\documentclass[useAMS,usenatbib,usegraphicx]{mn2e}  
\usepackage{longtable}
\usepackage{subfigure}
\newcommand{\ion}[2]{#1\,{\mdseries\textsc{#2}}}
\bibpunct{(}{)}{;}{a}{}{,}

   \title[Testing for Evolution Using SNe Ia Spectra from SNLS]{Testing for Evolution Using New Type Ia Supernovae \\ Spectra from the Supernova Legacy Survey}

    \author[E.S. Walker et al.]
{E.S.~Walker, $^{1,2}$\thanks{E-mail address emma.walker@sns.it}, I.M.~Hook $^{1,3}$, M.~Sullivan $^1$, D.A.~Howell $^{4,5}$, P.~Astier $^6$, C.~Balland $^{7,8}$, 
\newauthor S.~Basa $^8$, T.J.~Bronder $^9$, R.~Carlberg $^{10}$, A.~Conley $^{11}$, D.~Fouchez $^{12}$, J.~Guy $^6$, 
\newauthor  D.~Hardin $^6$, R.~Pain $^6$, K.~Perrett $^{10,13}$, C.~Pritchet $^{14}$, N.~Regnault $^6$, J.~Rich $^{15}$,
\newauthor G.~Aldering $^{16}$, H.~K. Fakhouri$^{16}$, T.~Kronborg $^6$,N.~Palanque-Delabrouille $^{15}$,
\newauthor S.~Perlmutter  $^{16}$,V.~Ruhlmann-Kleider $^{15}$ and T.~Zhang $^8$\\
$^1$ University of Oxford, Astrophysics, Denys Wilkinson Building, Keble Road, Oxford OX1 3RH, UK\\
$^2$ Scuola Normale Superiore di Pisa, Piazza dei Cavalieri 7, 56126 Pisa, Italy\\
$^3$ INAF-Osservatorio Astronomico di Roma, via di Frascati 33, 00040 Monteporzio Catone, Italy \\
$^{4}$ Las Cumbres Observatory Global Telescope Network, 6740 Cortona Dr., Suite 102, Goleta, CA 93117, USA \\
$^{5}$ Department of Physics, University of California, Santa Barbara, Broida Hall, Mail Code 9530, Santa Barbara, CA 93106-9530, USA \\
$^6$ LPNHE, Universit\'{e} Pierre et Marie Curie Paris 6, Universit\'{e} Paris Diderot Paris 7, CNRS-IN2P3, 4 place Jussieu, 75252 Paris Cedex 05, France\\
$^7$ University Paris 11, 91405 Orsay, France \\
$^8$ LAM, CNRS, BP8, P\^{o}le de l’\'{e}toile, Site de Ch\^{a}teau-Gombert, 38 rue Fr\'{e}d\'{e}ric Joliot-Curie, 13388 Marseille Cedex 13, France\\
$^9$ Advanced Electric Lasers Branch, Air Force Research Laboratory, 3550 Aberdeen SE, Kirtland Air Force Base, New Mexico 87117, USA\\
$^{10}$ Department of Astronomy and Astrophysics, 50 St. George Street, Toronto, ON M5S 3H4, Canada\\
$^{11}$ Department of Astrophysical and Planetary Sciences, University of Colorado, Boulder, CO 80309-0391, USA\\
$^{12}$ CPPM, CNRS-Luminy, Case 907, 13288 Marseille Cedex 9, France\\
$^{13}$ Netowrk Information Operations, DRDC Ottawa, 3701 Carling Avenue, Ottawa, ON, K1A 0Z4, Canada \\
$^{14}$ Department of Physics and Astronomy, University of Victoria, PO Box 3055, Victoria, BC V8W 3P6, Canada\\
$^{15}$ CEA/Saclay, DSM/Irfu/Spp, 91191 Gif-sur-Yvette Cedex, France\\
$^{16}$ Lawrence Berkely National Laboratory, Mail Stop 50-232, Lawrence Berkley Nation Laboratory, 1 Cyclotron Road, Berkeley, CA 94720, USA}
\begin{document}

\date{Released 2002 Xxxxx XX}

\pagerange{\pageref{firstpage}--\pageref{lastpage}} \pubyear{2002}

\maketitle

\label{firstpage}

\begin{abstract}
GMOS optical long-slit spectroscopy at the Gemini-North telescope was used to classify targets from the Supernova Legacy Survey (SNLS) from July 2005 and May 2006 -- May 2008.  During this time, 95 objects were observed.  Where possible the objects' redshifts ($z$) were measured from narrow emission or absorption features in the host galaxy spectrum, otherwise they were measured from the broader supernova features.  We present spectra of 68 confirmed or probable SNe Ia from SNLS with redshifts in the range $0.17 \leq z \leq 1.02$.  In combination with earlier SNLS Gemini and VLT spectra, we used these new observations to measure pseudo-equivalent widths (EWs) of three spectral features -- \ion{Ca}{ii} H\&K, \ion{Si}{ii} and \ion{Mg}{ii} -- in 144 objects and compared them to the EWs of low-redshift SNe Ia from a sample drawn from the literature.  No signs of evolution are seen for the \ion{Ca}{ii} H\&K and \ion{Mg}{ii} features.  Systematically lower EW \ion{Si}{ii} is seen at high redshift, but this can be explained by a change in demographics of the SNe Ia population within a two-component model combined with an observed correlation between EW \ion{Si}{ii} and photometric lightcurve stretch.
\end{abstract}


%

\section{Introduction}

Since the discovery of cosmic acceleration \citep[e.g.][]{R98,P99}, Type Ia supernovae have played a vital role as distance measures in the universe, up to and beyond $z=1$ through surveys such as the Supernova Legacy Survey \citep[SNLS][]{Astier06} and ESSENCE \citep{WoodVasey:2007p25}.  This is made possible by the fact that SNe Ia appear to be ``standardisable'' candles, a property first noted by \citet{P93} and later parameterised in a number of different ways \citep[for example][]{Riess:1996p2804,P97,Wang:2003p3195,Guy:2005p1610,Jha:2007p2805}.  However, the use of the same lightcurve calibration, e.g.~stretch, a measure of the relative width of the lightcurve, and colour, at high and low redshift relies on the fact that there is no evolution in the properties of the supernovae with time that cannot be accounted for by the calibration itself.

The properties of high- and low-z objects have been compared in a number of ways.  One way is to compare the line profiles and ejection velocities of features in the spectra.  \citet{Hook:2005p60} measured the ejection velocity from the blueshifted \ion{Ca}{ii} H\&K feature in 11 high-z supernovae with $z>0.38$ and compared this to low-z objects.  They found that at high-z the ejection velocities and phase evolution of the spectra are consistent with the low-z sample and concluded that there exists a sample at high-z with similar properties to SNe at low-z.  \citet{B06} measured the wavelength of maximum absorption and peak emission for the \ion{Ca}{ii} H\&K, \ion{Si}{ii} 6150 and  \ion{S}{ii} ``W'' features and found them to be similar in high-z and low-z objects.  They also noted evidence of a double absorption feature in the \ion{Ca}{ii} H\&K region which is seen to occur occasionally in low-z supernovae and has been attributed to \ion{Si}{ii} \citep{Howell:2006p95}.

Since the advent of large surveys producing high numbers of spectra, it has been possible to use these to make composite spectra, which are representative of the average supernovae spectrum in certain redshift and phase bins.  This has been done using Keck spectra of SNLS supernovae by \defcitealias{Ellis08}{E08}\citet[hereafter E08]{Ellis08} with $\overline{z} = 0.5$, and for ESSENCE spectra by \defcitealias{Foley:2008p1684}{F08}\citet[hereafter F08]{Foley:2008p1684}.  The studies used different methods to account for host galaxy contamination, but neither found definitive evidence for an evolving supernova population.  
\citetalias{Ellis08} created a composite in order to study possible effects of metallicity variations in progenitors in the rest-frame UV spectra.  This is where the absorption is dominated by metal line-blanketing effects and the metallicity differences are expected to be the most prominent, although the theoretical predictions are not in agreement \citep[for example]{Hoflich:1998p2194, Lentz:2000p2196}.   \citetalias{Ellis08} saw a large amount of variation in the spectrum in the UV region, more than is predicted and can be accounted for by dust, and also changes with phase that are not predicted.  However, due to a lack of local UV spectra of supernovae they are unable to draw any firm conclusions as to evolution.

\defcitealias{Folatelli:2004thesis}{F04}
\citetalias{Foley:2008p1684} do not perform galaxy subtraction, but instead contaminate their low-z spectra with a known galaxy spectrum in an attempt to match the host galaxy contamination of their high-z sample.  They find evidence of changes in the \ion{Fe}{ii} spectral feature which may indicate lower $^{54}$Fe production in a lower temperature environment brought on by a lower metallicity.  They attribute this not to evolution, but to changing demographics as the universe at higher redshift was less metal-rich compared to the universe today; however, again their conclusions are limited by the low-redshift sample of spectra.

The composite spectra from \citetalias{Ellis08} was included in \citet{Sullivan:2009p1769} which compared their medium redshift composite to a new low-z composite ($z<0.05$) and a high-z composite ($z>0.9$) based on data from \citet{Riess07}.  They observe a clear decrease in the size of spectral features of intermediate mass elements as redshift increases, but this is a change which is consistent with a change in demographics, rather than an evolution of SNe Ia.

Other studies have focussed on measuring the pseudo-equivalent widths of spectral features in the high-redshift spectra and comparing these to trends with phase in the low-redshift sample.  This method was first described by \citet[hereafter F04]{Folatelli:2004thesis} and used successfully on a small sample of spectra to compare low- and high-redshift objects for 5 distinct features in the spectrum. The blueshift of the absorption minimum in \ion{Ca}{ii} H\&K was also measured and compared to nearby supernovae and no evidence for any significant differences was found between the two samples \citep[see also][]{Garavini:2007p1645}.

An EW analysis of the first two and a half years of Gemini SNLS spectra was published in \defcitealias{Bronder08}{B08}\citet[hereafter B08]{Bronder08}.  Rest-frame EW measurements were made for the \ion{Ca}{ii} H\&K, \ion{Si}{ii} 4000 (hereafter referred to as just \ion{Si}{ii}) and \ion{Mg}{ii} features as well as the \ion{Ca}{ii} H\&K ejection velocity.  The spectra of 55 high-z supernovae were compared to 167 spectra from 24 low-z objects encompassing all three SNe Ia sub-types (normal, overluminous and underluminous).  No significant difference was seen between the two samples for the equivalent widths of \ion{Ca}{ii} H\&K\ and \ion{Si}{ii} or the ejection velocity.  For the \ion{Mg}{ii} feature, a slight difference was noted, but it was not significant enough to confirm evolution and could just have been a statistical effect.  This is discussed further in Section \ref{sec:low-redshift-trends}.  \citetalias{Bronder08} also found a relation between the EW \ion{Si}{ii} and $M_B$, the absolute B-band magnitude at maximum,  which could be used to calibrate the supernovae in the same way as lightcurve stretch and colour.

With the well-measured light curves of the SNLS supernovae it has also been possible to compare photometric properties.  In \citet{Conley:2006p96}, the rise-times of low- and high-z supernovae are compared and they find no compelling evidence that there is a difference between the two populations.

However, as was concluded by \citetalias{Foley:2008p1684}, there is to be expected some change in photometric properties due to changing demographics.  Due to the two-channel progenitor model \citep{Scannapieco:2005p2518,Mannucci:2005p2616,Mannucci:2006p2615,Sullivan:2006p3}, the ``prompt'' component which are brighter supernovae with broader light curves should dominate at higher redshift, whereas at $z=0$, the ``delayed'' component will be more important.  \citet{Howell:2007p1767,Sullivan:2009p1769} calculate that this should result in a 6\% increase in light curve width between $z=0$ and $z=1.5$.  Using observational data from \citet{Sullivan:2006p3}, they measure a percentage change in stretch of $8.1 \pm 2.7$\% between $0.03 <z<1.12$ which is consistent with changing demographics.  The spectral evolution observed in \citet{Sullivan:2009p1769} is consistent with this.

In summary, no firm evidence of evolution in the spectral properties of supernovae has been observed, but the studies are being hampered by the lack of local spectra to compare with.  In this paper we present an analysis of EWs of the \ion{Ca}{ii} H\&K, \ion{Si}{ii} and \ion{Mg}{ii} spectral features.  This study significantly expands on the work of \citetalias{Bronder08} using new data from SNLS with a different method of accounting for the effects of host galaxy light, as well as an expanded sample of low-z SNe from the literature for comparison.

The remainder of this paper is organised as follows.  In section 2 we describe the new observations.  In section 3 we detail the methods and measurements applied to the high-z sample of SNe Ia.  Section 4 describes the low-z sample drawn from the literature and used for comparison.  Section 5 details the results of the comparison study and an investigation into the possible cosmological uses for SNe Ia spectra.  These are discussed further in section 6 and our conclusions are summarised in section 7.

\section{Observations}\label{sec:observations}

The targets were selected from real-time SNLS imaging which took place at the Canada-France-Hawaii Telescope.  Fainter objects were usually observed spectroscopically at Gemini, with brighter candidates at either VLT or Keck (see Perrett et al. (in prep.) for more details on the real-time pipeline).

\subsection{Observational Set-up}

The new spectra presented here were taken in queue mode on the Gemini-North telescope with ToO status\footnote{Observing programmes: GN-2003B-Q-9, GS-2003B-Q-8, GN-2004A-Q-19, GS-2004A-Q-11, GN-2004B-Q-16, GS-2004B-Q-31, GN-2005A-Q-11, GS-2005A-Q-11, GN-2005B-Q-7, GS-2005B-Q-6, GN-2006A-Q-7, GN-2006B-Q-10, GN-2007A-Q-8, GN-2007B-Q-17, GN-2008A-Q-24.  PI Hook}.  The observational set-ups and data reduction methods used for SNLS-Gemini spectroscopy using GMOS \citep{Hook_GMOS} have been described in detail previously \citep{Howell05,Bronder08}.  In summary, a 'classical' setup with
  central wavelength of 680nm was used for observations of
  brighter targets which were expected to be at a lower redshift,
  whereas nod-and-shuffle mode \citep[N\&S,][]{Glazebrook:2001p1903}
  with a redder central wavelength of 720nm was used for
  fainter targets.  The only change to the methods of observational set-up since \defcitealias{Howell05}{H05}\citet[][hereafter H05]{Howell05} and \citetalias{Bronder08} are that from March 2007, first N\&S and then classical observations were taken at two wavelength settings in order to cover the gaps in the spectra caused by the spacing of the three GMOS CCDs.  In these cases, for the R400 grating and with $2\times2$ binning, the change in central wavelength used was 50\,nm.  Due to the unavailability of the R400 grating, three targets were observed with the R150 grating which has a lower dispersion.  For these objects, a change in central wavelength of 100\,nm was used.

The position angle of the slit was primarily determined by aligning the target and the centre of the host galaxy in order to try to observe narrow emission/absorption features in the host spectrum.  The second constraint was the availability of guide stars in the field.  Occasionally, less optimum position angles were used or the object was moved from the centre of the slit.  All of these considerations mean that the observations were not taken at parallactic angle, but this is taken into account in the analysis that follows.

The new targets observed as part of the Gemini SNLS spectroscopic follow-up programme are listed in Table \ref{obs_tab}.


\subsection{Data Reduction}

The data were reduced using standard IRAF\footnote{http://iraf.noao.edu/} and Gemini-specific\footnote{http://www.gemini.edu/} tasks to perform bias-subtraction, flat-fielding, sky-subtraction and chip mosaicing. A wavelength solution was applied to the 2D spectrum.  When combining the 2D frames, bad-pixel masks were used to take account of the ``missing data'' in the gaps between the CCDs.  A 1D spectrum was extracted from the 2D spectrum and corrected for telluric absorption features.  It was generally not possible to extract the target light separately from the host due to the small angular separation of the objects on the sky.  A sky spectrum was also extracted from the 2D frames, and an error spectrum was estimated from this by assuming that the noise is dominated by sky emission.  Where appropriate in some of the earlier data which were observed at only one central wavelength, a linear interpolation was used to remove the chip gaps and the error spectra set to a high value in these regions to deweight them.

\subsection{Redshift Determination and Object Classification}\label{sec:redshift-classification}

The redshifts of the supernova and its host are ideally measured using narrow host features such as [\ion{O}{iii}], [\ion{O}{ii}], H$\beta$ and  H$\alpha$ emission, and H\&K absorption.  This typically gives an error of $z\pm 0.001$, but if many lines are present then the error can be reduced further.  If no host features are present then the redshift measurement relies on the wider supernova features.  These are less precise due to the higher velocities involved and typically measure $z\pm0.01$ depending on the quality of the spectrum.

A high-z SN Ia is classified by identifying, typically, the \ion{Si}{ii} feature at $\lambda\approx 4000$\,\AA\ in the spectra.  Other features used are \ion{Si}{ii} $\lambda6150$ and \ion{S}{ii}.  To aid classification, all spectra are also run through the SUPERFIT program \citepalias{Howell05} which performs a $\chi^2$-minimisation to best-fit the observed spectra to a low-z supernova spectrum combined with host galaxy light.  Objects are classified  as SNIa (certain or highly probable Ia), SNIa? (probable Ia), SNII (SN Type II), SN (supernova of unknown type) or SN? (probable SN of unknown type).  To go alongside this, spectra are assigned a confidence index (CI) from 0--5 with 5 being a definite Ia and 0 definitely not.  Objects of CI = 0 are typically core-collapse supernovae.  The definitions of the CIs are given in \citetalias{Howell05}.  

The redshifts, their errors and CIs of all the objects observed in this study are given in Table \ref{res_tab}.


\subsection{Sub-Type Identification}

A number of SNe Ia were observed to have spectral features similar to known peculiar objects from their best-fit \texttt{SUPERFIT} matches \citepalias{Howell05}.  Identification of these objects is important so they can be flagged and removed from later studies.  The objects which were identified as having unusual spectral features during this period of observation are given in Table \ref{subtypes}.   

There are several key features which are distinct to each of the comparison spectra.  SNLS-03D3bb \citep[SN\,2003fg][]{Howell:2006p95} was an overluminous SNe Ia whose progenitor system is debated, but it is thought to be of super-Chandrasekhar mass.  The spectrum showed low ejection velocities, as well as an unburnt \ion{C}{ii} feature at 4200\AA.  SN\,2001ay \citep{2004cetd.conf..151H} has a high stretch value ($s = 1.6$), but appeared spectroscopically normal around maximum light, except for high-velocity \ion{Ca}{ii} and \ion{Si}{ii} features.  SN\,1999aa \citep{Garavini:2007p1645} has been seen as a transitional object between normal SNe Ia and the overluminous objects typified by SN\,1991T \citep{F1991T}, showing spectral features of both types of object.  SN\,1999aa spectra at maximum contained smaller amounts of intermediate mass elements and larger amounts of iron-group elements, and appeared peculiar until maximum light after which it appeared to be spectroscopically normal.

\begin{table}
\caption{Unusual SNe Ia identified in this sample from their spectral features only.  The stretch measurements are made using the SiFTO \citep{sifto} light curve fitter.  The references for the comparative objects are SNLS-03D3bb \citep{Howell:2006p95}, SN\,2001ay \citep{2004cetd.conf..151H} and SN\,1999aa \citep{Garavini:2004p1035}.}
\label{subtypes}
\begin{tabular}{l r@{ $\pm$ }l l}
\hline
Object & \multicolumn{2}{c}{Stretch} & Comments \\
\hline
07D3af & 0.943 & 0.017 & 03D3bb-like \\
&\multicolumn{2}{c}{}& \ion{C}{ii} feature at 4200\AA \\
07D3ap & 0.955 & 0.018 & SN\,2001ay-like \\
&\multicolumn{2}{c}{}& High-velocity \ion{Si}{ii} \\
07D3cr & 0.872 & 0.043 & SN\,2001ay-like \\
&\multicolumn{2}{c}{}& Broad \ion{Ca}{ii} \\
07D3dj & 1.118 & 0.018 & 03D3bb- or SN\,1999aa-like \\
&\multicolumn{2}{c}{}& Little \ion{Ca}{ii} and probable \ion{C}{ii} \\
08D2aa & 1.128 & 0.016 & SN\,1999aa- or SN\,2001ay-like \\
&\multicolumn{2}{c}{}& \ion{C}{ii} feature at 4200\AA \\
08D2ad & 1.155 & 0.021 & 03D3bb-like \\
&\multicolumn{2}{c}{}&\ion{C}{ii} feature at 4200\AA \\
\hline
\end{tabular}
\end{table}

\section{Methods and Measurements}

For the use of SNe Ia as cosmological distance indicators, it is important to verify that the objects we observe in the high-z universe have similar properties to those at low z, or, if there is some change in the mean properties \citep[for example]{Howell:2007p1767}, that these differences can be calibrated out.  All previous work on high-z SNe Ia spectra \citep[e.g.][]{Hook:2005p60,B06,Ellis08,Bronder08,Foley:2008p1684} have found no definitive evidence for evolution beyond what is expected from known changes in demographics.

In this paper out of the 95 objects observed, 68 were confirmed as Type Ia or probably Type Ia events.  Previous Gemini SNLS samples have published a further 41 \citepalias{Howell05} and 46 \citepalias{Bronder08}.

\subsection{Sample Selection}

For the study here, all SNLS supernovae observed with Gemini, including objects published in \citetalias{Howell05} and \citetalias{Bronder08}, with confidence indices of 4 or 5 were considered.  A cut was then made based on the availability of final lightcurves and measurements of conditions on the night of observation, specifically hour angle (HA) and mean seeing.  The HA and seeing are required for light-loss correction described in Section \ref{sec:correcting-adr}.  In the case of observations spread over more than one night, the larger value of the seeing was used.  

A comparison sample of spectra drawn from the first three years of SNLS-VLT spectra analysed in real-time (Basa et al.~in prep) was also used.  These objects were observed using the FORS1 spectrograph \citep{Appenzeller:1998p2422} on the VLT.  Again, these objects were required to have CI=4 or 5 and have final lightcurves and a measurement of seeing for the EW study.  In this case, the measurements of the hour angle on the night of observation were not required as FORS1 employs an atmospheric dispersion correction (ADC) which minimises the amount of light lost when not observing at the parallactic angle.

\begin{figure}
\begin{center}
\includegraphics[width=8cm]{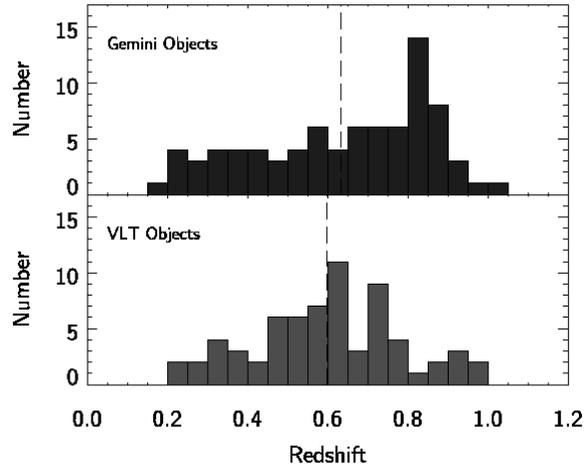}
\end{center}
\caption{Redshift histogram for the Gemini and VLT samples.  The dotted line represents the mean redshifts of the two samples.}
\label{fig:redshift_hist}
\end{figure}

Removing objects classified as peculiar (see Table \ref{subtypes}), the sample for the EW study contains 149 objects, 82 from Gemini and 67 from the VLT.  A histogram of redshift is shown for the final Gemini and VLT samples in Figure \ref{fig:redshift_hist}.  It shows that the Gemini sample peaks at a higher mean redshift, as expected since Gemini was used to observe the candidates that were expected to be at higher redshift, but also contains a larger number of lower redshift objects than the VLT sample.  These are objects from the D3 field which is not visible in the southern sky.  The mean redshifts illustrated by the dotted line are $\overline{z}_{Gem} = 0.63$ and $\overline{z}_{VLT} = 0.60$.  The median of the VLT sample is similar to its mean, but for the Gemini sample, due to the large number of points at $z>0.8$ has a median value of $0.70$.

\subsection{Photometric Data and Host Galaxy Template Fitting}\label{sec:photometric-data}

For this study, SNLS multicolour photometry of the supernovae and their host galaxies was used for two main purposes: firstly for subtraction of the host galaxy contamination from the spectra; and secondly, for comparison of spectral features with photometric properties of the supernovae.

The photometric lightcurve data for the high-z SNLS sample above, and the low-z comparison sample described in Section \ref{sec:low-z-comparison}, was fitted using the SiFTO lightcurve-fitter \citep{sifto} to ensure homogeneity in the sample.  This provided measurements of lightcurve stretch, colour and date of maximum.  It also permitted the interpolation of supernova magnitudes in $g'r'i'z'$-bands to the date of spectroscopic observation.

From stacked images, the host galaxy surface brightness in $u^*g'r'i'z'$-bands at the position of the supernova was measured.  These were used to find a best-fitting galaxy template spectrum.  The same template library as in \citetalias{Ellis08} was used, which was drawn from the P\'{E}GASE programme \citep{Fioc:1997p509,Fioc:1999p507}, and a star-burst template (SB1) from \citet{Calzetti94} was added.  Emission-line features were interpolated over.

A $\chi^2$ minimisation was used to compare synthetic photometry  from the templates in the $u^*g'r'i'z'$-bands, $f_{temp,i}$ to the measured values $f_{gal,i}$ by minimising
\begin{equation}
  \label{eq:1}
  \chi ^2  =  \sum_i{\frac { (f_{gal,i} - (A\lambda +B)f_{temp,i})^2}{\sigma_{gal,i}^2}}
\end{equation}

\noindent where $\sigma_{gal,i}$ is the error on the galaxy flux in band $i$.  The template is normalised by the function $A\lambda + B$ to allow for reddening in the host galaxy.  The best-fit galaxy template $(A\lambda + B)f_{temp}$ is denoted below as $f_g^{bf}(\lambda)$.

\subsection{Galaxy Subtraction Method}\label{sec:methods-measurements}

Here we use a galaxy subtraction method similar to the one described in \citetalias{Ellis08}, rather than the correction factor method in \citetalias{Bronder08}, but with one large difference.  The objects observed as part of the \citetalias{Ellis08} study were all observed at parallactic angle where supernova light-loss from the slit was minimised whereas in this study the observations were not taken at parallactic angle.  In order to carry out a galaxy subtraction, we assume the galaxy is a flat background and the supernova is modelled as a point-source with a FWHM equal to the seeing in the band used for object acquisition (either r$'$ or i$'$), and scaling with wavelength as $\lambda^{-0.2}$.  Under this assumption, light from the supernova will be lost due to atmospheric differential refraction (ADR).  We account for this in Section \ref{sec:correcting-adr}.

The galaxy subtraction method is summarised in the following sections.

\subsubsection{Correcting To Compensate For Different Observing Conditions}

Starting with the raw spectrum, $f_{tot}^{raw}(\lambda)$, which is the sum of supernova flux, $f_{SN}^{raw}(\lambda)$, and galaxy flux, $f_g^{raw}(\lambda)$,  the aim is to produce a final corrected supernova spectrum, $f_{corr}(\lambda)$.  

The SNLS observations were not spectrophotometric, therefore before the galaxy template (which is normalised to SNLS host photometry) can be subtracted, the observed spectrum must be placed on the same flux scale.  This is done by multiplying by a correction factor $A_{sens}$ calculated by comparing photometry of the supernova and host from the imaging, correctly scaled to match observing conditions and observational set-up, to the synthetic $r'$- and $i'$-band flux in the observed spectrum.  $A_{sens}$ is the mean of the corrections needed to match the $r'$- and $i'$-band flux.
\begin{eqnarray*}
f_{tot}(\lambda) & = & f_{tot}^{raw}(\lambda) \times A_{sens}
\end{eqnarray*}

\noindent We denote the photometrically-normalised SN and host galaxy spectra as
\begin{eqnarray*}
f_{SN}^s(\lambda) & = & f_{SN}^{raw}(\lambda) \times A_{sens}\\
f_{g}^{s}(\lambda) & = & f_{g}^{raw}(\lambda) \times A_{sens}.
\end{eqnarray*}

\subsubsection{Template Subtraction}

The fractional contamination (FC) of galaxy light in the total spectrum is calculated for both the $r'$- and $i'$ bands using the galaxy and SN photometry and accounting for the effects of the observing conditions.  Any object with fractional contamination of galaxy light in the total spectrum $FC \geq 0.1$ in either band has the best-fitting galaxy template subtracted-off.  The FC is calculated in two bands instead of one because as it is measured in the observed frame, there will be a redshift-dependence on the value of FC as the filters will fall at different points in the rest-frame of the object.  In situations with $z<0.7$, the $i'$-band filter falls in the redder part of the spectrum where there is less supernova flux relative to galaxy; and at $z>0.7$, the reverse is true.  The FC will also be phase-dependent as the supernova spectrum becomes redder with time.  The value of $0.1$ was chosen because at this level of contamination, the error on the EW measurements will be $\approx$10\%, which is comparable to the size of the error introduced by misidentifying the galaxy host type (see Section \ref{sec:error-evaluation}).  Also, below this level of contamination, the risk of subtracting too much galaxy template is high.

Before subtraction can occur, the best-fit galaxy template, $f_g^{bf}(\lambda)$, must be scaled to account for differences between the photometry apertures and the spectroscopy apertures.  The galaxy photometry is in units per square arcsecond (or effectively in an aperture of size $1'' \times 1''$).  The aperture size of the spectroscopy is given by the slit width, which is always $0.75''$ for GMOS observations, but varies for the VLT spectra, multiplied by the size of the aperture window used during extraction, and is denoted by $A_{ap}$ in units of square arcseconds.  Under the assumption of a flat host galaxy background, the flux of the host transmitted through the slit, $f_g^t(\lambda)$, is then
\begin{eqnarray*}
f_g^{t}(\lambda)& =& f_g^{bf}(\lambda) \times A_{ap}
\end{eqnarray*}

\noindent The galaxy template is then interpolated to the same wavelength scale as the observed spectrum and subtracted off leaving the intermediate step
\begin{eqnarray*}
f_{corr}^{int1}(\lambda) & = & f_{tot}(\lambda) - f_g^t(\lambda) \\
& = & f_{SN}^s(\lambda) + f_g^s(\lambda) - f_g^{t}(\lambda).
\end{eqnarray*}

\noindent In an ideal situation, where the galaxy template matches the observed galaxy perfectly, $f_g^s = f_g^{t}$ leaving a spectrum containing only supernova flux.

\subsubsection{Correcting for ADR}\label{sec:correcting-adr}

The spectrum now contained only supernova light, but the colours were not correct due to the loss of light at the wavelengths not used for acquisition.  See \citet{Filippenko82} for the equations governing light loss as a function of wavelength, hour-angle, seeing, airmass etc..  A correction function, $c_{ADR}(\lambda)$, for the supernova spectrum was found by fitting a second-order polynomial to the inverse of the fraction of SN light which is transmitted through the slit at the effective wavelength of each band, i.e. multiplication by $c_{ADR}(\lambda)$ should have corrected the spectrum so it represented the total supernova flux.  The fitting was calculated using \texttt{MPFITFUN}, part of \texttt{MPFIT} \citep{Markwardt:2009p2371}.  The final, subtracted, ADR-corrected spectrum, $f_{corr}(\lambda)$, was given by:\
\begin{eqnarray*}
f_{corr}^{int2}(\lambda) & = & (f_{SN}^s(\lambda) + f_g^s(\lambda) - f_g^{t}(\lambda))c_{ADR}(\lambda).
\end{eqnarray*}

\noindent Note that this correction was not applied to the VLT spectra.

\subsubsection{Extinction Correction}

The final step is to correct for extinction in both the host galaxy and the Milky Way.  In the rest-frame, the SALT2 colour law \citep{salt2}, a function of wavelength, redshift and SN colour, is applied to each object to account for the host extinction.  This correction is small and has almost no effect on the EW measurements.  In the observed frame, the Cardelli, Clayton and Mathis (\citeyear{Cardelli:1989p920}) extinction law is applied to account for Milky Way extinction using the dust maps of \citet{Schlegel:1998p1609}.  The SNLS fields are selected to be in regions with low galactic extinction so the Milky Way correction is not large either.  These corrections can be summarised as $c_{ext}(\lambda,z,c)$ where $c$ is the supernova colour.
\begin{eqnarray*}
f_{corr}(\lambda) & = & (f_{SN}^s(\lambda) + f_g^s(\lambda) - f_g^{t}(\lambda))c_{ADR}(\lambda)c_{ext}(\lambda,z,c).
\end{eqnarray*}

\noindent The final spectrum contains only supernova light and can have the equivalent widths of the features measured.

\begin{figure}
\begin{center}
\includegraphics[width=8cm]{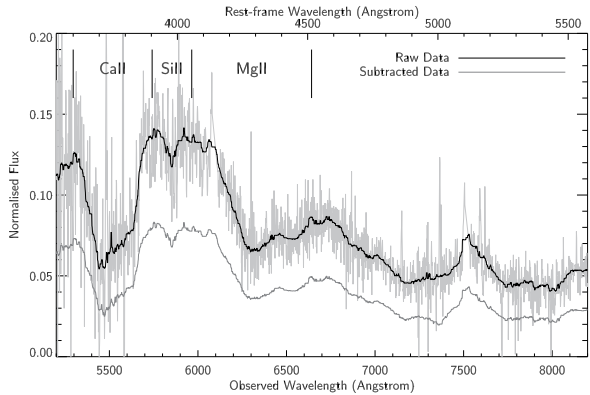}
\end{center}
\caption[Galaxy subtraction example]{Example of host galaxy subtraction from a high-z SNLS supernova spectrum.  The raw data (for object 07D3ea at z=0.471) is shown in light grey, with a smoothed version over the top in black.  The SN spectrum after subtraction is shown as the smoothed light grey line underneath.  07D3ea was best-fit by an star-burst galaxy template and has $FC = 0.19$ in the $i'$-band.}
\label{fig:galsub}
\end{figure}

The same scaling treatment is applied to the error spectrum, but no subtractions are done because it is assumed to contain only sky noise.  As an example, the galaxy-subtracted spectrum for 07D3ea is illustrated in Figure \ref{fig:galsub}.

\subsection{Equivalent Width Measurements}\label{sec:equiv-width-meas}

The equivalent widths were measured in the same way as in \citetalias{Bronder08}, fitting a pseudo-continuum to the feature and calculating the EW between these limits.  The pseudo-continuum was defined by the highest point in a region either side of the feature.  In general, an EW and its associated statistical error are given by:
\begin{eqnarray}
\label{eweq}
EW &=& \sum_{i=1}^N \Bigg (1 - \frac{f_{\lambda}(\lambda_i)}{f_c(\lambda_i)}\Bigg)\Delta\lambda_i \\
\label{ewe}
\sigma^2_{EW} &=& \sum_{i=1}^N \Bigg( \Big( \frac{\sigma_f(\lambda_i)}{f_c(\lambda_i)} \Big)^2 + \Big( \frac{f_{\lambda}(\lambda_i)}{f_c^2(\lambda_i)}\Big)^2\sigma^2_{c_i}\Bigg)\Delta\lambda_i^2 
\end{eqnarray}

\noindent where $f_{\lambda}$ is the flux of the spectrum, $f_c$ is the flux of the pseudo-continuum, $\sigma_f$ is the uncertainty in the flux from the error spectrum and $\sigma_{c_i}$ is the uncertainty in the pseudo-continuum.  However, as error spectra were available for the SNLS observations, it was possible to weight each point in the calculation and account for uncertainties in the flux due to sky noise.  Equation \ref{eweq} was modified to:
\begin{eqnarray}
\label{ewSNLS}
EW_{\mathrm{SNLS}}&=& \Bigg(\sum_{i=1}^N \Bigg (1 - \frac{f_{\lambda}(\lambda_i)}{f_c(\lambda_i)}\Bigg)\Delta\lambda_iw_i\Bigg)\frac{N}{\sum_{i=1}^N(w_i)}
\end{eqnarray}

\noindent where $w_i$ is the weighting of each point given by $w_i=\frac{1}{\sigma_f(\lambda_i)^2}$. The statistical error was still calculated as given in Equation \ref{ewe}.

The same EW code for measuring the features as in \citetalias{Bronder08} was used with changes to the method used to select the limits of the pseudo-continuum to make the process more automatic.  The ranges of wavelength for the blue and red bounds of the SNe Ia features were first defined in \citetalias{Folatelli:2004thesis} and were the ones used in \citetalias{Bronder08}.  In this study, we used the \ion{Ca}{ii} H\&K, \ion{Si}{ii} and \ion{Mg}{ii} features and the boundaries for these features were redefined according to observations of low-z and high-z objects using trial and error to find the optimum values in order to make the process more automated.  The new ranges are given in Table \ref{tab:spec_ranges}.

\begin{table}
\caption[New spectral ranges]{The spectral ranges defined in \protect\citet{Folatelli:2004thesis} compared to the new ranges defined here.}
\label{tab:spec_ranges}
\begin{tabular}{c c c c c}
\hline
Feature & Folatelli Blue& New Blue& Folatelli Red& New Red\\
& Limit (\AA)& Limit (\AA)& Limit (\AA)& Limit (\AA)\\
\hline
\ion{Ca}{ii} & 3500 -- 3800 & 3500 -- 3700 & 3900 -- 4100 & 3900 -- 4100\\
\ion{Si}{ii} & 3900 -- 4000 & 3900 -- 4000 & 4000 -- 4150 & 4050 -- 4125\\
\ion{Mg}{ii} & 3900 -- 4150 & 4050 -- 4200 & 4450 -- 4700 & 4450 -- 4700\\
\hline
\end{tabular}
\end{table}

As the definitions of the SNe Ia spectral features in \citetalias{Folatelli:2004thesis} overlap, the code could over-estimate the \ion{Ca}{ii} feature or the \ion{Mg}{ii} feature by including the \ion{Si}{ii} feature.  The code now automatically adjusts the ranges to compensate for this depending on the supernova ``colour''.  As a crude way of calculating this from the spectrum, the ratio of integrated flux in the region 50-100\,\AA\ above and below the minimum of the \ion{Si}{ii} feature was measured  and used to move the boundaries of the features so the over-estimation no longer occurs.  For a redder supernova, the red limit of the \ion{Ca}{ii} feature was decreased and for a blue supernova, the blue limit of the \ion{Mg}{ii} feature was increased. 

In the event that the automation still failed, the bounds for the pseudo-continuum had to be selected manually.  In this case an error to account for the uncertainty in the placement of the feature bounds was calculated by varying the boundaries within $\pm$10\,\AA\ at random and evaluating the EW fifty times.  The resulting EWs were fitted with a Gaussian and the pseudo-continuum contribution to the EW error was given by the standard deviation of the Gaussian.  This was added in quadrature to the statistical error.

The EW measurements for the 149 objects in the final sample are in Tables \ref{tab:walker_ew}, \ref{tab:bronder_ew} and \ref{tab:vlt_ew}.  At this point, five objects were removed from the sample as the galaxy subtraction method was believed to have failed for these objects.  This was reflected in either negative values of EWs or values of EW \ion{Mg}{ii} close to 0.  These objects tended to occur in the very centre or on the outskirts of the host where the assumption that the host contribution does not vary within the size of the spectroscopy aperture.  The objects are marked with an * in the results tables.  This left a total of 144 objects: 78 objects from the Gemini telescope and 66 from the VLT.

\defcitealias{Bronder08}{B08}
\subsection{Evaluation of Additional Sources of Error}\label{sec:error-evaluation}

\begin{figure}
\begin{center}
\includegraphics[width=8cm]{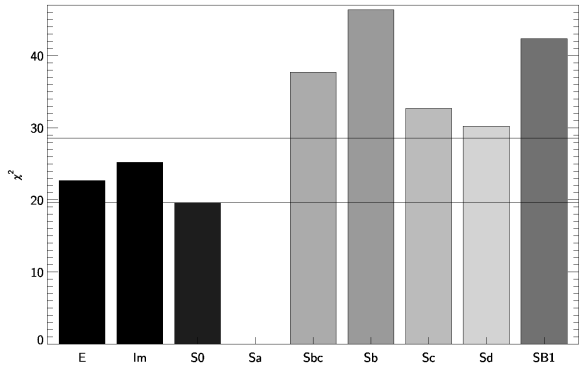}
\end{center}
\caption[Host galaxy error]{A histogram plotting the minimum $\chi^2$ values found for each of the galaxy templates for the photometry of 06D1bz.  The best fit is an S0 galaxy, but the elliptical and irregular templates cannot be ruled out.  The lower line marks the minimum $\chi^2$ and the higher line marks the $3\sigma$ region.  There is no $\chi^2$ value for the Sa template as this fit does not converge.}
\label{fig:galhist}
\end{figure}

Using the multicolour host galaxy subtraction method there are several extra terms in the error calculation as well as the statistical error.  Firstly, there is an uncertainty in the selection of the galaxy template.  Any galaxy template within a $3\sigma$ range of the best-fit galaxy is identified to rule out any galaxy template which is definitely not the host.  This is illustrated in Figure \ref{fig:galhist} for the example of 06D1bz above.  The best-fitting template is that of an S0 galaxy, but the elliptical and irregular galaxy spectrum lie within the $3\sigma$ range.

Any galaxy template fit falling within this $3\sigma$ range are appropriately scaled, subtracted from the raw spectrum and the EW measured.  The scatter in EW measurements is then used to produce an error due to misidentification of the galaxy type.  It is typically in the order of a few \AA\ (i.e.~10\% EW) and is added in quadrature to the $1\sigma$ statistical error from the EW measurement.

Another change from \citetalias{Bronder08} is that measurements of EW features where a GMOS chip gap has fallen in the boundary regions of the feature are no longer removed from the sample.  Instead, a linear interpolation is made across the gap.  The presence of the chip gap adds to the error on the EW measurement as there is no data in that region.  To compensate for this, the statistical error is increased by a multiplicative correction factor $A_{gap}$ to take into account the number of pixels which form the feature where there is essentially no information.
\begin{eqnarray*}
A_{gap} & = & \sqrt{ \frac{N_{tot}}{N_{tot} - N_{gap}} }
\end{eqnarray*}

\noindent where $N_{tot}$ is the total number of pixels in the feature and $N_{gap}$ is the number of pixels from the chip gap included in that feature.

\section{Low-z Comparison Sample}\label{sec:low-z-comparison}

\subsection{Spectroscopy Sample}

The equivalent widths of the \ion{Ca}{ii} H\&K, \ion{Si}{ii} and \ion{Mg}{ii} features change with the phase of the supernova so it is important to find a sample with good phase coverage.  In order to select a sample, the SUSPECT archive\footnote{\texttt{http://bruford.nhn.ou.edu/\~{}suspect/}} was used to search for published spectra with phases of $\pm11$ days from B-band maximum. This phase range was chosen to match the range of rest-frame phases in the high-z spectra.  Before $-11$\,days spectra are rare because of the time needed to discover the objects and arrange spectroscopy scheduling.  After $11$\,days, there are not many high-z spectra as the supernovae have faded.  The selection criteria applied to the spectra were: that there should be at least two spectra within the $\pm11$\,day range as we want to study the evolution of spectra features with time; the spectra should include all three of the \ion{Ca}{ii} H\&K, \ion{Si}{ii}\ and \ion{Mg}{ii}\  features so as to be certain the supernovae are normal in the whole region being investigated; all the objects should be classified as core-normal supernovae.

There are a number of problems with applying these criteria to the current low-z sample.  Firstly, a lot of published spectra do not have the wavelength coverage to include the \ion{Ca}{ii} H\&K\ feature as this is in the near-UV which is difficult to observe from the ground.  Secondly, the well-studied objects tended to be the spectroscopically or photometrically peculiar examples which are not useful for this study.  

Applying these criteria to the SUSPECT archive produced 104 spectra from 14 objects.  This was then combined with 49 spectra from 11 objects published in \citet{Matheson:2008p755}; however, the latter objects are excluded from the analysis of the \ion{Ca}{ii} H\&K\ feature.  This is because the spectrograph used has a blue limit of 3700\,\AA\ which would mean the feature would only be completely visible in spectra of objects with a narrow \ion{Ca}{ii} feature at certain redshifts and so bias the sample.  Table \ref{low-z_refs} details the 23 low-z objects used in this study.

\begin{table}
\caption{Low-z sample used to make the low-z trends.  Objects with published photometry are marked with a $^{\dagger}$ symbol.  Objects in galaxies with known distance moduli or within the Hubble flow ($z\geq0.2$) are marked with a $^{\ddagger}$.  Five objects have been used in the low-z cosmology sample in \citet{Astier06} and these are marked with an $\ast$.}
\label{low-z_refs}
\begin{tabular}{l p{2.1cm} l}
\hline
Supernova & Phases of Spectra Used\footnotemark[1] & Reference \\
\hline
SN\,1989B $^{\dagger}$$^{\ddagger}$ & $0$,$6$,$11$ & \citet{Barbon:1990p1227}\\
SN\,1990N $^{\dagger}$$^{\ddagger}$ & $-8$ & \citet{Leibundgut:1991p2426}\\
& $2$& \citet{Mazzali:1993p1235}\\
SN\,1994D$^{\dagger}$$^{\ddagger}$ & $-11$,$-10$,$-9$,$-8$,$-5$, $-4$,$-2$,$2$,$4.10$,$11$& \citet{Patat:1996p7}\\
SN\,1996X $^{\dagger}$ & $-2$,$0$,$1$,$7$ & \citet{Salvo:2001p1226}\\
SN\,1998aq $^{\dagger}$$^{\ddagger}$ & $-9$,$-8$,$1$,$2$,$3$,$4$,$5$,$6$,$7$ & \citet{Branch:2003p1101}\\
SN\,1998bu $^{\dagger}$$^{\ddagger}$ & $-4$,$-3$,$-2$,$8$,$9$,$10$,$11$ & \citet{Jha:1998p1202}\\
SN\,1999ee $^{\dagger}$ & $-11$,$-9$,$-4$,$-2$,$0$,$5$, $7$,$9$ & \citet{Hamuy:2002p1094}\\
SN\,2000E $^{\dagger}$ & $-9$,$-6$,$-5$,$5$ & \citet{Valentini:2003p1223}\\
SN\,2002bo $^{\dagger}$ & $-11$,$-6$,$-5$,$-4$,$-3$, $-2$,$-1$,$4$ & \citet{Benetti:2004p27}\\
SN\,2002er $^{\dagger}$ & $-11$,$-9$,$-8$,$-7$,$-6$, $-5$,$-4$,$-3$,$-2$,$-1$,$0$, $2$,$4$,$5$,$6$,$10$ & \citet{Kotak:2005p1106}\\
SN\,2003du $^{\dagger}$ & -11$,$-7 & \citet{Anupama:2005p716}\\
& $-8$,$-6$,$-4$,$-2$,$-1$,$0$, $1$,$3$,$4$,$7$,$8$,$9$,$10$ & \citet{Stanishev:2007p82}\\ 
SN\,2004S $^{\dagger}$ & $1$,$7$ & \citet{Krisciunas:2007p1103}\\
SN\,2004dt & $-10$,$-9$,$-7$,$-6$,$-4$, $-3$,$-2$,$-1$,$2$,$3$ & \citet{Altavilla:2007p17}\\
SN\,2004eo $^{\dagger}$ & $-6-3$,$2$,$7$,$11$ & \citet{Pastorello:2007p1224}\\
SN\,1997do $^{\dagger}$ & $-11$,$-10$,$-7$,$-6$,$9$,$11$ & \citet{Matheson:2008p755} \\
SN\,1998V $^{\dagger}$$\ast$ & $1$,$2$,$3$ &\citet{Matheson:2008p755} \\
SN\,1998dh $^{\dagger}$ & $-9$,$-8$,$-7$,$-5$,$-3$,$0$ &\citet{Matheson:2008p755} \\
SN\,1998ec $^{\dagger}$$^{\ddagger}$ & $-3$,$-2$ &\citet{Matheson:2008p755} \\
SN\,1998eg $^{\dagger}$$^{\ddagger}$$\ast$& $0$,$5$ &\citet{Matheson:2008p755} \\
SN\,1999cc $^{\dagger}$$^{\ddagger}$$\ast$ & $-3$,$-1$,$1$,$2$&\citet{Matheson:2008p755} \\
SN\,1999ej & $-1$,$ 3$,$5$,$9$ &\citet{Matheson:2008p755} \\
SN\,1999gd $^{\dagger}$ &$3$,$10$ &\citet{Matheson:2008p755} \\
SN\,1999gp $^{\dagger}$$^{\ddagger}$$\ast$ & $-5$,$-2$,$1$,$3$,$5$,$7$ &\citet{Matheson:2008p755} \\
SN\,2000fa  $^{\dagger}$$^{\ddagger}$$\ast$& $-10$,$2$,$3$,$5$,$10$ &\citet{Matheson:2008p755} \\
SN\,2001V  $^{\dagger}$ & $-11$,$-10$,$-9$,$-8$,$-7$, $-6$,$-4$,$9$,$10$ & \citet{Matheson:2008p755} \\
\hline
\end{tabular}
\end{table}

\subsection{Photometry Sample}\label{sec:low-z-cosmo}

The photometric properties of all 21 supernovae which have published photometry were refitted using the SiFTO light curve fitter to produce a consistently-fitted sample.  The lightcurves were fitted for stretch, colour, date of maximum and peak magnitude, which were useful for studying any possible relation between the spectroscopy features and residuals from the Hubble diagram.  The results are summarised in Table \ref{lowz_phot}, which also shows any independent distance modulus measurements available for the host. 

\subsection{Low-z Equivalent Width Measurements and Errors}

The EWs were evaluated using Equation \ref{eweq} and all measurements were made in the object's rest-frame. The values are given in Table \ref{tab:lowz-ew}.  Error spectra were not available for the low-z spectra as they were for the high-z SNLS data which is why Equation \ref{eweq} was used.  This meant that no statistical error on the EW could be measured.  The assumption in \citetalias{Bronder08} was that the main source of error was due to incorrectly placing the pseudo-continuum bounds around the features and this error was quantified in the same way as described for the high-z data in Section \ref{sec:equiv-width-meas}; however, the low-z spectra are of high signal-to-noise and, as the equivalent width measurement code had been made much more automated, these errors were negligible.  When one measurement of the EW was required per supernova, the mean over all epochs was measured and the error was assigned based on the RMS scatter about the mean.  This was a good estimation for the features being investigated here as the spectral features appear to be approximately constant with phase within the phase range investigated (see Figure \ref{fig:lowz_trends}).

\subsection{Low-Redshift Trends}\label{sec:low-redshift-trends}

\begin{figure}
\centering
\includegraphics[width=8cm]{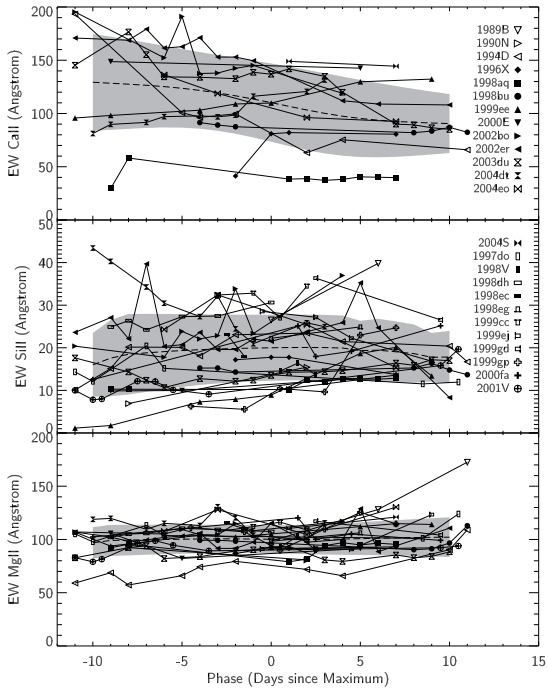}
\caption[Low-z Trends for the three spectral features]{Low-z Trends for \protect\ion{Ca}{ii} H\&K, \ion{Si}{ii} and \ion{Mg}{ii}.  The individual supernova measurements are marked by the different symbols.  The central thick dotted line is the trend and the grey shaded region represents the $1\sigma$ error distribution.  The legend applies to all 3 panels.}
\label{fig:lowz_trends}
\end{figure}

We calculate low-redshift trends as a measure of the variation of spectral features with phase relative to the date of maximum.  The trend is calculated by measuring a resistant median each day $\pm10$ from B-band maximum.  The resistant median removes outliers which are more than $3\sigma$ from the median.  The weights applied to each measurement is a Gaussian centered on that date with a FWHM of 4 days.  This time-frame was chosen as smaller time intervals result in the trend becoming affected by the scatter of the EW and any poor phase sampling.  A weighted RMS was calculated to produce the $1\sigma$ scatter for the sample.  The resulting trends are presented in Figure \ref{fig:lowz_trends}.

\begin{figure*}
\centering
\includegraphics[scale=0.35]{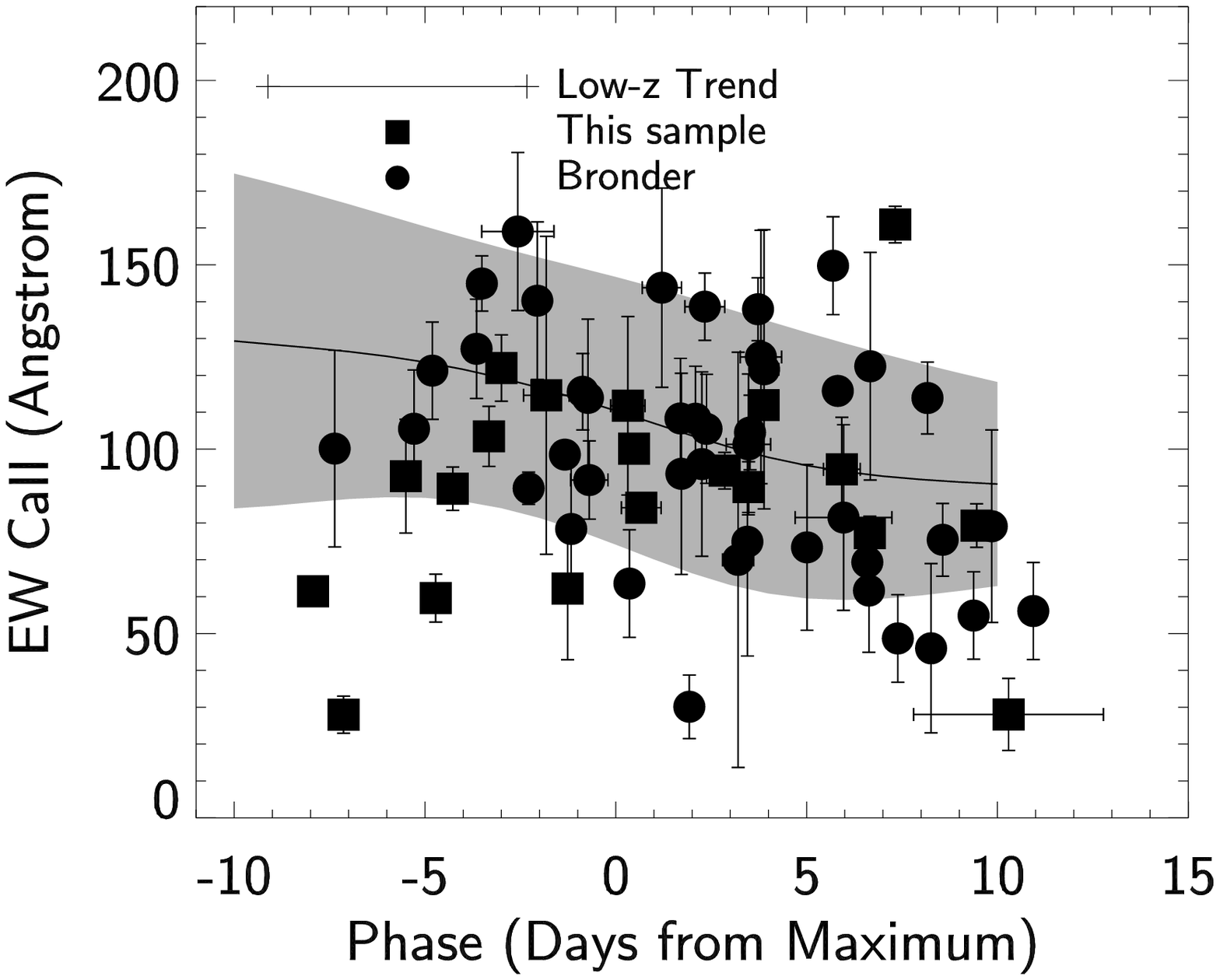}
\includegraphics[scale=0.35]{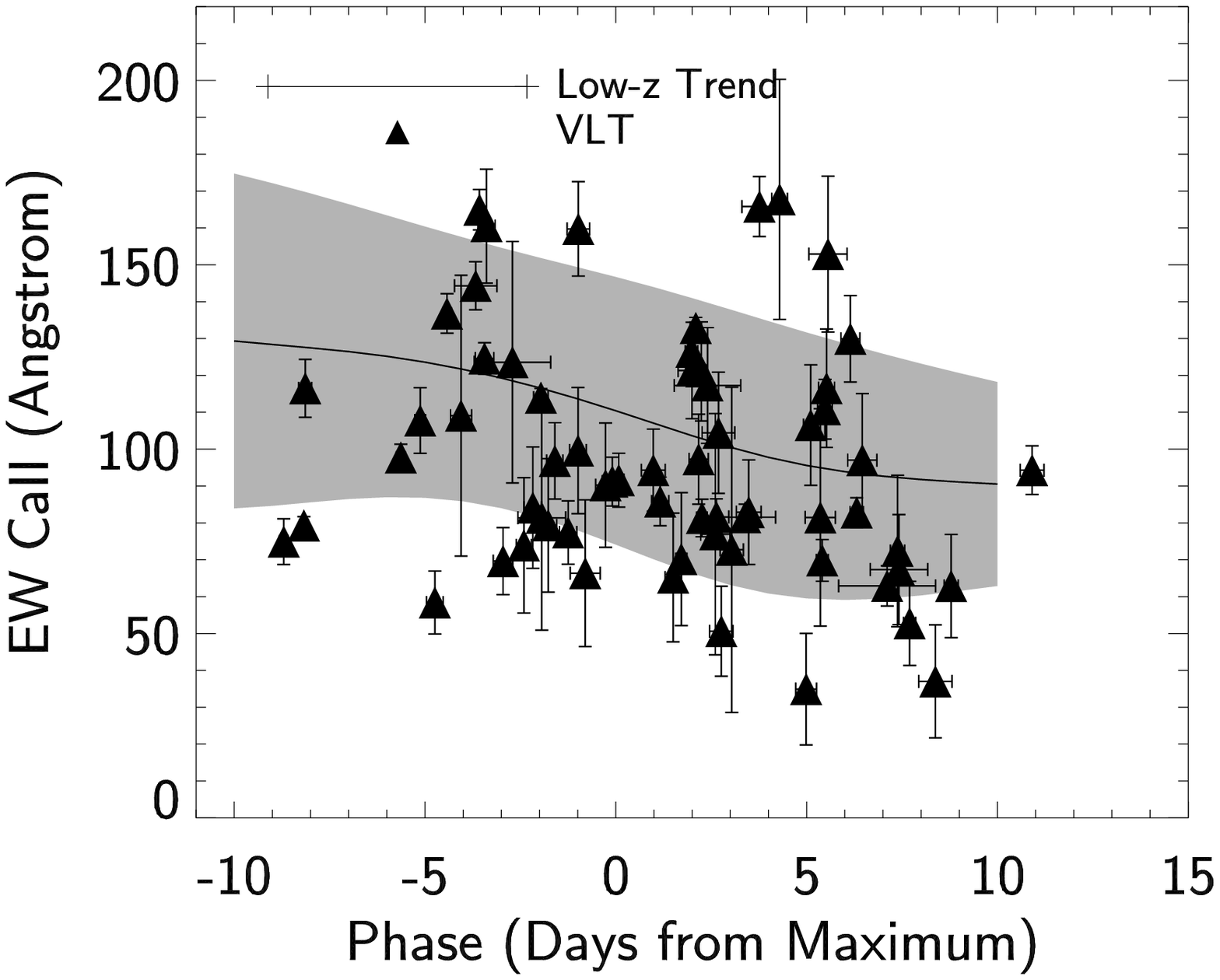}
\includegraphics[scale=0.35]{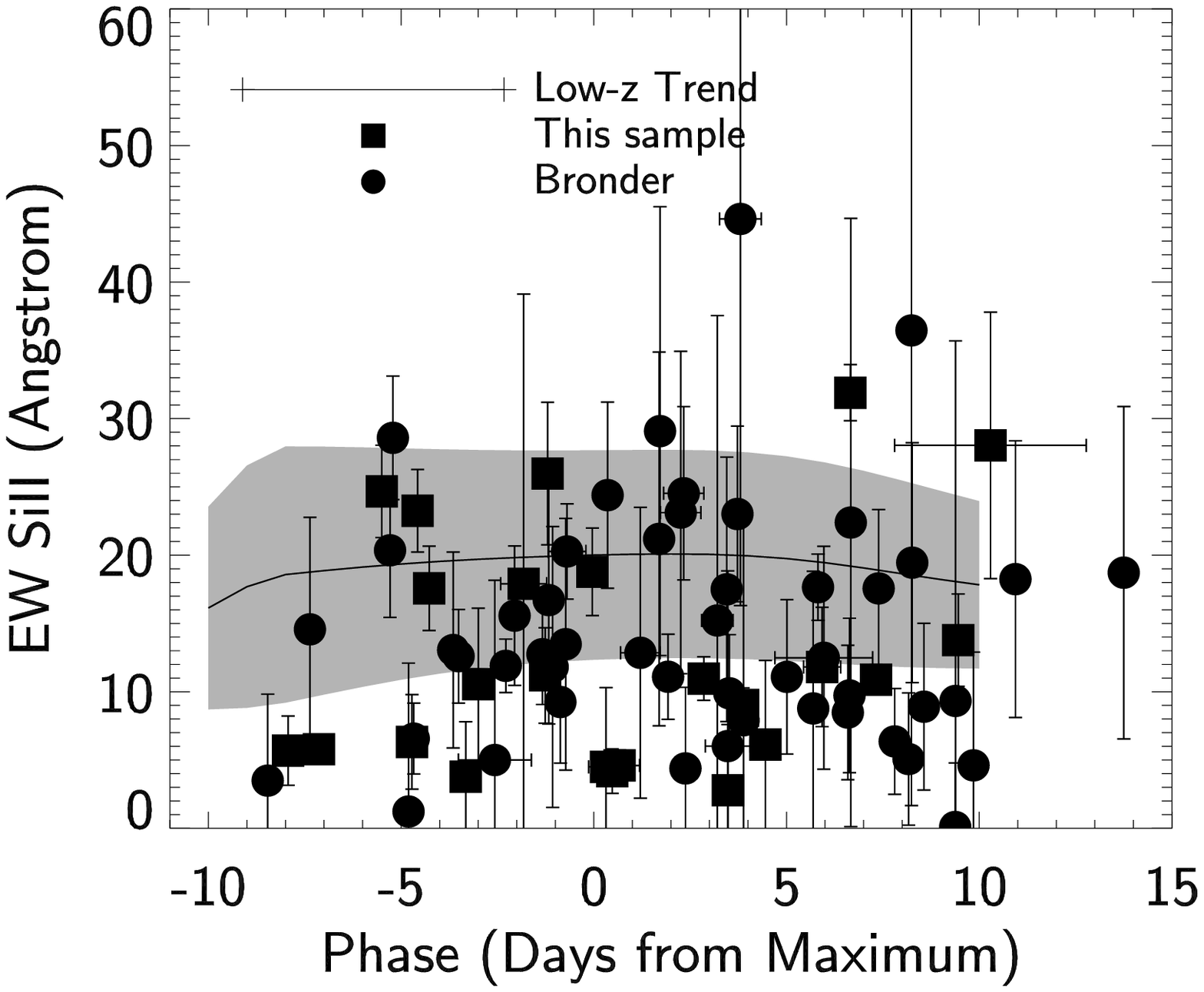}
\includegraphics[scale=0.35]{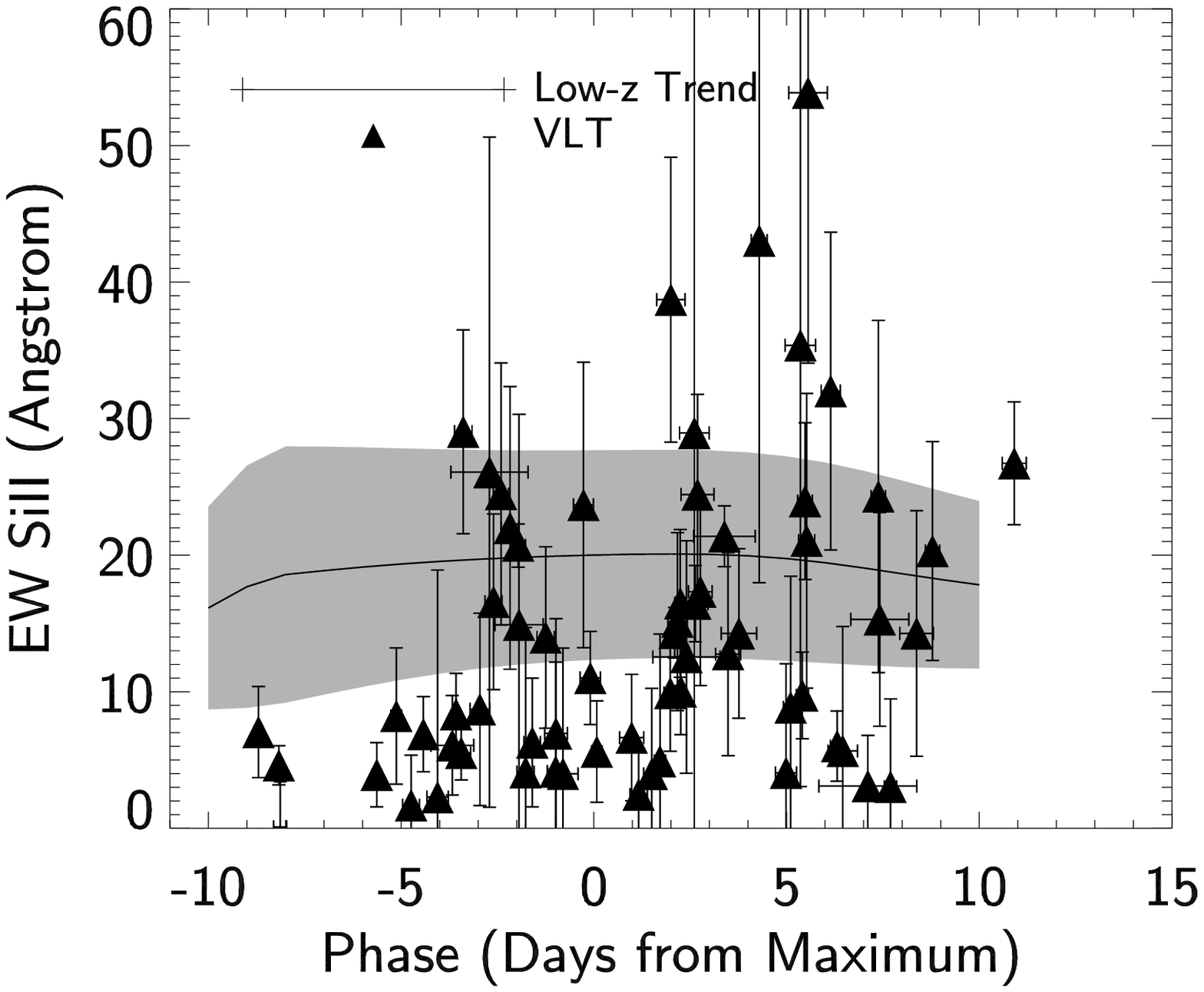}
\includegraphics[scale=0.35]{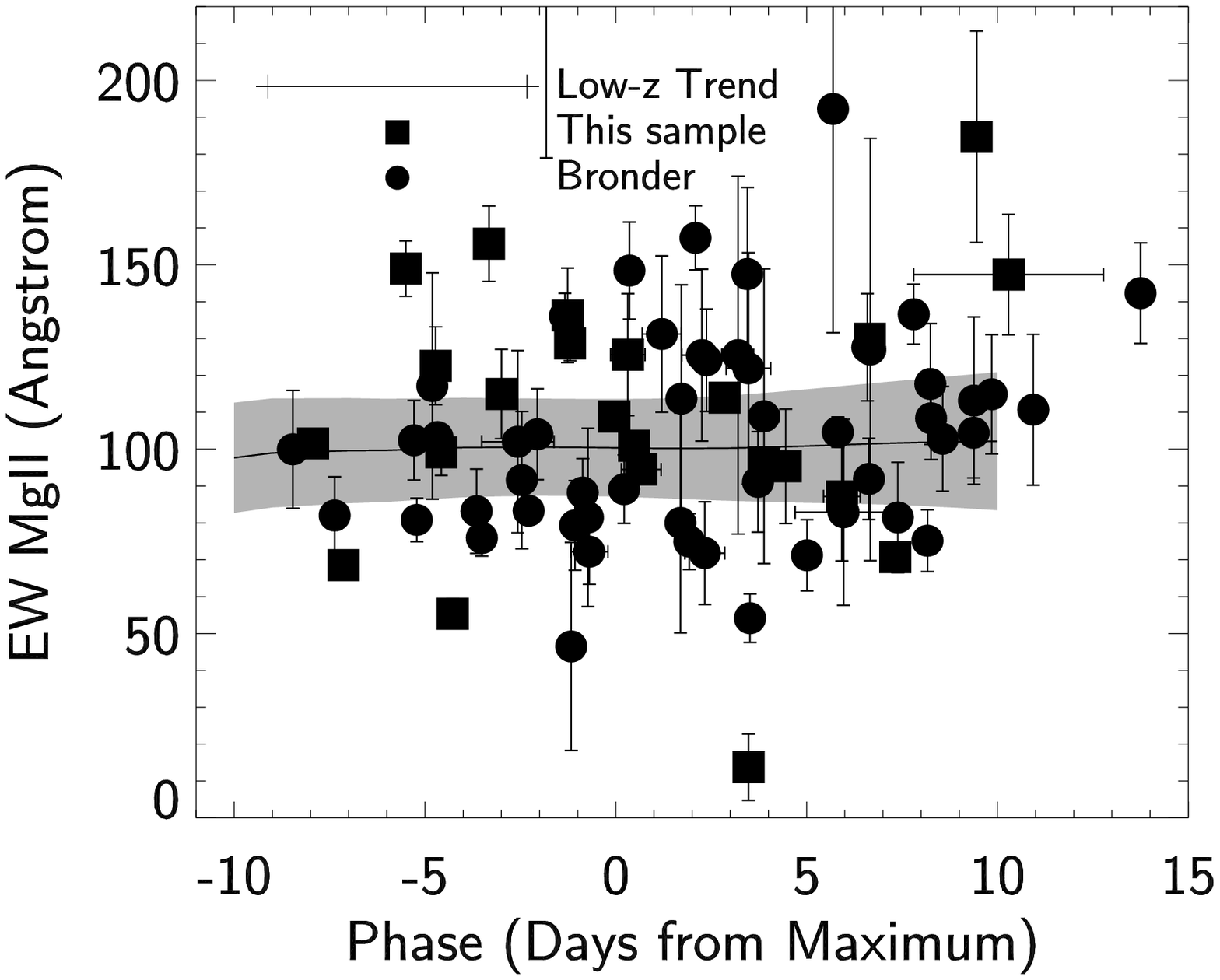}
\includegraphics[scale=0.35]{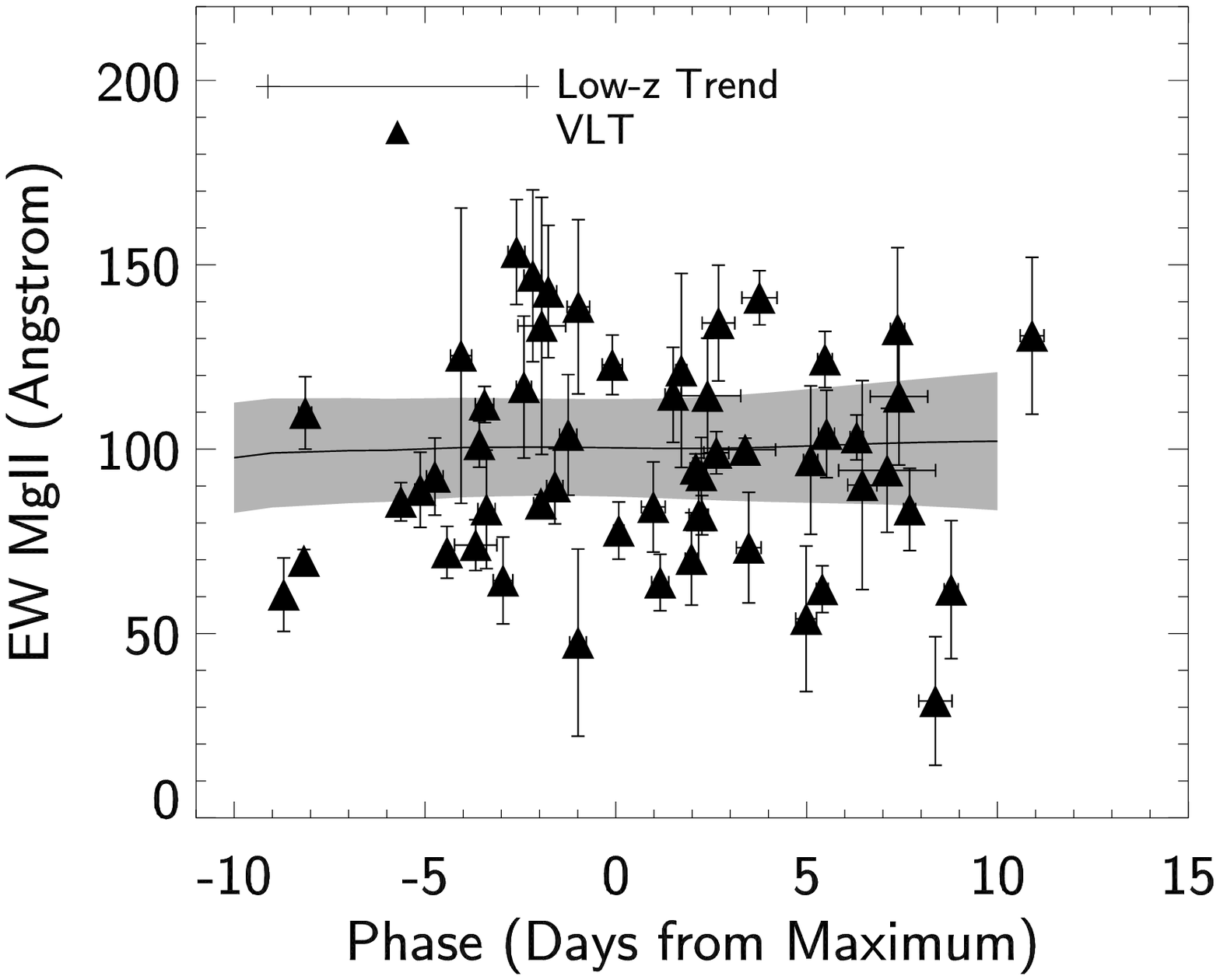}
\caption[High-redshift/Low-redshift Comparison]{Each figure shows the low-z trend of the EW of either \ion{Ca}{ii} H\&K, \ion{Si}{ii}\ or \ion{Mg}{ii}\ as a solid black line with the $1\sigma$ region in grey.  The high-z sample is shown as the points plotted over the top as squares and circles (Gemini) and triangles (VLT)}\label{fig:highz_lowz_comp}
\end{figure*}

The trends shown in Figure \ref{fig:lowz_trends} show that over the range of phases studied, all three features are approximately constant.  The \ion{Ca}{ii} H\&K feature decreases in strength slightly post-maximum, but only by 20\,\AA.  Both the \ion{Ca}{ii} H\&K\ and \ion{Mg}{ii}\ features show one supernova with a significantly reduced EW measurement.  In the case of \ion{Ca}{ii} H\&K, it is SN\,1998aq and for the \ion{Mg}{ii}\ feature is it SN\,1994D.  Both objects appear to have quite a blue colour (see Table \ref{lowz_phot}).  Due to the small size of the sample, especially for the \ion{Ca}{ii} H\&K\ feature, it is not possible to tell whether these are anomalous results, or important sub-classes of objects which have not yet been identified.  For example, if one object in the low-z sample has a low \ion{Ca}{ii} H\&K\ equivalent width then a number of objects like this should these should appear in the high-z sample.  This may explain some of the low EW values seen in Figure \ref{fig:highz_lowz_comp}.

These trends differ in a number of ways from those presented in \citetalias{Bronder08}.  Firstly, the method of calculating the trends is different.  The trends in \citetalias{Bronder08} are created by measuring EWs from the template spectra  of \citet{Nugent:2002p712}.  An error region was determined by adding an error to the trend and comparing to low-z measurements until a $\chi^2_r = 1$ was reached.  This method assumes that the low-z measurements are distributed symmetrically about the trend.

In this paper, the trends are generated directly from the measurements themselves and the $1\sigma$ scatter was determined directly from the EW measurements.  There are also a number of objects included in this study which were not in \citetalias{Bronder08} and vice versa due to the selection criteria or date of publication.  As the sample size is small, this has had a considerable effect on the trends.

The \citetalias{Bronder08} \ion{Si}{ii}\ trend is slightly lower that the one seen here.  This is primarily caused by the fact that new spectra introduced into the sample all had higher EW values than the \citetalias{Bronder08} mean.  The increase in EW to maximum followed by a decline is not seen here to such a large degree as in the \citetalias{Bronder08} trends.

The rapid rise in the \ion{Mg}{ii}\ EW seen after $+3$\,days in \citetalias{Bronder08} is not seen in this sample where the trend is flat from $-10$ to $+10$\,days.  This is likely due to the fact that the sample is different; however, SN\,1989B, which is in both samples, does show a larger increase in the \ion{Mg}{ii}\ EW between the measurements at +6 and +11 days.

Due to the lack of available spectra of SN\,1991T-like and SN\,1991bg-like objects, it was not possible to measure trends for these objects in a similar way to the core-normal supernovae, which is why spectroscopically peculiar events are removed from the EW analysis.

\defcitealias{Howell05}{H05}
\section{Results}

\subsection{Comparison between High-z Data and Low-z Trends}\label{sec:stat-comp}

The high-z sample is divided into the Gemini sample (objects presented in this paper, plus those published in \citetalias{Howell05} and \citetalias{Bronder08}) and VLT samples in order to identify any systematic differences between the samples: none are seen in this analysis.  Figure \ref{fig:highz_lowz_comp} shows the comparison for the three features in this analysis for both samples.  The figures show the low-redshift trends seen in Figure \ref{fig:lowz_trends} with the high-z points measured from the Gemini and VLT samples overplotted.

The high-z measurements for the \ion{Si}{ii}\ feature appear to be smaller than the low-z trend.  In order to look for systematic shifts, we make trends of the high-z data.  The trends are calculated in a similar way to the low-z ones, except that the points are weighted according to the size of the error on the measurement, rather than applying a Gaussian weight based on phase.  A rolling mean is applied with a bin-size of $\pm3$\,days.  The mean for each phase bin is constrained so that there must be more than 10 equivalent width measurements within $\pm3$\,days in order for the value to stand.  Consequentially, the trends are not evaluated at some early and later phases.

\begin{figure*}
\centering
\includegraphics[scale=0.35]{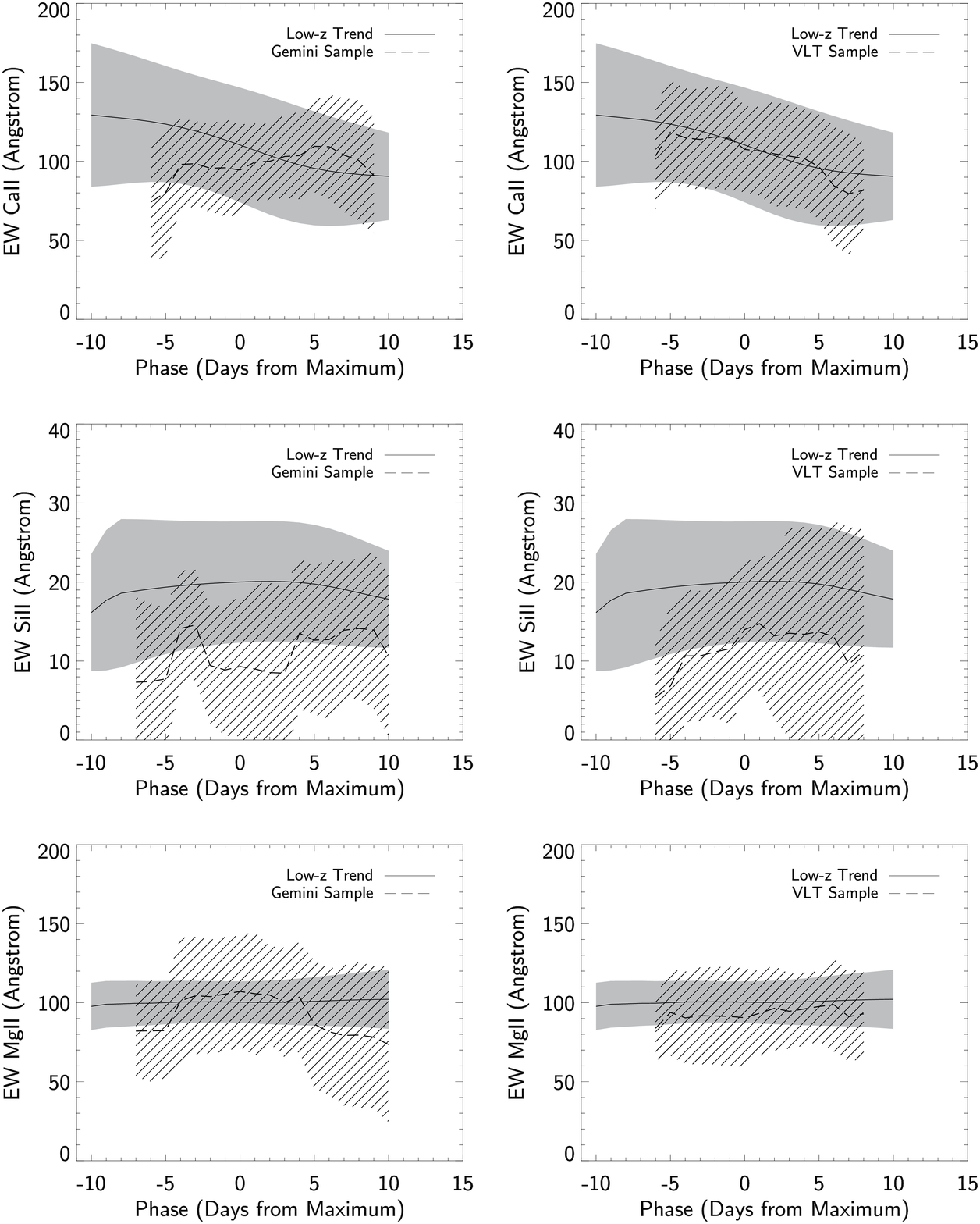}
\caption[High-redshift/Low-redshift Trends Comparison]{Each figure show the low-z trend of the EW of either \ion{Ca}{ii} H\&K, \ion{Si}{ii}\ or \ion{Mg}{ii} as a solid black line with the $1\sigma$ region in grey.  The high-z trends are overplotted in blue (Gemini) and green (VLT) with the shaded region representing the $1\sigma$ RMS scatter}\label{fig:highz_trends}
\end{figure*}

Figure \ref{fig:highz_trends} shows that for the \ion{Ca}{ii} H\&K\ and \ion{Mg}{ii}\ features, the low- and high-z trends are approximately consistent.  The \ion{Si}{ii}\ feature shows a high-z trend which is consistently below the low-z one, as expected from the low-EW points seen in Figure \ref{fig:highz_lowz_comp}.  This will be discussed later.  This is contrary to the observations of \citetalias{Bronder08} which saw no difference between the high- and low-z populations in the \ion{Si}{ii} feature.  The difference between the high- and low-z populations that may have been observed for the \ion{Mg}{ii}\ feature in \citetalias{Bronder08} is not seen in this analysis which uses a larger high-z sample and a different method for generating low-z trends.  

\subsection{Correlations Between Spectroscopic and Photometric Properties}\label{sec:corr-spec-phot}

One of the reasons why there might be differences between the low- and high-z samples is because the spectral properties will be sensitive to the explosion physics of the supernovae.  There is increasing evidence that there are two populations of supernova (e.g.\ \citet{Sullivan:2006p3}) and that the prompt supernovae tend to be brighter and bluer.  This will produce a drift in the mean photometric properties of SNe Ia as a function of redshift if the mix of prompt and delayed supernovae changes with redshift \citep{Howell:2007p1767}.  $M_B$, the absolute B-band magnitude, is known to correlate with lightcurve width in the sense that more luminous SNe have higher stretch.  We also know from \citetalias{Bronder08} that EW \ion{Si}{ii}\ is correlated with $M_B$, so brighter supernovae appear to contain less \ion{Si}{ii}.  These facts taken together imply that the mean spectral properties between the high- and low-z samples will be different in the sense that the high-z sample, where we expect a greater proportion of the prompt-type objects, should have a lower EW \ion{Si}{ii}.  This is indeed the effect seen in Figure \ref{fig:highz_lowz_comp}.  In the following we quantify the size of this effect and determine whether it explains the observed differences seen in Figure \ref{fig:highz_lowz_comp}.  First we quantify the correlations between spectral and photometric properties, stretch and colour, using a Spearman's Rank correlation method.

We combine the Gemini and VLT samples into one high-z sample as the correlations are likely to be small effects.  The high-z spectra in this study are also noisy and so there are large errors on the EW measurements.  As the correlation coefficient calculation does not deweight objects based on the size of their error bars, we have grouped all the high-z objects together and ranked the objects according to the size of their error and used only the lowest 75\% for this part of this analysis.The cuts for \ion{Ca}{ii} H\&K, \ion{Si}{ii}\ and \ion{Mg}{ii}\ correspond to error values of 18.1\,\AA, 9.7\,\AA\ and 19.3\,\AA\ respectively.  The correlations are evaluated for all the high-z objects with errors less than the 75-percentile cut.  The correlation coefficients are presented in Table \ref{tab:phot_corr}.

\begin{table}
\caption[Photometric Correlations]{Correlation coefficients between the measured EWs and the light curve stretch and colour for the low-z, high-z and combined samples.  The number of standard deviations from a null hypothesis is given in the final column.}
\label{tab:phot_corr}
\begin{tabular}{c c c c c c}
\hline
Feature & Variable & Sample & $\rho$ & $\sigma$ from null \\
\hline
\ion{Ca}{ii} H\&K & Stretch & Low-z & 0.21 & -0.66\\
&& High-z & 0.10 & -1.0\\
&& All & 0.09 & -0.91\\
& Colour & Low-z & 0.44 & -1.38 \\
&& High-z & 0.12  & -1.20 \\
&& All & 0.21 &  -2.15 \\
\ion{Si}{ii} & Stretch & Low-z & -0.76 & 3.58 \\
&& High-z & -0.24 & 2.50 \\
&& All & -0.33 & 3.78 \\
& Colour & Low-z & 0.19 & -0.89 \\
&& High-z & 0.08 & -0.80 \\
&& All & 0.19 & -2.12 \\
\ion{Mg}{ii} & Stretch & Low-z & -0.31 & 1.47 \\
&& High-z & 0.09 & -0.93 \\
&& All & 0.03 & -0.38 \\
& Colour & Low-z & 0.47 & -2.22 \\
&& High-z & 0.09 & -0.85 \\
&& All & 0.17 & -1.87 \\
\hline
\end{tabular}
\end{table}

The correlation analysis shows that the strongest correlations exist between stretch and EW \ion{Si}{ii}, and this is seen in all samples at the 2.5--3.5$\sigma$ level.  There are also weak correlations between the EW \ion{Si}{ii}\ and colour for the combined sample, and for the low-z sample between EW \ion{Mg}{ii}\ and colour.  There also appears to be a slight correlation between the EW \ion{Ca}{ii} H\&K\ and colour.  There appears to be no correlation between EW \ion{Ca}{ii} H\&K\ and stretch.  We now examine the two most promising relations: EW \ion{Si}{ii} and stretch; and EW \ion{Mg}{ii} and colour.

\subsubsection{EW \ion{Si}{ii} and Stretch}\label{sec:ew-stretch}

\begin{figure}
\centering
\includegraphics[width=8cm]{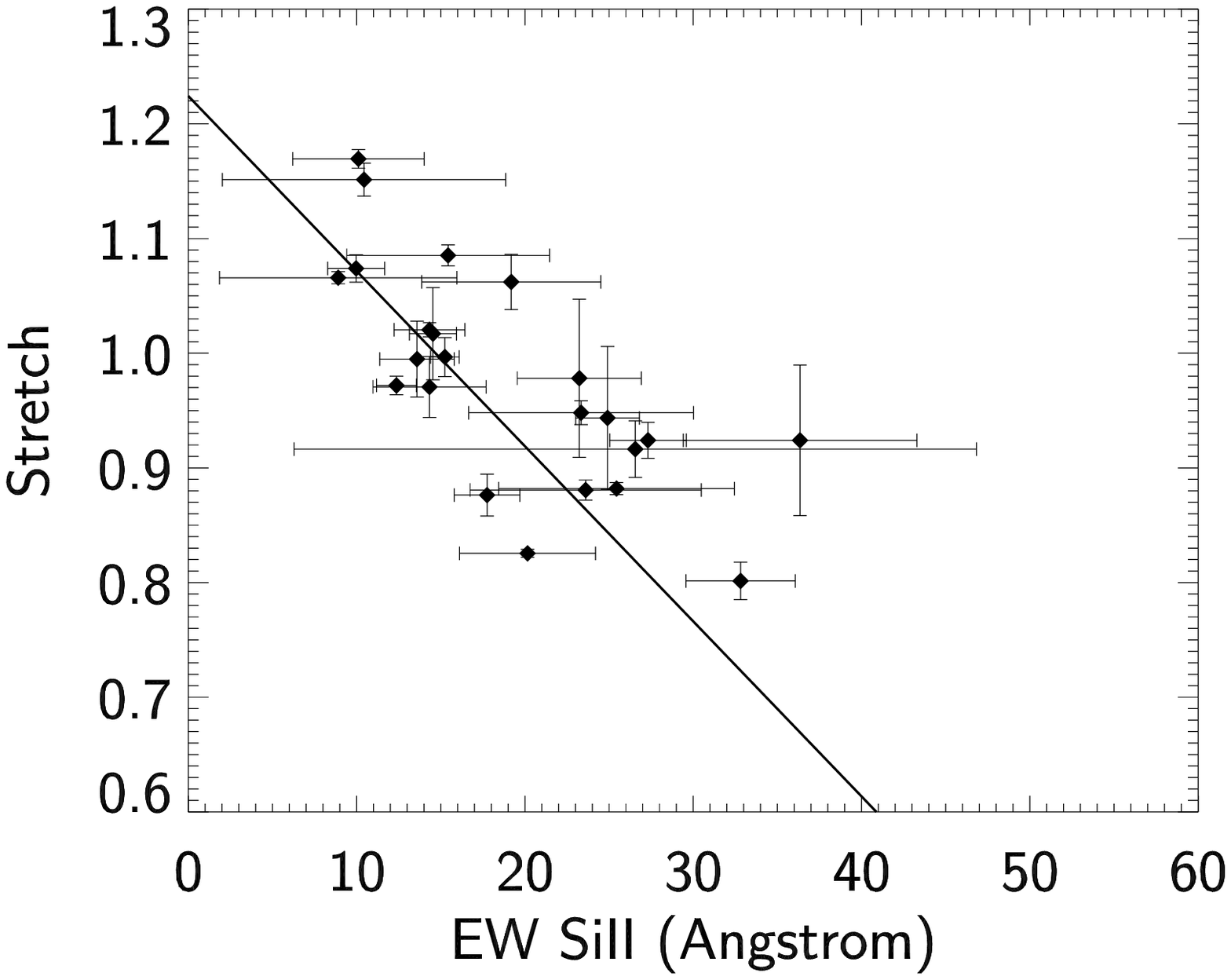}
\includegraphics[width=8cm]{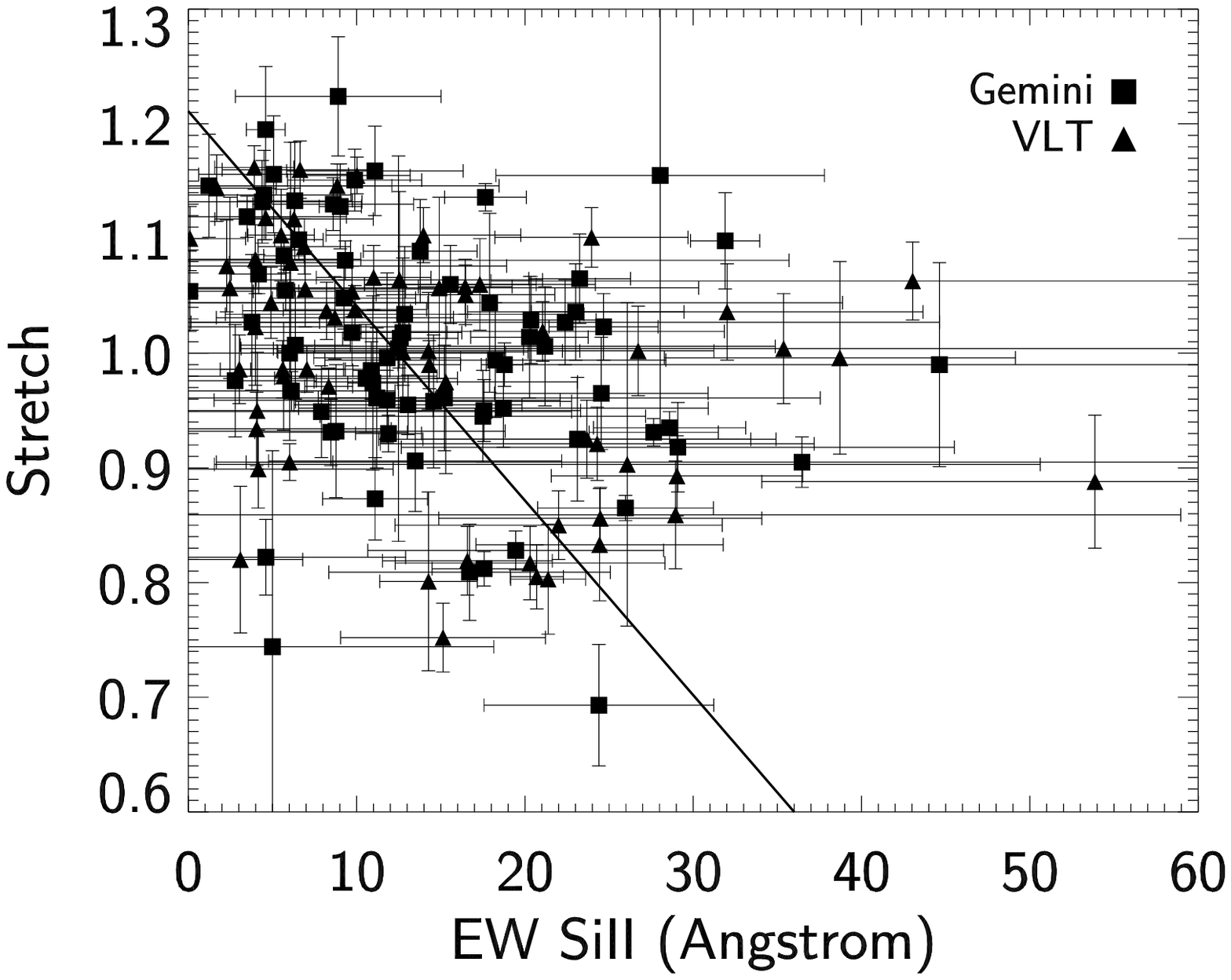}
\caption[EW \ion{Si}{ii}-Stretch Plots]{Top: Stretch against EW \ion{Si}{ii}\ for the low-z data.  Bottom: Stretch against EW \ion{Si}{ii}\ for the high-z Gemini (blue squares) and VLT (red triangles) data.  In both plots, the black line represents the best-fit straight line to the data.  The fits are given in Equations (\ref{eq:low-z_ew_s}) and (\ref{eq:high-z_ew_s})}
\label{fig:ew_strech}
\end{figure}

The stretch-EW \ion{Si}{ii}\ measurements are plotted in Figure \ref{fig:ew_strech} for the low-z (diamond) and high-z (square and triangle) points.  Here the high-z Gemini and VLT samples are plotted separately in order to identify any possibly systematic offsets between the samples: none are seen.  The best-fit was found using \texttt{LINFITEX}, part of \texttt{MPFIT}, which takes account of errors on both the variables.  The low- and high-z relationships shown by the black lines are are:
\begin{eqnarray}
s & = & (1.22 \pm 0.03) + (-0.015 \pm 0.002)\times EW_{SiII}\textnormal{\quad low-z}\label{eq:low-z_ew_s}\\
s & = & (1.21 \pm 0.01) + (-0.017 \pm 0.001)\times EW_{SiII}\textnormal{\quad high-z}\label{eq:high-z_ew_s}
\end{eqnarray}

\noindent The reduced $\chi^2$ for the fits are $\chi^2_r = 2.2$ (low-z) and $\chi^2_r = 2.7$ (high-z).  The fits are consistent to within 1$\sigma$, although the errors on the gradient and intercept are correlated.  We would expect the fits to be the same if the high- and low-z populations behave in the same way; however this does not mean that they must have the same distribution at high and low redshift.

\begin{figure}
\begin{center}
\includegraphics[width=8cm]{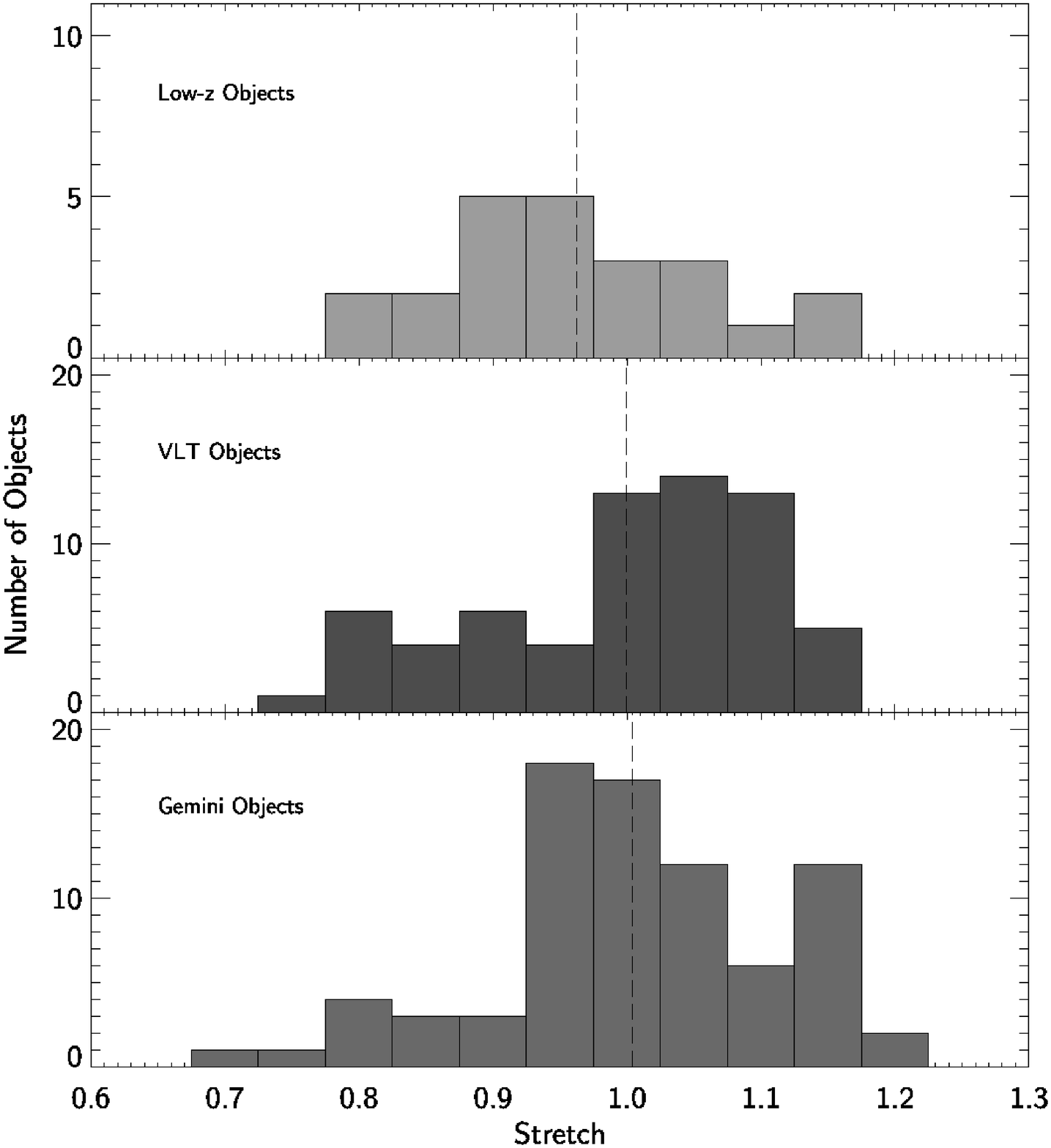}
\end{center}
\caption[Stretch Histogram]{Histograms of the stretch distribution for the low-z (green), VLT (red) and Gemini (blue) samples.  The dotted line marks the mean stretch for each sample.  The low-z sample has a lower mean stretch than the two high-z samples.}
\label{fig:stretch_hist}
\end{figure}

It has already been established in \citetalias{Bronder08} that there is a relationship between EW \ion{Si}{ii}\ and $M_B$, the absolute magnitudes of the individual supernovae: the brighter the supernova, the smaller EW \ion{Si}{ii}.  This has been extended here to use lightcurve stretch instead of magnitude, as is shown in Figure \ref{fig:ew_strech}, as $M_B$ is known to correlate with stretch.  This relationship can potentially explain the difference seen between the low-z and high-z objects in Figures \ref{fig:highz_lowz_comp} and \ref{fig:highz_trends} above.  Figure \ref{fig:stretch_hist} shows that the stretch distribution is shifted to higher stretch for the higher redshift samples compared to the low-z one.  The relationships seen in Figure \ref{fig:ew_strech} show that more higher stretch objects would mean more low-EW \ion{Si}{ii}\ measurements.

To test whether the difference in mean stretch between the high- and low-z samples can be attributed as the cause of all the systematic shift in EW \ion{Si}{ii} as seen in Figure \ref{fig:highz_lowz_comp}, a predicted high-z mean EW was calculated based on the mean high-z stretch and the low-z relation from Equation (\ref{eq:low-z_ew_s}).  This predicted high-z mean EW \ion{Si}{ii} can then be compared to the measured value.

From Figure \ref{fig:stretch_hist}, the mean stretches for the high- and low-z samples are $\overline{s}_{\textnormal{\footnotesize{low-z}}} = 0.96$, $\overline{s}_{\textnormal{\footnotesize{Gem}}} = 1.00$ and $\overline{s}_{\textnormal{\footnotesize{VLT}}} = 1.00$, with standard deviations $\sigma_{\textnormal{\footnotesize{low-z}}} = 0.10$,  $\sigma_{\textnormal{\footnotesize{Gem}}} = 0.11$ and $\sigma_{\textnormal{\footnotesize{VLT}}} = 0.10$.  Using the high-redshift standard deviation to produce upper and lower bounds and ignoring the errors on the fit in Equation (\ref{eq:low-z_ew_s}), the predicted mean high-z EW \ion{Si}{ii}\ values are 14.67\,\AA\ with a standard deviation of 7.34\,\AA\ (Gemini), and 14.67\,\AA\ with a standard deviation of  6.67\,\AA\ (VLT).  Taking the weighted mean and standard deviations of the observed high-z trends in Figure \ref{fig:highz_trends} yields observed mean EW \ion{Si}{ii}\ of 11.23\,\AA\ with a standard deviation of 4.58\,\AA\ (Gemini) and 11.05 with a standard deviation of 6.64\,\AA\ (VLT).  Although the errors on the quantities are large, the predicted and observed high-z EW \ion{Si}{ii}\ are in good agreement with each other.

Thus it appears that the difference in observed EW \ion{Si}{ii} between the low- and high-z samples can be explained by the difference in the stretch distributions of the two samples, i.e.~the differences are due to changing demographics rather than a new, unexplained form of evolution.

\subsubsection{EW \ion{Mg}{ii} and Colour}

\begin{figure}
\centering
\includegraphics[width=8cm]{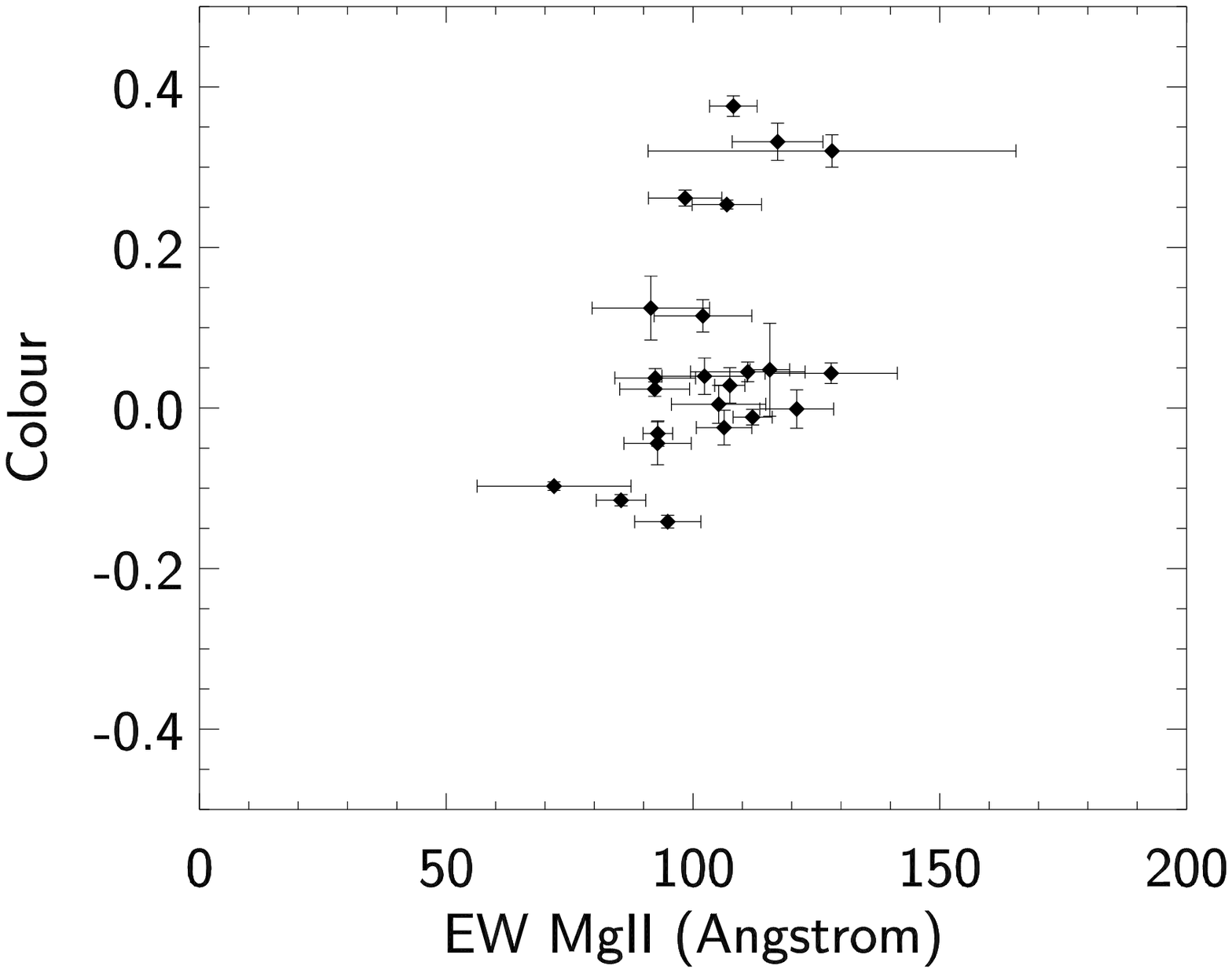}
\includegraphics[width=8cm]{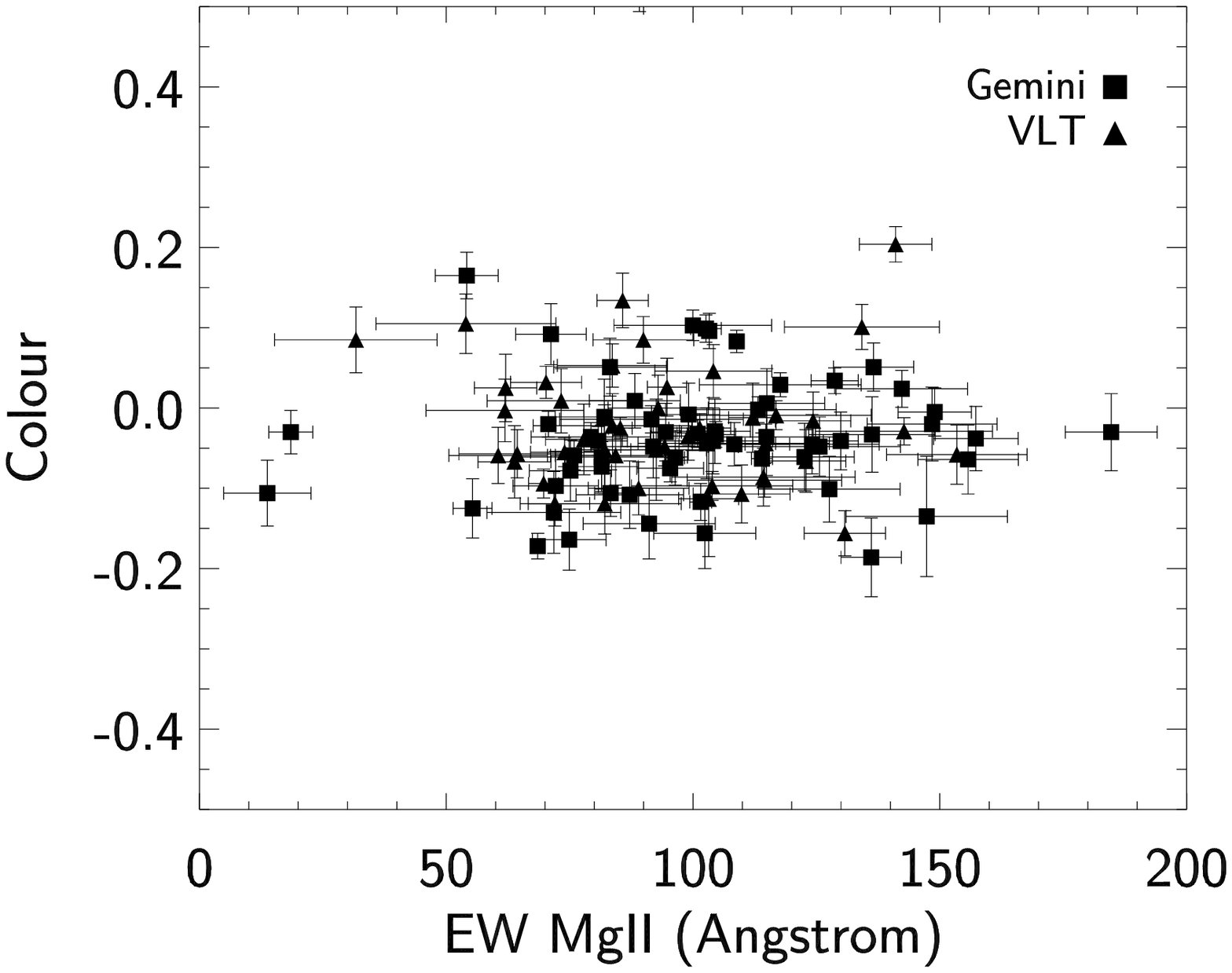}
\caption[EW \ion{Mg}{ii}-Colour Plots]{Top: Colour against EW \ion{Mg}{ii}\ for the low-z data.  Bottom: Colour against EW \ion{Mg}{ii}\ for the high-z Gemini (square) and VLT (triangle) data.}
\label{fig:ew_colour}
\end{figure}

The relationships between colour and EW \ion{Mg}{ii}\ in the low- and high-z data are plotted in Figure \ref{fig:ew_colour}.  Figure \ref{fig:ew_colour} highlights the differences in properties between the high- and low-z samples.  The range of EW \ion{Mg}{ii}\ values in the high-z data is much broader than for the low-z sample, but the reason is unclear.  This could be because the low-z sample is too small to be representative of the SNe Ia population as a whole.  The high-z sample is suffering from selection bias as a rough colour-cut is applied during high-z target selection and we do not see the range of supernova colours observed in the low-z sample. 

\begin{figure}
\centering
\includegraphics[width=8cm]{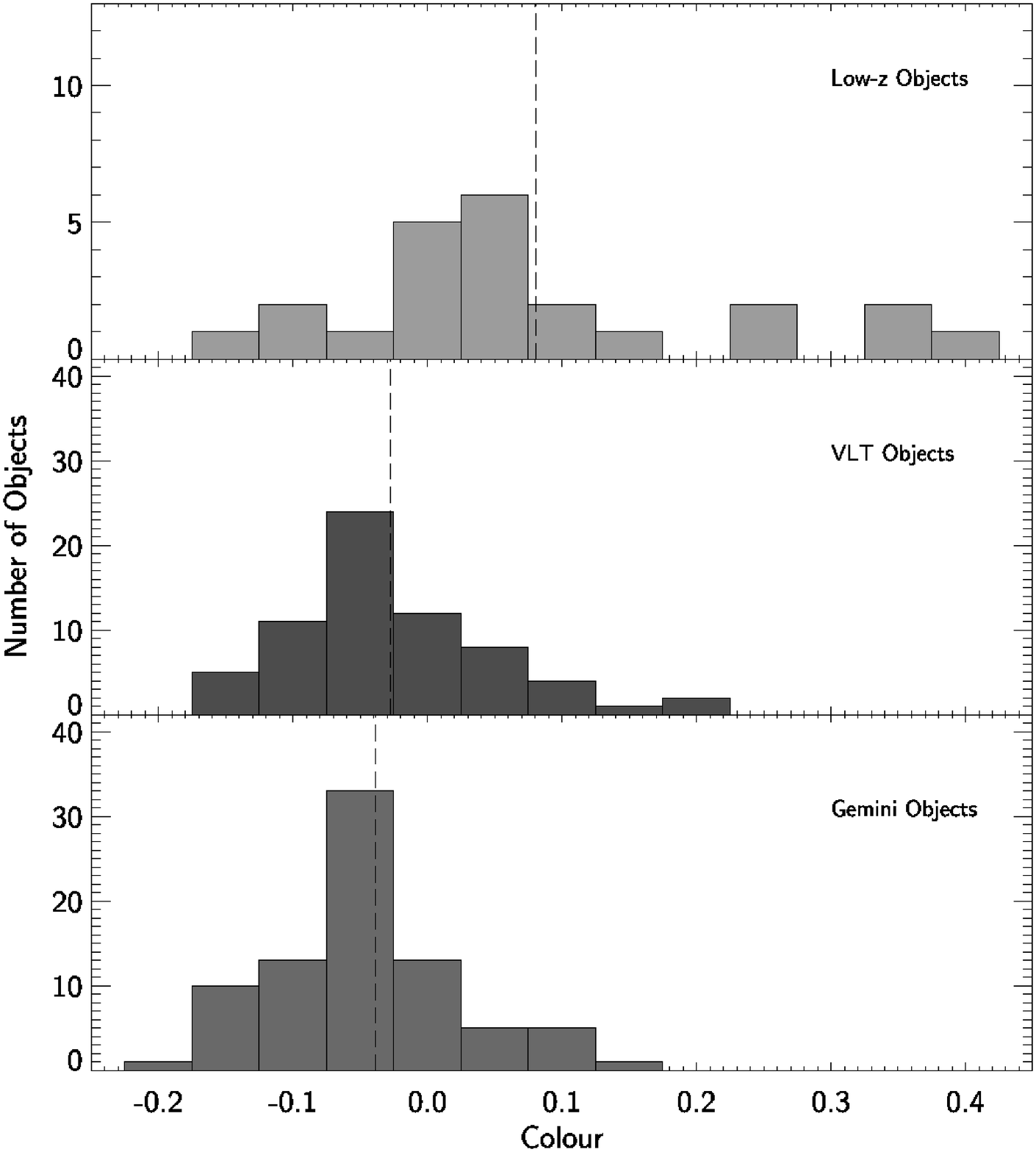}
\caption[Colour Histogram]{Histograms of the colour distribution for the low-z (green), VLT (red) and Gemini (blue) samples.  The dotted line marks the mean stretch for each sample.  The low-z sample has a redder mean than the two high-z samples.}
\label{fig:colour_hist}
\end{figure}

Figure \ref{fig:colour_hist} shows that the colour distributions for all three samples are very different.  The low-z sample contains a number of redder objects which aren't seen in the high-z data.  These are not underluminous, redder supernovae, but objects classed as core-normal.The Gemini distribution shows a strongly-peaked distribution around $c=0$
whereas the VLT distribution is more spread out.  The mean colours for the samples are \mbox{$\overline{c}_{\textnormal{\footnotesize{low-z}}} = +0.08$,} $\overline{c}_{\textnormal{\footnotesize{Gemini}}} = -0.03$ and $\overline{c}_{\textnormal{\footnotesize{VLT}}} = -0.04$.    

It is not possible to state anything conclusive about any relationship between colour and EW \ion{Mg}{ii} as the data is very noisy at high-z and the sample at low-z is very small.  This would be interesting to investigate further with better data as the relation could potentially reveal more about the supernova progenitor and be another test for an evolving population, but it is not possible at this time.

\subsection{Applications of SNe Ia Spectra in Cosmological Studies}\label{sec:cosmological-uses}

One of the aims of this study is to investigate potential uses for supernova spectra in the calibration of the objects for cosmological studies.  This could be either as an alternative to the current method of photometric stretch and colour, or as a third calibrator, with the goal of reducing the scatter around the best-fit line on the Hubble diagram and hence improve the accuracy of cosmological parameter estimation.  The large, spectroscopic dataset, with accompanying photometry lends itself well to this type of study.

\subsubsection{Cosmology Fitter}

In order to measure a scatter in the Hubble diagram, we must first fit the cosmology.  In this section we detail the cosmological fitter used for this purpose.  Throughout, we assume a flat universe with $w=-1$.

The supernova cosmology is fitted using a \texttt{MINUIT}-based, lighter-weight version of \texttt{SIMPLE COSFITTER}\footnote{\texttt{http://qold.astro.utoronto.ca/conley/simple\_cosfitter/}} as opposed to a full Bayesian analysis used in \citet{Astier06}. We refit the data rather than using the cosmology found in \citet{Astier06} because the samples used in the studies are different.  A best-fit method is faster and sufficiently precise for this study.

\texttt{SIMPLE COSFITTER} takes the supernova apparent B-band magnitudes at maximum, $m_B$; the redshift, $z$; the supernova lightcurve stretch, $s$; and supernova lightcurve colour $c$ and performs a least-squared fit to find the best-fitting values of $\Omega_M$, $\alpha$, $\beta$ and $\mathcal{M}$ by minimising the $\chi^2$ where
\begin{eqnarray}
\chi^2& = &\sum_{\textnormal{objects}}\frac{(\mu - 5\log (d_L(z,\theta)/10\textnormal{\,pc}))^2}{\sigma^2(\mu) + \sigma^2_{int}}
\end{eqnarray}

\noindent and 
\begin{eqnarray}\label{eq:cosmo_fit}
\mu & = & m_B - \mathcal{M} + \alpha(s-1) - \beta c
\end{eqnarray}

\noindent $\theta$ stands for the cosmological parameters that define the fitted model, which in this case is simply $\Omega_M$; $\sigma(\mu)$ is the error on the observed distance modulus; and $\sigma_{int}$ is the intrinsic dispersion in the supernova absolute magnitudes \citep{Astier06}.  A good fit is obtained by varying the parameter $\sigma_{int}$ manually until the $\chi^2_r = 1$; the other parameters are varied by \texttt{COSFITTER} itself.  By assuming a value of $H_0$ then the mean supernova absolute magnitude $\overline{M_B}$ can be calculated from the fitted value of $\mathcal{M}$.  

The code can be exploited to fit alternatives to stretch and colour, e.g.\ equivalent width.  It was also adapted to fit for a fourth variable which has a linear contribution to the distance modulus.  For instance, the fit could be performed using equivalent width as a third possible calibrator:
\begin{eqnarray}
\mu & = &  m_B - \mathcal{M} + \alpha(s-1) - \beta c +\gamma EW\label{eq:cosfitter-gamma}
\end{eqnarray}
  
\noindent The code also enables evaluation of $\mathcal{M}$, $\alpha$, $\beta$ and $\gamma$ within a fixed cosmology.

\subsubsection{Cosmological Parameters Derived From Photometric Data}\label{sec:cosmology-whole-data}

In this section, we fit the cosmology to the data sample in order to have a baseline to compare against when testing for spectroscopic calibrators.

As stated in Section \ref{sec:low-z-cosmo}, the low-z sample used for the spectroscopy study is not sufficient for cosmology-fitting.  This is because most of the objects are not in the Hubble flow and their peculiar velocities affect their redshifts and hence their distances are very poorly constrained.  To account for this, we use the low-z sample from \citet{Astier06} with photometry fitted using SiFTO, but removing the objects marked with $\ast$ in Table \ref{low-z_refs}, to keep the cosmology and spectroscopy samples entirely separate.  We also make stretch and colour cuts on the sample to remove any possible outliers which may not be fully described by the width-brightness or colour relations.  The cuts are $0.75\leq s \leq 1.25$ and $-0.25 \leq c \leq 0.25$ which leaves a sample of 32 low-z objects, 75 Gemini objects and 66 VLT objects.  This is similar to the cuts that would be applied for a full cosmology analysis.

For the first fit no stretch or colour correction was applied to the data.  This is designated Fit A in Table \ref{tab:cosmo-fits}.  The fit shows that the intrinsic dispersion ($\sigma_{int}$) must be set to be quite high (0.255\,mag) in order to obtain a $\chi^2_r \approx 1$.  The value of $\Omega_M = 0.462$ is higher than current measurements which employ stretch and colour correction (e.g.\ \citet{Astier06}), but this is to be expected because at higher redshift the supernovae tend to be brighter and bluer which causes the fit to favour a higher matter fraction.

In Fit B, stretch and colour corrections are applied and the resulting best-fit value of $\Omega_M$ is much lower.  The value is within 2 standard deviations of that published in \citet{Astier06}.  The latest full analysis of SNLS data finds, $\Omega_M = 0.268 \pm 0.020$ (Sullivan et al. (in prep.)), and so the value calculated in Fit B is within 3 standard deviations.  The value of $\alpha$ is very similar to the one achieved from the full analysis, and  $\beta$ is very different from the value published in \citet{Astier06}, but in close agreement with the latest value from the SNLS analysis ($\beta = 2.9$).  This is due to the fact that the treatment of the $\beta$ parameter in the fitting methods has changed since \citet{Astier06}.  Fit B shows a reduced RMS compared to Fit A which shows that the calibration applied using stretch and colour corrections are important.  Applying these corrections also reduces the intrinsic scatter required for a good fit.  It is a result similar to this which is the goal of this part of the study.

\begin{table*}
\begin{minipage}{15cm}
\caption[Cosmology fits]{Cosmology fits table.  The fit from \citet{Astier06} is included for reference.  Fits C, D and E were made without the low-z sample and fixing the cosmology to the values from Fit B.  Fit F was made without the use of the low-z sample, but the cosmology was not fixed.  Parameters which were fixed are marked with a *.  The quoted values of $\sigma_{int}$ are those used to obtain the best fits.}
\label{tab:cosmo-fits}
\centering
\begin{tabular}{c c c c r@{$\pm$}l r@{$\pm$}l r@{$\pm$}l r@{$\pm$}l r@{$\pm$}l}
\hline
Fit & Extra Calibrator & $\sigma_{int}$ & RMS & \multicolumn{2}{c}{$\Omega_M$} & \multicolumn{2}{c}{$\alpha$} & \multicolumn{2}{c}{$\beta$}  & \multicolumn{2}{c}{$\gamma$}  & \multicolumn{2}{c}{$\mathcal{M}$} \\
&& (mag) &(mag) & \multicolumn{2}{c}{}& \multicolumn{2}{c}{}& \multicolumn{2}{c}{}&\multicolumn{2}{c}{}&\multicolumn{2}{c}{(mag)} \\
\hline
Astier & -- & -- & -- & 0.263 & 0.053 & 1.52 & 0.14 & 1.57 & 0.15 & \multicolumn{2}{c}{--} & 23.85 & 0.03 \\
A& -- & 0.255 &  0.260 & 0.462 & 0.065 & \multicolumn{2}{c}{--} &  \multicolumn{2}{c}{--} &  \multicolumn{2}{c}{--} & 24.091 & 0.045\\
B & -- & 0.12 &  0.181 & 0.197 & 0.031 & 1.28 & 0.14 & 3.21 & 0.19 & \multicolumn{2}{c}{--} & 23.999 & 0.025 \\
C & \ion{Ca}{ii} H\&K & 0.10 & 0.176 & \multicolumn{2}{c}{0.197*} & \multicolumn{2}{c}{1.28*} & \multicolumn{2}{c}{3.21*} & 0.001 & 0.001 & 24.050 & 0.028 \\
D & \ion{Si}{ii} & 0.10 & 0.183 & \multicolumn{2}{c}{0.197*} & \multicolumn{2}{c}{1.28*} & \multicolumn{2}{c}{3.21*} & 0.006 & 0.002 & 24.084 & 0.030 \\
E & \ion{Mg}{ii} & 0.10 & 0.177 & \multicolumn{2}{c}{0.197*} & \multicolumn{2}{c}{1.28*} & \multicolumn{2}{c}{3.21*} & 0.000 & 0.001 & 24.044 & 0.039 \\
F & \ion{Si}{ii} & 0.10 & 0.193 & 0.214 & 0.053 & 1.50 & 0.18 & 3.49 & 0.24 & 0.007 & 0.002 & 24.123 & 0.054 \\
\hline
\end{tabular}
\end{minipage}
\end{table*}

\begin{figure}
  \centering
  \includegraphics[width=8cm]{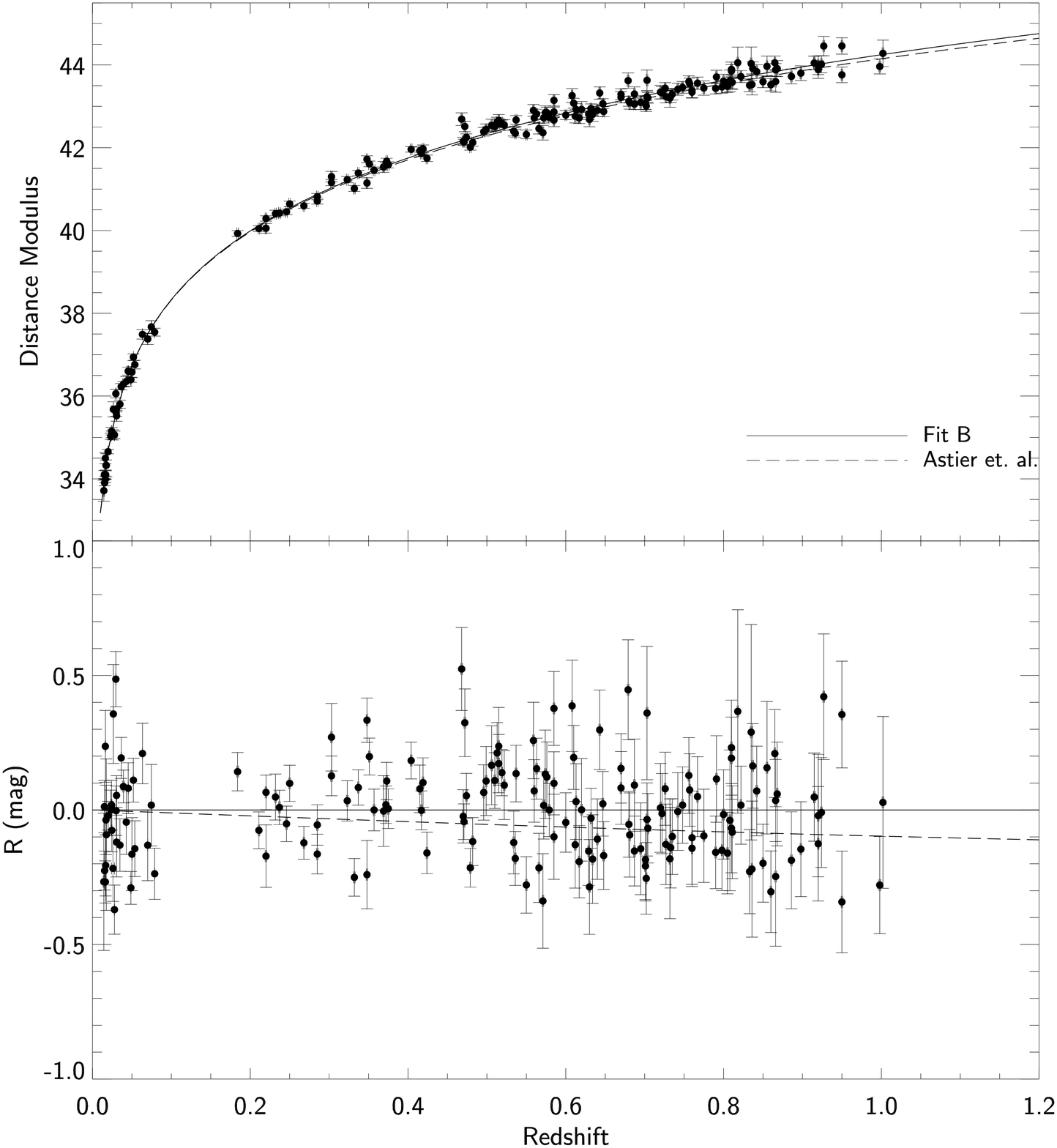}
  \caption{Hubble diagram for Fit B (solid line) which uses stretch and colour corrections, compared to the cosmology fit from \citet[][dashed]{Astier06}.  The data used are the low-z cosmology sample from \citet{Astier06}, and all the VLT and Gemini objects from the samples defined in Section \ref{sec:observations}.  Stretch and colour cuts are applied.  In both plots, the black points are calculated using the distance moduli from the apparent magnitude of the supernova fitted by SiFTO and the cosmological parameters from Fit B, correcting for stretch and colour.  The lines represent the distance moduli calculated from Equation (\ref{eq:cosmo_fit}).  The bottom plot shows the residuals from Fit B (see Section \ref{sec:hubble-res-corr} for definition) compared to the Astier cosmology.}
  \label{fig:hubble-diag}
\end{figure}

\subsubsection{Hubble Residual Correlations}\label{sec:hubble-res-corr}

Now the cosmology has been fitted to the data, residuals from the fit can be plotted against spectral features and possible alternative calibrators found.  The Hubble residuals, $R$, as shown in Figure \ref{fig:hubble-diag}, are defined as
\begin{eqnarray}
R & = & m_B - m_{\textnormal{\footnotesize{cosmo}}}(z,\Omega_M)
\end{eqnarray}

\noindent where $m_B$ is the observed B-band apparent peak magnitude and $m_{\textnormal{\footnotesize{cosmo}}}$ is derived from the cosmological parameters.  The assumed cosmology is that measured using Fit A i.e.\ $\Omega_M = 0.462$ in a flat universe.  The correlations with the photometric quantities stretch and colour are also presented as a guide as, naturally, strong correlations are expected for these variables.  As in Section \ref{sec:corr-spec-phot}, we use only the high-z EW measurements from the least noisy 75\% of the sample.  

\begin{figure*}
  \centering
  \includegraphics[width=12cm]{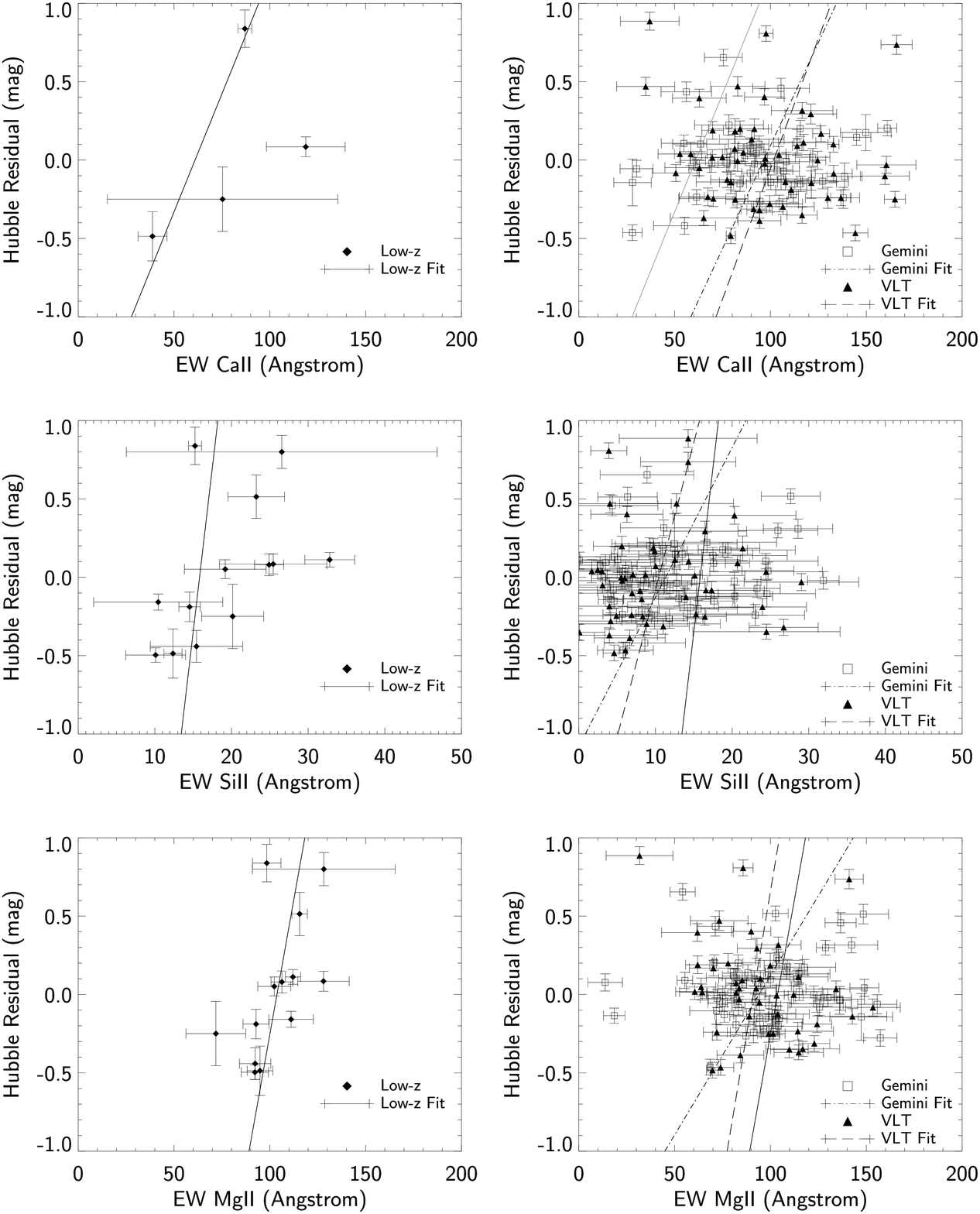}
  \caption[Hubble Residual - EW Plots]{Hubble Residual - EW Plots for  the low-z (left column) and the high-z (right column) objects.  The  best-fits to the lines are also shown.  Solid line -- low-z; dot-dashed -- Gemini; dashed -- VLT.}
  \label{fig:res-ew-plot}
\end{figure*}

For the low-z sample, the method of calculating the observed distance modulus is slightly different as not all the objects are in the Hubble flow.  For five very nearby objects (see Table \ref{lowz_phot}) which are not in the Hubble flow we use distance moduli measured from their host galaxies either using Cepheid variable stars \citep{Saha:2006p2423} or galaxy surface brightness fluctuations \citep{Ajhar:2001p2373}.  These values minus the value of $\overline{M_B}$ determined from the best-fit value of $\mathcal{M}$ from Fit A and assuming $H_0 = 70$kms$^{-1}$Mpc$^{-1}$, are used as $m_{\textnormal{\footnotesize{cosmo}}}$.  For 5 other objects, they are situated in host galaxies with $z \geq 0.02$ and can be assumed to be in the Hubble flow and so the same method as the high-z sample to calculate $m_{\textnormal{\footnotesize{cosmo}}}$ is used.  Plots of the low-z and high-z data are shown in Figure \ref{fig:res-ew-plot}.  The correlations are calculated and tabulated in the same way as Table \ref{tab:phot_corr} and presented in Table \ref{tab:res-corr}.

\begin{table}
\caption[Hubble residual correlations]{Correlations between photometric and spectral properties and the Hubble residuals.  There are not enough low-z objects with EW \ion{Ca}{ii} H\&K\ measurements to measure a correlation.}
  \label{tab:res-corr}
  \begin{tabular}{l l c c c}
    \hline
Variable & Sample & $\rho$ & $\sigma$ from null \\
\hline
Stretch & Low-z & -0.45 & 1.62 \\
& Gemini & -0.45 & 4.04 \\
& VLT & -0.51 & 4.10 \\
Colour & Low-z & 0.79 & -2.73 \\
& Gemini & 0.51 & -4.48 \\
& VLT & 0.58 & -4.67 \\
EW \ion{Ca}{ii} H\&K & Low-z & -- & -- \\
& Gemini & 0.09 & -0.60 \\
& VLT & -0.13 & 0.97 \\
EW \ion{Si}{ii} & Low-z & 0.72 & -2.59 \\
& Gemini & 0.13 & -0.94 \\
& VLT & 0.10 & -0.71 \\
EW \ion{Mg}{ii} & Low-z & 0.78 & -2.80 \\
& Gemini & 0.05 & -0.38 \\
& VLT & -0.29 & 1.87 \\
\hline
  \end{tabular}
  \end{table}

The strong correlations between stretch and colour and Hubble residual in Table \ref{tab:res-corr} show that the current low- and high-z spectral features should not be used as direct replacements for either of the photometric properties as the relations are much weaker.  The strongest correlation between one of the high-z spectral features and the residuals is for EW \ion{Mg}{ii}\ for the VLT sample, but this is not seen in the Gemini data - the correlation goes in the opposite direction.  However, the linear fits to the data in Figure \ref{fig:res-ew-plot} where points are weighted according to their errors show the best-fit lines all have positive gradients, implying that the negative correlation coefficient seen in the VLT data is spurious.  By comparing Table \ref{tab:res-corr} and Figure \ref{fig:res-ew-plot}, we can see that even applying a cut to remove the most noisy points from the high-z data, the correlation coefficient does not necessarily represent the best way to look for a relation between two quantities as it is affected by outliers.

Instead of looking to replace stretch or colour with one of the EW measurements, it might be more instructive to look for a correlation which exists between the EW and the Hubble residual after the application of stretch and colour correction.  To do this, we fix the cosmology and values of $\alpha$ and $\beta$ to $\Omega_M = 0.197$, $\alpha = 1.28$ and $\beta = 3.21$ as in Fit B, and use the adapted version of \texttt{COSFITTER} to fit Equation (\ref{eq:cosfitter-gamma}).

To see just how great an improvement can be made, the instrinsic dispersion was allowed to vary and Equation (\ref{eq:cosfitter-gamma}) fitted allowing $\mathcal{M}$, $\sigma_{int}$ and $\gamma$ as the free parameters.  The cosmology must be fixed as the objects in the low-z cosmology sample do not have accompanying spectra and so could not be used to anchor the Hubble diagram for these fits.  The results of this process are given by Fits C, D and E in Table \ref{tab:cosmo-fits}.

The results in Table \ref{tab:cosmo-fits} show that by adding in a third calibrator, the intrinsic dispersion necessary for a good fit can be decreased, as expected.  The \ion{Ca}{ii} H\&K\ and \ion{Mg}{ii}\ features show values of $\gamma$ consistent with 0 and so they are ruled out as possible calibrators.  However, for the \ion{Si}{ii}\ feature $\gamma$ is three standard deviations from zero and produces a value of $\mathcal{M}$\ consistent with the one in Fit B.  The RMS is comparable to the one obtained in Fit B as well.

In order to test whether this result is meaningful, the constraints on $\alpha$, $\beta$ and $\Omega_M$ were relaxed so they were free parameters.  Even without a low-z sample to anchor the Hubble diagram at low redshift, the fit using EW \ion{Si}{ii}\ as a third calibrator still gave reasonable values for the parameters, denoted as Fit F in Table \ref{tab:cosmo-fits}.  However, the RMS scatter was increased slightly to 0.193\,mag.

From this is is clear that \ion{Si}{ii}\ could have a role to play in SNe Ia calibration for cosmology.  However, with the large errors on the high-z EW measurements, the RMS scatter around the best-fit cosmology do not improve and it is not possible to draw any firm conclusions as to the value of $\gamma$.  Instead, in Section \ref{sec:spectral-stretch}, we demonstrate that the relationship between EW \ion{Si}{ii}\ and stretch in the low-z sample can be exploited to reduce the residuals on the Hubble diagram by reducing the scatter on $M_B$ measurements.

\subsection{EW \ion{Si}{ii}\ as a Cosmological Calibrator}\label{sec:spectral-stretch}

Drawing on the fact that we have seen that EW \ion{Si}{ii}\ and photometric stretch are correlated, and that there is a strong correlation between the Hubble residuals and the low-z EW \ion{Si}{ii}, we will explore the possible use of EW \ion{Si}{ii}\ as a calibrator for the cosmology in place of stretch for the low-z sample.  Here the errors on the EW are smaller so it is possible to be more definite about the potential of the calibrator.

\begin{figure}
  \centering
  \includegraphics[width=8cm]{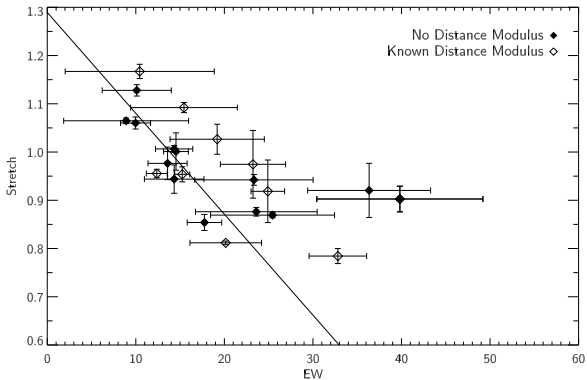}
  \caption[Refitting spectral stretch]{The two sub-samples of the low-z data are shown.  Those with closed circles do not have known distance moduli and are used to define the spectral-stretch relations (black line).  Those with open circles will be used to test the relation and are not included in the fit.}
  \label{fig:spec-stretch-sample}
\end{figure}

\begin{figure*}
  \centering
  \includegraphics[width=12cm]{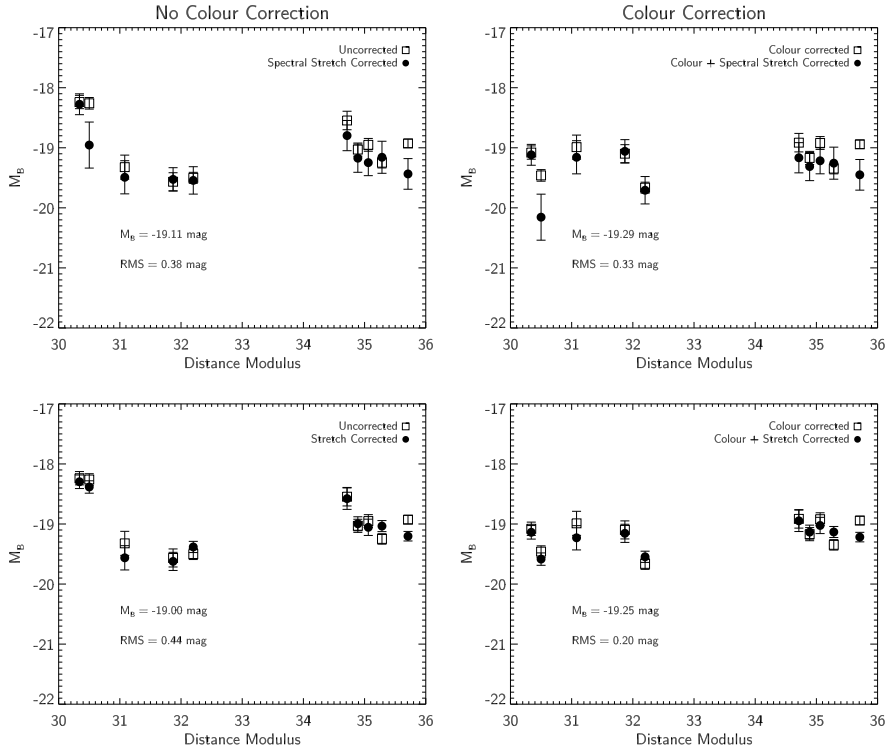}
  \caption[Spectral stretch in use for low-z data]{Comparing spectral stretch correction to photometric stretch correction with and without colour correction.  The value of $\overline{M_B}$ given on the plot is the weighted mean of the points after the stretch correction.  Without colour correction (left column) the RMS before any stretch correction is 0.45\,mag and the mean absolute magnitude $\overline{M_B}=-18.94$.  With colour correction but no stretch correction (right column), $\overline{M_B} = -19.19$\,mag and the RMS is 0.24\,mag}
  \label{fig:spec-stretch-comp}
\end{figure*}

We divide the objects in the low-z spectroscopy sample which have photometry into two independent sub-sets in order to have a sample to define the relation and a separate one to test its use.  The first sample (i) contains the objects with no reliable absolute distance measures, i.e.\ those without Cepheid or surface brightness fluctuation distances and not in the Hubble flow (those marked with a $^{\dagger}$ only in Table \ref{low-z_refs}); and the second sample (ii) containing the rest (those marked as $^{\dagger}$$^{\ddagger}$).  The two samples are illustrated in Figure \ref{fig:spec-stretch-sample} as open circles for the objects with distance measurements and closed circles for those objects without.  Using sample (i), the relation between stretch and EW \ion{Si}{ii}\ is refitted to derive an expression for ``spectral stretch'', $s_{\textnormal{\footnotesize{spec}}}$, using \texttt{LINFITEX}. The best-fit is given by
\begin{eqnarray}
  \label{eq:spectral-stretch}
  s_{\textnormal{\footnotesize{spec}}} & = & (-0.021\pm 0.005)EW_{\textnormal{\footnotesize{SiII}}} +(1.29 \pm 0.07)
\end{eqnarray}

\noindent where $\chi^2_r = 1.55$.  This fit is different from the ones in Equations (\ref{eq:low-z_ew_s}) and (\ref{eq:high-z_ew_s}) as fewer objects are used.  The remaining 10 low-z objects were corrected for stretch and also spectral stretch according to the relation
\begin{eqnarray*}
  M_B & =& m_B - \mu + \alpha(s - 1)
\end{eqnarray*}

\noindent where $s$ can either be the normal photometric stretch or spectral stretch as fitted in Equation (\ref{eq:spectral-stretch}).  We then plot $M_B$ against $\mu$ for illustration purposes, and measure the weighted-mean, $\overline{M_B}$, and the RMS scatter about the mean.  This is shown in Figure \ref{fig:spec-stretch-comp}.  The values of $\overline{M_B}$ and the RMS are shown on each plot and summarised in Table \ref{tab:spec-stretch-fits}.  For the purposes of this exercise we take the values of $\Omega_M$ and $\alpha$ from Fit B.  The right-hand column of Figure \ref{fig:spec-stretch-comp} shows the effect of also adding a correction for supernova colour. 

\begin{table}
\caption[$\overline{M_B}$ and the RMS scatter results]{A summary of the vaules of $\overline{M_B}$ and the RMS scatter about the mean for the fits in Figure \ref{fig:spec-stretch-comp}.}
  \label{tab:spec-stretch-fits}
  \begin{tabular}{l c c}
    \hline\hline
Corrections Applied & $\overline{M_B}$ & RMS \\
 & (mag) & (mag) \\
\hline
None & -18.94 & 0.45 \\
Spectral stretch & -19.11 & 0.38 \\
Photometric stretch & -19.00 & 0.44 \\
Colour & -19.19 & 0.24 \\
Colour + Spectral stretch & -19.29 & 0.33 \\
Colour + Photometric stretch & -19.25 & 0.20 \\
\hline   
    \end{tabular}
 \end{table} 

Table \ref{tab:spec-stretch-fits} shows that the lowest RMS is obtained when using photometric stretch and colour together, but when no colour correction is made, spectral stretch correction produces a lower RMS than photometric stretch.

It is not possible to apply this correction to the high-z supernovae and see any improvement in the data due to the large errors on EW \ion{Si}{ii}, but in the future with a sample of well-measured equivalent widths it is possible that spectral stretch could replace photometric stretch.  This would mean that once the supernova was identified fewer photometry points would be needed to fit the light curve, only the B-band magnitude at maximum and the redshift would be needed: supernova colour could be measured from synthetic photometry of a spectrum obtained around maximum light.  However, the stretch-EW \ion{Si}{ii}\ relation needs better calibration first using, for example, a large sample of low-z supernovae either with known host galaxy distances or within the Hubble flow.  As seen from the $\gamma$ parameter analysis in the noisy high-z data, as shown in Table \ref{tab:cosmo-fits}, if corrections for photometric stretch, colour and spectral stretch/EW \ion{Si}{ii}\ can all be combined then there is the potential that calibration could be improved.

\section{Discussion}

\subsection{Tests for Evolution}

The comparison between the high-z and low-z trends seen in Figure \ref{fig:highz_trends} shows that for the \ion{Ca}{ii} H\&K\ and \ion{Mg}{ii}\ features, the high-z sample is consistent with the low-z trends, but the scatter on the data is quite large.  For the \ion{Si}{ii}\ feature, a large number of low equivalent width measurements are seen in the high-z data and the high-z trend is consistently below the low-z one.  This can be explained as a change in the demographics.  In Section \ref{sec:ew-stretch} we see that light curve stretch is correlated with EW \ion{Si}{ii}.  At higher redshift, due to the ``A+B'' model we expect more of the prompt component which are brighter and bluer as more $^{56}$Ni is thought to be produced during the explosion.  This would produce less IME which is reflected in the lower EW \ion{Si}{ii}.

To the level of current precision, the variation in mean EW \ion{Si}{ii}\ between high and low redshift can be understood in terms of the combination of a change of demographics within a two-component SN model and an observed correlation between photometric stretch and EW \ion{Si}{ii}.  Therefore, we see no evidence for an additional bias that is not already accounted for by the standard photometric stretch and colour corrections.  However, in future experiments which may go to even higher redshift to look for $w = w(z)$ (models where dark energy is not a cosmological constant) will need a higher level of precision.   Future cosmological studies observing to $z=1.7$ will need a precision of 2\% in the standardisation of supernova magnitudes if they are to keep systematic errors smaller than statistical errors, and it not yet clear whether the current methods of calibration will be suitable for this \citep{Howell:2007p1767}.  We also know that the width-brightness relation may not be valid for supernovae with large stretch such as SN\,2001ay-like objects.

It will only be possible to make a strong statement regarding the evolution of SNe Ia when there is a large, unbiased sample of low-z spectra.  In Figure \ref{fig:stretch_hist} we have only 3 low-z objects in the $s=1$ bin, compared to 14 in the VLT sample and 17 in the Gemini sample.  Future surveys such as Palomar Transient Factory \citep{Law:2009p2424} and SkyMapper \citep{2007PASA...24....1K} will largely address this problem.

\subsection{Implications for Cosmology}\label{sec:impl-cosm}

The origin of the width-luminosity relation is well-understood to first order: a variation in the amount of $^{56}$Ni alters the amount of gamma-rays emitted due to radioactive decay to $^{56}$Co and $^{56}$Fe which are absorbed and thermally reemitted by the expanding ejecta \citep{Colgate:1969p2731,Arnett:1979p2732}.  The more $^{56}$Ni produced, the wider the light curve \citep{Arnett:1982p2622}.  This also means the production of less IME, which is reflected in Figure \ref{fig:ew_strech} \citep[see also Figure 2 in][]{Sullivan:2009p1769}.  What causes the varying amount of $^{56}$Ni is not understood although it may be due in part to the way that burning of carbon and oxygen occurs in the supernova, or the metallicity of the progenitor.  The relation between stretch and EW \ion{Si}{ii}\ from Figure \ref{fig:ew_strech} is exploited in Section \ref{sec:spectral-stretch} for the low-z dataset to reduce the scatter of the objects on the Hubble plot by correcting $\overline{M_B}$ using spectral stretch.

The origins of supernova colour are more complex than stretch and not as well-understood.  At high redshift, colours may be affected by the environment surrounding the supernova simply due to the fact that it is not possible to separate the supernova light from the surrounding interstellar and circumstellar medium \citep[see][for a discussion of the effect of dust and the colour parameter on cosmology]{Conley:2007p93}.  In addition those effects, from studies in the local universe, there is known to be an intrinsic difference in supernova colours as well \citep{P93}.  In Table \ref{tab:res-corr} we see that supernova colour has a stronger correlation with Hubble residual than stretch, implying that it is more important for calibration purposes.

Another application for spectral stretch is that it may allow more supernovae to be added to the low-z sample for cosmology.  Currently, most SNe Ia have either good photometric coverage or good spectroscopic coverage, but not both.  For instance, in the spectroscopic sample used for this study, only 5 objects overlap with the \citet{Astier06} cosmology sample.  By taking the objects that have good spectroscopic coverage, instead of using photometrically-derived properties to calibrate them, spectral stretch could be used.  This would also require a B-band magnitude at maximum, or interpolated from other data, and a known distance modulus or redshift.  There is not yet a spectral replacement for colour, but the EW \ion{Mg}{ii}-colour relation seen in the low-z sample could be exploited for this purpose when the relationship is better quantified.

For future studies out to $z=1.7$, \citet{Howell:2007p1767} have shown that supernova calibration needs to be better than 2\% in magnitude in order to keep systematic errors below the level of statistical errors.  By introducing a third calibrator based on EW \ion{Si}{ii} into the cosmology in Section \ref{sec:hubble-res-corr} we find a non-zero value for the constant $\gamma$ whilst achieving a similar RMS scatter around the best-fit cosmology as when using standard photometric stretch and colour corrections.  The estimate of $\Omega_M$ is also closer to the current value from the full Bayesian analysis of SNLS data.  In the future with higher signal-to-noise spectra, the EW \ion{Si}{ii}\ may provide a further method of cosmological calibration.  At the moment, the best fit is not approaching the 2\% level.  This can be done with more observations with high signal-to-noise spectra and good photometric follow-up of discoveries. 

\subsection{Improvements}

\subsubsection{Galaxy Subtraction Method}

The methods presented in Section \ref{sec:methods-measurements} detail the way host galaxy contamination is removed from the spectra.  This is made somewhat more difficult for the Gemini objects which were not observed at parallactic angle and hence the spectra suffer from slit-losses.  The objects from the VLT sample were observed using an ADC and so this is less of a problem.  Extraction of spectra from the host galaxy and the supernovae separately was not usually possible due to the small angular size of the galaxy on the sky.  In order to subtract off host galaxy light, we must make an assumption about the nature of the background light, and we assume that it is a flat background in the slit.  This assumption is reasonable in most cases, but it can fail when the supernovae are situated in the very centre of the host, or on the outskirts. 

Another limitation is the number of galaxy template spectra available for the subtraction.  In order to make a better subtraction, ideally we would reobserve the host galaxies after the supernova light has faded with an identical set-up.  However, there will be slight variations in the seeing and airmass so the correction would still not be perfect.  Another method would be to use the CFHT imaging directly to aid subtraction.  Photometry-Assisted Spectral Extraction \citep[PHASE;][]{Baumont:2008p1659} has been developed and used on SNLS-VLT spectra to extract the supernova spectrum with an extraction window determined from the position of the target in the imaging data \citep{Balland:2009p2990}.  A $\chi^2$-minimisation is used to evaluate the contribution of galaxy and supernova in each pixel to produce a spectrum composed of only supernova light.  This method produces clean spectra for approximately 70\% of observations, but when the supernova is close to the centre of the host, or the fractional contamination of the the spectrum is high, the method cannot separate the two components.  These are two of the occasions when the multicolour host galaxy subtraction method also fails so PHASE could be used as an alternative to the method used in this study, but not as a replacement.

\subsubsection{Low-z Trends}

In comparison to the large number of high-z objects included in this analysis, there are only 14 objects used to make the low-z trends for the \ion{Ca}{ii} H\&K\ feature, and 25 for the \ion{Si}{ii}\ and \ion{Mg}{ii}\ features.  The sample is biased towards supernovae in bright, elliptical hosts which will tend to host redder supernovae with lower stretch (the delayed component from the two-component model).  This can be seen in Figure \ref{fig:stretch_hist}.  The sample is also drawn from many sources where the photometric coverage is different.  To do a study of this kind in the future to an increased level of precision we would require a large homogeneous sample of low-z spectra and photometry.

The trends shown in Figure \ref{fig:lowz_trends} reflect the diversity seen in the low-z objects, an advantage of using the EW measurements directly; however, there is a much larger scatter on the high-z data.  Some of this scatter will be due to a lower signal-to-noise ratio in the high-z data, but some might be intrinsic to the supernova populations that we do not observe in the low-z sample due to its limited size.  The lack of a well-calibrated set of supernova spectra which cover the blue \ion{Ca}{ii} H\&K\ feature limits this part of the study particularly.  There might be additional subtypes in the supernovae population, or transitional objects, which have not yet been identified.  It is not yet clear whether there is a continual class of objects from underluminous SN\,1991bg-like supernovae \citep{F1991bg} to overluminous SN\,1991T-like objects \citep{F1991T}, or whether the progenitors are different \citep[for example][]{Benetti:2005p1566,Branch:2006p1746}.  Indeed, unusual objects objects falling outside these classes are now observed.

One of the most interesting results to come out of the low-z trend of the \ion{Mg}{ii}\ feature is the fact that this feature seems to be constant with time and with very little scatter between objects.  The feature is not made up of solely \ion{Mg}{ii}, but also contains \ion{Fe}{ii} and \ion{Fe}{iii} which are thought to be sensitive to temperature variations in the supernova and drive the relation between $\mathcal{R}_{642/443}$ and colour described in \citet{Bailey:2009p2202}.  Furthermore the recombination from \ion{Fe}{iii} to \ion{Fe}{ii} changes the opacity of the ejecta and causes the secondary maximum in the supernova lightcurves \citep{Kasen:2006p909}.  We see weak correlations between EW \ion{Mg}{ii}\ and colour in Figure \ref{fig:ew_colour}, but there is a large variation in the relations between the low- and high-z data.  As there is a larger range of colours observed in the low-z sample compared to the high-z sample, we would expect the any diversity to be seen in the low-z trend.  A larger data-set at low-z would enable this to be investigated further.

\section{Conclusions}

The aims of this study were to observe spectroscopically potential Type Ia supernovae and confirm their type and redshift in order for them to be used for cosmology; to study the spectra of the objects themselves to verify their use as cosmological standard candles; and to investigate new ways the spectral information can be used to calibrate the objects for cosmology.

The first aim is easily satisfied.  Here we present observations of 95 targets, 68 of which were confirmed as SNe Ia or probable SNe Ia events.  8 events were confirmed as being core-collapse in nature, leaving 19 observations as probably not Ia in nature or of unknown type.  The spectra are presented in Appendix \ref{sec:high-z-spectra}.  The spectroscopic observations of such a large dataset, especially when combined with other SNLS observations, provide an opportunity to study supernovae covering billions of years of cosmic look-back time.  SNLS provides not only spectroscopic coverage, but complete light curves in four bands to fit the photometric properties such as apparent magnitude, stretch and colour; and deep images of the host galaxy in five bands.  In total, spectral analysis was possible for 144 objects drawn from SNLS survey work at the Gemini and VLT telescopes, covering a redshift range of $0.184 \leq z \leq 1.002$.

The second aim was investigated in Section \ref{sec:stat-comp} where high-z equivalent width measurements of the \ion{Ca}{ii} H\&K, \ion{Si}{ii}\ and \ion{Mg}{ii}\ spectral features are compared to trends seen in a low-z sample.  We find that for the \ion{Ca}{ii} H\&K\ feature, the scatter on the measurements at low- and high-z is large, but the features appear to be consistent with each other.  The \ion{Mg}{ii}\ feature has a much lower scatter at low redshift than with high-z measurements and the statistical comparison between the two sample yields $\chi^2_r >1$; however, there does not appear to be a systematic shift in the mean value of EW \ion{Mg}{ii}\ between the two samples. The \ion{Si}{ii}\ feature has a lower trend at high redshift compared to low redshift.  This is explained by the fact that EW \ion{Si}{ii}\ is seen to be anticorrelated with photometric stretch.  At high redshift we see more of the prompt component from the ``A+B'' model which are brighter and have a broader light curve due to greater $^{56}$Ni production.  These supernovae have larger stretches and hence smaller EW \ion{Si}{ii}.  We also observe different correlations between EW \ion{Mg}{ii}\ and photometric colour for the low-z and high-z samples.  This difference is probably due to a difference in sample selection and the colour-cut introduced in the high-z target selection in particular.  The origin of any possible relationship between colour and EW \ion{Mg}{ii}\ is not understood, but may be connected to the balance between \ion{Fe}{ii}/\ion{Fe}{iii} which are also present in the feature.  It does not appear to systematically affect the comparison of EW trends despite the high-z sample being bluer than the low-z sample.  We find nothing within the dataset to introduce doubt as to the use of SNe Ia as standardisable candles for cosmology.

The final aim, to exploit spectral properties to aid cosmological calibration is studied in Section \ref{sec:cosmological-uses}.  We find that the spectral features do not correlate with Hubble diagram residuals as strongly as photometric stretch or colour, but significant relationships are seen between the residuals and EW \ion{Si}{ii}\ and EW \ion{Mg}{ii}.  This is to be expected as we see relations between EW \ion{Si}{ii}\ and EW \ion{Mg}{ii}, and stretch and colour, respectively.  The high-z data are too noisy to be used to produce a spectroscopically-calibrated Hubble diagram.  However, we demonstrate that such a calibration may be possible by defining a spectral stretch from a sub-set of the low-z sample based on the observed relationship between EW \ion{Si}{ii}\ and photometric stretch.  This is then applied to other low-z supernovae with known distance moduli or which are within the Hubble flow and we demonstrate that this reduces the scatter on the measurement of the absolute magnitude.  The spectral stretch correction reduces the RMS on the measurement of $\overline{M_B}$ to comparable, in fact slightly smaller value than that obtained by photometric stretch correction, although this is demonstrated for a small sample of objects.

\section*{Acknowledgements}
We gratefully acknowledge the Gemini queue observers and support staff for taking all the SNLS Gemini data.  This work is based on observations obtained at the Gemini Observatory, which is operated by the Association of Universities for Research in Astronomy, Inc., under a cooperative agreement with the NSF on behalf of the Gemini partnership: the National Science Foundation (United States), the Science and Technology Facilities Council (United Kingdom), the National Research Council (Canada), CONICYT (Chile), the Australian Research Council (Australia), Ministério da Ciência e Tecnologia (Brazil) and Ministerio de Ciencia, Tecnología e Innovación Productiva  (Argentina).  This work is also based in part on observations obtained with MegaPrime/MegaCam, a joint project of CFHT and CEA/DAPNIA, at the CFHT which is operated by the National Research Council (NRC) of Canada, the Institut National des Sciences de l'Universe of the Centre National de la Recherche Scientifique (CNRS) of France, and the University of Hawaii.  The work also makes use of SUSPECT, the Online Supernova Spectrum Archive.  ESW acknowledges the support of the Science and Technology Facilities Council and ASI contract I/016/07/0; TJB acknowledges the support of the Alberta-Bart Holaday scholarship; MS acknowledges support from th Royal Society.

\bibliographystyle{mn2e}

\appendix

\section{Data Tables}

\begin{table*}
\begin{minipage}{150mm}
\caption{Table of observations. $^1$ Measured using GEMSEEING, $^2$ Measured using IMEXAM, $^3$ R150 observation, $^4$ Second night of data only used $^5$ First night of data only used}\label{obs_tab}
\begin{tabular}{c c c c c c c c}
\hline
SN & RA & Dec &UT Date& Exposure Time &Mode & $\lambda_{cent}$ & Seeing \footnotemark[1]\\
(SNLS name) & (J2000) & (J2000) & & (s) & &(nm) & ($''$)\\
\hline
05D4bo & 22:16:44.630 & -17:20:12.37 & 2005-07-14  & 2700 & N\&S & 720 & 0.43\\
06D1bt & 02:24:59:57 & -04:01:05.43 & 2006-08-31 & 5400 & N\&S & 720 & 0.45 \\
06D1bz & 02:25:07.62 & -04:42:06.58 & 2006-08-22 & 5400 & N\&S & 720 & 0.52 \\
06D1ck & 02:26:25.37 & -04:40:43.80 & 2006-08-31 & 3600 & N\&S & 720 & 0.40 \\
 &  &  &  2006-09-01 & 1800 & N\&S & 720 & 1.03\\
06D1dt & 02:27:15.20 & -04:35:44.88 & 2006-09-23 & 3600 & C & 680 & 0.51 \\
06D1gg & 02:26:28.56 & -04:44:54.29 & 2006-09-23 & 2400 & C  & 680 & 0.66\\
06D1gu & 02:24:33.91 & -04:17:37.80 & 2006-11-18 & 5400 & N\&S & 720 & 0.37\\
06D1hl & 02:26:42.95 & -04:18:22.34 & 2006-12-11 & 5400 & N\&S & 720 & 0.36\\
06D1jp & 02:25:48.07 & -04:17:56.02 & 2006-12-22 & 7200 & N\&S & 720 & 0.61\\
06D2iz & 10:00:34.76 & 02:19:20.17 & 2006-12-21 & 5400 & N\&S & 720 & 0.64\\
06D2ja & 10:00:33.51 & 02:28:01.01 & 2006-12-21 & 3600 & N\&S & 720 & 0.46 \\
06D2ji & 09:59:51.03 & 01:51:50.14 & 2006-12-19 & 7200 & N\&S & 720 & 0.66 \\
 &  &  &2006-12-20 & 1800 & N\&S & 720 & 0.38\\
06D2js & 10:02:15.70 & 02:35:06.62 & 2006-12-25 & 7200 & N\&S & 720 & 1.10\footnotemark[2] \\
06D2ju & 10:01:34.65 & 02:26:42.72 & 2006-12-24 & 5400 & N\&S & 720 & 0.83 \\
06D3ed & 14:18:47.82 & 52:45:23:28 & 2006-05-27 & 2400 & N\&S & 720 & 0.82 \\
06D3el & 14:17:01.06 & 52:13:56.84 & 2006-05-27 & 5400 & C & 680 & 0.72 \\
06D3et & 14:22:04.76 & 52:20:18.30 & 2006-05-31 & 5400 & N\&S & 720 & 0.56 \\
06D3fp & 14:20:18.62 & 52:53:49.62 & 2006-06-18 & 2600 & C & 680 & 0.55 \\
06D3fr & 14:19:20.24 & 52:31:50.11 & 2006-06-29 & 3600 & C & 680 & 1.07 \\
06D3fv & 14:21:03.31 & 53:00:52.77 & 2006-06-25 & 1800 & N\&S & 720 & 0.73 \\
 &  &  & 2006-06-29 & 5400 & N\&S & 720 & 0.65\\
06D3gg & 14:20:17.36 & 53:10:42.27 & 2007-07-20 & 3600 & N\&S & 720 &0.51 \\
06D3gh & 14:18:31.38 & 52:26:34.69 & 2006-06-30 & 5400 & N\&S & 720 & 0.41 \\
06D3gn & 14:17:44.61 & 52:21:40:42 & 2006-07-06 & 2700 & C & 680 & 0.40 \\
06D3gx & 14:17:03.36 & 52:56:21.82 & 2007-07-27 & 3600 & N\&S & 720 & 0.52 \\
06D3hd & 14:20:42.76 & 52:14:20.27 & 2006-07-18 & 3600 & C & 680 &0.72 \\
06D4dh & 22:14:19.70 & -17:35:06:23 & 2006-07-30 & 5400 & C & 680 & 0.51 \\
06D4dr & 22:16:46.00 & -17:20:45.49 & 2006-08-19 & 4800 & N\&S & 720 & 0.44 \\
06D4fc & 22:13:52.30 & -17:19:30.31 & 2006-09-23 & 5400 & N\&S & 720 & 0.39 \\
06D4fy & 22:14:06.22 & -17:34:19.34 & 2006-09-16 & 5400 & N\&S & 720 & 0.39 \\
06D4gn & 22:15:14.81 & -17:23:01.36 & 2006-09-22 & 7200 & N\&S & 720 & 0.42 \\
06D4hc & 22:15:40.52 & -18:09:08.71 & 2006-10-13 & 5400 & N\&S & 720 & 0.96 \\
07D1cx & 02:24:58.48 & -04:01:54.23 & 2007-11-10 & 3600 & N\&S & 720 & 0.44 \\
&&&&3600&  & 715\\
07D1ds & 02:25:24.44 & -04:18:49.29 & 2007-11-10 & 3600 & N\&S & 720 & 0.41 \\
&&&&3600&& 720\\
07D1dv & 02:26:51.27 & -04:47:18.29 & 2007-11-10 & 1800 & N\&S & 720 & 1.06 \\
&&&2007-11-11 & 1800 &&&0.51\\
&&&& 3600 &&715\\
07D1ea & 02:25:14:77 & -04:55:39.80 & 2007-11-13 & 3600 & N\&S & 720 & 0.45 \\
&&&& 3600 && 715\\
07D2bu & 09:59:27.590 & 02:33:03.48 & 2007-03-15 & 5400 &N\&S & 720 & 0.93 \\
&&& 2007-03-16 & 1800 & & & 0.63\\
&&& 2007-03-17 & 3600 & & & 0.77\\
07D2cy & 10:01:49.486  & 02:40:59.09  & 2007-03-22 & 3600 & N\&S & 720 & 0.82 \\
&&& 2007-03-22 & 5400 & & 715 & 0.51\\
07D2kc & 10:01:07.19 & 02:02:28.65 & 2007-12-20 & 900 & C & 680 & 0.67 \\
&&&& 1800&& 675\\
07D2ke & 10:01:04.73 & 01:59:31.81 & 2007-12-20 & 900 & C & 680 & 0.80 \\
&&&&900 && 675\\
07D2kh & 10:00:28.94 & 02:13:39.65 & 2008-01-04 & 3600 & N\&S & 720 & 0.63 \\
&&&& 1800 && 715\\
&&& 2008-01-06 & 1800 & && 0.32\\
07D2ki & 10:00:26.69 & 01:56:48.87 & 2008-01-05 & 3600 & N\&S & 720 & 0.54 \\
&&&& 1800 && 715\\
&&&2008-01-08 & 3600 &&&0.54\\
07D2kl & 09:59:16.95 & 02:31:28.13 & 2008-01-4 & 3600 & N\&S & 720 & 0.48 \\
&&&&3600 && 715\\
\end{tabular}
\end{minipage}
\end{table*}

\begin{table*}
\contcaption{}
\begin{minipage}{15cm}
\begin{tabular}{c c c c c c c c}
07D3ae & 14:17:33.70 & 52:31:27.17 & 2007-01-25 & 2400 & C & 680 & 0.57 \\
07D3af & 14:19:05.01 & 53:06:08.98 & 2007-01-27 & 2400 & C & 680 & 1.12\footnotemark[2] \\
07D3ai & 14:19:06.711 & 52:40:11.69 & 2007-02-14 & 3600 & C & 680 & 0.49 \\
07D3am & 14:22:36.55 & 52:20:26.26 & 2007-01-23 & 3600 & C & 680 & 0.70 \\
07D3ap & 14:18:10.26 & 52:23:07.38 & 2007-01-23 & 3600 & N\&S & 720 & 0.54 \\
07D3bb & 14:18:10.31 & 52:18:26.38 & 2007-01-25 & 7200 & N\&S & 720 & 0.46 \\
07D3bo & 14:20:58.408& 52:20:37.22  & 2007-02-14 & 5400 & N\&S & 720 & 0.54 \\
&&& 2007-02-15 & 1800 \\
07D3bp & 14:21:01.32 & 52:13:37.19 & 2007-01-27 & 7200 & B\&S & 720 & 0.75 \\
07D3bt & 14:21:25.719  & 52:58:13.67  & 2007-02-22 & 9000 & N\&S & 720 & 0.55 \\
07D3cn & 14:16:17.255 & 52:32:22.60  & 2007-03-24 & 3600 & N\&S & 720 & 0.50 \\
&&& 2007-03-25 & 5400 & & 715 & 0.41\\
07D3cr & 14:18:12.821 &52:19:04.57 & 2007-03-20 & 7200 & N\&S & 720 & 0.41 \\
07D3cu & 14:20:56.531 &52:15:13.36  & 2007-03-24 & 2400 & C & 680 & 0.49 \\
&&&& 2400 &&675 \\
07D3da & 14:20:06.236 & 52:16:16.64  & 2007-03-21 & 3600 & N\&S & 720 & 0.36 \\
&&&& 3600 & &715\\
07D3dj & 14:22:40.795 & 53:02:14.69  & 2007-04-12 & 3600 & C & 680 & 0.44 \\
07D3do & 14:20:35.683 & 52:44:48.58 & 2007-04-10 & 3600 & N\&S & 720 & 0.54\\
&&&& 3600 & &715\\
07D3ea & 14:21:16.116 &  52:40:07.88  & 2007-04-12 & 3600 & C & 680 & 0.47\\
&&& 2007-04-14 & 1200 & & &0.66\\
07D3ey\footnotemark[3] & 14: 19:18.944 & 53:04:35.92  & 2007-05-14 & 3600 & N\&S & 720 & 0.47 \\
&&&& 3600&&710\\
07D3fi\footnotemark[3] & 14:21:24.800 & 53:10:40.10  & 2007-04-23 & 3600 & N\&S & 720 & 0.77 \\
&&&& 3600 & &710\\
07D3gm\footnotemark[3] & 14:18:32.321 & 52:43:29.62 & 2007-05-19 & 3600 & N\&S & 720 & 0.47 \\
&&&& 3600&&710\\
07D3hl & 14:16:25.147 & 53:08:33.66& 2007-06-09 & 3600 & N\&S & 720 & 0.72 \\
&&&&3600&&715\\
07D3hu & 14:21:14.560 & 52:47:53.41 & 2007-06-11 & 3600 & N\&S & 720 & 0.37\\
&&&&1800&&715\\
07D3hv & 14:21:05.942 & 52:53:36.62 & 2007-06-11 & 3600 & C & 680 & 0.43 \\
07D3hw & 14:21:48.410 & 52:30:33.64 & 2007-06-08 & 3600 & N\&S & 720 & 0.53\\
&&&&3600&&715\\
07D3hz & 14:20:29.569 & 52:23:56.79 & 2007-06-10 & 3600 & C & 680 & 0.54 \\
07D3ib & 14:17:10.791 & 53:04:46.91 & 2007-06-16 & 3600 & N\&S & 720 & 0.74\\
&&&&1800&&715\\
&&&2007-06-18 & 1800 &&&0.89\\
07D3it & 14:18:08.730 & 52:25:04.06 & 2007-07-16 & 3600 & N\&S & 720 & 0.50 \\
&&&&3600&&715\\
07D4fl & 22:16:13.16 & -18:11:46.00 &2007-11-13 & 1800 & N\&S & 720 & 0.47 \\
&&&&3600 && 715\\
08D1aa \footnotemark[4]& 02:27:37.13 & -04:03:02.64 & 2008-01-27 & 1800 & N\&S & 720 & 0.69 \\
&&&&1800 && 715\\
&&&2008-02-13 & 1800 &&720 & 0.98\\
08D2aa \footnotemark[5]& 10:01:52.77 & 02:15:57.74 & 2008-01-15 & 1800 & N\&S & 720 &0.76\\
&&&2008-01-17 & 1800&&& 0.95\\
&&&&1800 && 715\\
08D2ad & 10:00:59.19 & 01:45:27.31 & 2008-01-09 & 1800 & N\&S & 720 & 0.60 \\
&&& 2008-01-10 & 3600\\
08D2au & 10:02:01.84 & 02:15:03.76 & 2008-01-17 & 1800 & N\&S & 720 & 0.60\\
&&&&1800 && 715\\
08D2bj \footnotemark[4]& 09:58:51.48 & 02:07:07.38 & 2008-01-10 & 3600 & N\&S & 720 & 0.42\\
&&&&3600 && 715 & 0.47\\
&&&2008-01-12 & 1800 &&715 \\
08D2ch & 10:01:39.61 & 02:06:57.56 & 2008-02-12 & 1800 & N\&S & 720 & 0.68 \\
&&&& 1800 && 715\\
08D2cl & 09:59:28.55 & 02:29:49.02 & 2008-02-12 & 3600 & N\&S & 720 & 0.94 \\
&&&& 1800 && 715\\
08D2dr & 10:01:54.13 & 02:04:01.96 & 2008-02-28 & 1200 & C & 680 & 0.66 \\
&&&& 1200 && 675\\
08D2dz & 09:58:53.07 & 01:44:19.21 & 2008-03-05 & 3600 & N\&S & 720 & 0.54 \\
&&&&3600 && 715\\
08D2eo & 10:02:03.85 & 02:33:44.74 & 2008-03-07 & 1200 & C & 680 & 0.51 \\
&&&&1200 && 675\\
\end{tabular}
\end{minipage}
\end{table*}

\begin{table*}
\contcaption{}
\begin{minipage}{15cm}
\begin{tabular}{c c c c c c c c}
08D2fj & 10:00:57:55 & 02:02:30.68 & 2008-03-09 & 1200 & C & 680 & 0.61 \\
&&&&1200 && 675\\
08D2gw & 10:01:05.95 & 02:31:19.93 & 2008-03-08 & 3600 & N\&S & 720 & 0.63\\
&&&& 1800 && 715\\
08D2hw & 10:00:06.70 & 02:33:08.66 & 2008-03-11 & 1800 & N\&S & 720 & 0.69 \\
&&&&3600 && 715\\
08D2id & 10:02:01.98 & 02:13:13.38 & 2008-04-02 & 1800 & N\&S & 720 & 0.60 \\
&&&&3600 && 715\\
08D2iq & 09:58:56.05 & 02:34:26.99 & 2008-04-08 & 1800 & N\&S & 720 & 0.45 \\
&&&&3600 && 715\\
08D2kj & 09:59:13.37 & 02:25:04.45 & 2008-04-08 & 3600 & N\&S & 720 & 0.43 \\
&&&& 3600 && 715\\
08D3au & 14:19:43.14 & 53:06:31.22 & 2008-02-15 & 1800 & N\&S & 720 & 0.61 \\
&&& 2008-02-17 & 3600 && 715 & 0.51\\
08D3bh & 14:17:06.50 & 53:01:37.38 & 2008-02-15 & 1800 & N\&S & 720 & 0.69 \\
&&&&1800 && 715\\
08D3dc & 14:20:44.79 & 52:46:48.15 & 2008-03-08 & 3600 & N\&S &720 & 0.51 \\
&&& 2008-03-10 & 3600 && 715 & 0.55\\
08D3dx & 14:17:47.82 & 52:23:02.76 & 2008-03-08 & 3600 & N\&S & 720 & 0.58 \\
&&&& 3600 && 715\\
08D3fp & 14:16:28.53 & 53:06:10.23 & 2008-04-08 & 3600 & N\&S & 720 & 0.64 \\
&&&& 3600 && 715\\
08D3fu & 14:22:35.86 & 53:00:36.98 & 2008-04-02 & 1800 & N\&S & 720 & 0.52 \\
&&&& 1800 && 715\\
08D3gb & 14:21:53.41 & 52:11:47.89 & 2008-04-03 & 1200 & C & 680 & 0.67 \\
&&&& 1200 && 675\\
08D3gf & 14:18:32.27 & 53:04:57.26 & 2008-04-12 & 1200 & C & 680 & 0.88 \\
&&&&1200 && 675\\
08D3gu & 14:21:24.10 & 53:05:35.12 & 2008-05-08 & 1800 & N\&S & 720 & 0.68 \\
&&&& 3600 && 715\\
08D3hh & 14:19:51.20 & 52:38:00.25 & 2008-05-12 & 1800 & N\&S & 720 & 0.49 \\
&&&& 3600 && 715\\
\hline
\end{tabular}
\end{minipage}
\end{table*}

\newpage

\begin{table*}
\begin{minipage}{12cm}
\caption{Classifications and redshifts.  Objects types are defined in Section \ref{sec:redshift-classification}}\label{res_tab}
\begin{tabular}{c c c c c c}
\hline
Target & Type & Redshift & Redshift Error & CI & Redshift Source \\
\hline
05D4bo & SN & ... & ... & 2 & ... \\
06D1bt & SNIa & 0.81 & 0.01 & 5 & target \\
06D1bz & SNIa & 0.833 & 0.001 & 4 & host \\
06D1ck & SNIa? & 0.90 & 0.02 & 3 & target \\
06D1dt & SN & 0.298 & 0.001 & 0 & target \\
06D1gg & SN & ... & ... & 1 & ... \\
06D1gu & SN & 0.87 & 0.01 & 2 & host \\
06D1hl & SN? & ... & ... & 2 & ... \\
06D1jp & SN & ... & ... & 2 & ... \\
06D2iz & SNIa? & 0.85 & 0.01 & 3 & target \\
06D2ja & SNIa & 0.726 & 0.001 & 4 & host \\
06D2ji & SNIa? & 0.90 & 0.01 & 3 & host \\
06D2js & SNIa & 0.60 & 0.01 & 5 & target \\
06D2ju & SNIa? & 0.927 & 0.001 & 3 & host \\
06D3ed & SNIa & 0.404 & 0.001 & 5 & host \\
06D3el & SNIa & 0.519 & 0.001 & 5 & host \\
06D3et & SNIa & 0.5755 & 0.0002 & 4 & host \\
06D3fp & SNIa & 0.268 & 0.001 & 5 & host \\
06D3fr & SNII & 0.275 & 0.001 & 0 & host \\
06D3fv & SN & ... & ... & 2 & ... \\
06D3gg & SNII & 0.2665 & 0.0003 & 0 & host \\
06D3gh & SNIa & 0.720 & 0.005 & 4 & host \\
06D3gn & SNIa & 0.2501 & 0.0002 & 5 & host \\
06D3gx & SNIa & 0.76 & 0.02 & 5 & target \\
06D3hd & SN & 0.2409 & 0.0002 & 2 & host \\
06D4dh & SNIa & 0.3027 & 0.0004 & 5 & host \\
06D4dr & SNIa & 0.76 & 0.01 & 5 & target \\
06D4fc & SNIa? & 0.677 & 0.001 & 3 & host \\
06D4fy & SNIa & 0.88 & 0.02 & 4 & target \\
06D4gn & SN & ... & ... & 2 & ... \\
06D4hc & SNII? & 0.38 & 0.01 & 1 & target \\
07D1cx & SNIa & 0.74 & 0.01 & 5 & target \\
07D1ds & SNIa & 0.706 & 0.001 & 4 & host \\
07D1dv & SNIa & 0.887 & 0.001 & 4 & host \\
07D1ea & SNIa & 0.775 & 0.001 & 5 & host \\
07D2bu & SN & 0.733 & 0.001 & 2 & host \\
07D2kc & SNIa & 0.354 & 0.001 & 4 & host \\
07D2ke & SNII & 0.114 & 0.001 & 0 & host \\
07D2kh & SNIa & 0.731 & 0.001 & 4 & host \\
07D2ki & SN? & 0.673 & 0.001 & 2 & host \\
07D2kl & SN? & 1.023 & 0.001 & 2 & host \\
07D2cy & SNIa & 0.886 & 0.001 & 4 & host \\
07D3ae & SNIa & 0.237 & 0.001 & 5 & host \\
07D3af & SNIa & 0.356 & 0.001 & 5 & host \\
07D3ai & SNII & 0.198 & 0.001 & 0 & host \\
07D3am & SNII & 0.214 & 0.001 & 0 & host \\
07D3ap & SNIa & 0.451 & 0.001 & 5 & host \\
07D3bb & SN? & ... & ... & 2 & ... \\
07D3bo & SNIa? & 0.92 & 0.02 & 3 & target \\
07D3bp & SN? & 0.769 & 0.001 & 2 & host \\
07D3bt & SNIa? & 0.91 & 0.01 & 3 & target \\
07D3cn & SNIa & 0.898 & 0.001 & 4 & host \\
07D3cr & SNIa & 0.746 & 0.001 & 4 & host \\
07D3cu & SNIa & 0.512 & 0.001 & 5 & host \\
07D3da & SNIa & 0.837 & 0.001 & 4 & host \\
07D3dj & SNIa & 0.444 & 0.001 & 5 & host \\
07D3do & SNIa? & 1.02 & 0.01 & 3 & target \\
07D3ea & SNIa & 0.471 & 0.001 & 5 & host \\
07D3ey & SNIa & 0.740 & 0.001 & 4 & host \\
07D3fi & SN & ... & ... & 2 & ... \\
07D3gm & SNIa & 0.83 & 0.02 & 5 & target \\
07D3hl & SNIa & 0.67 & 0.01 & 5 & target \\
\end{tabular}
\end{minipage}
\end{table*}

\begin{table*}
\contcaption{}
\begin{minipage}{12cm}
\begin{tabular}{c c c c c c}
07D3hu & SNIa & 0.572 & 0.001 & 4 & host \\
07D3hv & SNIa & 0.351 & 0.001 & 5 & host \\
07D3hw & SNIa & 0.748 & 0.001 & 5 & host \\
07D3hz & SNIa & 0.506 & 0.001 & 5 & host \\
07D3ib & SNIa & 0.681 & 0.001 & 4 & host \\
07D3it & SNIa & 0.835 & 0.001 & 4 & host \\
07D4fl & SNIa & 0.503 & 0.001 & 5 & host \\
08D1aa & SNIa? & 0.54 & 0.01 & 3 & target \\
08D2aa & SNIa & 0.538 & 0.001 & 5 & host \\
08D2ad & SNIa & 0.554 & 0.001 & 5 & host \\
08D2au & SN? & 0.563 & 0.001 & 2 & host \\
08D2bj & SNIa & 0.84 & 0.01 & 4 & target \\
08D2ch & SNIa? & 0.474 & 0.001 & 3 & host \\
08D2cl & SNIa? & 0.831 & 0.001 &3 & host \\
08D2dr & SNIa & 0.355 & 0.001 & 4 & host \\
08D2dz & SNIa & 0.65 & 0.01 & 4 & target \\
08D2eo & SNII & 0.125 & 0.001 & 0 & host \\
08D2fj & SN? & 0.507 & 0.001 & 1 & host \\
08D2gw & SNIa & 0.715 & 0.001 & 5 & host \\
08D2hw & SNIa & 0.746 & 0.001 & 4 & host \\
08D2id & SNIa & 0.833 & 0.001 & 4 & host \\
08D2iq & SNIa & 0.709 & 0.001 & 5 & host \\
08D2kj & SNIa & 0.702 & 0.001 & 5 & host \\
08D3au & SNII & ... & ... & 0 & ... \\
08D3bh & SNIa & 0.52 & 0.01 & 4 & target \\
08D3dc & SNIa & 0.799 & 0.001 & 5 & host \\
08D3dx & SNIa? & 0.928 & 0.001 & 3 & host \\
08D3fp & SN? & ... & ... & 2 & ... \\
08D3fu & SN? & 0.426 & 0.01 & 2 & target \\
08D3gb & SNIa & 0.17 & 0.01 & 5 & target \\
08D3gf & SNIa & 0.352 & 0.01 & 5 & host \\
08D3gu & SNIa & 0.767 & 0.001 & 5 & host \\
08D3hh & SNIa & 0.452 & 0.001 & 5 & host \\
\hline
\end{tabular}
\end{minipage}
\end{table*}

\newpage
\begin{table*}
\begin{minipage}{8cm}
\caption{EW results for objects presented in this paper.  Objects marked * are removed from further analysis.}
\label{tab:walker_ew}
\begin{tabular}{c r@{$\pm$}l r@{$\pm$}l r@{$\pm$}l}
\hline
Object & \multicolumn{2}{c}{\protect\ion{Ca}{ii}} & \multicolumn{2}{c}{\protect\ion{Si}{ii}} & \multicolumn{2}{c}{\protect\ion{Mg}{ii}}\\
 & \multicolumn{2}{c}{(\AA)}&\multicolumn{2}{c}{(\AA)}&\multicolumn{2}{c}{(\AA)}\\
\hline
06D3ed & \multicolumn{2}{c}{--} & 6.13 & 6.16 & 95.37 & 15.52 \\
06D3el & 61.45 & 4.05 & 5.68 & 2.54 & 101.55 & 2.20 \\
06D3et & 89.26 & 5.87 & 17.57 & 3.09 & 55.34 & 4.05 \\
06D3fp & \multicolumn{2}{c}{--} & 18.78 & 3.21 & 108.85 & 0.89 \\
06D3gh & 103.48 & 8.15 & 3.78 & 4.02 & 155.73 & 10.23 \\
06D3gn & \multicolumn{2}{c}{--} & 27.65 & 4.91 & 102.63 & 6.63 \\
06D3gx & 94.55 & 14.06 & 11.82 & 4.37 & 87.19 & 17.48 \\
06D4dh & 27.98 & 5.04 & 5.79 & 0.53 & 68.54 & 0.68 \\
06D4dr & 106.78 & 8.31 & 23.25 & 3.03 & 99.13 & 6.25 \\
06D1bz & 122.01 & 9.01 & 10.56 & 5.56 & 114.96 & 12.12 \\
06D1bt & 89.71 & 6.89 & 2.78 & 4.82 & 13.76 & 8.99 \\
06D4fy* & 103.96 & 9.72 & 17.24 & 5.99 & -2.83 & 7.76 \\
06D2ja & 111.78 & 24.19 & 4.50 & 5.80 & 125.62 & 16.54 \\
06D2js & 92.66 & 15.41 & 24.67 & 3.38 & 148.98 & 7.53 \\
07D3ae & \multicolumn{2}{c}{--} & 25.97 & 5.22 & 128.72 & 4.77 \\
07D3da & 59.61 & 6.49 & 6.33 & 3.46 & 122.60 & 10.56 \\
07D2cy & 62.22 & 19.31 & 11.19 & 3.50 & 136.26 & 12.91 \\
07D3cn & 79.27 & 5.90 & 13.78 & 3.38 & 184.74 & 28.68 \\
07D3ea & 100.15 & 3.50 & 4.18 & 1.63 & 100.93 & 2.15 \\
07D3hw & 84.12 & 3.35 & 4.60 & 1.15 & 94.49 & 3.14 \\
07D3hl & 77.24 & 4.57 & 31.9 & 2.06 & 129.93 & 3.53 \\
07D3hv & 160.93 & 4.96 & 10.88 & 1.10 & 70.7 & 1.05 \\
07D3hu & 111.96 & 2.64 & 9.03 & 1.25 & 96.4 & 1.87 \\
07D3hz & 94.18 & 4.96 & 10.96 & 1.60 & 113.97 & 2.14 \\
07D3ib & 114.62 & 43.12 & 17.9 & 21.22 & 156.82 & 43.91 \\
07D3it & 49.31 & 14.42 & 28.04 & 9.76 & 147.35 & 16.35 \\
\hline
\end{tabular}
\end{minipage}
\end{table*}

\begin{table*}
\begin{minipage}{8cm}
\caption[EW results for the Bronder sample]{EW results for the Bronder sample.  Objects marked * are removed from further analysis}
\label{tab:bronder_ew}
\begin{tabular}{c r@{$\pm$}l r@{$\pm$}l r@{$\pm$}l}
\hline
Object & \multicolumn{2}{c}{\protect\ion{Ca}{ii}} & \multicolumn{2}{c}{\protect\ion{Si}{ii}} & \multicolumn{2}{c}{\protect\ion{Mg}{ii}}\\
 & \multicolumn{2}{c}{(\AA)}&\multicolumn{2}{c}{(\AA)}&\multicolumn{2}{c}{(\AA)} \\
\hline
03D1ax & 89.39 & 4.35 & 11.89 & 1.96 & 83.30 & 3.32 \\
03D1bk & 105.53 & 15.93 & 20.36 & 4.92 & 102.4 & 10.78 \\
03D1ew & 138.65 & 9.13 & 24.53 & 6.35 & 71.81 & 13.92 \\
03D1fq & 78.36 & 17.97 & 16.71 & 9.08 & 46.51 & 28.21 \\
03D4cn & 81.45 & 25.18 & 12.48 & 8.16 & 82.89 & 25.22 \\
03D4cy & 149.75 & 13.28 & 8.77 & 10.05 & 192.29 & 60.66 \\
03D4cz & 79.09 & 26.11 & 4.61 & 8.30 & 114.89 & 16.16 \\
03D4fd & 91.63 & 10.63 & 20.27 & 3.49 & 72.20 & 8.86 \\
03D4gl & 1.15 & 26.67 & 14.57 & 8.21 & 82.03 & 10.49 \\
04D1hd* & 138.83 & 47.54 & 9.4 & 10.61 & 1.62 & 14.68 \\
04D1hy & 140.26 & 21.38 & 15.57 & 5.10 & 104.05 & 12.32 \\
04D1ow & 137.96 & 8.54 & 23.02 & 6.43 & 91.12 & 13.65 \\
04D2ae & 159.04 & 21.44 & 5.0 & 13.16 & 102.02 & 24.69 \\
04D3dd & 124.99 & 34.38 & 44.63 & 28.33 & \multicolumn{2}{c}{--} \\
04D3fq & 93.32 & 27.30 & 29.09 & 16.44 & 113.69 & 30.91 \\
04D3kr & 115.8 & 3.07 & 17.65 & 2.43 & 104.62 & 4.12 \\
04D3lu & 69.35 & 8.48 & 8.47 & 4.92 & 127.63 & 14.53 \\
04D3ml & 98.52 & 3.5 & 12.75 & 3.68 & 136.11 & 6.12 \\
04D3nq & \multicolumn{2}{c}{--} & 9.32 & 26.37 & 113.18 & 22.69 \\
04D3ny & 108.34 & 16.28 & 21.19 & 13.69 & 80. & 29.84 \\
04D3oe & 30.14 & 8.61 & 11.10 & 3.11 & 74.93 & 7.59 \\
04D4dm & 121.67 & 37.87 & 7.91 & 14.98 & 108.92 & 39.95 \\
04D4gg & \multicolumn{2}{c}{--} & 3.49 & 6.33 & 99.97 & 15.97 \\
04D4hu & 61.65 & 16.73 & 9.73 & 5.65 & 91.93 & 11.00 \\
04D4ic & 69.98 & 56.30 & 15.22 & 22.33 & 125.52 & 48.55 \\
04D4ii & 121.3 & 13.19 & 1.22 & 10.87 & 117.13 & 30.69 \\
04D4im* & 108.27 & 14.24 & -0.23 & 2.99 & 157.3 & 8.69 \\
04D2mj & 104.52 & 10.11 & 9.88 & 4.30 & 54.16 & 6.55 \\
05D1az & 75.43 & 9.87 & 8.9 & 6.11 & 102.81 & 14.20 \\
05D1by & \multicolumn{2}{c}{--} & \multicolumn{2}{c}{--} & 89.15 & 9.26 \\
05D1cc & 48.66 & 11.90 & 17.57 & 5.77 & 81.48 & 14.97 \\
05D1ee & 54.87 & 11.85 & 0.09 & 4.69 & 104.43 & 12.22 \\
05D1em & 122.48 & 30.87 & 22.39 & 22.27 & 127.02 & 57.28 \\
05D2ab & \multicolumn{2}{c}{--} & 11.81 & 10.28 & 79.32 & 12.16 \\
05D2ah & \multicolumn{2}{c}{--} & \multicolumn{2}{c}{--} & 91.60 & 18.60 \\
05D2ck & 63.55 & 14.59 & 24.40 & 6.82 & 148.44 & 13.16 \\
05D2ja & \multicolumn{2}{c}{--} & 6.36 & 3.87 & 136.59 & 8.12 \\
05D2nt & 105.5 & 14.77 & 4.38 & 5.94 & 124.12 & 13.89 \\
05D2ob & 101.28 & 19.1 & 6.02 & 12.83 & 121.92 & 31.37 \\
05D3ax & 113.87 & 9.76 & 5.07 & 4.84 & 75.17 & 8.36 \\
05D3cf & \multicolumn{2}{c}{--} & 18.72 & 12.17 & 142.31 & 13.63 \\
05D3ci & 73.37 & 22.47 & 11.09 & 5.65 & 71.22 & 9.64 \\
05D3cx & 56.10 & 13.18 & 18.24 & 10.13 & 110.71 & 20.46 \\
05D3jq & 55.17 & 16.18 & 8.61 & 5.30 & 18.52 & 5.63 \\
05D3kt & 127.22 & 13.46 & 13.05 & 7.17 & 83.19 & 11.43 \\
05D3lb & 115.57 & 10.36 & 9.25 & 4.47 & 88.22 & 9.21 \\
05D3mh & 74.90 & 31.00 & 17.50 & 9.68 & 147.49 & 23.48 \\
05D3mq & \multicolumn{2}{c}{--} & 36.46 & 34.80 & 117.68 & 16.37 \\
05D3mx & 45.99 & 22.97 & 19.45 & 8.79 & 108.38 & 11.20 \\
05D4bm & 144.93 & 7.48 & 12.60 & 3.43 & 75.91 & 4.910 \\
05D4dt* & \multicolumn{2}{c}{--} & 2.13 & 2.64 & 6.38 & 3.17 \\
05D4dy & 95.96 & 25.01 & 23.12 & 11.81 & 125.51 & 23.31 \\
05D4fo & \multicolumn{2}{c}{--} & 28.59 & 4.52 & 80.82 & 5.90 \\
05D4gw & 143.8 & 27.02 & 12.85 & 10.66 & 131.23 & 21.20 \\
06D3bz & 113.9 & 21.40 & 13.48 & 9.22 & 81.48 & 24.17 \\
06D3cn & \multicolumn{2}{c}{--} & 6.57 & 2.6 & 103.29 & 0.93 \\
\hline
\end{tabular}
\end{minipage}
\end{table*}

\begin{table*}
\begin{minipage}{8cm}
\caption[EW results for the VLT sample]{EW results for the VLT sample.  Objects marked * are removed from further analysis.}
\label{tab:vlt_ew}
\begin{tabular}{c r@{$\pm$}l r@{$\pm$}l r@{$\pm$}l}
\hline
Object & \multicolumn{2}{c}{\protect\ion{Ca}{ii}} & \multicolumn{2}{c}{\protect\ion{Si}{ii}} & \multicolumn{2}{c}{\protect\ion{Mg}{ii}}\\
 & \multicolumn{2}{c}{(\AA)}&\multicolumn{2}{c}{(\AA)}&\multicolumn{2}{c}{(\AA)} \\
\hline
03D1bf & 81.38 & 30.45 & 14.90 & 15.42 & 133.44 & 34.85 \\
03D1co & 109.09 & 38.07 & 2.278 & 16.64 & 125.35 & 40.03 \\
03D1dt & 67.35 & 14.91 & 15.30 & 7.83 & 114.28 & 18.58 \\
03D1fc & 164.95 & 5.49 & 8.34 & 3.00 & 101.37 & 6.28 \\
03D1fl & 94.31 & 11.13 & 6.64 & 4.63 & 84.33 & 12.22 \\
03D4ag & 79.31 & 2.41 & 4.61 & 1.43 & 69.74 & 3.06 \\
03D4at & 97.01 & 18.06 & 5.67 & 9.12 & 90.27 & 28.34 \\
03D4au & 37.01 & 15.31 & 14.27& 9.00 & 31.70 & 17.43\\
03D4dy & 106.50 & 16.34 & 8.85 & 9.61 & 97.03 & 20.14 \\
04D1dc & 113.95 & 2.49 & 20.70 & 1.58 & 85.26 & 2.40 \\
04D1ff & 167.75 & 32.56 & 43.04 & 25.06 & \multicolumn{2}{c}{--} \\
04D1hx & 116.53 & 16.05 & 21.05 & 10.80 & 104.12 & 11.85 \\
04D1iv & 72.72 & 44.10 & \multicolumn{2}{c}{--} & \multicolumn{2}{c}{--} \\
04D1kj & 107.76 & 8.91 & 8.22 & 4.99 & 88.98 & 10.20 \\
04D1ks & 159.73 & 12.80 & 6.95 & 8.40 & 138.59 & 23.63 \\
04D1ow & 129.94 & 11.71 & 32.01 & 11.64 & \multicolumn{2}{c}{--} \\
04D1pd & 76.94 & 32.72 & 28.95 & 32.36 & \multicolumn{2}{c}{--} \\
04D1pg & 96.83 & 10.35 & 6.28 & 4.71 & 89.95 & 10.21 \\
04D1pp & 97.28 & 12.21 & 15.13 & 6.52 & 82.12 & 15.50 \\
04D1qd & 90.23 & 16.87 & 23.68 & 10.45 & \multicolumn{2}{c}{--} \\
04D1rh & 85.81 & 6.60 & 2.47 & 3.25 & 63.84 & 7.66 \\
04D1sa & \multicolumn{2}{c}{--} & 16.58 & 6.43 & 153.46 & 14.23 \\
04D1si & 84.13 & 16.46 & 22.00 & 10.35 & 147.02 & 23.31\\
04D2ac & 165.80 & 8.14 & 14.27 & 6.21 & 141.06 & 7.35 \\
04D2al & 123.58 & 32.75 & 26.08 & 24.54 & \multicolumn{2}{c}{--} \\
04D2an & 144.31 & 6.52 & 6.08 & 3.64 & 73.99 & 6.89 \\
04D2bt & 81.54 & 3.54 & 21.38 & 2.23 & 99.86 & 3.11 \\
04D2cf & 62.91& 5.47 & 3.09 & 3.71 & 94.26 & 16.80 \\
04D2fp & 126.43 & 4.39 & 9.90 & 4.26 & 70.24 & 12.53 \\
04D2fs & 132.86 & 2.88 & 14.32 & 1.85 & 94.73 & 3.98 \\
04D2gc & 58.41 & 8.54 & 1.69 & 3.66 & 92.56 & 10.47 \\
04D2gp & 104.44 & 16.43 & 24.44 & 7.34 & 134.22 & 15.70 \\
04D2mc* & \multicolumn{2}{c}{--} & -2.73 & 4.52 & 309.42 & 13.62 \\
04D4an & 62.89 & 13.99 & 20.31 & 8.01 & 61.90 & 18.72 \\
04D4bq & 110.81 & 8.08 & 23.95 & 5.74 & 124.34 & 7.62 \\
04D4fx & 116.50 & 7.85 & 0.10 & 3.30 & 109.84 & 9.82 \\
04D4ib & 65.19 & 17.47 & 3.98 & 6.26 & 114.75 & 12.89 \\
04D4jr & 97.73 & 3.62 & 3.92 & 2.35 & 85.73 & 5.19\\
04D4ju & 69.66 & 9.09 & 8.71 & 7.05 & 64.38 & 11.79 \\
05D1cb & 34.92 & 15.12 & 4.06 & 8.00 & 53.98 & 19.76 \\
05D1ck & 73.91 & 18.36 & 24.48 & 9.60 & 116.83 & 19.26 \\
05D1dn & 132.98 & 9.63 & 8.00 & 5.59 & 210.53 & 11.24 \\
05D2ac & 81.34 & 5.11 & 10.02 & 3.16 & 82.08 & 5.34 \\
05D2ay & 152.92 & 21.13 & 53.87 & 19.80 & \multicolumn{2}{c}{--} \\
05D2bt & 70.17 & 18.04 & 4.91 & 9.31 & 121.36 & 26.28 \\
05D2bv & 91.60 & 7.29 & 5.61 & 3.71 & 77.89 & 7.76 \\
05D2bw & 121.34 & 13.08 & 38.71 & 10.43 & \multicolumn{2}{c}{--} \\
05D2ci & 72.39 & 20.53 & 24.30 & 12.90 & 132.56 & 22.12 \\
05D2dt & 79.33 & 18.10 & 4.10 & 10.61 & 142.75 & 17.93 \\
05D2dw & 136.80 & 5.35 & 6.89 & 2.75 & 72.02 & 7.04 \\
05D2dy & 81.61 & 4.93 & 16.45 & 2.79 & 99.04 & 5.74 \\
05D2eb & 124.60 & 4.28 & 5.51 & 1.98 & 112.12 & 4.90 \\
05D2ec & 82.90 & 14.20 & 12.75 & 7.43 & 73.30 & 15.01 \\
05D2fq & 50.61 & 12.19 & 17.32 & 6.87 & \multicolumn{2}{c}{--} \\
05D2he & 117.27 & 15.70 & 12.54 & 8.51 & 114.48 & 15.64 \\
05D2ie & 74.94 & 6.19 & 7.05 & 3.34 & 60.53 & 9.98 \\
05D4af & 94.30 & 6.61 & 26.73 & 4.50 & 130.77 & 21.27 \\
05D4be & 69.84 & 5.65& 9.73 & 3.17 & 62.03 & 6.33 \\
05D4bi & 66.36 & 19.90 & 3.99 & 9.21 & \multicolumn{2}{c}{--} \\
05D4bj & 99.63 & 17.11 & 4.15 & 8.02 & 47.54 & 25.38 \\
05D4cq & 91.23 & 6.62 & 11.01 & 3.41 & 122.89 & 8.08 \\
05D4cs & 77.39 & 8.59 & 13.97 & 6.65 & 103.87 & 16.34 \\
05D4cw & 82.81 & 4.03& 6.02 & 2.57 & 103.19 & 6.08 \\
05D4dw & 81.49 & 29.46 & 35.37 & 32.32 & \multicolumn{2}{c}{--} \\
05D4ej & 52.77 & 11.41 & 3.03 & 6.44 & 83.65 & 11.16 \\
05D4ek & 121.09 & 13.43 & 16.47 & 5.41 & 92.95 & 10.24 \\
05D4ev & 160.49 & 15.43 & 29.04 & 7.47 & 83.67 & 16.07 \\
\hline
\end{tabular}
\end{minipage}
\end{table*}


\begin{table*}
\begin{minipage}{12cm}
\caption[Low-z Photometry]{Photometric properties of the low-z sample fitted using SiFTO}
\label{lowz_phot}
\begin{tabular}{c c r@{$ \pm $}l r@{$ \pm $}l r@{$ \pm $}l c}
\hline 
SN & Redshift\footnotemark[1] & \multicolumn{2}{c}{Stretch} & \multicolumn{2}{c}{Colour} & \multicolumn{2}{c}{B-band\footnotemark[2]}& Independent Distance \\ 
& &\multicolumn{2}{c}{} &\multicolumn{2}{c}{}& \multicolumn{2}{c}{(mag)}\\
\hline
SN\,1989B & 0.004 & 0.903 &  0.027 &  0.374 &  0.029 &  12.25 & 0.04 & 30.50 $\pm$ 0.09$^a$\\
SN\,1990N & 0.004 & 1.092 &  0.010 &  0.051 &  0.013 & 12.69 & 0.02 & 32.20 $\pm$ 0.09$^a$\\
SN\,1994D & 0.003 & 0.812 &  0.004 & -0.104 &  0.006 & 11.75 & 0.02 & 31.08 $\pm$ 0.20$^b$\\
SN\,1996X & 0.008 & 0.854 &  0.017 & -0.023 &  0.016 & 12.99 & 0.04 & -- \\
SN\,1998aq & 0.004 & 0.955 &  0.009 & -0.146 &  0.009  & 12.31 & 0.01 & 31.87 $\pm$ 0.15$^a$\\
SN\,1998bu & 0.004 & 0.954 &  0.016 &  0.262 &  0.011  & 12.11 & 0.02 & 30.34 $\pm$ 0.11$^a$\\
SN\,1999ee & 0.011 & 1.065 &  0.006 &  0.261 &  0.006  &  14.85 & 0.01 & -- \\
SN\,2000E & 0.004 & 1.060 &  0.012 &  0.128 &  0.035 & 12.85 & 0.15 & -- \\
SN\,2002bo & 0.005 & 0.942 &  0.011 &  0.379 &  0.014  & 13.95 & 0.02 & -- \\
SN\,2002er & 0.009 & 0.876 &  0.009 &  0.124 &  0.019  & 14.25 & 0.06 & -- \\
SN\,2003du & 0.007 & 1.006 &  0.007 & -0.112 &  0.008  & 13.46 & 0.01 & -- \\
SN\,2004eo & 0.015 & 0.869 &  0.005 &  0.048 &  0.013  & 15.07 & 0.05 & -- \\
SN\,2004S & 0.010 & 0.976 &  0.034 &  0.011 &  0.025 & 14.14 & 0.05 & -- \\
SN\,1997do & 0.010 & 0.944 &  0.029 &  0.002 &  0.027  & 14.27 & 0.05 & -- \\
SN\,1998V & 0.017 & 1.001 &  0.039 & -0.027 &  0.025 & 15.07 & 0.08 & -- \\
SN\,1998dh & 0.008 & 0.907 &  0.017 &  0.029 &  0.025  & 13.85 & 0.04 & -- \\
SN\,1998ec & 0.020 & 0.975 &  0.070 &  0.115 &  0.058  & 16.16 & 0.12 & -- \\
SN\,1998eg & 0.024 & 0.919 &  0.065 & -0.010 &  0.022  & 16.11 & 0.05 & -- \\
SN\,1999cc & 0.032 & 0.784 &  0.016 &  0.005 &  0.010  & 16.78 & 0.02 & -- \\
SN\,1999gd & 0.019 & 0.920 &  0.056 &  0.365 &  0.024  & 16.94 & 0.04 & -- \\
SN\,1999gp & 0.026 & 1.167 &  0.015 &  0.031 &  0.012  & 16.04 & 0.03 & -- \\
SN\,2000fa & 0.022 & 1.027 &  0.031 &  0.043 &  0.027  & 15.85 & 0.05 & -- \\
SN\,2001V & 0.016 & 1.128 &  0.012 &  0.050 &  0.015  & 14.62 & 0.02 & -- \\
\hline
\end{tabular}
\medskip
$^1$ Relative to CMB, $^2$ At date of maximum.\\
References: $^a$ \protect\citet{Saha:2006p2423} $^b$ \protect\citet{Ajhar:2001p2373}
\end{minipage}
\end{table*}

\begin{table*}
\begin{minipage}{8cm}
\caption{EW measurements for the low-z spectra.}\label{tab:lowz-ew}
\begin{tabular}{c c c c c}
\hline
Object & Phase &\ion{Ca}{ii} & \ion{Si}{ii} & \ion{Mg}{ii}\\
 & (Days) & (\AA) & (\AA) & (\AA)\\
\hline
SN\,1989B & 0 & -- & 26.56 & 98.58 \\
SN\,1989B & 6 & -- & 39.82 & 128.16 \\
SN\,1989B & 11 & -- & -- & 172.61 \\
SN\,1990N & -8 & -- & 6.92 & 92.35 \\
SN\,1990N & 2 & -- & 15.44 & 80.78 \\
SN\,1994D & -11 & 193.27 & 10.09 & 59.06 \\
SN\,1994D & -9 & -- & 15.13 & 68.78 \\
SN\,1994D & -8 & -- & 20.16 & 57.26 \\
SN\,1994D & -5 & 100.77 & 20.10 & 65.99 \\
SN\,1994D & -4 & 96.37 & 18.09 & 74.21 \\
SN\,1994D & -2 & 99.63 & 23.05 & 79.46 \\
SN\,1994D & 2 & 63.20 & 23.45 & 71.85 \\
SN\,1994D & 4 & 75.34 & 21.47 & 65.94 \\
SN\,1994D & 10 & -- & 20.39 & 90.39 \\
SN\,1994D & 11 & 65.92 & 16.71 & 108.98 \\
SN\,1996X & -2 & 41.29 & 17.18 & 86.65 \\
SN\,1996X & 0 & 81.01 & 17.77 & 92.88 \\
SN\,1996X & 1 & 82.05 & 17.75 & 91.66 \\
SN\,1996X & 7 & 80.56 & 13.71 & 93.10 \\
SN\,1998aq & -9 & 29.61 & 10.29 & 92.22 \\
SN\,1998aq & -8 & 58.20 & 10.27 & 95.58 \\
SN\,1998aq & 1 & 38.40 & 10.18 & 79.06 \\
SN\,1998aq & 2 & 38.79 & 12.37 & 81.56 \\
SN\,1998aq & 3 & 37.22 & 12.01 & 94.22 \\
SN\,1998aq & 4 & 38.15 & 12.78 & 96.98 \\
SN\,1998aq & 5 & 40.50 & 12.70 & 94.88 \\
SN\,1998aq & 6 & 40.15 & 12.61 & 96.50 \\
SN\,1998aq & 7 & 39.76 & 12.88 & 95.41 \\
SN\,1998bu & -4 & 91.48 & 15.24 & 99.23 \\
SN\,1998bu & -3 & 89.16 & 15.24 & 99.23 \\
SN\,1998bu & -2 & 87.53 & 14.30 & 98.41 \\
SN\,1998bu & 8 & 81.97 & 15.27 & 90.43 \\
SN\,1998bu & 9 & 83.49 & 16.36 & 91.26 \\
SN\,1998bu & 10 & 86.99 & 14.78 & 96.61 \\
SN\,1998bu & 11 & 82.38 & 13.68 & 112.96 \\
SN\,1999ee & -11 & 95.76 & 1.14 & 106.85 \\
SN\,1999ee & -9 & 98.16 & 1.75 & 93.78 \\
SN\,1999ee & -4 & 103.10 & 7.29 & 104.13 \\
SN\,1999ee & -2 & 108.93 & 7.92 & 102.26 \\
SN\,1999ee & 0 & 109.82 & 8.91 & 105.23 \\
SN\,1999ee & 5 & 129.76 & 18.67 & 112.07 \\
SN\,1999ee & 7 & -- & 20.11 & 114.87 \\
SN\,1999ee & 9 & 132.32 & 13.38 & 114.11 \\
SN\,2000E & -9 & 148.74 & 9.88 & 105.79 \\
SN\,2000E & -5 & -- & 9.97 & 82.18 \\
SN\,2000E & 5 & 142.57 & 12.86 & 91.46 \\
SN\,2002bo & -11 & 195.34 & 20.38 & 107.06 \\
SN\,2002bo & -6 & 152.29 & 17.78 & 105.13 \\
SN\,2002bo & -5 & 191.09 & 23.85 & 109.45 \\
SN\,2002bo & -4 & 137.57 & 22.08 & 104.52 \\
SN\,2002bo & -3 & 137.96 & 22.85 & 104.29 \\
SN\,2002bo & -2 & 142.47 & 33.86 & 114.94 \\
SN\,2002bo & -1 & 145.12 & 23.34 & 108.19 \\
SN\,2002bo & 4 & 118.94 & 36.97 & 117.07 \\
SN\,2002er & -11 & 170.87 & 23.61 & 107.01 \\
SN\,2002er & -9 & -- & 27.09 & 102.27 \\
SN\,2002er & -8 & 168.78 & 22.12 & 98.25 \\
SN\,2002er & -7 & 179.28 & 39.70 & 99.45 \\
SN\,2002er & -6 & 161.45 & 20.32 & 102.01 \\
SN\,2002er & -5 & 162.72 & 20.72 & 103.20 \\
SN\,2002er & -4 & 171.20 & 23.61 & 105.90 \\
SN\,2002er & -3 & 152.98 & 19.58 & 91.61 \\
SN\,2002er & -2 & 152.98 & 19.58 & 91.61 \\
\end{tabular}
\end{minipage}
\end{table*}

\begin{table*}
\contcaption{}
\begin{minipage}{8cm}
\begin{tabular}{c c c c c}
SN\,2002er & -1 & 149.57 & 22.05 & 97.47 \\
SN\,2002er & 0 & -- & 21.51 & 115.32 \\
SN\,2002er & 2 & -- & 23.78 & 97.44 \\
SN\,2002er & 4 & 111.98 & 25.90 & 99.99 \\
SN\,2002er & 5 & -- & 35.25 & 128.83 \\
SN\,2002er & 6 & 108.88 & 24.68 & 88.57 \\
SN\,2002er & 10 & 108.03 & 8.29 & 110.61 \\
SN\,2003du & -11 & 145.34 & 17.62 & 82.77 \\
SN\,2003du & -8 & 176.57 & 14.33 & 92.39 \\
SN\,2003du & -7 & 155.24 & 12.39 & 93.97 \\
SN\,2003du & -6 & 134.42 & 10.66 & 82.10 \\
SN\,2003du & -4 & 133.97 & 12.75 & 83.85 \\
SN\,2003du & -2 & 132.71 & 12.15 & 89.28 \\
SN\,2003du & -1 & 138.70 & 12.40 & 90.29 \\
SN\,2003du & 0 & 136.87 & 13.52 & 93.04 \\
SN\,2003du & 1 & 141.85 & 14.25 & 93.59 \\
SN\,2003du & 3 & 134.86 & 14.39 & 80.77 \\
SN\,2003du & 4 & 119.98 & 14.63 & 79.39 \\
SN\,2003du & 7 & 89.99 & 16.42 & 85.42 \\
SN\,2003du & 8 & 89.31 & 15.90 & 82.90 \\
SN\,2003du & 9 & 86.63 & 17.18 & 84.06 \\
SN\,2003du & 10 & 85.06 & 16.99 & 88.40 \\
SN\,2004dt & -10 & 81.19 & 43.41 & 118.9 \\
SN\,2004dt & -9 & 89.76 & 40.28 & 119.82 \\
SN\,2004dt & -7 & 91.64 & 34.27 & 109.54 \\
SN\,2004dt & -6 & 97.04 & 30.46 & 115.12 \\
SN\,2004dt & -4 & 97.79 & 27.33 & 113.16 \\
SN\,2004dt & -3 & 96.84 & 32.15 & 130.73 \\
SN\,2004dt & -2 & 98.46 & 24.45 & 120.35 \\
SN\,2004dt & -1 & 116.19 & 20.96 & 104.04 \\
SN\,2004dt & 2 & 120.69 & 25.87 & 99.30 \\
SN\,2004dt & 3 & 131.45 & 24.94 & 115.57 \\
SN\,2004eo & -6 & 136.35 & 24.30 & 107.61 \\
SN\,2004eo & -3 & 118.77 & 32.41 & 128.00 \\
SN\,2004eo & 2 & 96.18 & 25.44 & 104.67 \\
SN\,2004eo & 7 & 92.51 & 15.36 & 130.34 \\
SN\,2004S & 1 & 148.99 & 10.46 & 110.43 \\
SN\,2004S & 7 & 144.40 & 13.59 & 121.01 \\
SN\,1997do & -11 & -- & 14.33 & 105.19 \\
SN\,1997do & -10 & -- & 12.12 & 97.86 \\
SN\,1997do & -7 & -- & 20.44 & 113.20 \\
SN\,1997do & -6 & -- & 15.18 & 103.68 \\
SN\,1997do & 8.5 & -- & 11.52 & 101.07 \\
SN\,1997do & 10.5 & -- & 11.94 & 123.93 \\
SN\,1998V & 0.5 & -- & 14.54 & 92.82 \\
SN\,1998V & 1.5 & -- & 16.38 & 103.11 \\
SN\,1998V & 2.5 & -- & 13.64 & 90.24 \\
SN\,1998dh & -9 & -- & 24.87 & 102.04 \\
SN\,1998dh & -8 & -- & 26.35 & 105.01 \\
SN\,1998dh & -7 & -- & 24.21 & 105.78 \\
SN\,1998dh & -5 & -- & 27.31 & 110.46 \\
SN\,1998dh & -3 & -- & 30.62 & 107.46 \\
SN\,1998dh & 0 & -- & 27.46 & 109.28 \\
SN\,1998ec & -2.5 & -- & 23.23 & 115.56 \\
SN\,1998ec & -1.5 & -- & 18.02 & 109.87 \\
SN\,1998eg & 0 & -- & 22.23 & 98.36 \\
SN\,1998eg & 5 & -- & 24.91 & 106.30 \\
SN\,1999cc & -3 & -- & 32.39 & 107.29 \\
SN\,1999cc & -1 & -- & 32.82 & 112.07 \\
SN\,1999cc & 0.5 & -- & 26.97 & 116.27 \\
SN\,1999cc & 2 & -- & 34.46 & 108.90 \\
SN\,1999ej & -0.5 & -- & 28.54 & 91.06 \\
SN\,1999ej & 2.5 & -- & 27.16 & 91.83 \\
SN\,1999ej & 4.5 & -- & 22.25 & 90.77 \\
SN\,1999ej & 9 & -- & 16.85 & 123.35 \\
\end{tabular}
\end{minipage}
\end{table*}

\begin{table*}
\contcaption{}
\begin{minipage}{8cm}
\begin{tabular}{c c c c c}
SN\,1999gd & 2.5 & -- & 36.35 & 117.13 \\
SN\,1999gd & 9.5 & -- & 26.53 & 104.11 \\
SN\,1999gp & -4.5 & -- & 6.28 & 111.09 \\
SN\,1999gp & -1.5 & -- & 5.55 & 107.13 \\
SN\,1999gp & 0.5 & -- & 10.44 & 89.89 \\
SN\,1999gp & 3 & -- & 9.64 & 109.73 \\
SN\,1999gp & 5 & -- & 22.98 & 125.49 \\
SN\,1999gp & 7 & -- & 24.69 & 114.61 \\
SN\,2000fa & -10 & -- & 12.55 & 102.34 \\
SN\,2000fa & 1.5 & -- & 25.13 & 120.30 \\
SN\,2000fa & 2.5 & -- & 17.96 & 100.17 \\
SN\,2000fa & 4.5 & -- & 19.19 & 105.09 \\
SN\,2000fa & 9.5 & -- & 25.16 & 99.18 \\
SN\,2001V & -11 & -- & 10.01 & 82.89 \\
SN\,2001V & -10 & -- & 7.81 & 79.05 \\
SN\,2001V & -9.5 & -- & 7.98 & 81.46 \\
SN\,2001V & -7.5 & -- & 12.15 & 96.19 \\
SN\,2001V & -6.5 & -- & 12.07 & 98.80 \\
SN\,2001V & -5.5 & -- & 10.11 & 94.55 \\
SN\,2001V & -3.5 & -- & 9.10 & 89.31 \\
SN\,2001V & 9.5 & -- & 15.82 & 92.23 \\
SN\,2001V & 10.5 & -- & 19.66 & 93.89 \\
\hline
\end{tabular}
\end{minipage}
\end{table*}

\clearpage

\section{High-z Spectra}\label{sec:high-z-spectra}

Here we present all the SNLS supernova candidates observed at Gemini  between  May 2006 -- May 2008.  The spectra show the raw data in light blue with a smoothed version in dark blue.  The smoothing is done using a Savitzky-Golay smoothing filter \citep{citeulike:4226570}.  The spectra are grouped according to their classifications.  In the data plots, if any separate host galaxy extraction was possible it is  plotted in green.  Any narrow lines from host emission/absorption are marked.  The results of galaxy template subtraction from the method presented in Section \ref{sec:methods-measurements} are plotted in red.

\newpage

\begin{figure*}
\begin{center}
\subfigure[06D1bt $z=0.81$]{
\includegraphics[scale=0.40]{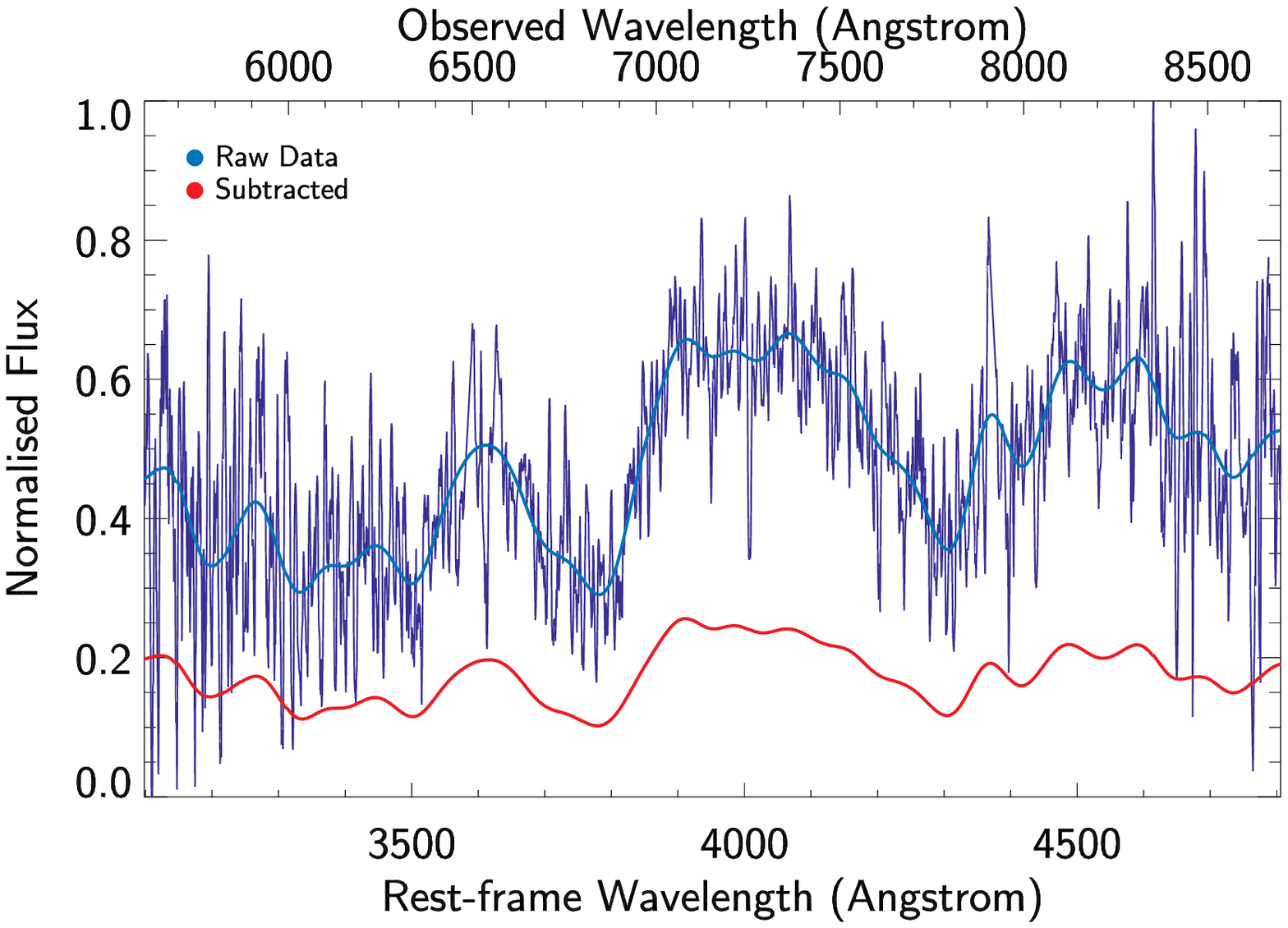}
\label{subfig:06D1bt_data2}
}
\subfigure[06D1bz $z=0.833$]{
\includegraphics[scale=0.40]{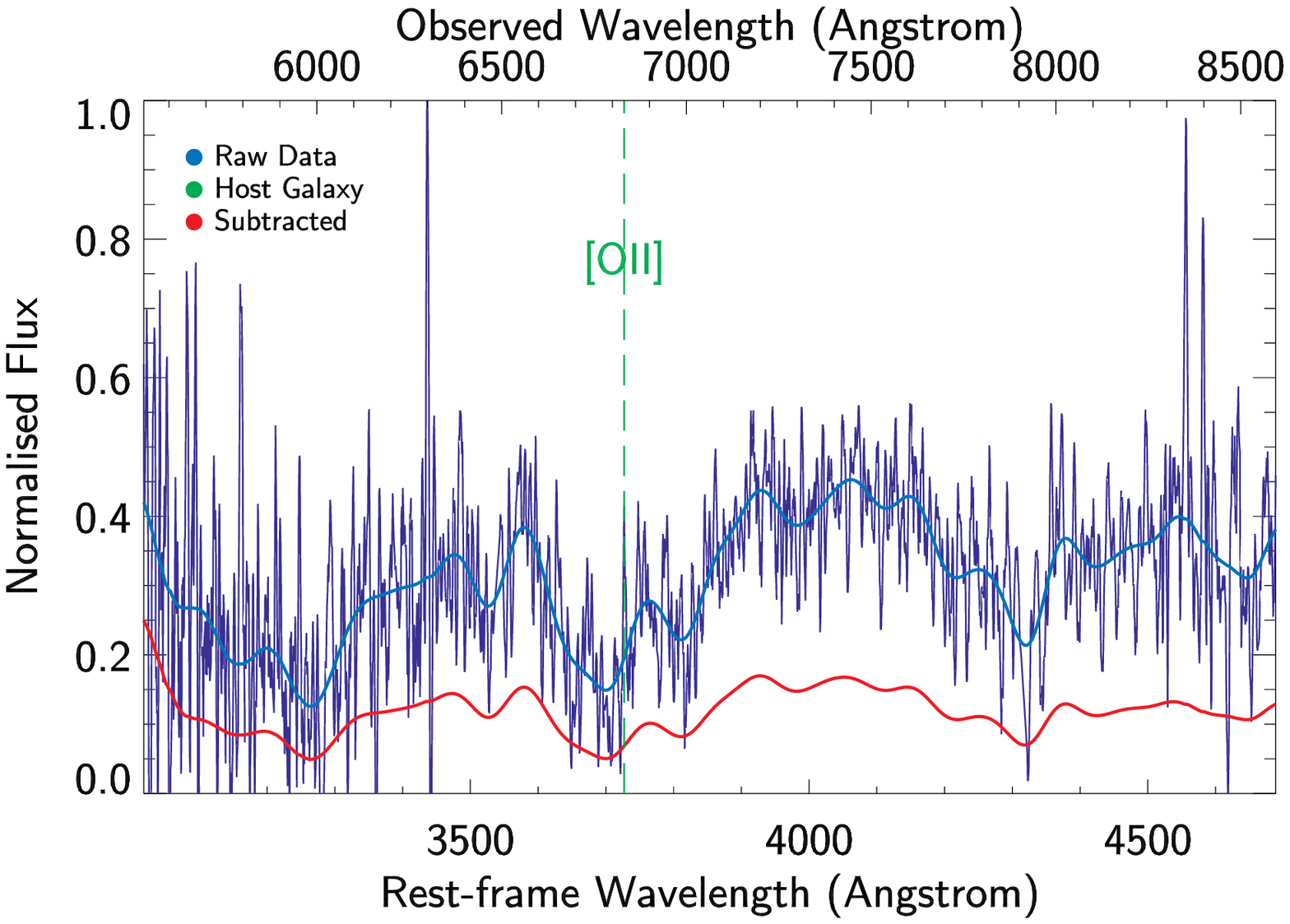}
\label{subfig:06D1bz_data2}
}
\subfigure[06D2ja $z=0.726$]{
\includegraphics[scale=0.40]{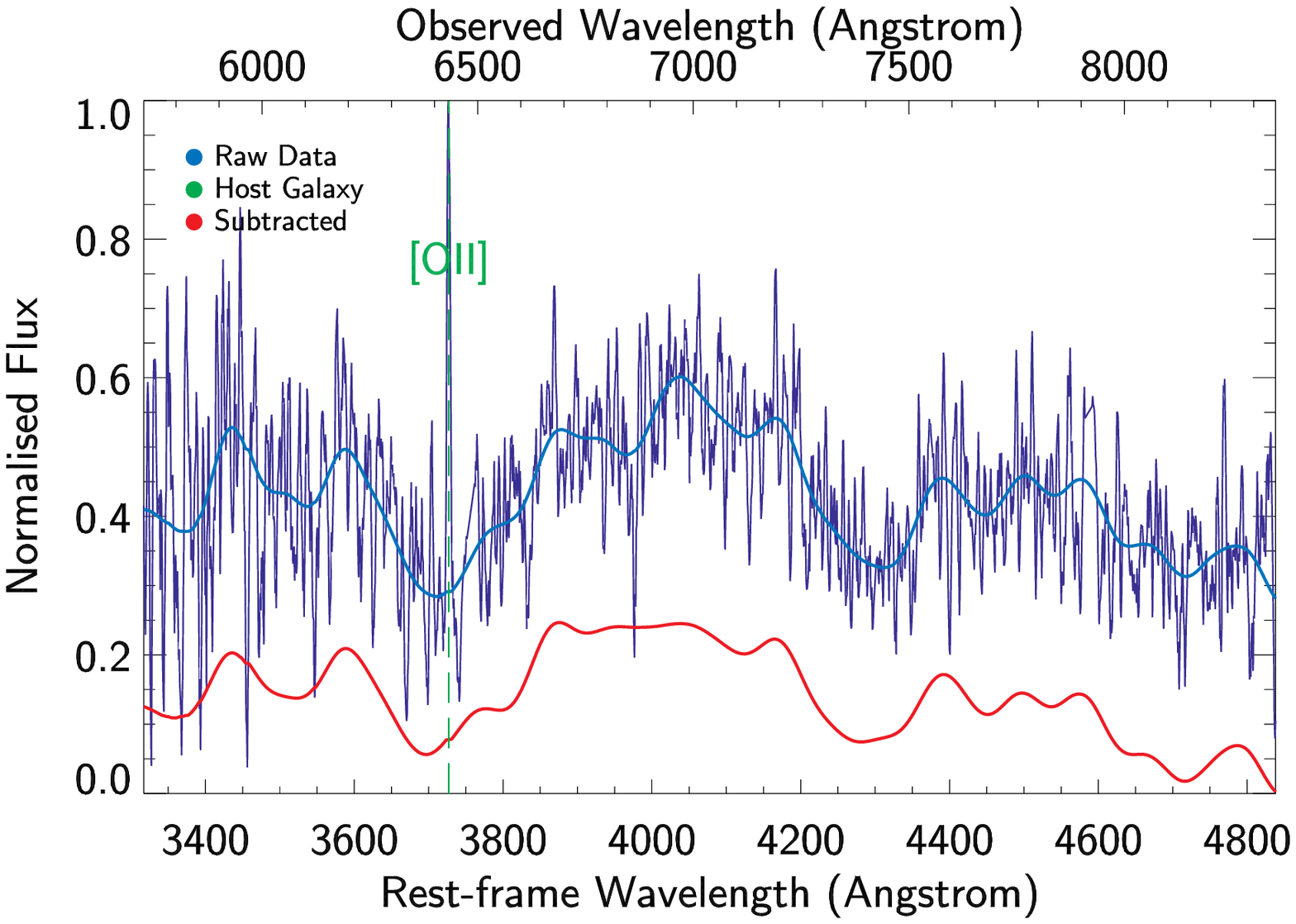}
\label{subfig:06D2ja_data2}
}
\subfigure[06D2js $z=0.60$]{
\includegraphics[scale=0.40]{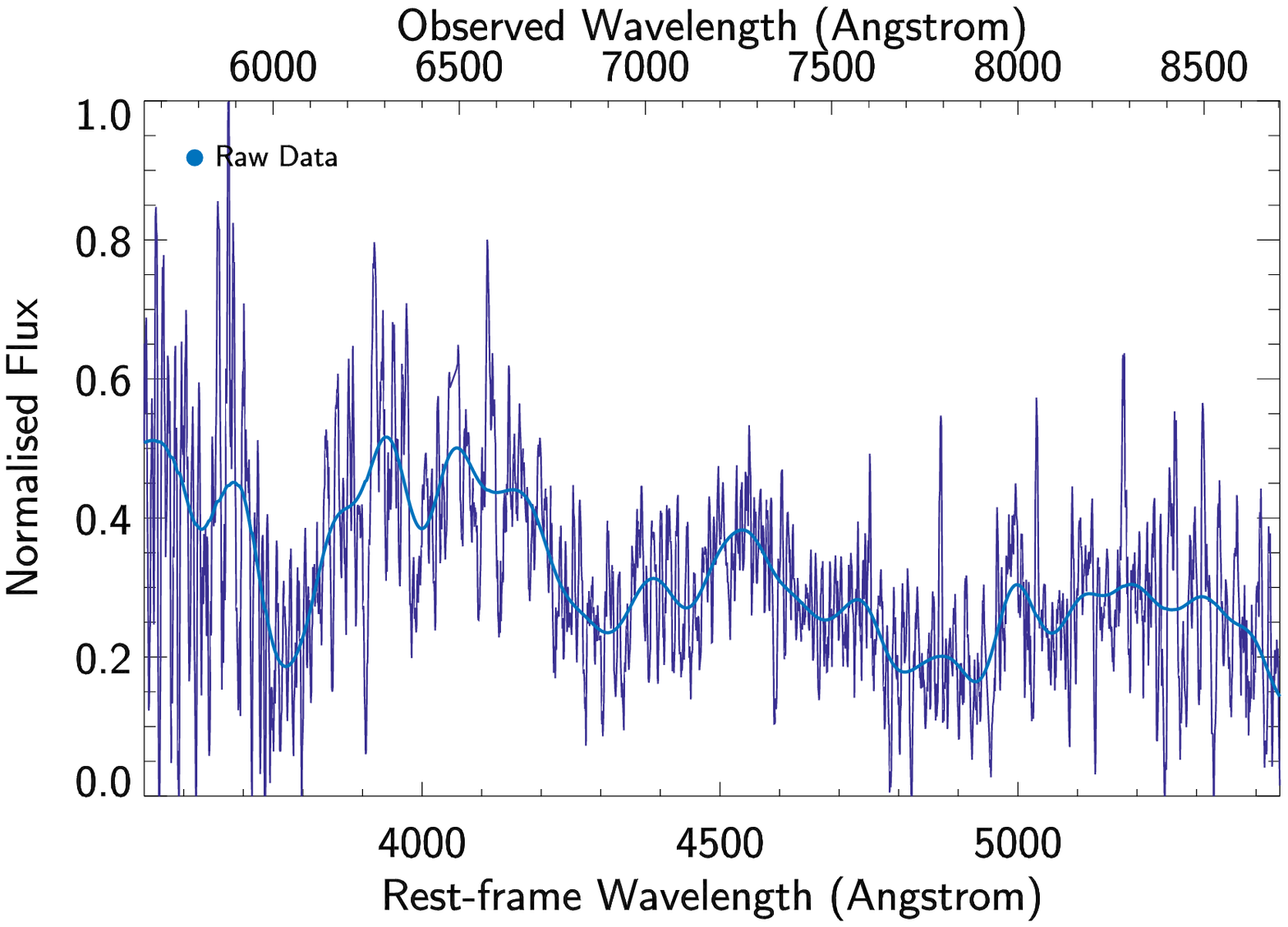}
\label{subfig:06D2js_data2}
}
\subfigure[06D3ed $z=0.404$]{
\includegraphics[scale=0.40]{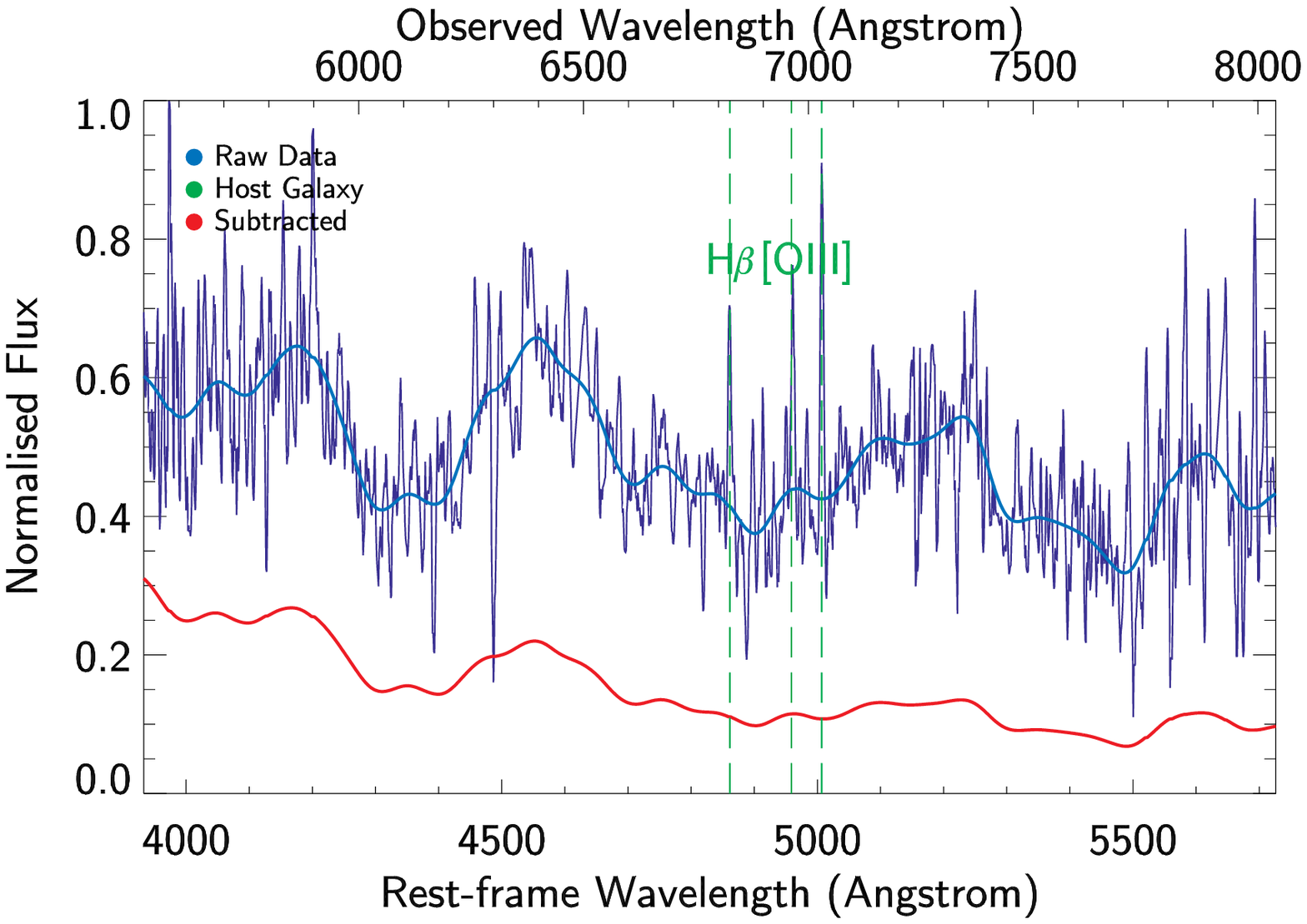}
\label{subfig:06D3ed_data2}
}
\subfigure[06D3el $z=0.519$]{
\includegraphics[scale=0.40]{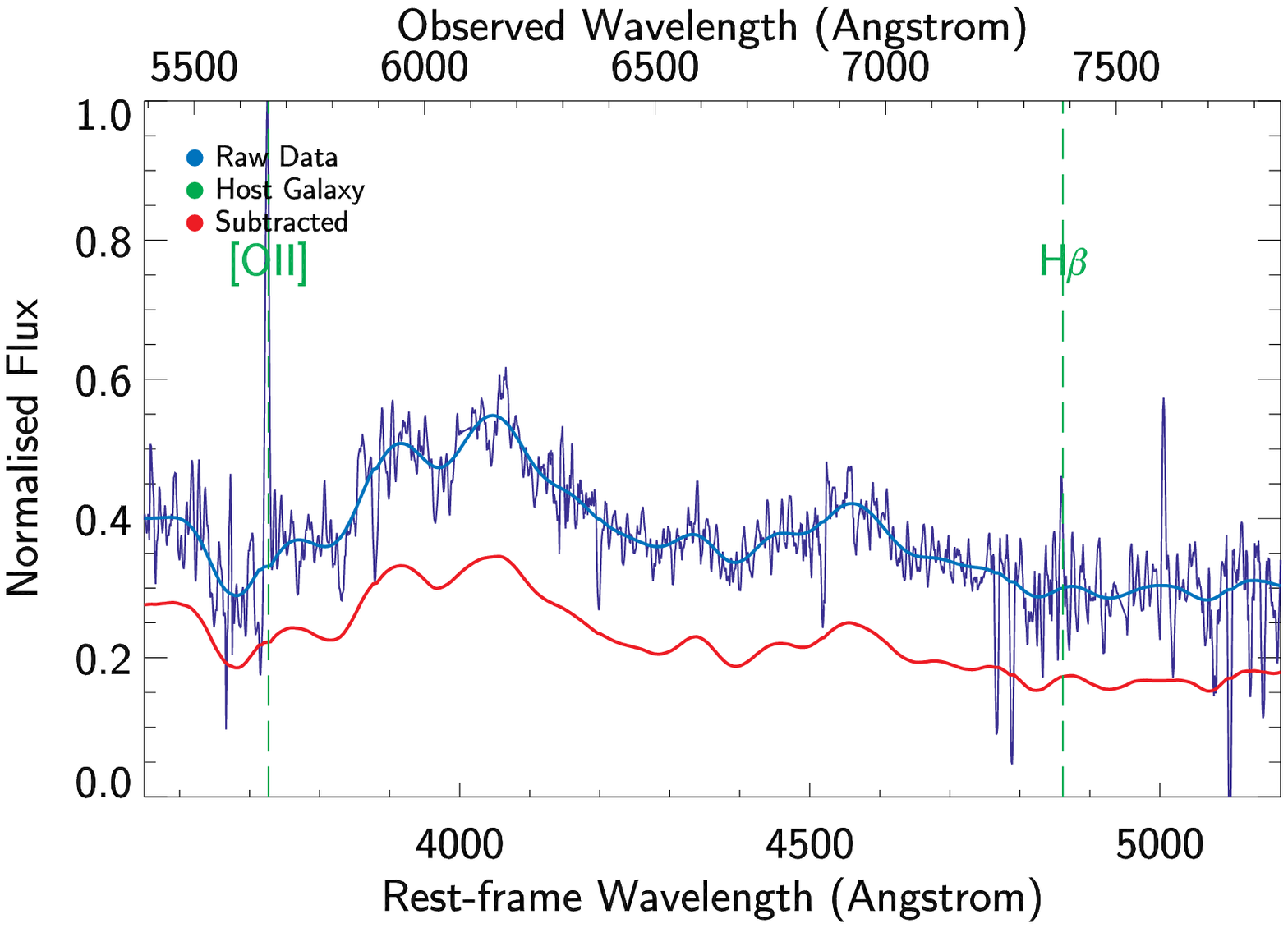}
\label{subfig:06D3el_data2}
}
\subfigure[06D3et $z=0.5755$]{
\includegraphics[scale=0.40]{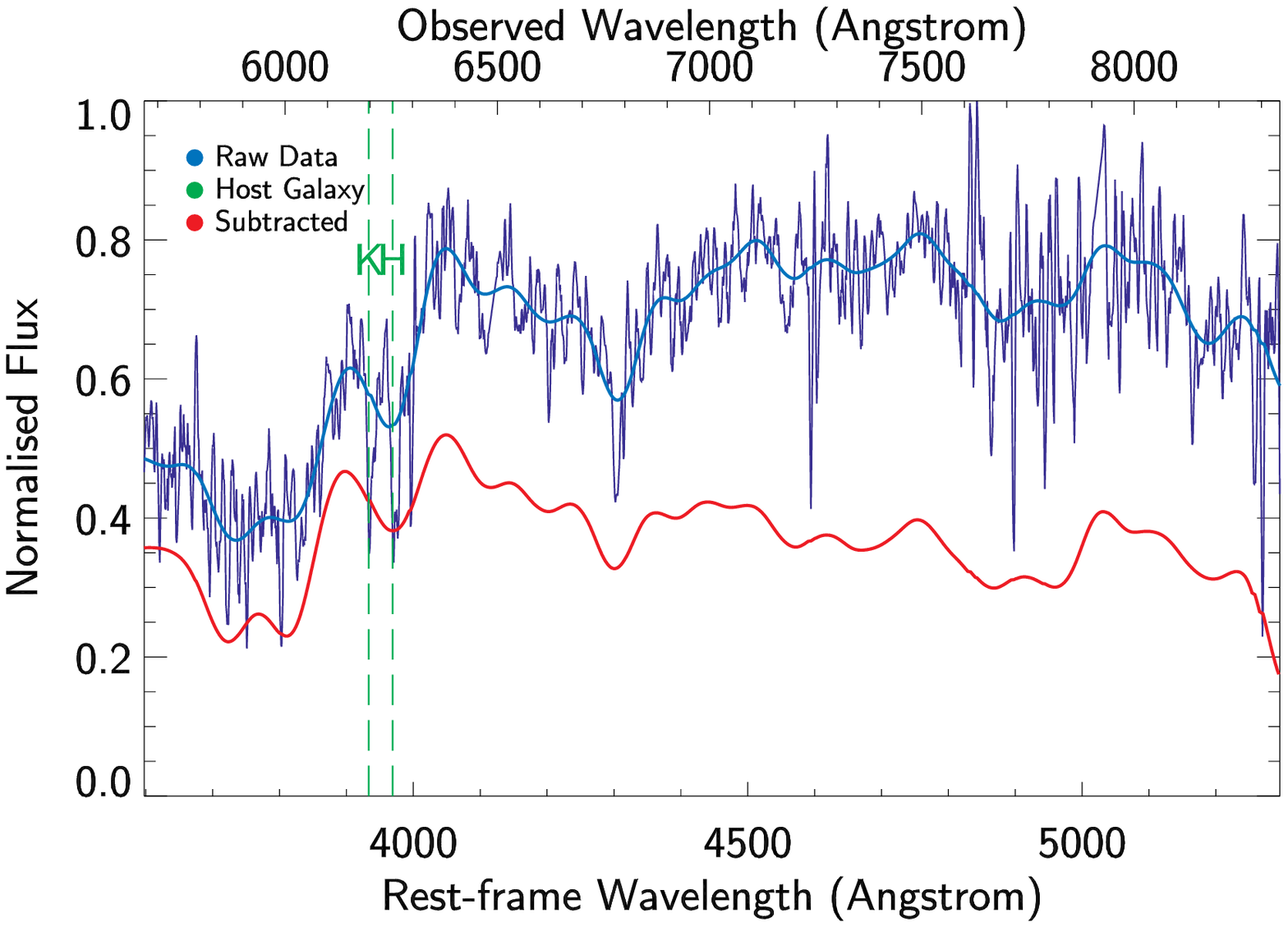}
\label{subfig:06D3et_data2}
}
\subfigure[06D3fp $z=0.268$]{
\includegraphics[scale=0.40]{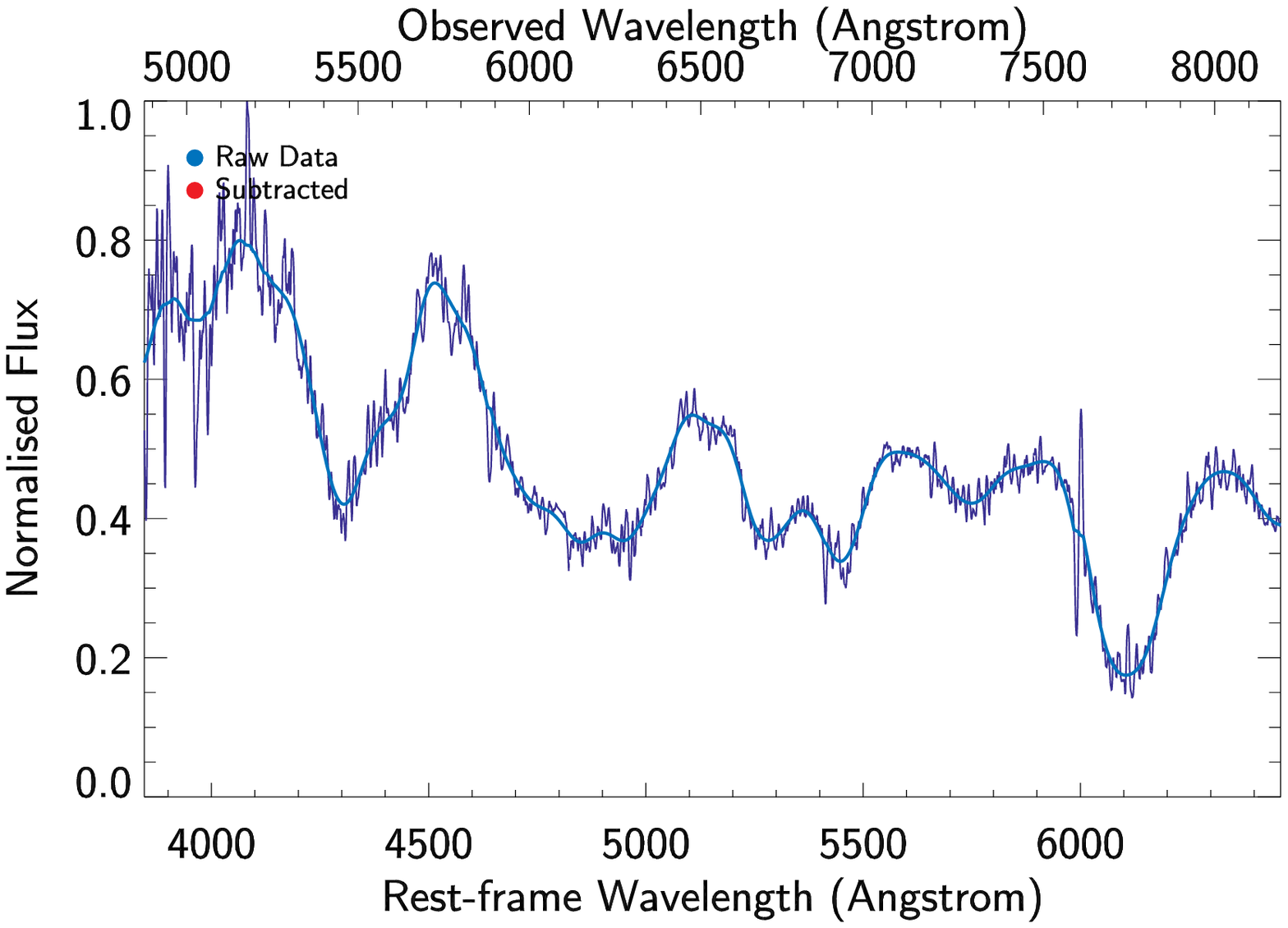}
\label{subfig:06D3fp_data2}
}
\end{center}
\caption{Type Ia Supernovae (CI = 4 or 5)}
\end{figure*}

\begin{figure*}
\begin{center}
\subfigure[06D3gh $z=0.720$]{
\includegraphics[scale=0.40]{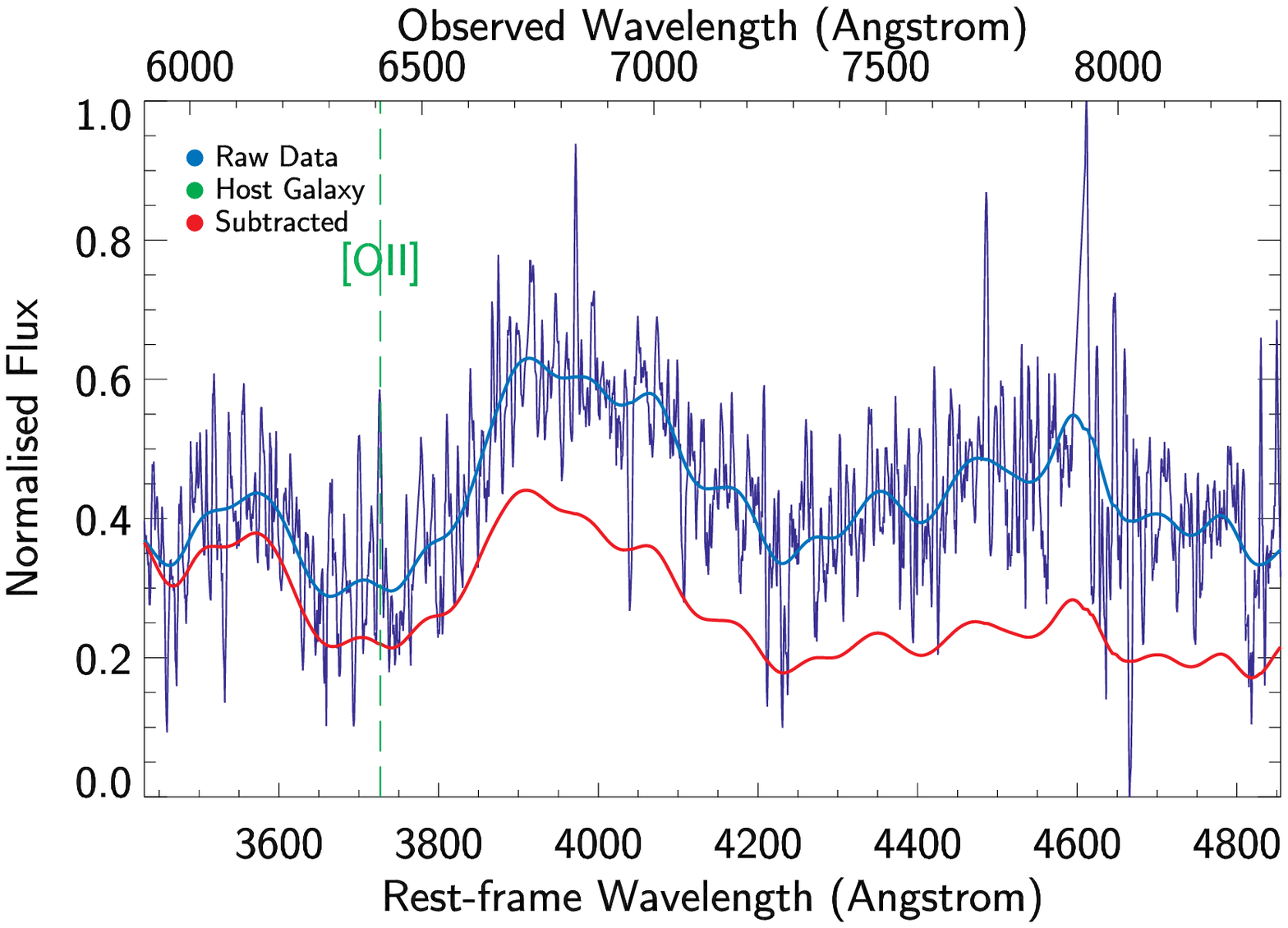}
\label{subfig:06D3gh_data2}
}
\subfigure[06D3gn $z=0.2501$]{
\includegraphics[scale=0.40]{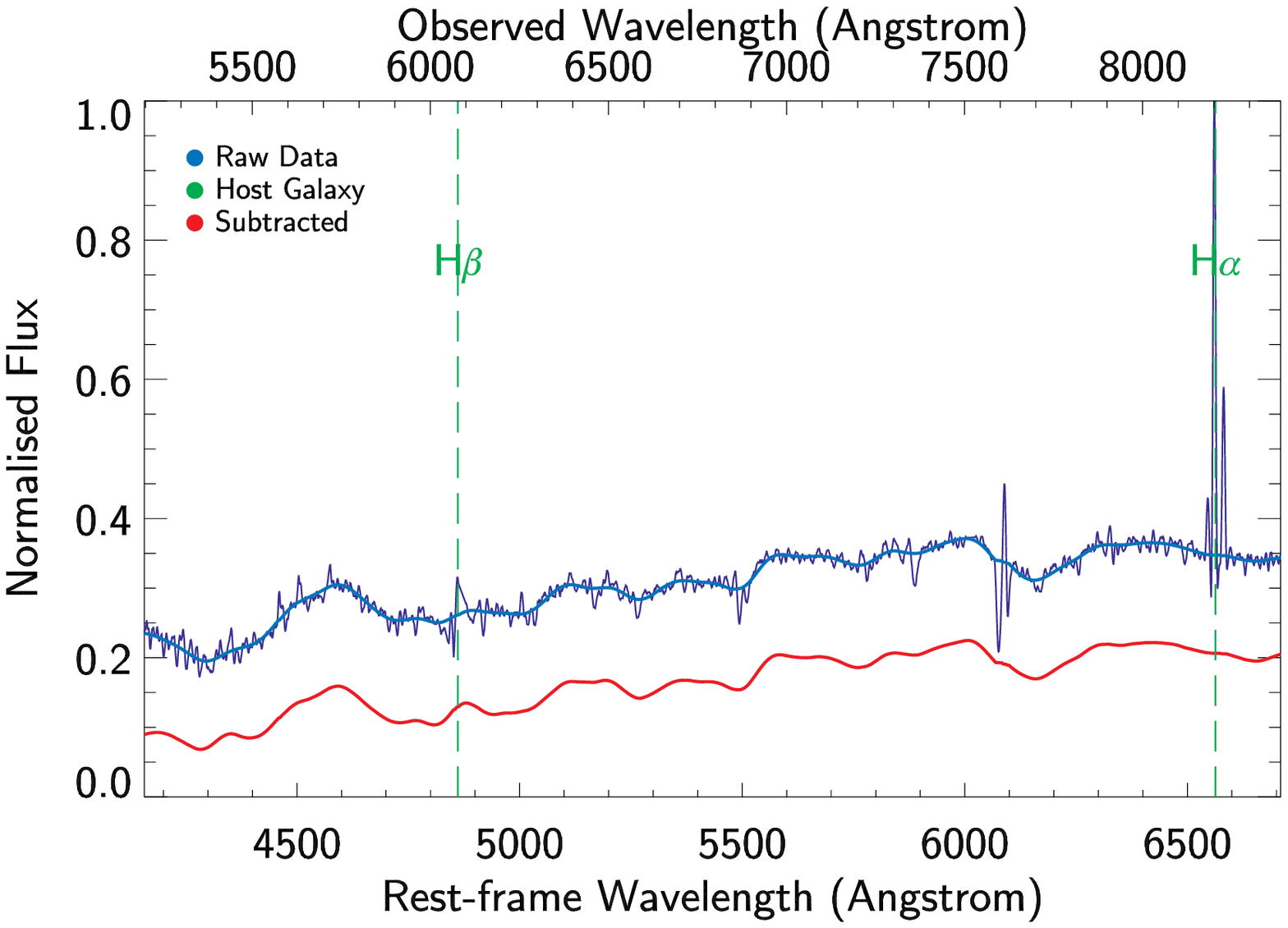}
\label{subfig:06D3gn_data2}
}
\subfigure[06D3gx $z=0.76$]{
\includegraphics[scale=0.40]{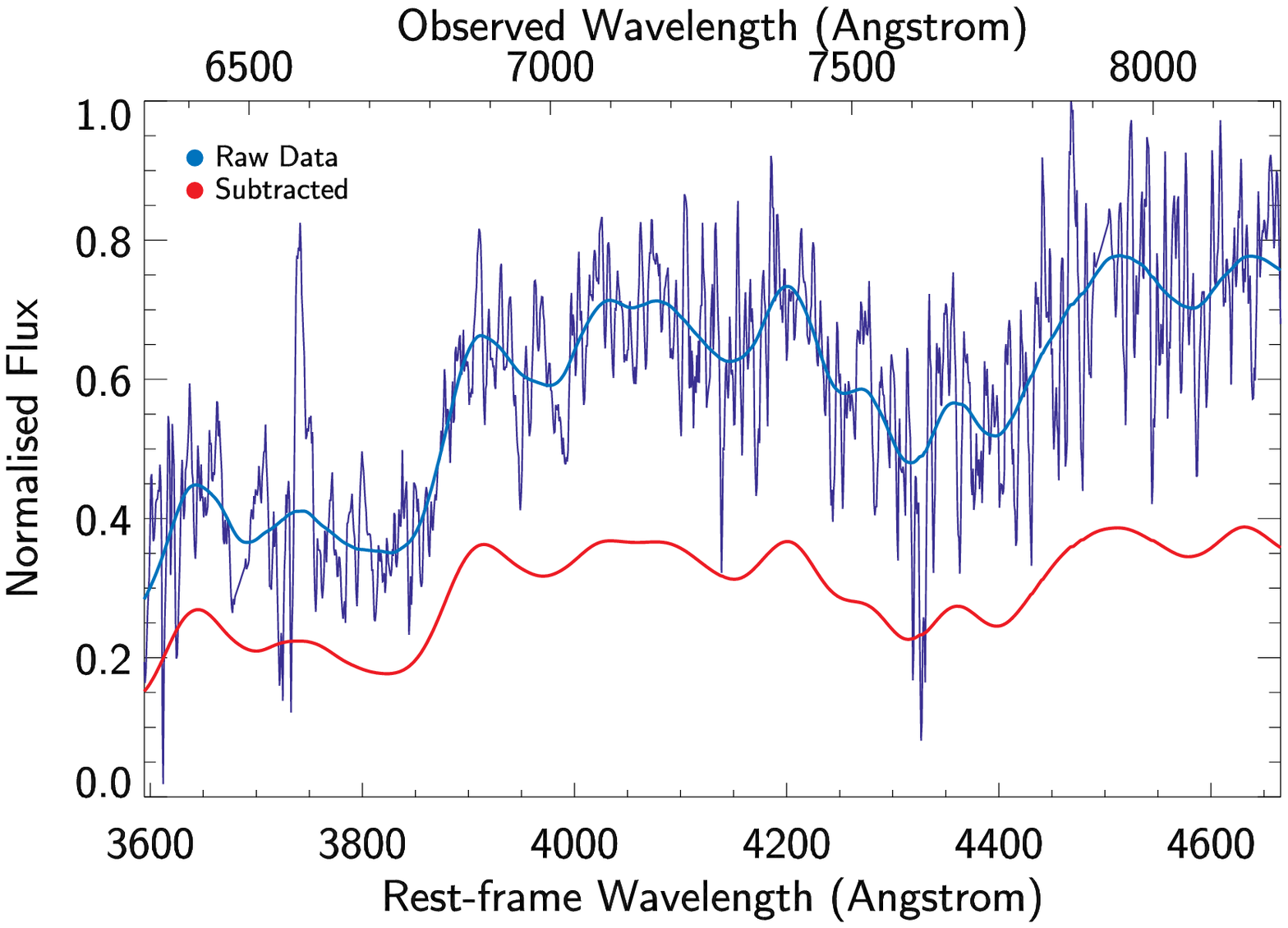}
\label{subfig:06D3gx_data2}
}
\subfigure[06D4dh $z=0.3027$]{
\includegraphics[scale=0.40]{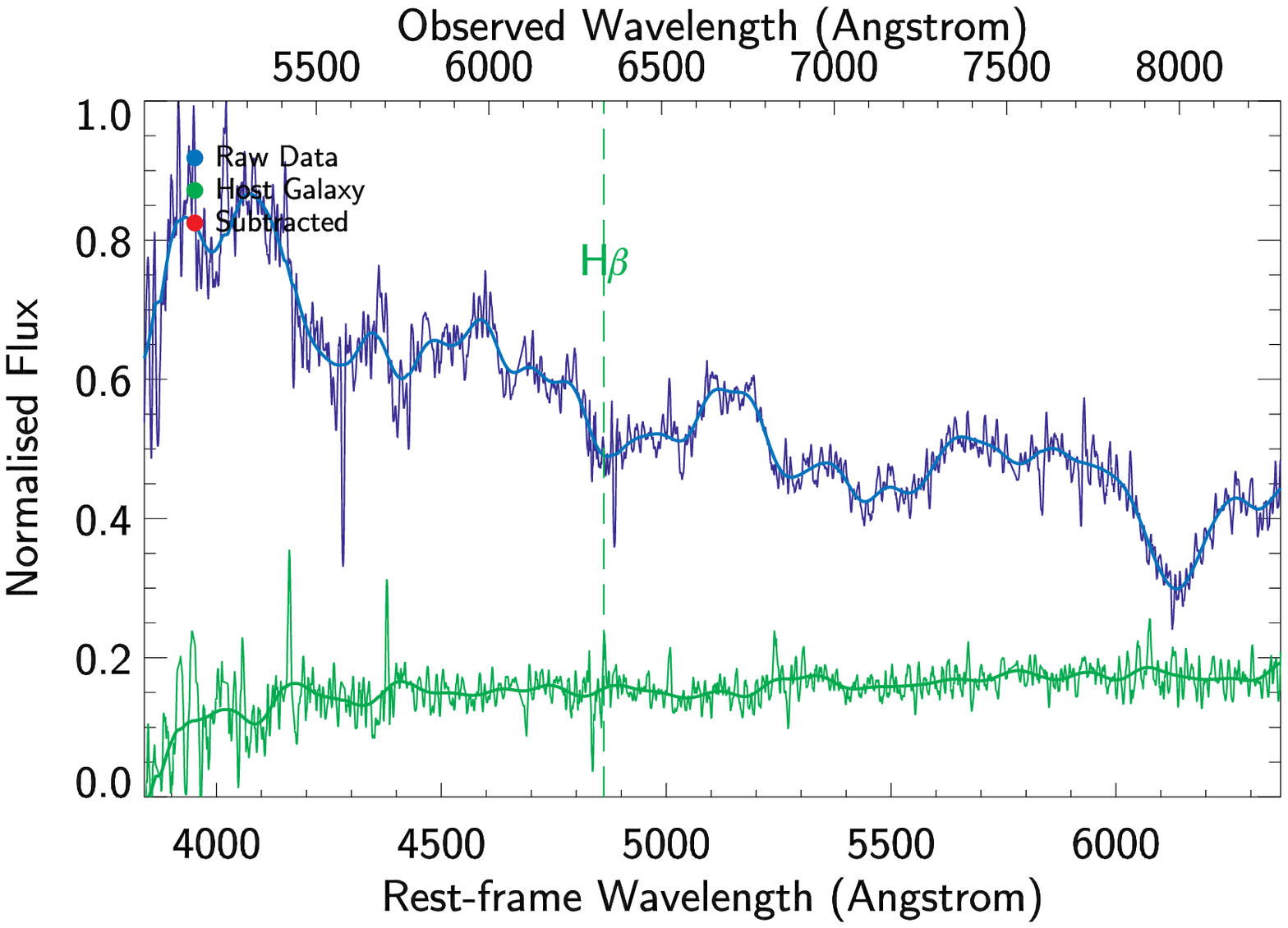}
\label{subfig:06D4dh_data2}
}
\subfigure[06D4dr $z=0.76$]{
\includegraphics[scale=0.40]{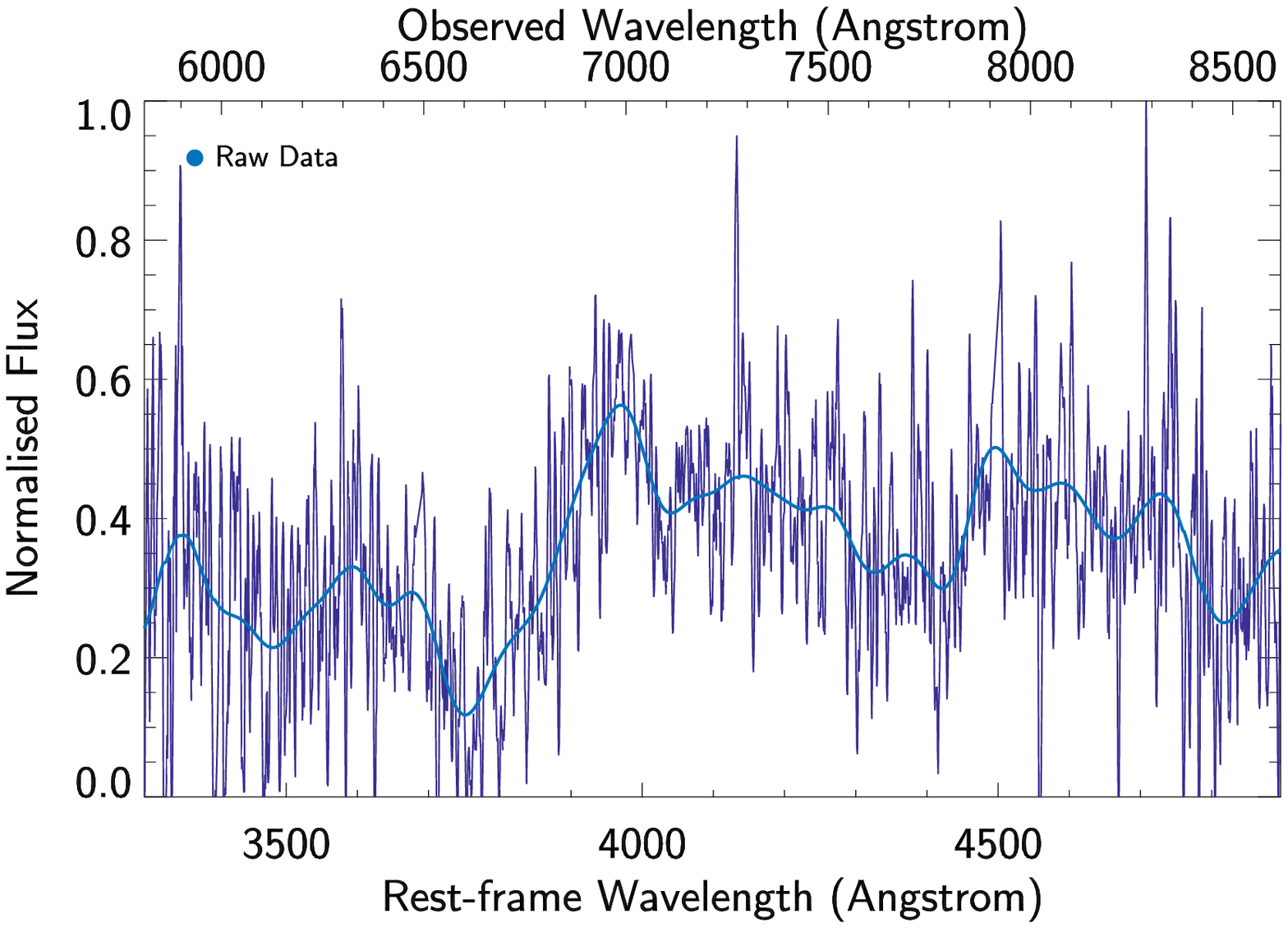}
\label{subfig:06D4dr_data2}
}
\subfigure[06D4fy $z=0.88$]{
\includegraphics[scale=0.40]{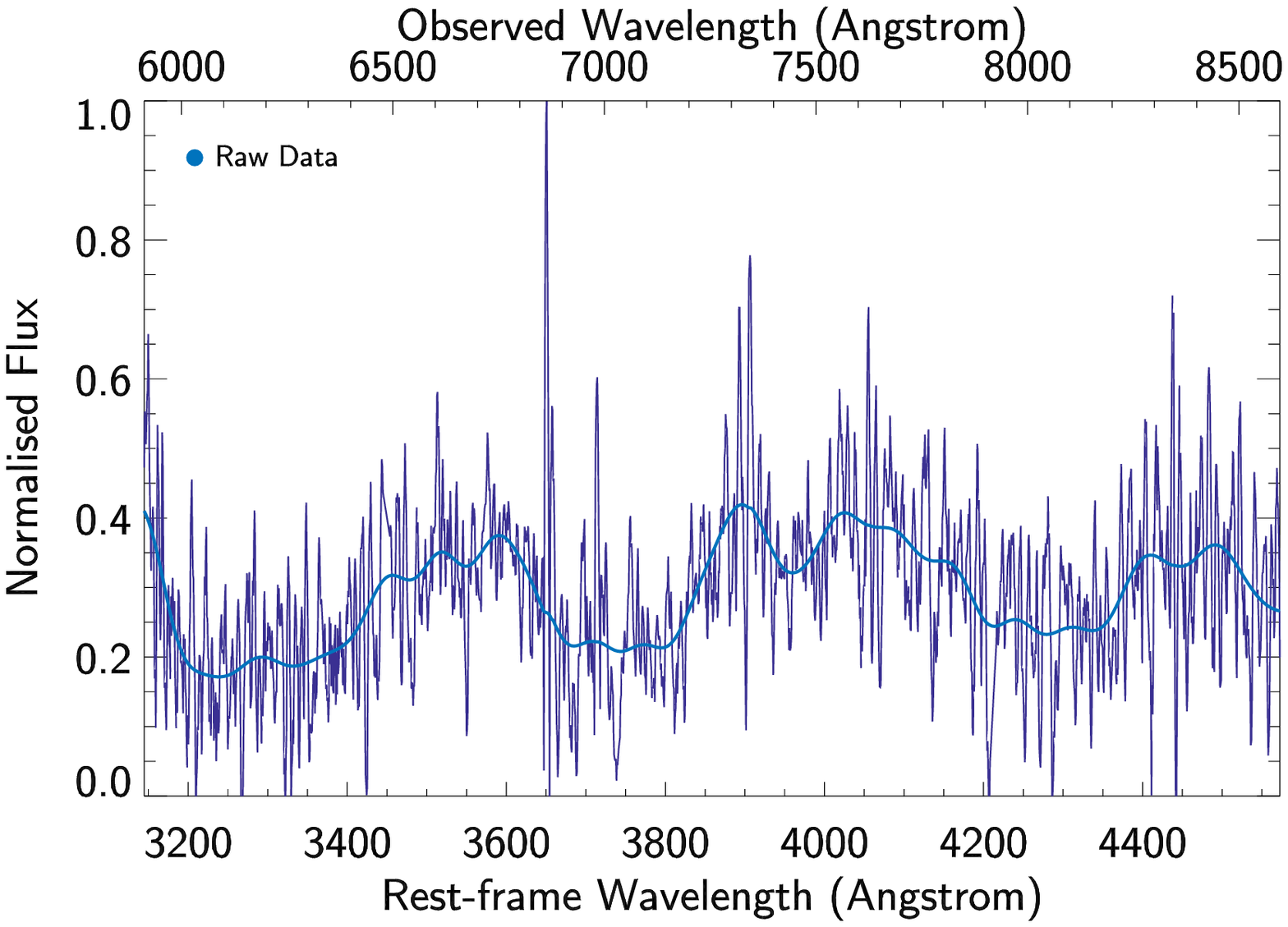}
\label{subfig:06D4fy_data2}
}
\subfigure[07D1cx $z=0.74$]{
\includegraphics[scale=0.40]{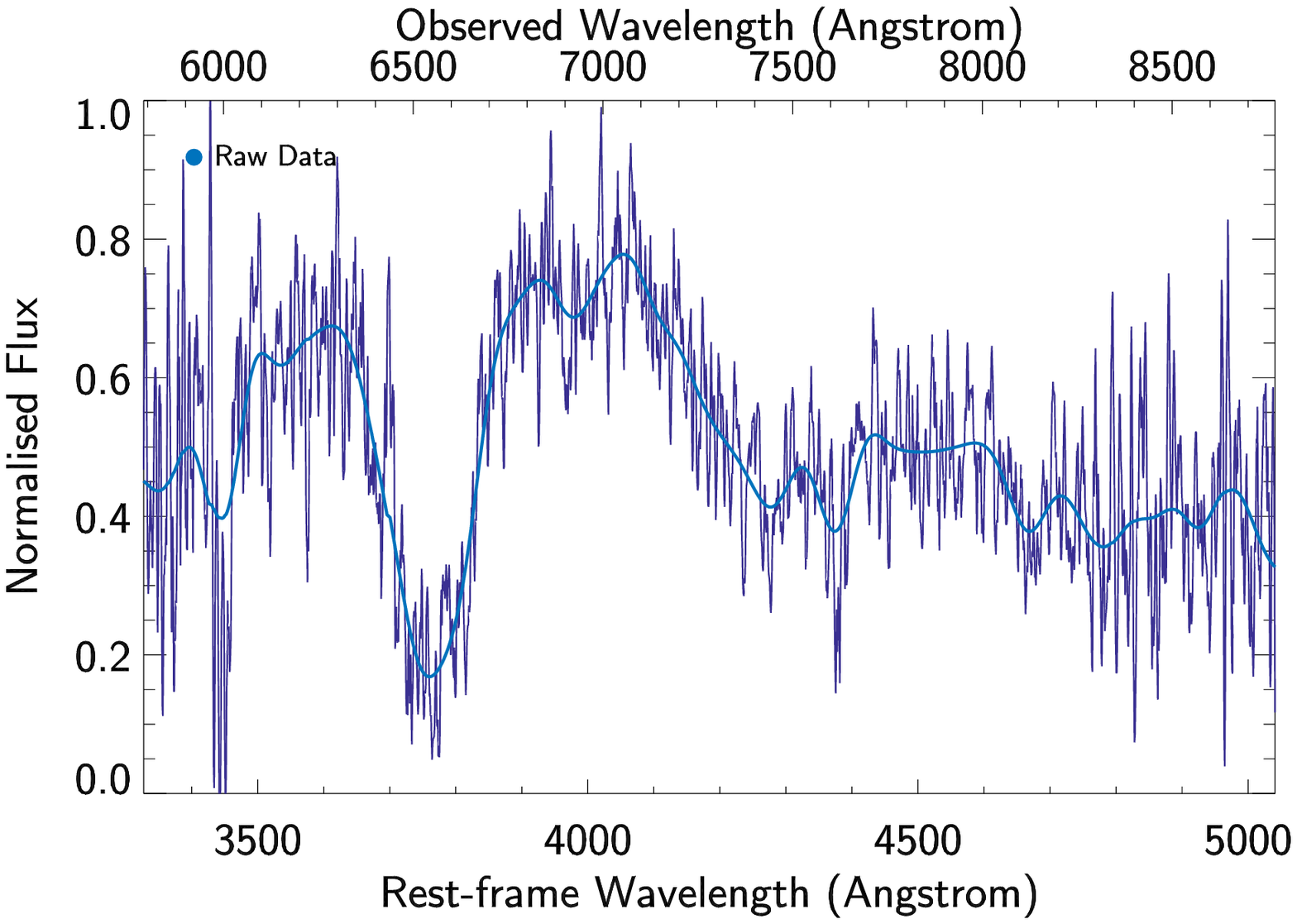}
\label{subfig:07D1cx_data2}
}
\subfigure[07D1ds $z=0.706$]{
\includegraphics[scale=0.40]{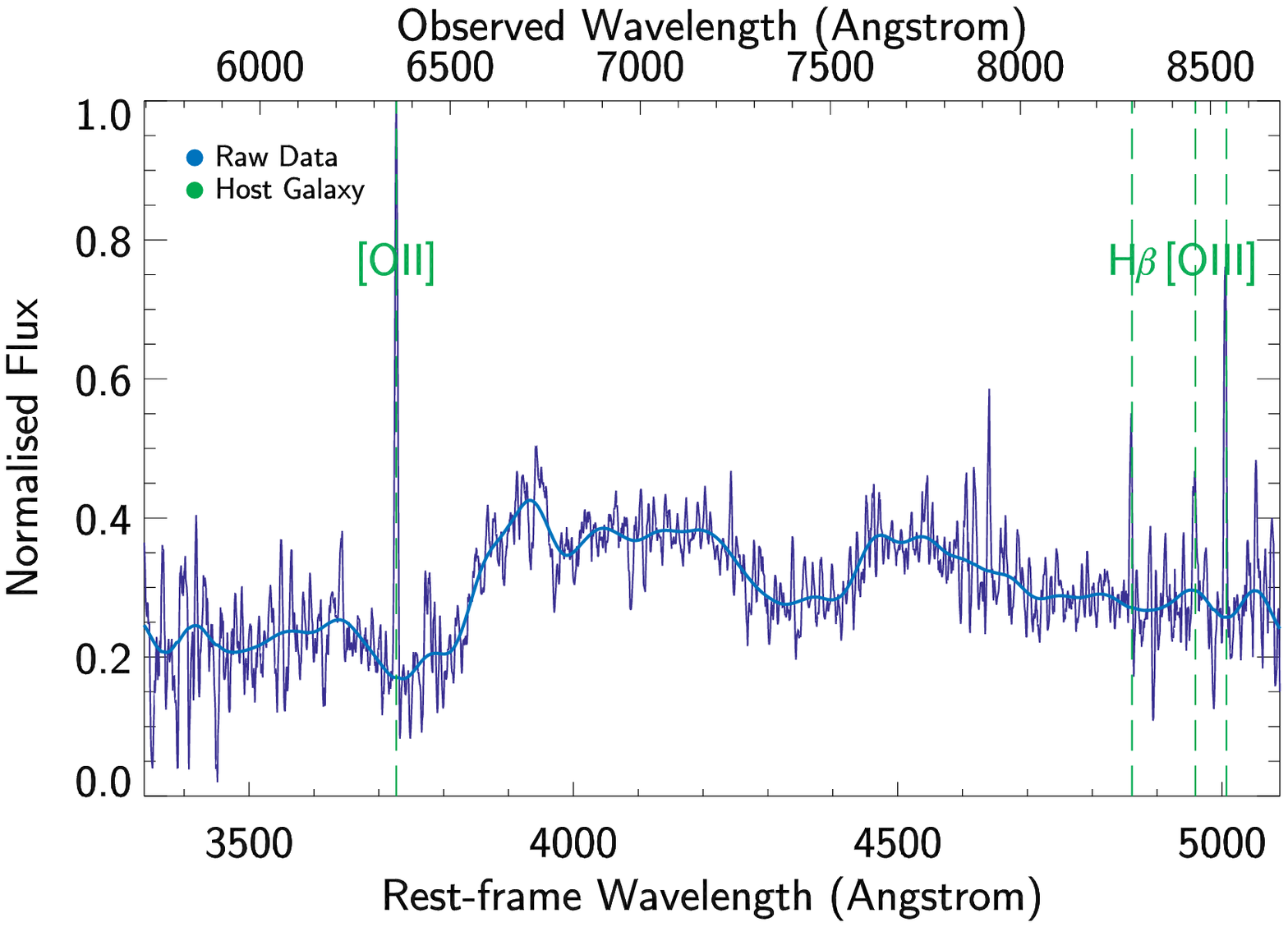}
\label{subfig:07D1ds_data2}
}
\end{center}
\caption{Type Ia Supernovae (CI = 4 or 5)}
\end{figure*}

\begin{figure*}
\begin{center}
\subfigure[07D1dv $z=0.887$]{
\includegraphics[scale=0.40]{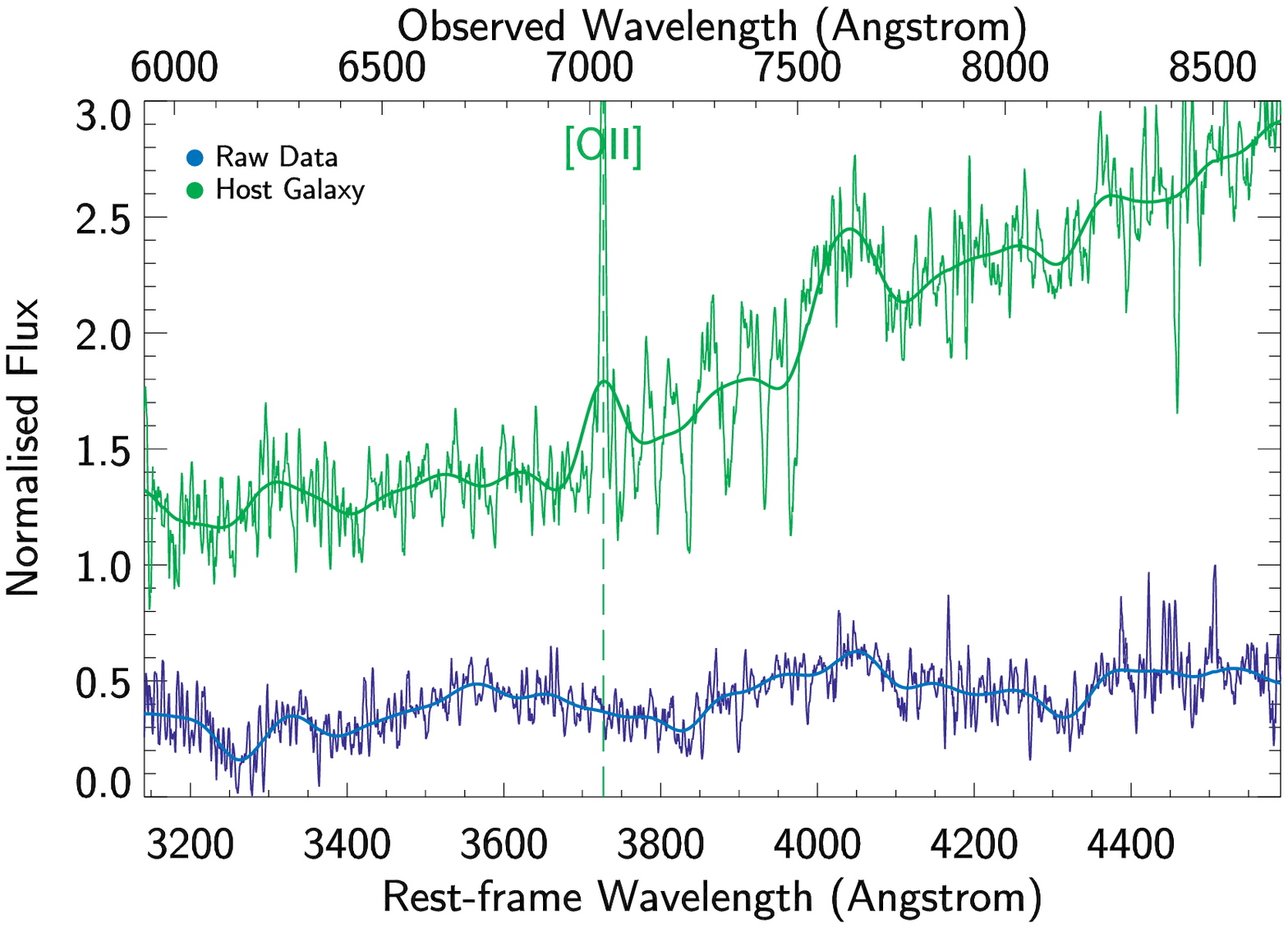}
\label{subfig:07D1dv_data2}
}
\subfigure[07D1ea $z=0.775$]{
\includegraphics[scale=0.40]{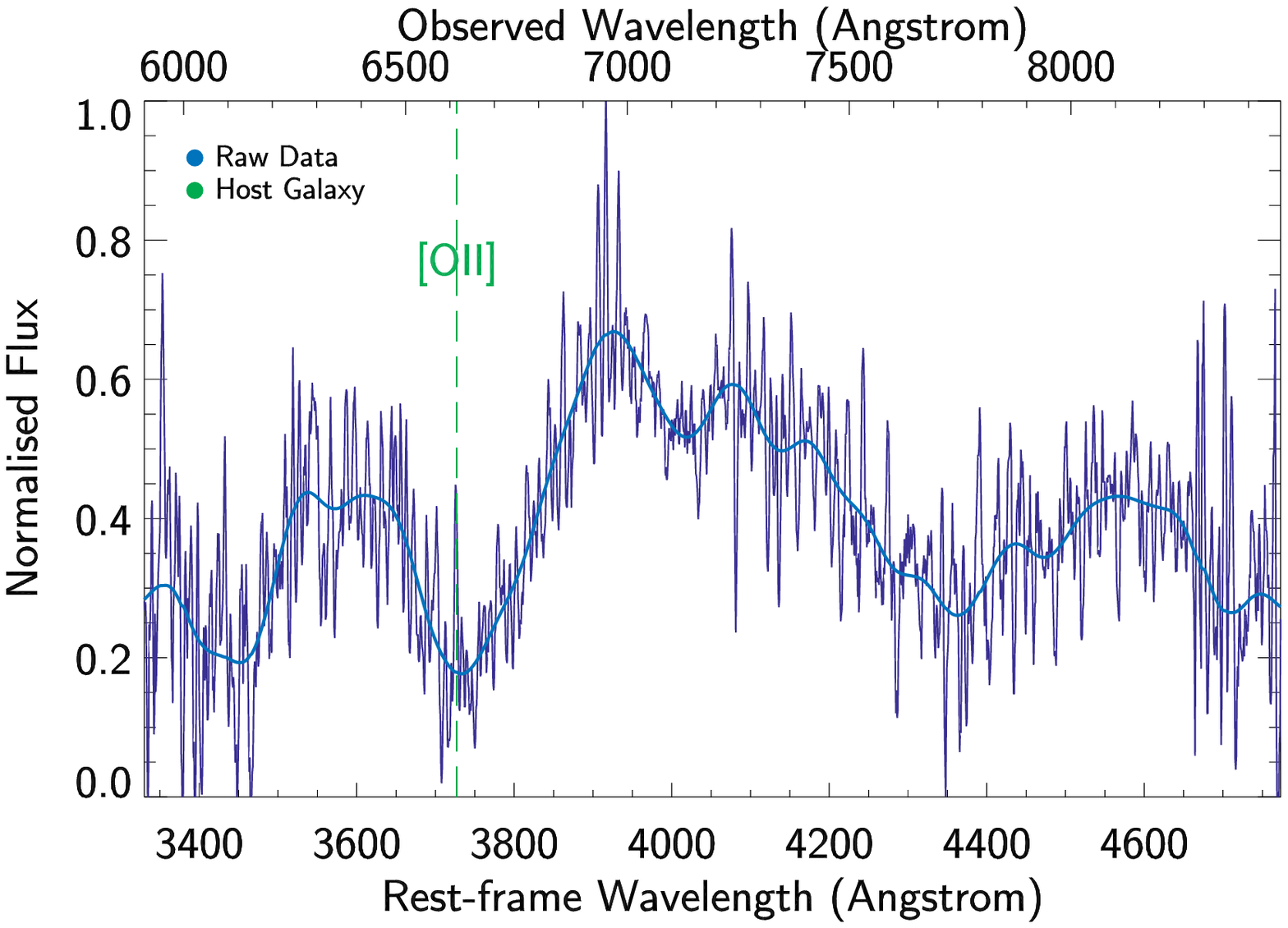}
\label{subfig:07D1ea_data2}
}
\subfigure[07D2kc $z=0.354$]{
\includegraphics[scale=0.40]{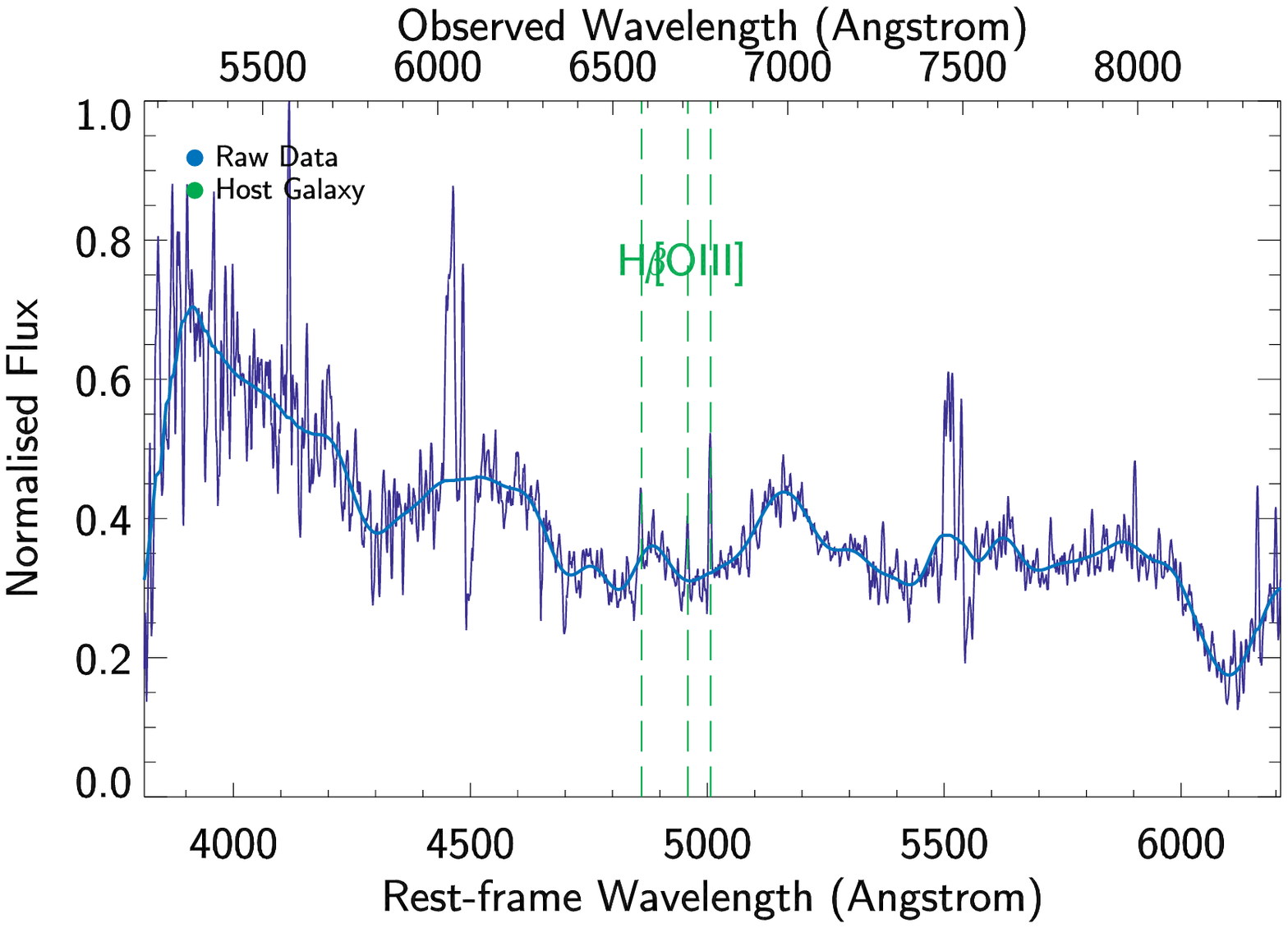}
\label{subfig:07D2kc_data2}
}
\subfigure[07D2kh $z=0.731$]{
\includegraphics[scale=0.40]{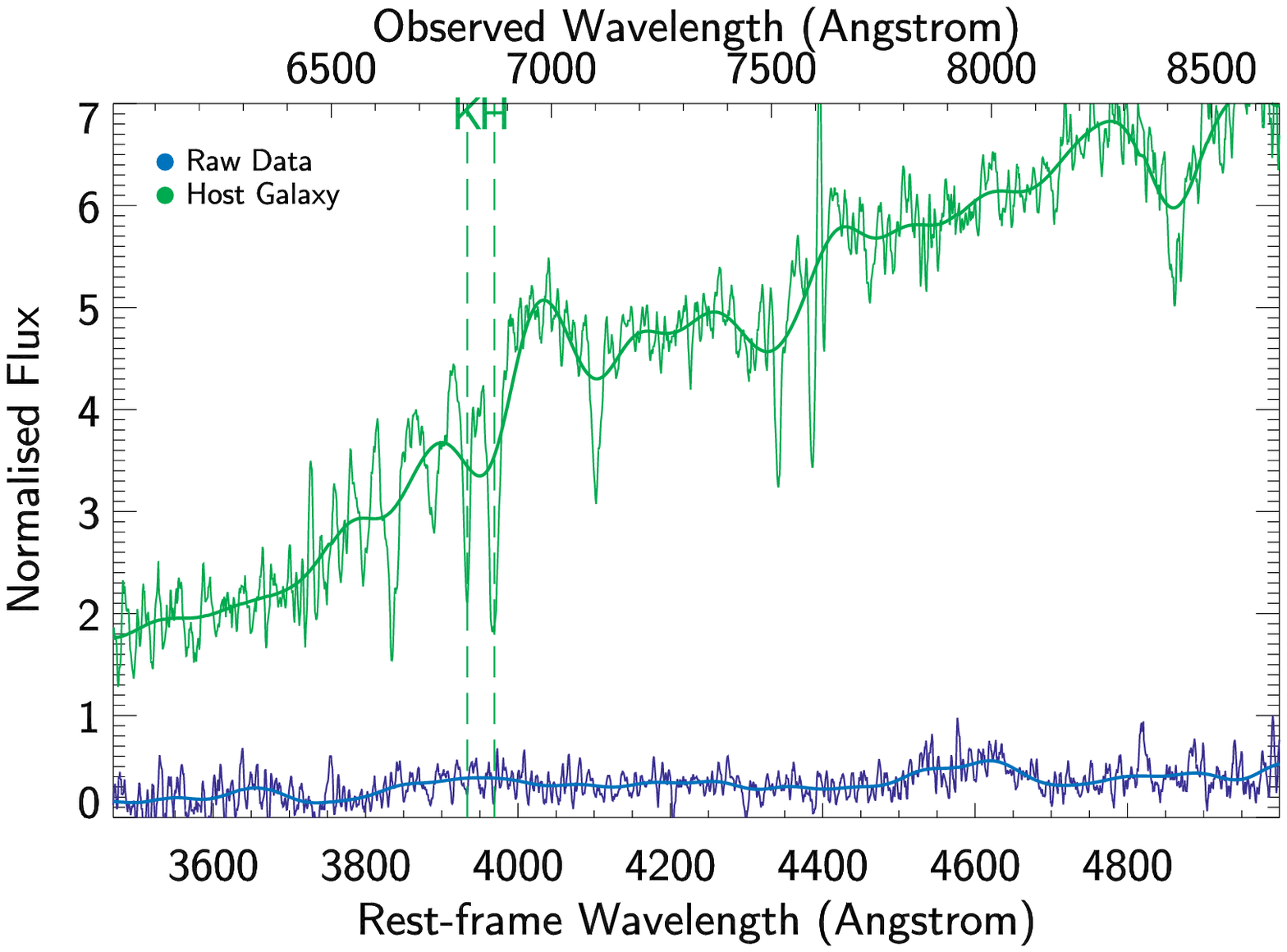}
\label{subfig:07D2kh_data2}
}
\subfigure[07D2cy $z=0.886$]{
\includegraphics[scale=0.40]{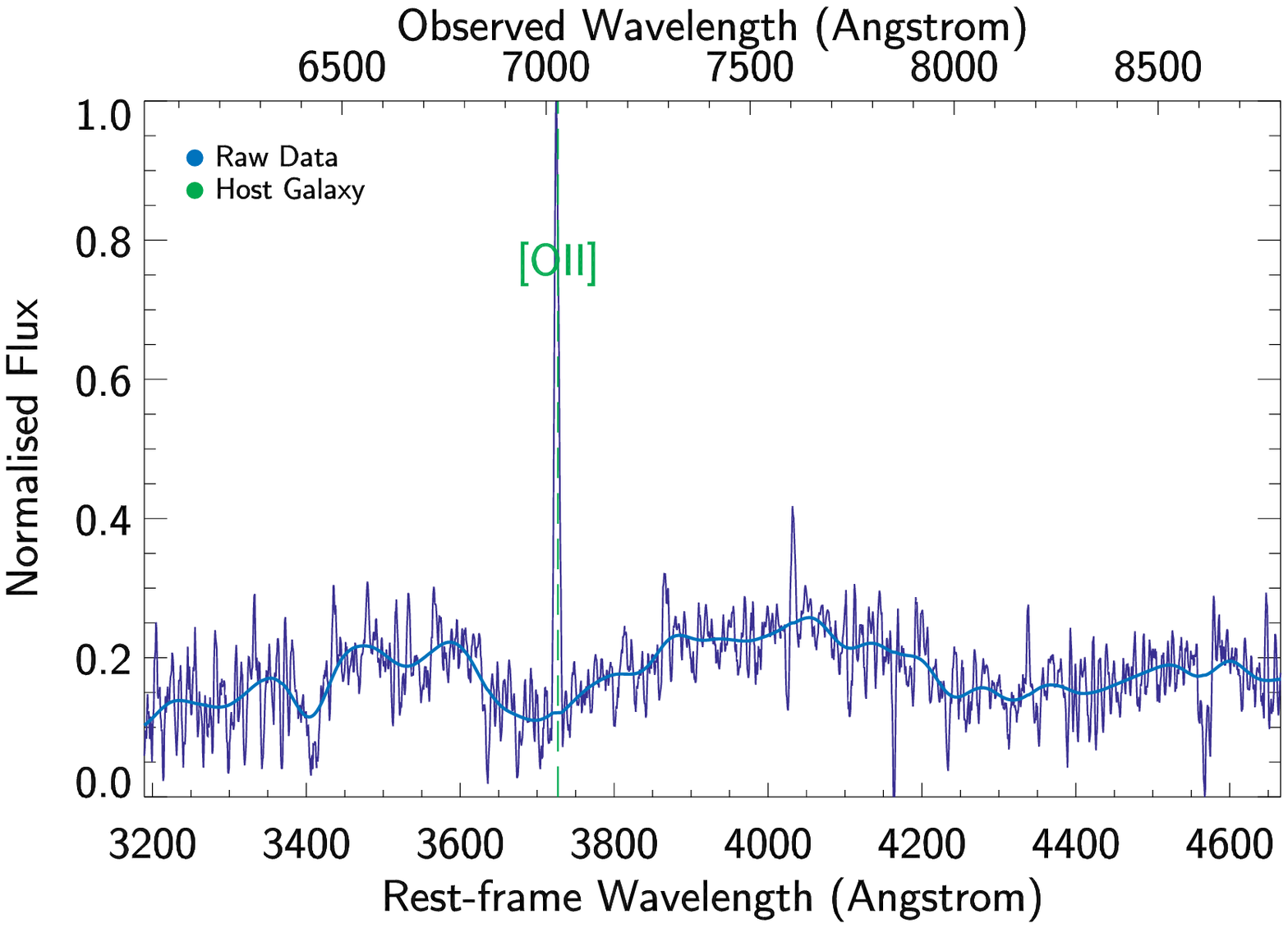}
\label{subfig:07D2cy_data2}
}
\subfigure[07D3ae $z=0.237$]{
\includegraphics[scale=0.40]{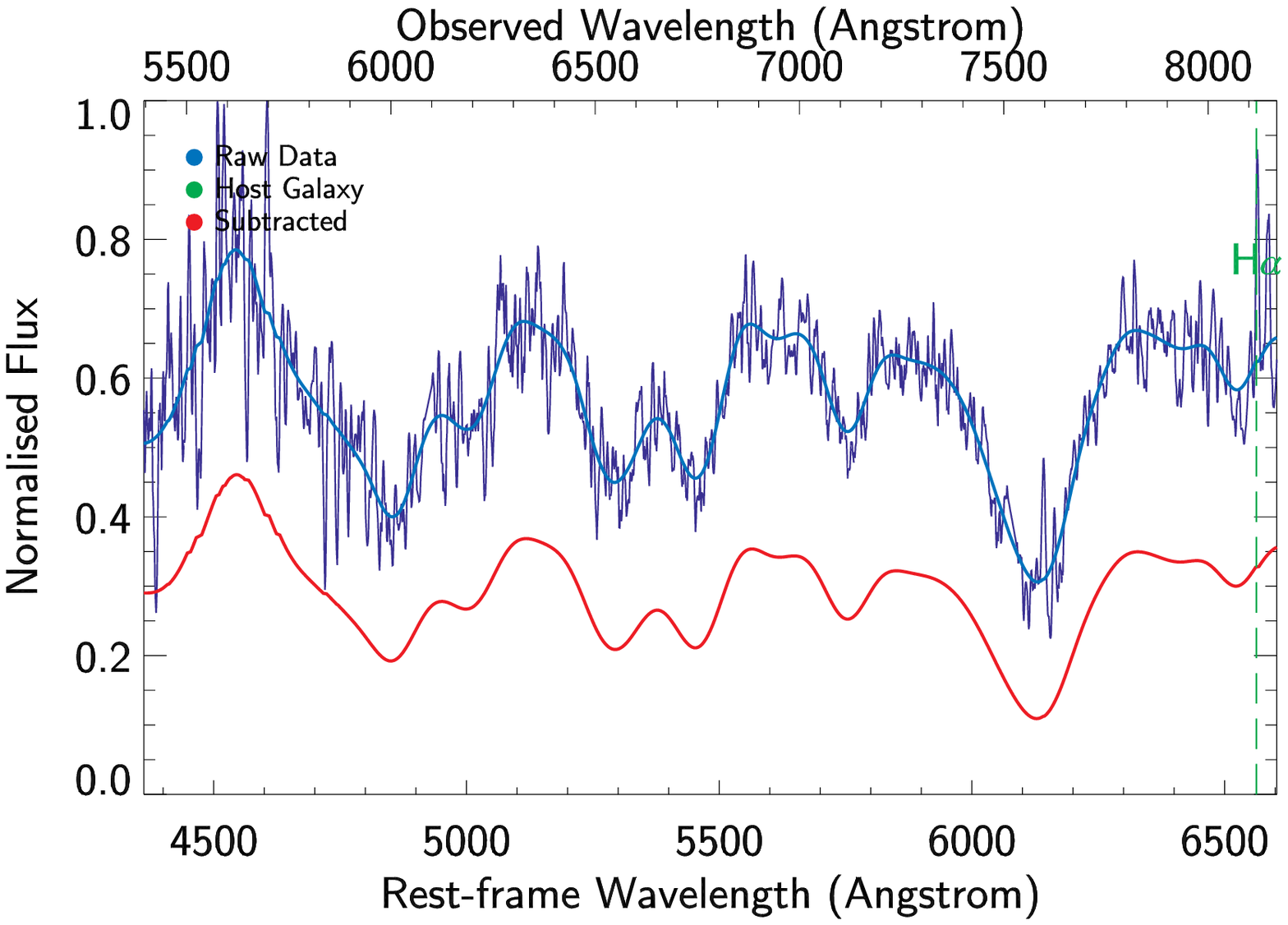}
\label{subfig:07D3ae_data2}
}
\subfigure[07D3af $z=0.237$]{
\includegraphics[scale=0.40]{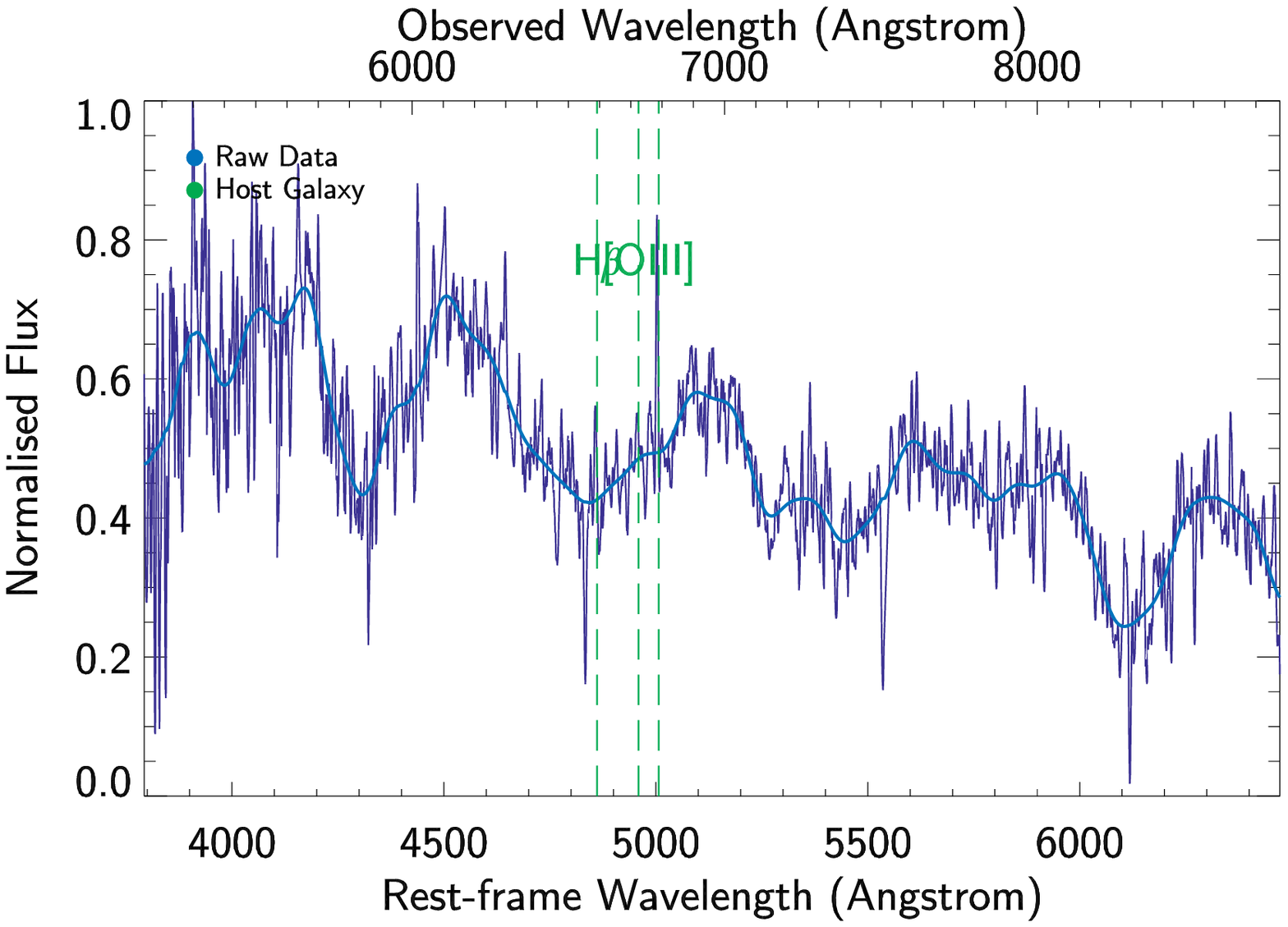}
\label{subfig:07D3af_data2}
}
\subfigure[07D3ap $z=0.451$]{
\includegraphics[scale=0.40]{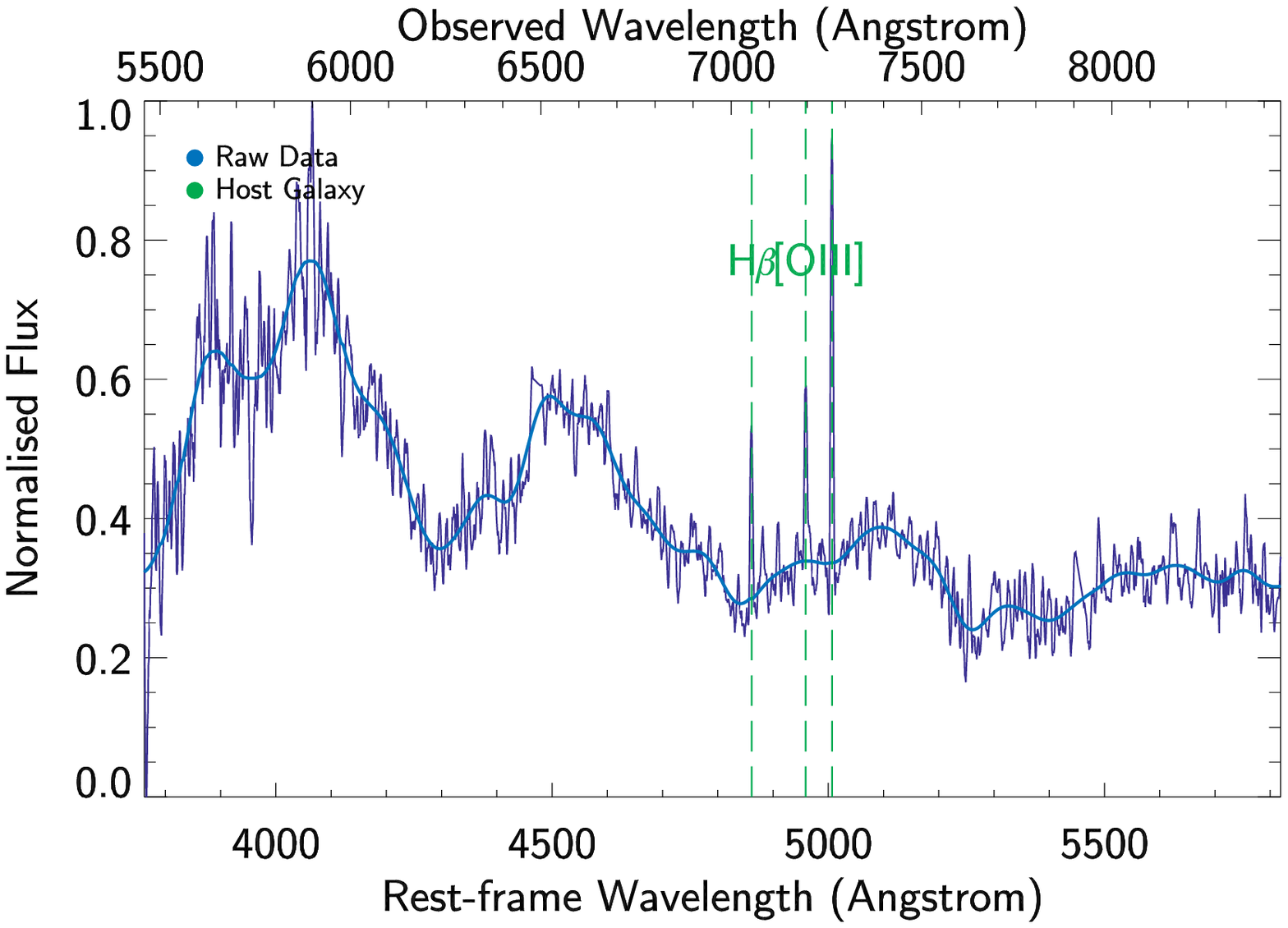}
\label{subfig:07D3ap_data2}
}
\end{center}
\caption{Type Ia Supernovae (CI = 4 or 5)}
\end{figure*}

\begin{figure*}
\begin{center}
\subfigure[07D3cn $z0.898$]{
\includegraphics[scale=0.40]{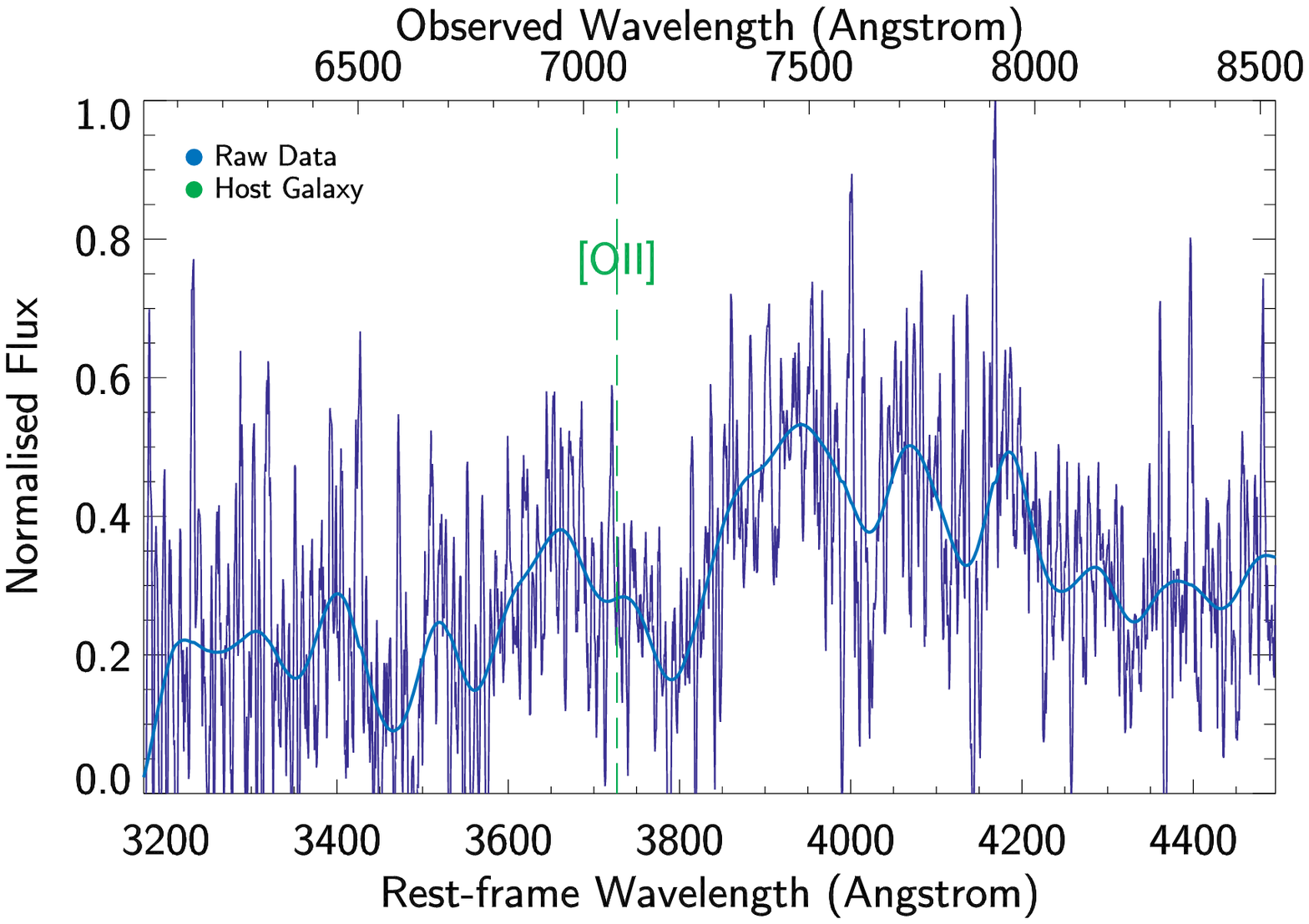}
\label{subfig:07D3cn_data2}
}
\subfigure[07D3cr $z-0.746$]{
\includegraphics[scale=0.40]{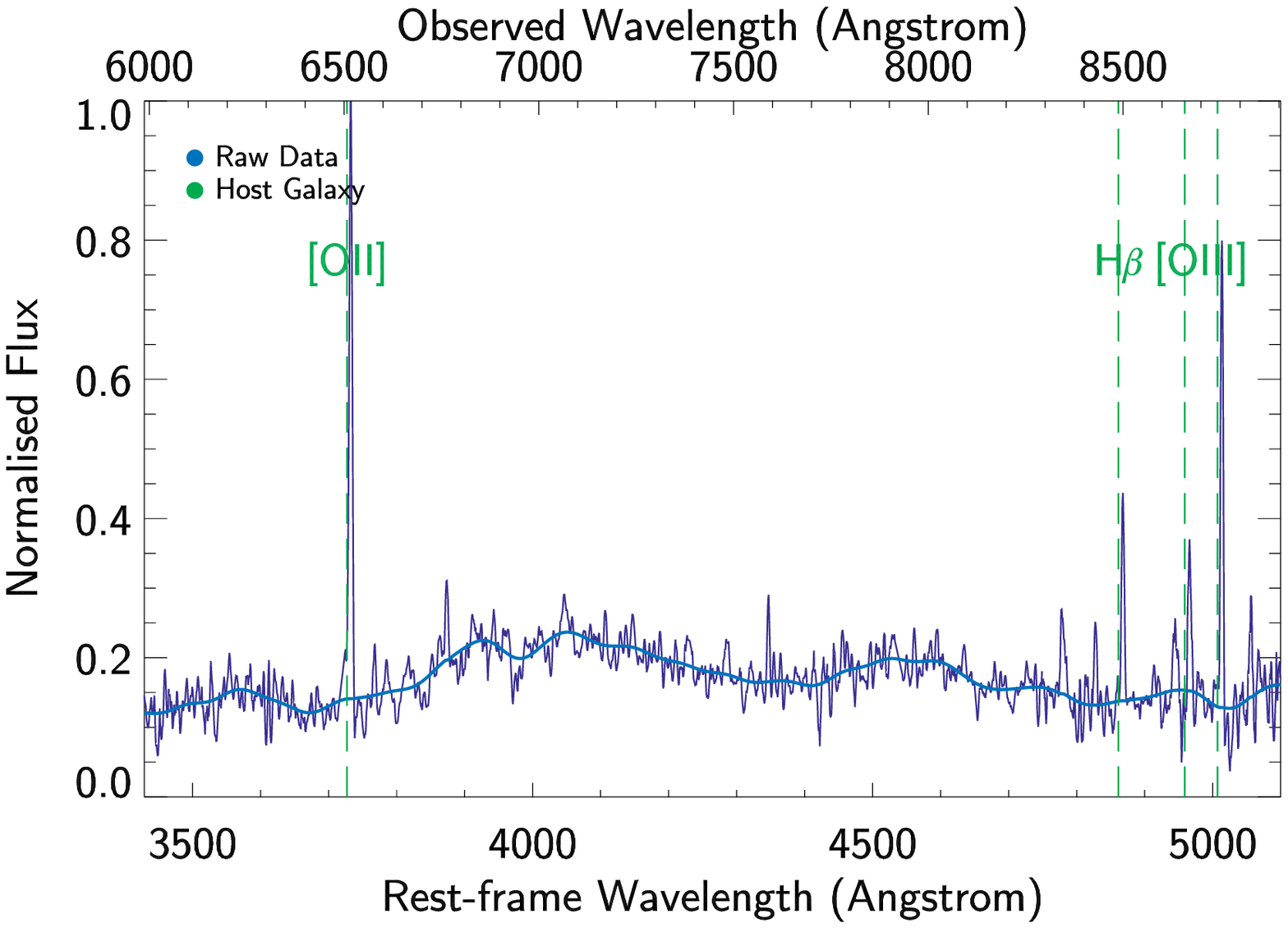}
\label{subfig:07D3cr_data2}
}
\subfigure[07D3cu $z=0.512$]{
\includegraphics[scale=0.40]{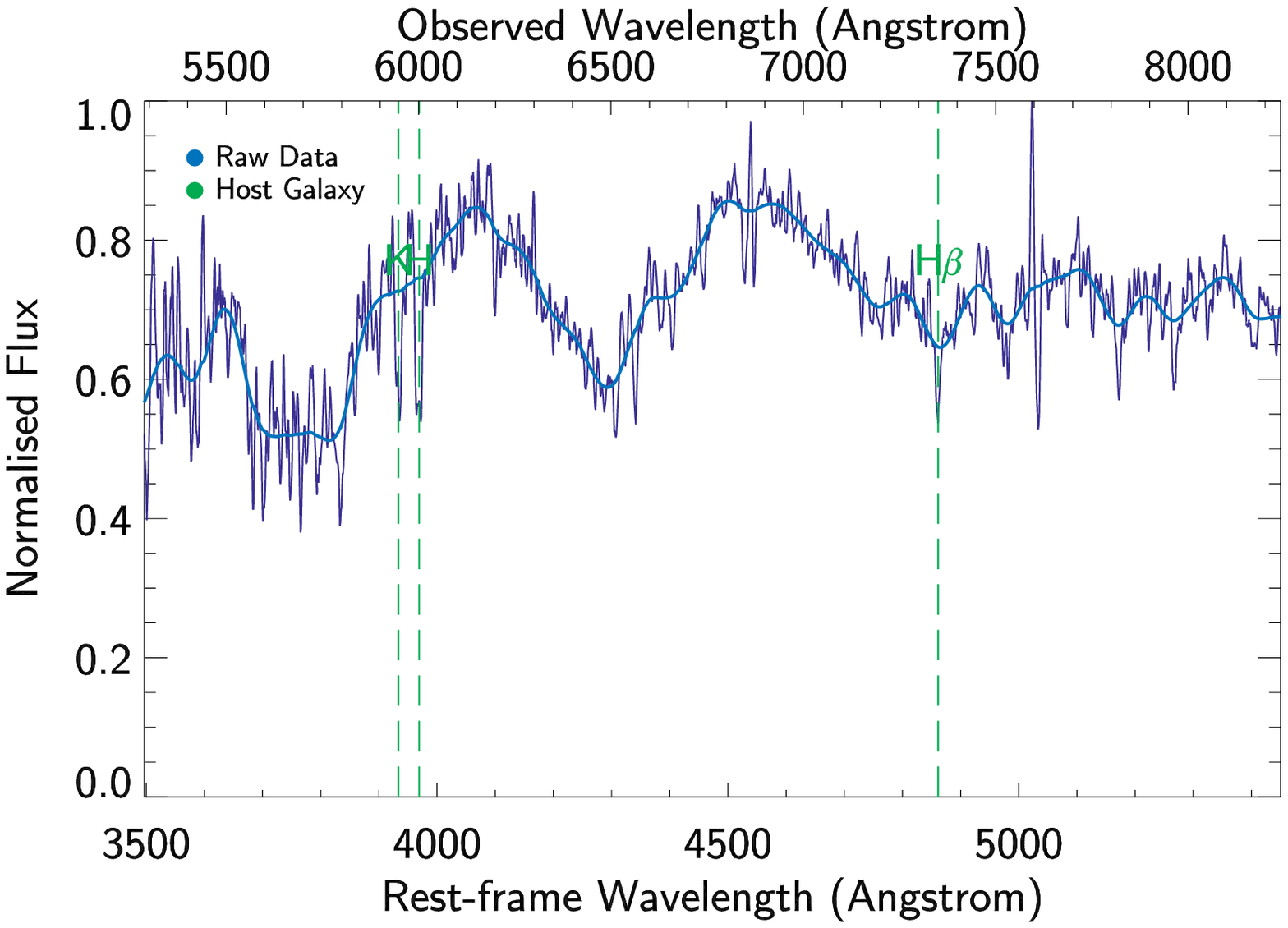}
\label{subfig:07D3cu_data2}
}
\subfigure[07D3da $z=0.837$]{
\includegraphics[scale=0.40]{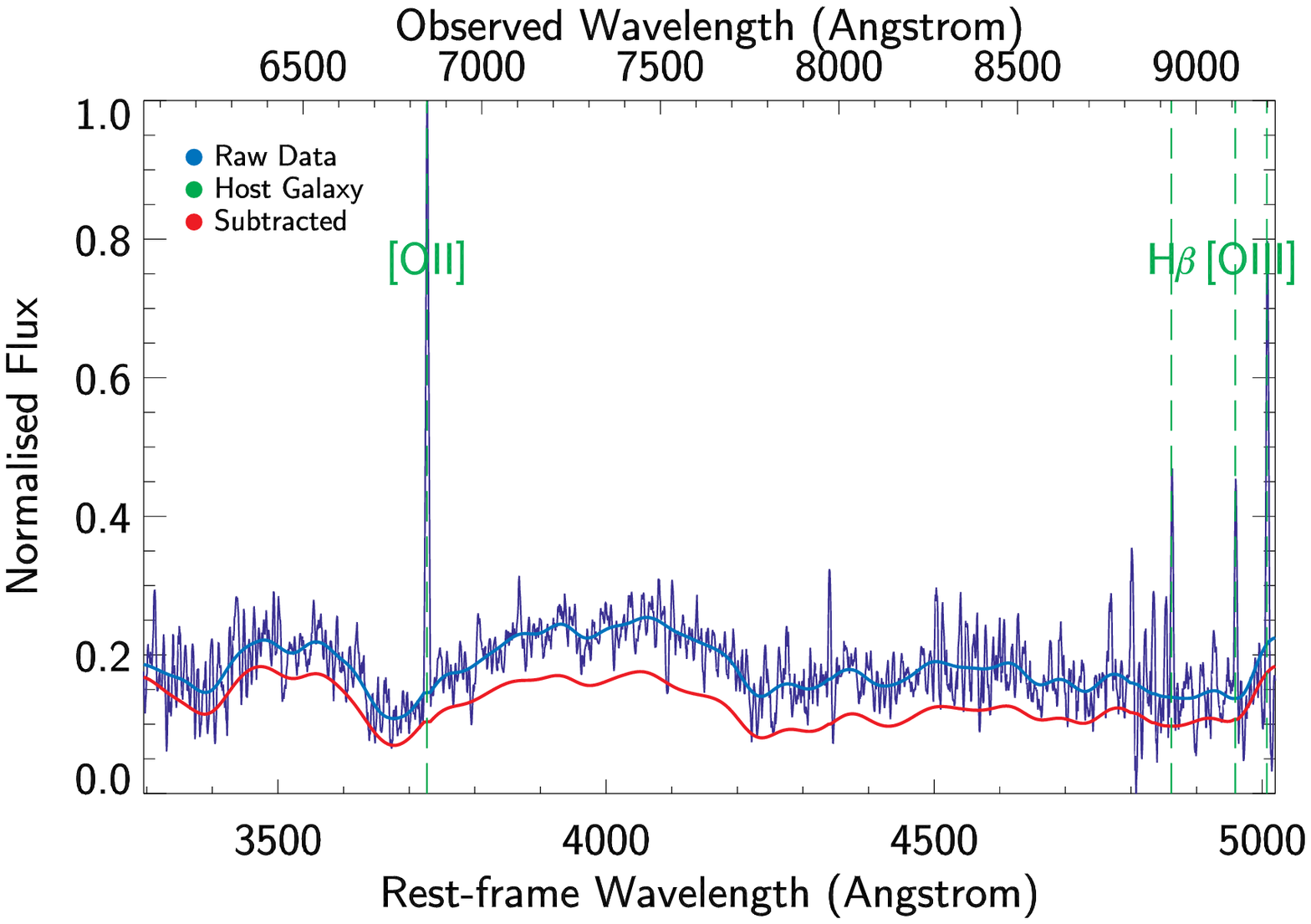}
\label{subfig:07D3da_data2}
}
\subfigure[07D3dj $z=0.444$]{
\includegraphics[scale=0.40]{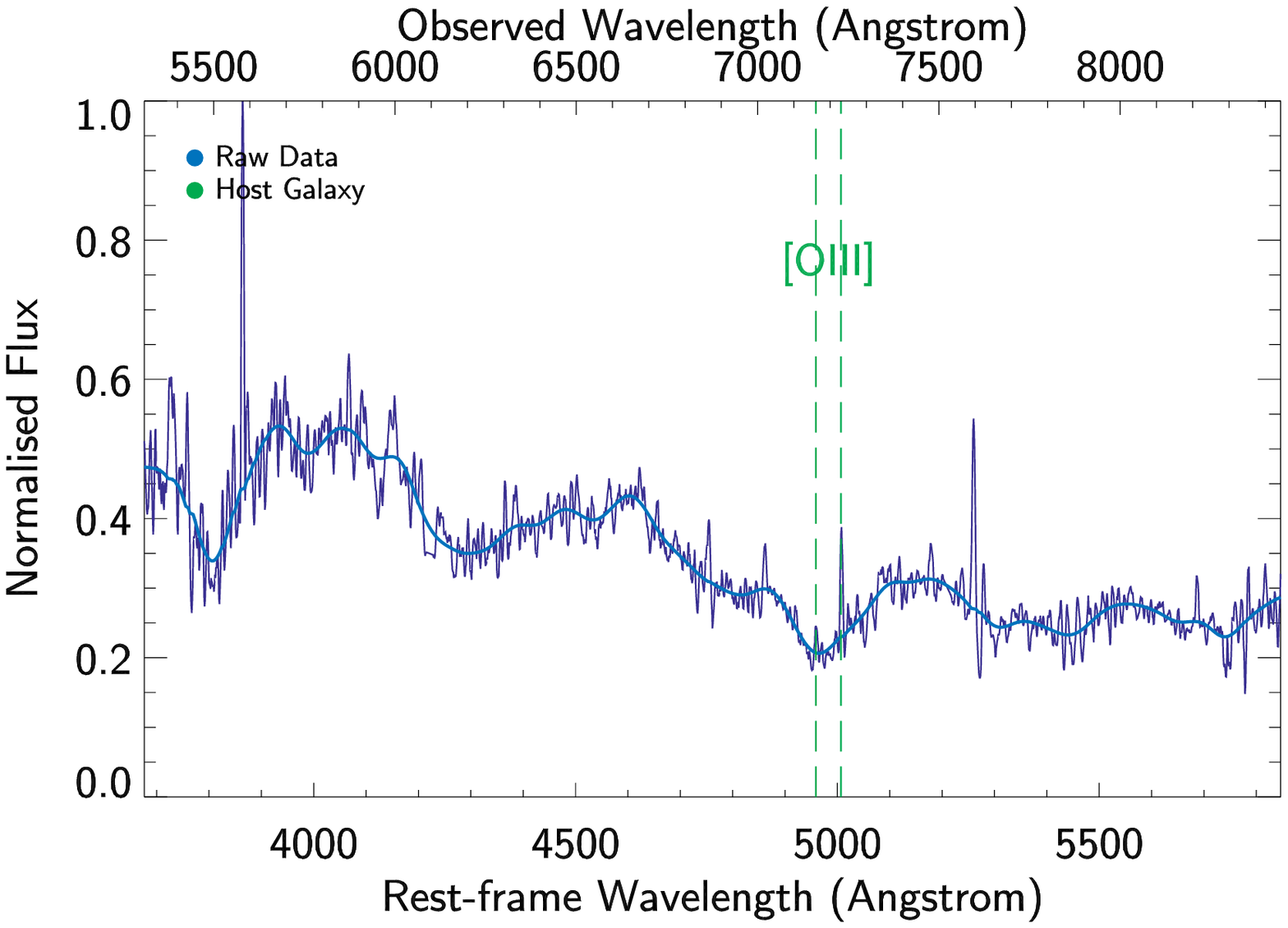}
\label{subfig:07D3dj_data2}
}
\subfigure[07D3ea $z=0.471$]{
\includegraphics[scale=0.40]{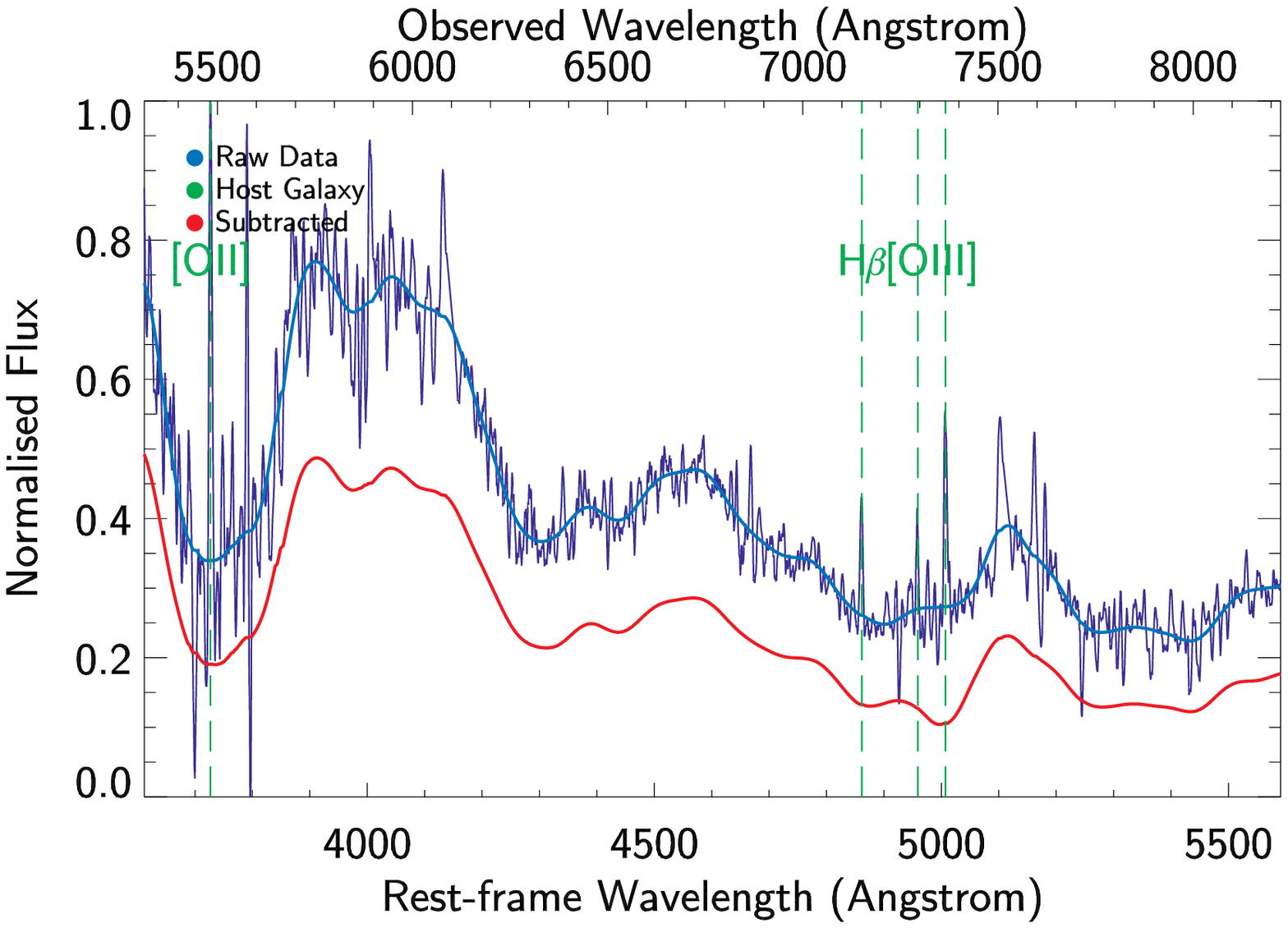}
\label{subfig:07D3ea_data2}
}
\subfigure[07D3ey $z=0.740$]{
\includegraphics[scale=0.40]{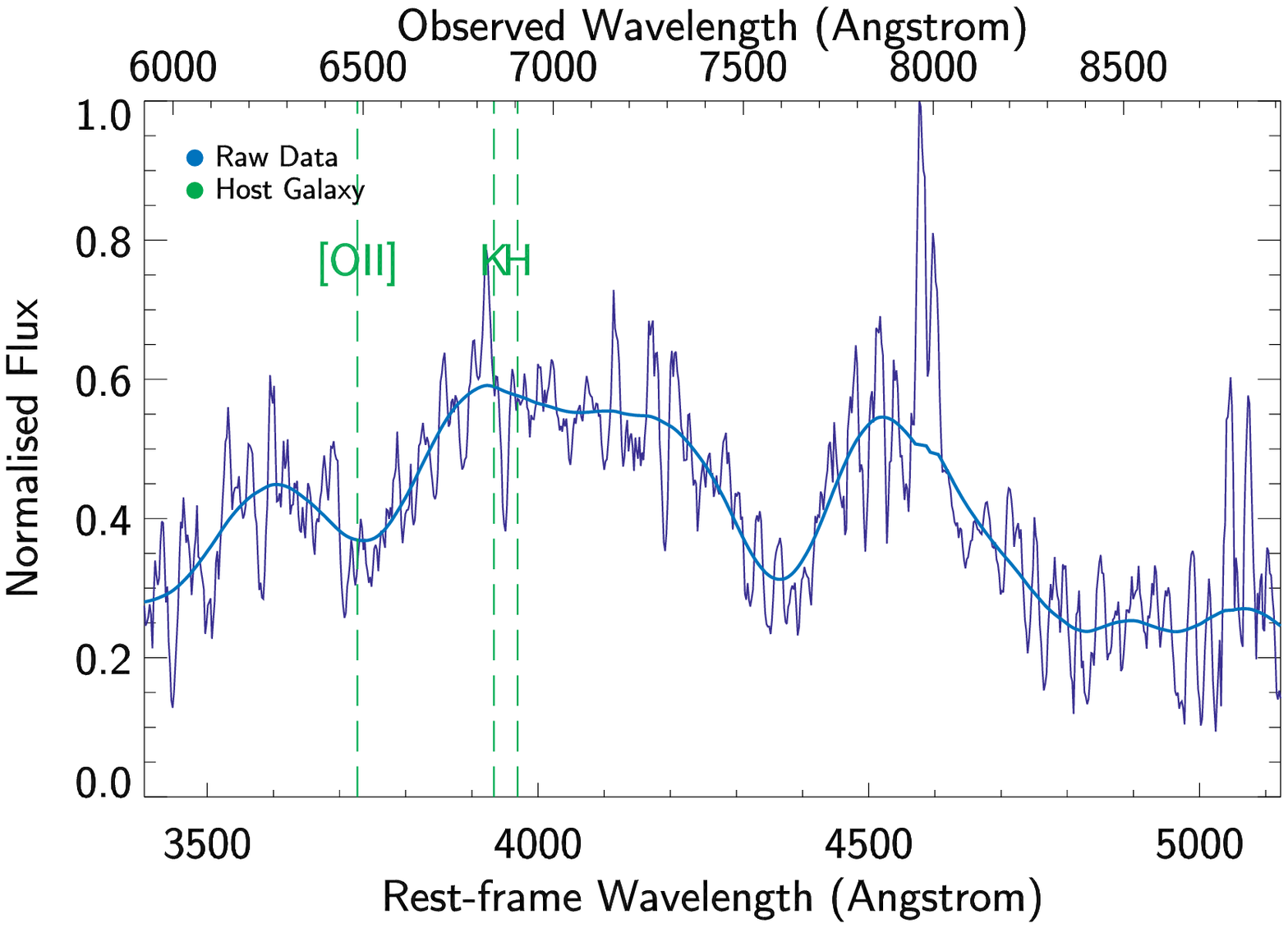}
\label{subfig:07D3ey_data2}
}
\subfigure[07D3gm $z=0.83$]{
\includegraphics[scale=0.40]{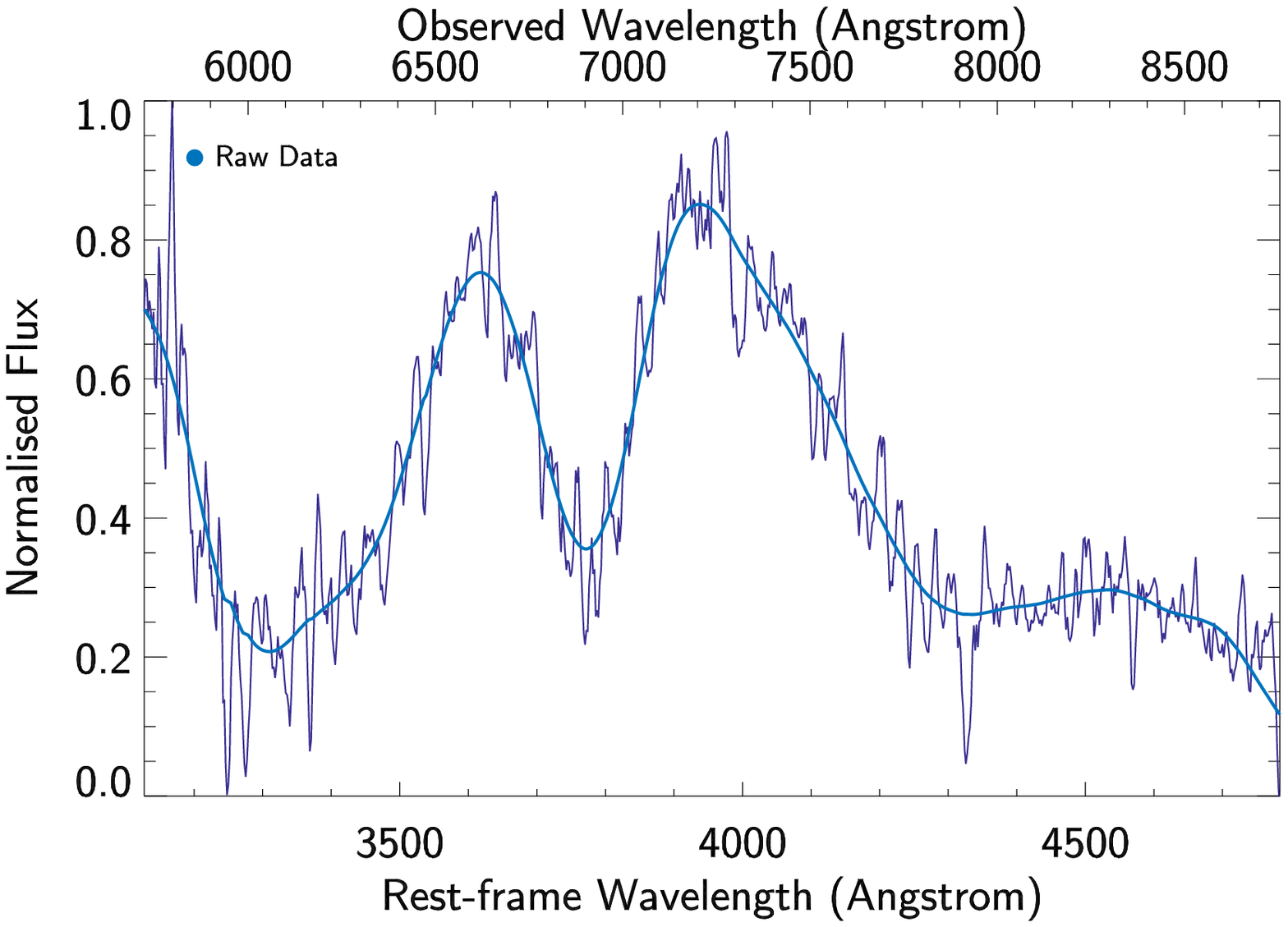}
\label{subfig:07D3gm_data2}
}
\end{center}
\caption{Type Ia Supernovae (CI = 4 or 5)}
\end{figure*}

\begin{figure*}
\begin{center}
\subfigure[07D3hl $z=0.67$]{
\includegraphics[scale=0.40]{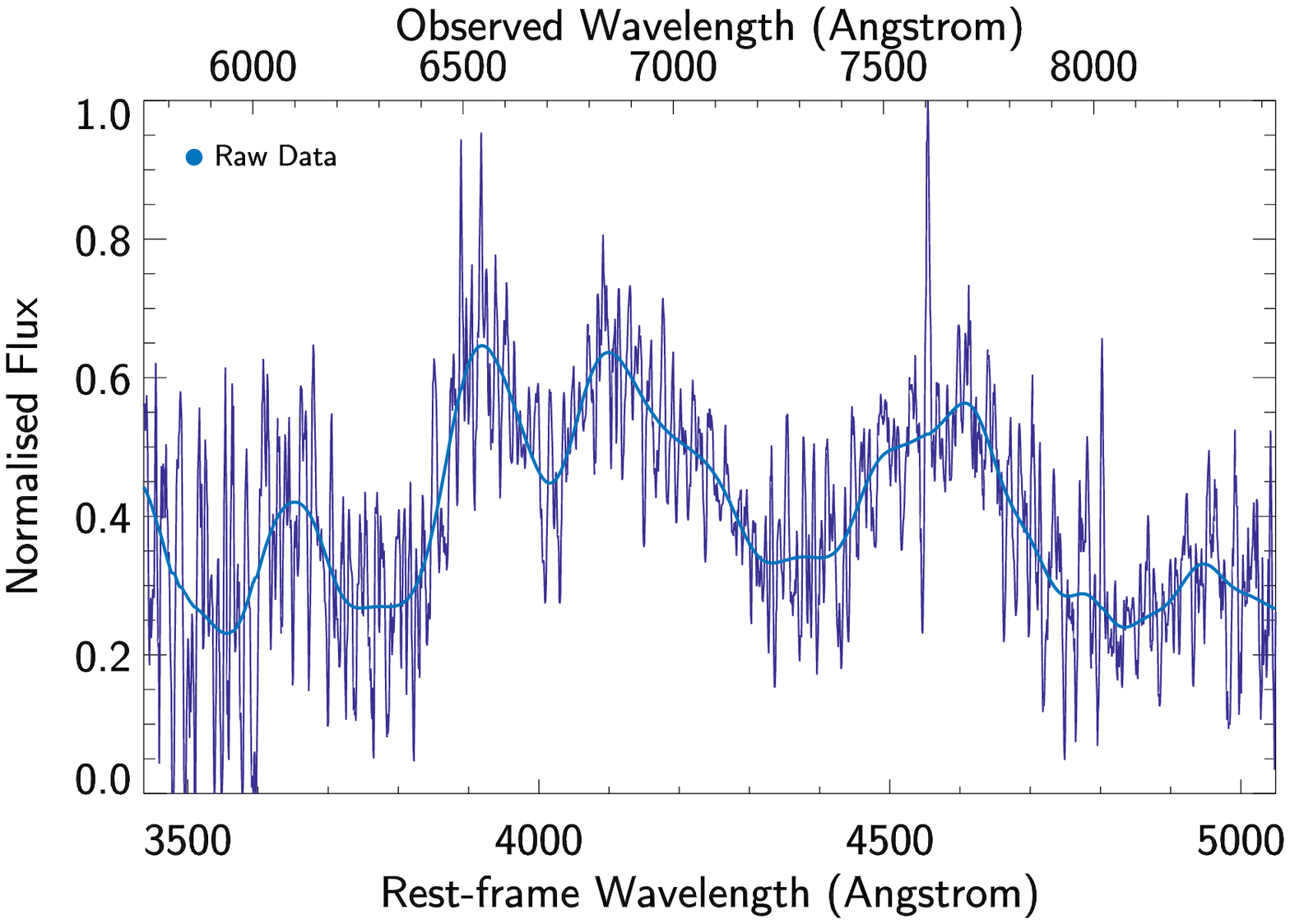}
\label{subfig:07D3hl_data2}
}
\subfigure[07D3hu $z=0.572$]{
\includegraphics[scale=0.40]{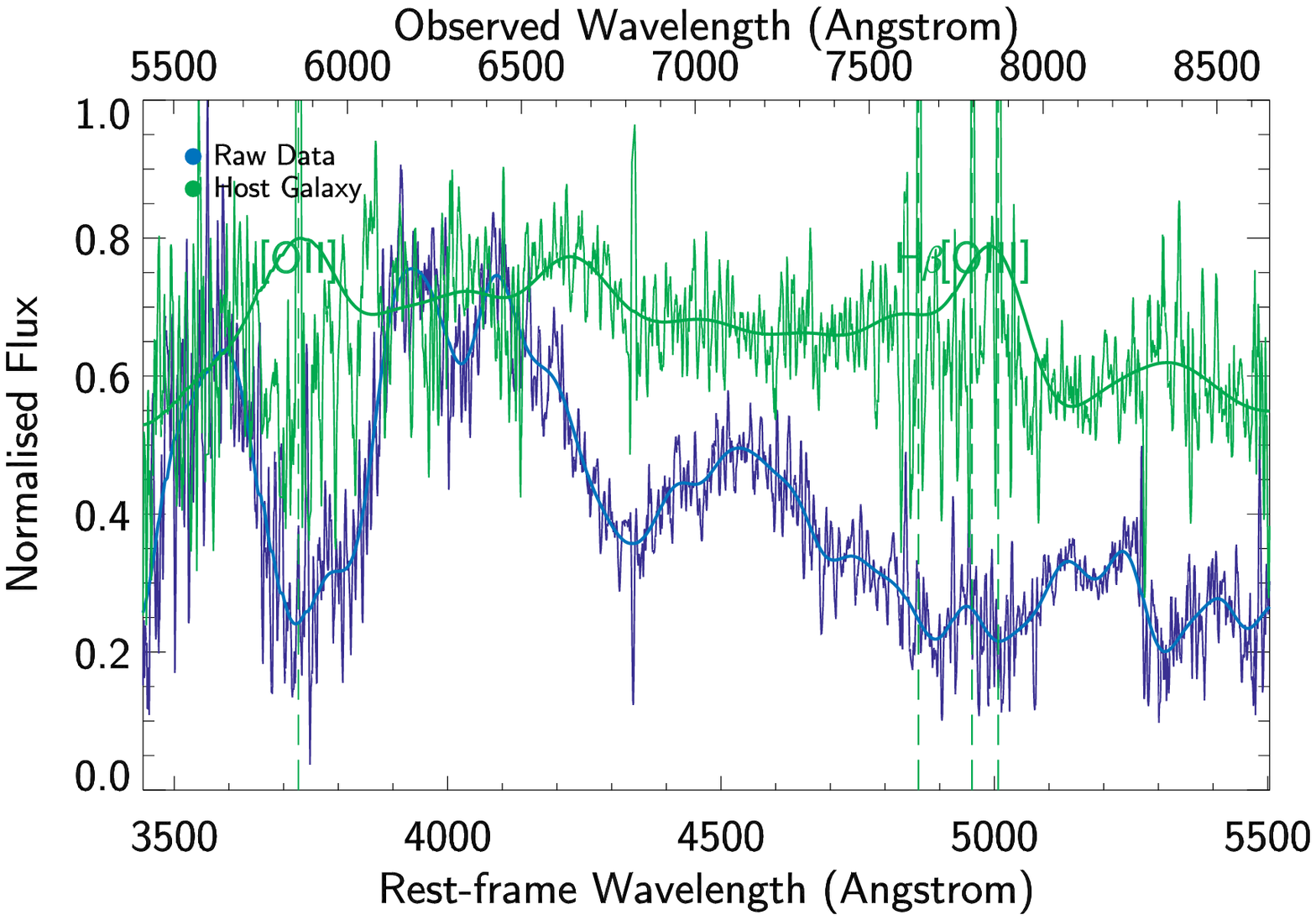}
\label{subfig:07D3hu_data2}
}
\subfigure[07D3hv $z=0.351$]{
\includegraphics[scale=0.40]{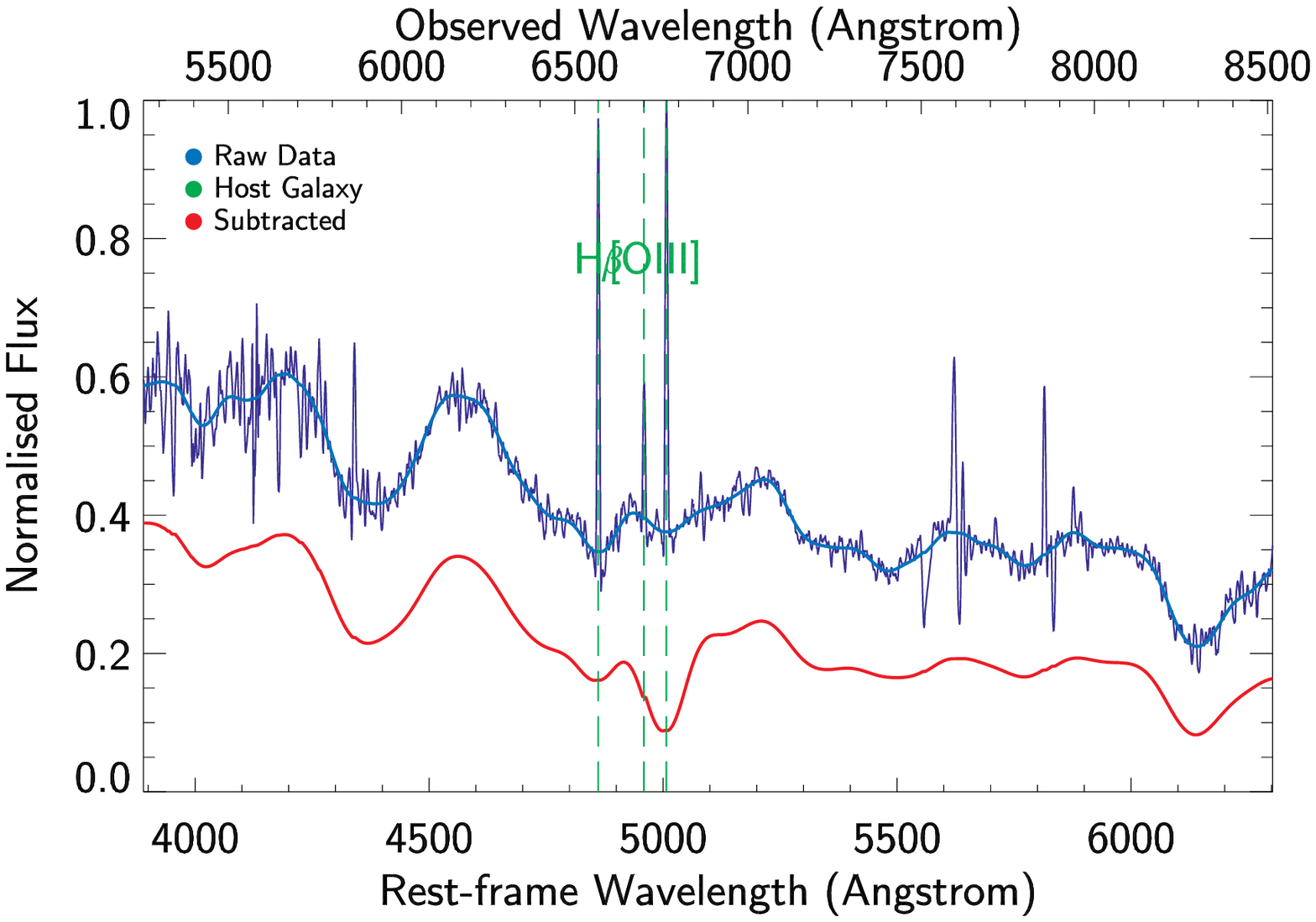}
\label{subfig:07D3hv_data2}
}
\subfigure[07D3hw $z=0.748$]{
\includegraphics[scale=0.40]{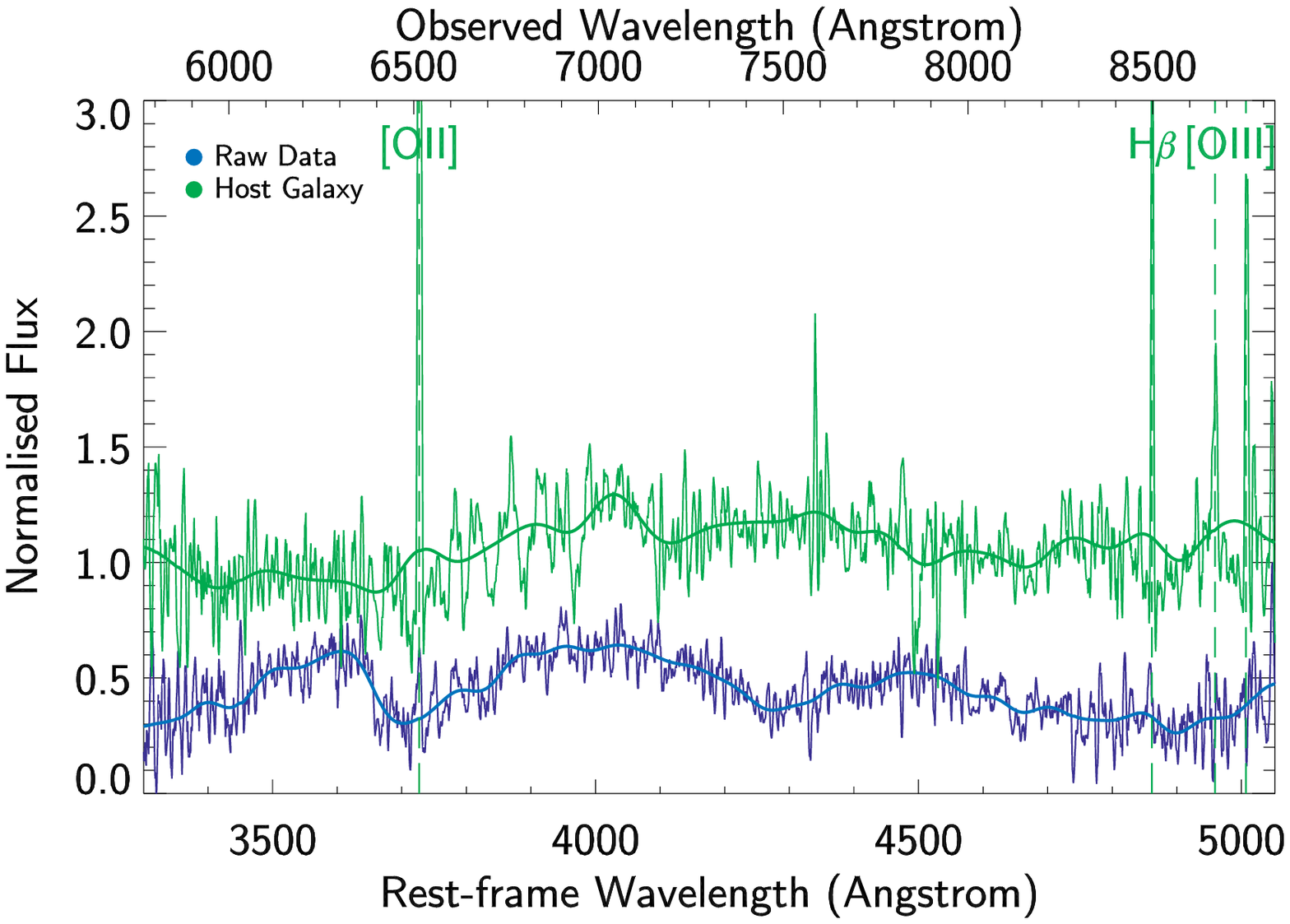}
\label{subfig:07D3hw_data2}
}
\subfigure[07D3hz $z=0.506$]{
\includegraphics[scale=0.40]{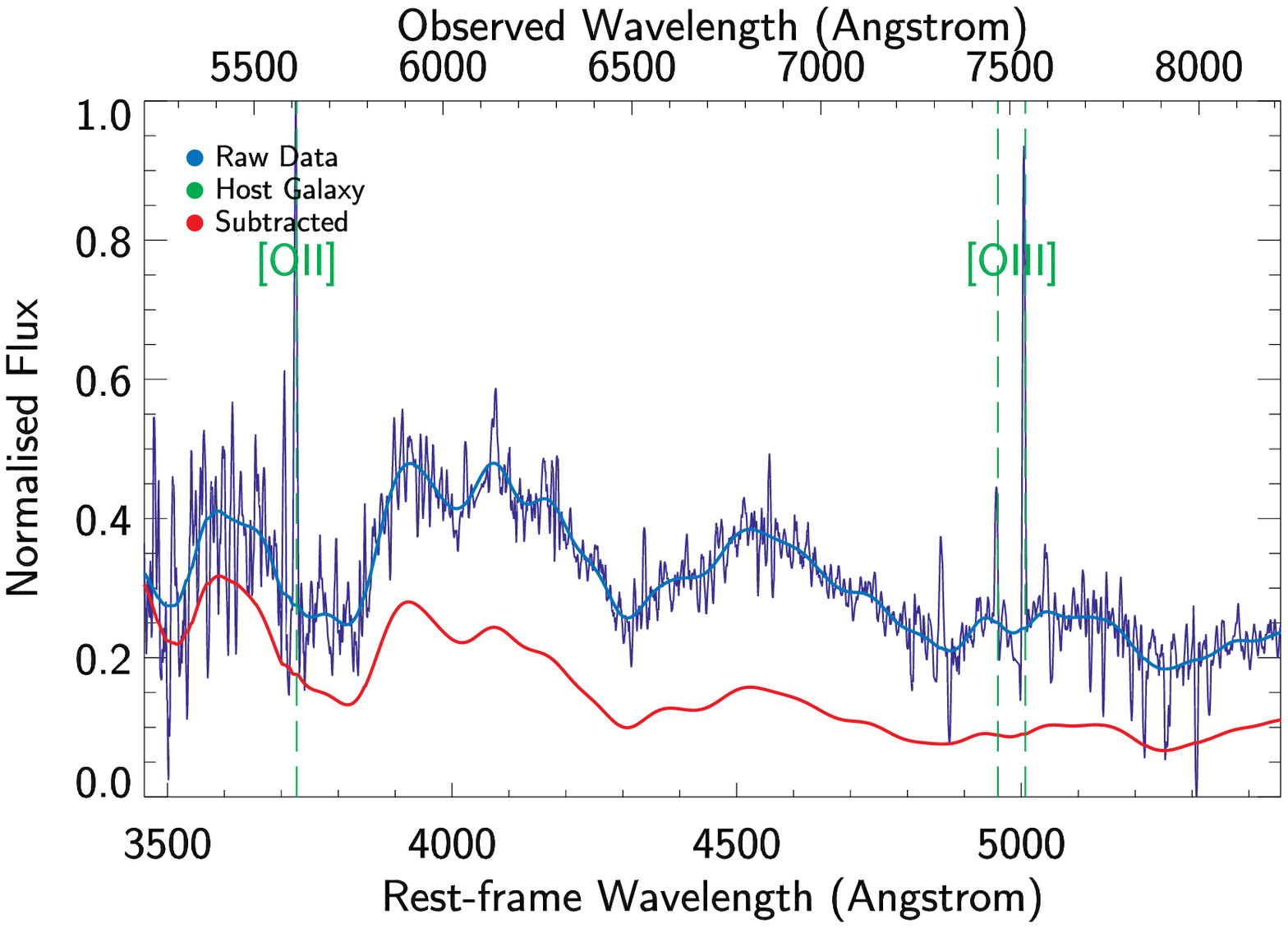}
\label{subfig:07D3hz_data2}
}
\subfigure[07D3ib $z=0.681$]{
\includegraphics[scale=0.40]{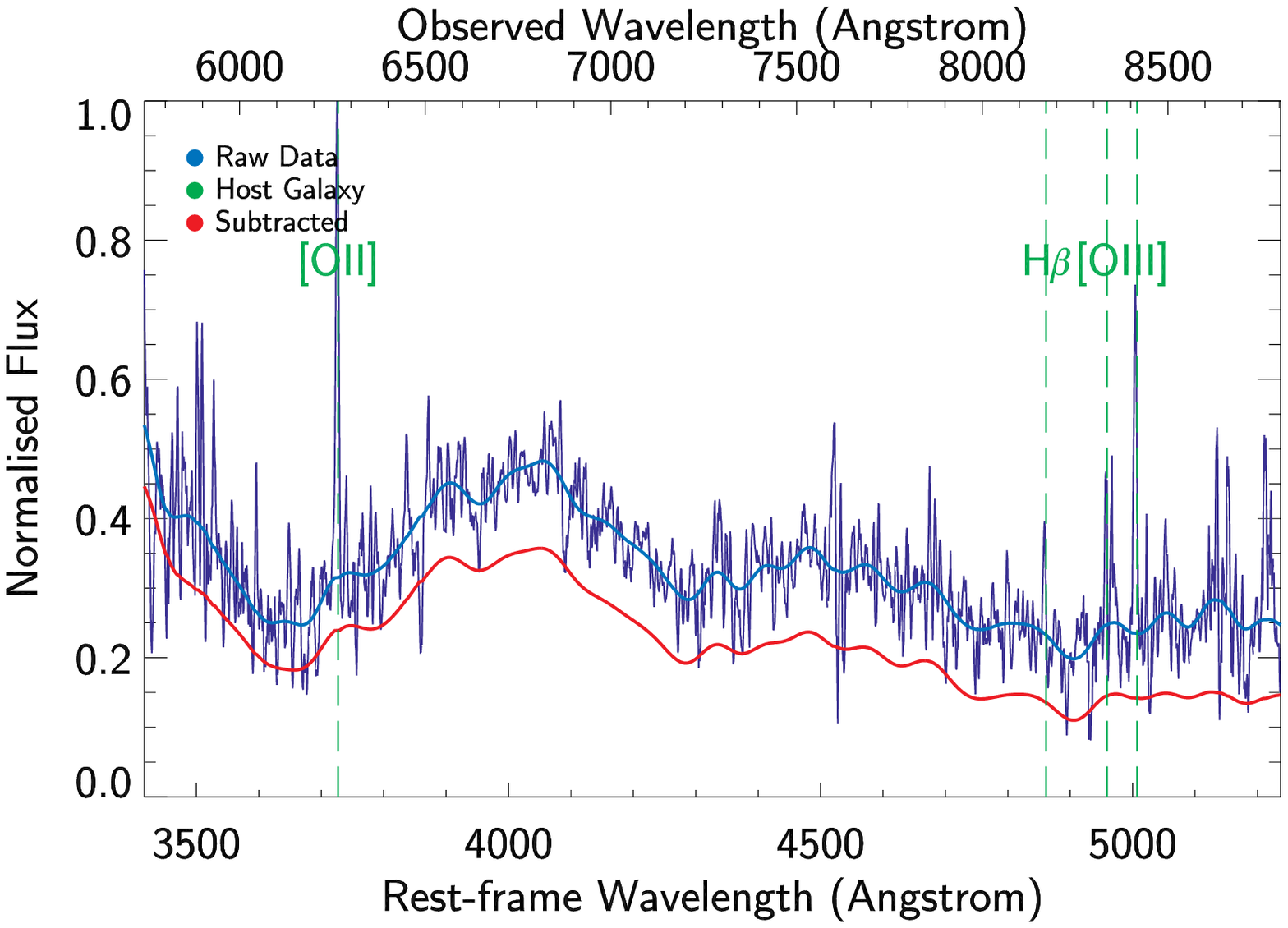}
\label{subfig:07D3ib_data2}
}
\subfigure[07D3it $z=0.835$]{
\includegraphics[scale=0.40]{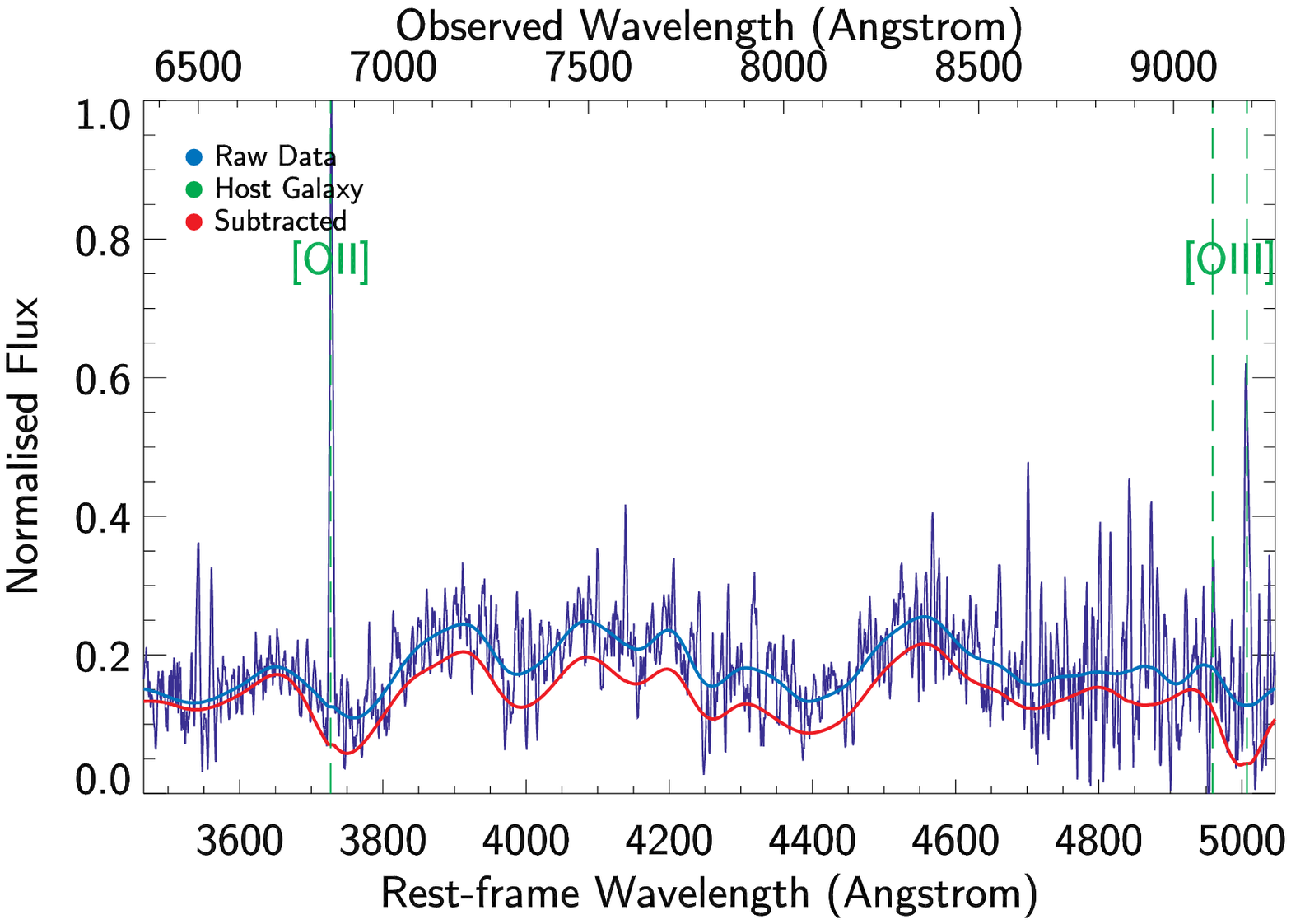}
\label{subfig:07D3it_data2}
}
\subfigure[07D4fl $z=0.503$]{
\includegraphics[scale=0.40]{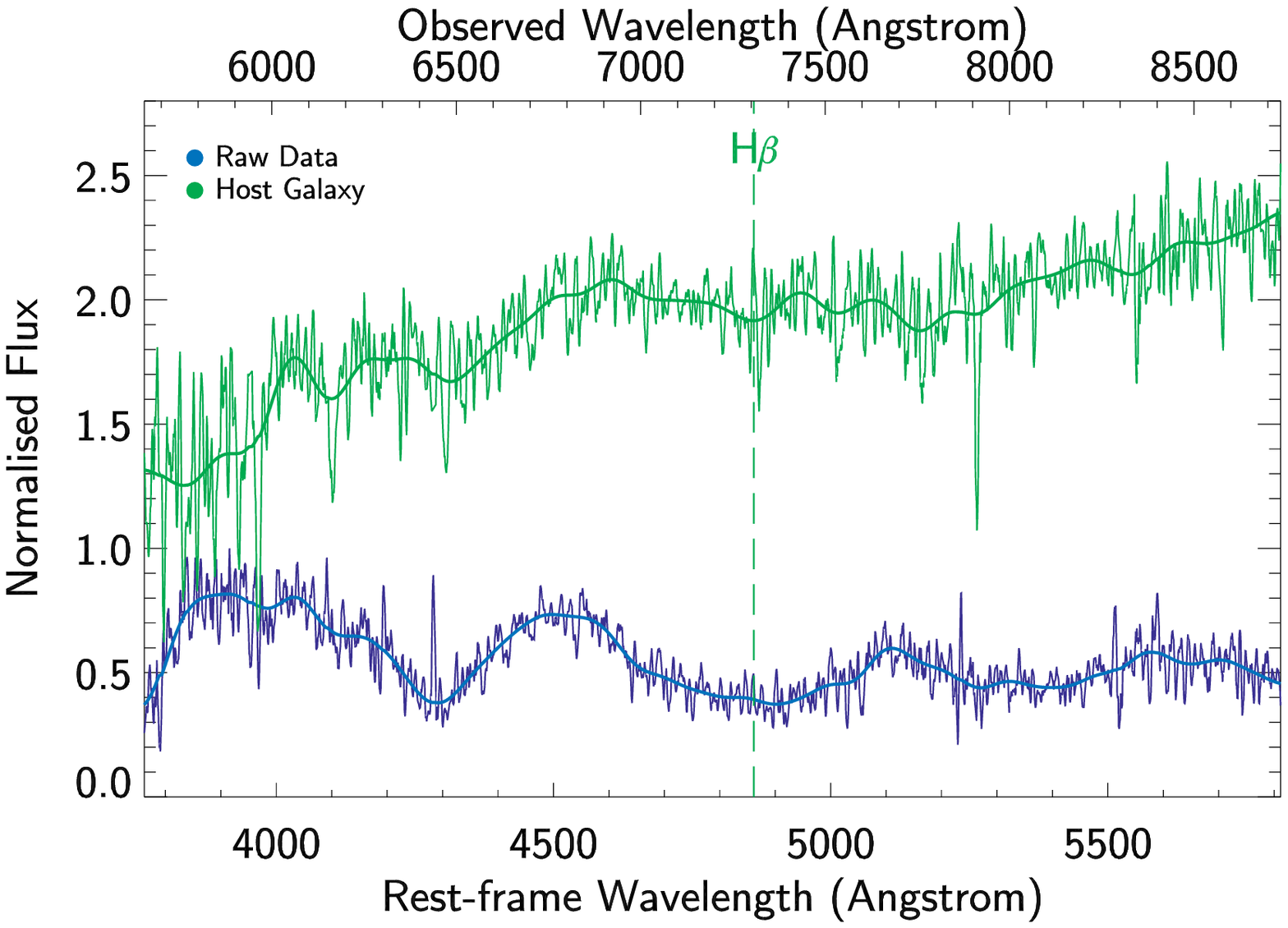}
\label{subfig:07D4fl_data2}
}
\end{center}
\caption{Type Ia Supernovae (CI = 4 or 5)}
\end{figure*}

\begin{figure*}
\begin{center}
\subfigure[08D2aa $z=0.538$]{
\includegraphics[scale=0.40]{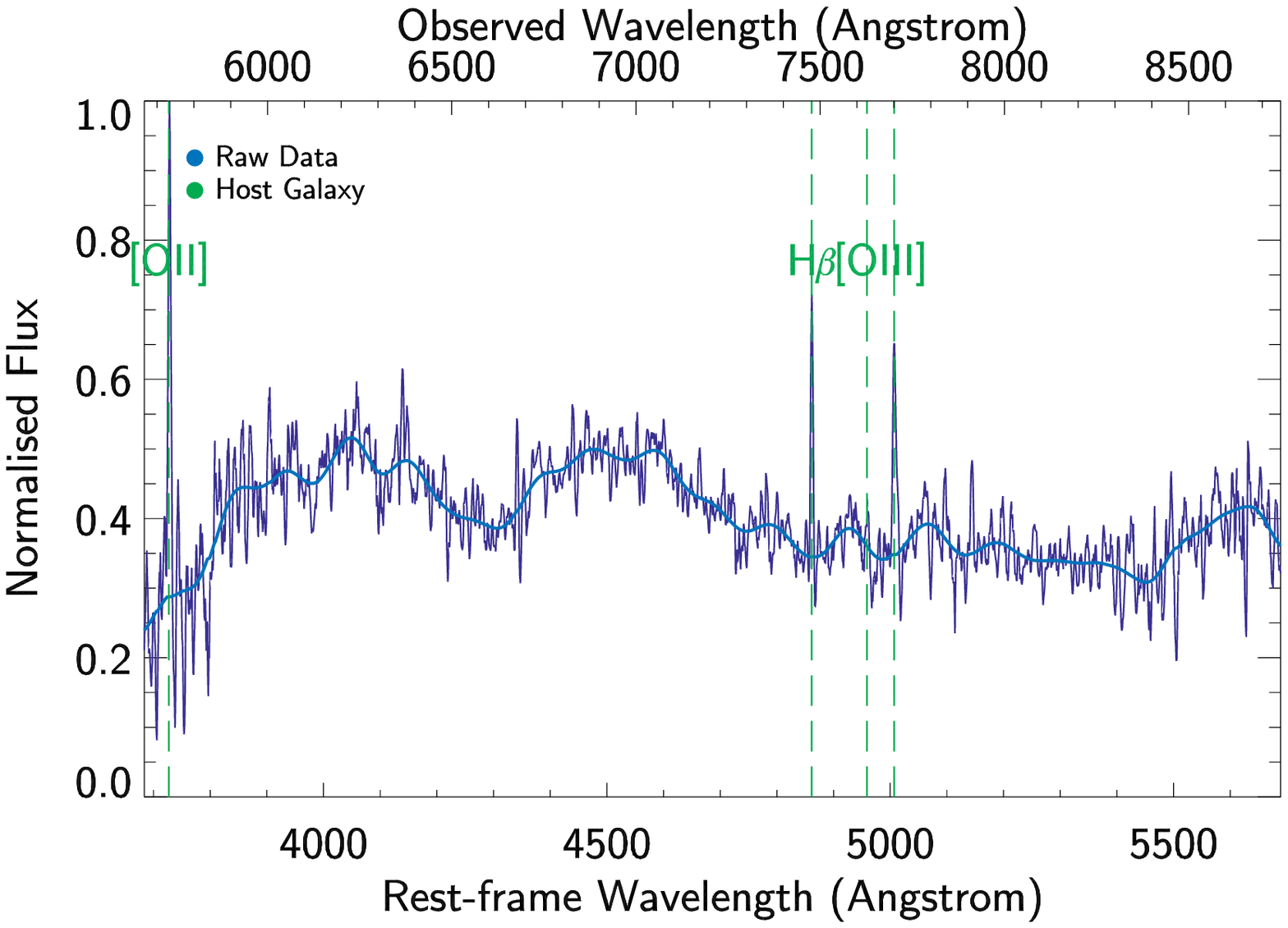}
\label{subfig:08D2aa_data2}
}
\subfigure[08D2ad $z=0.554$]{
\includegraphics[scale=0.40]{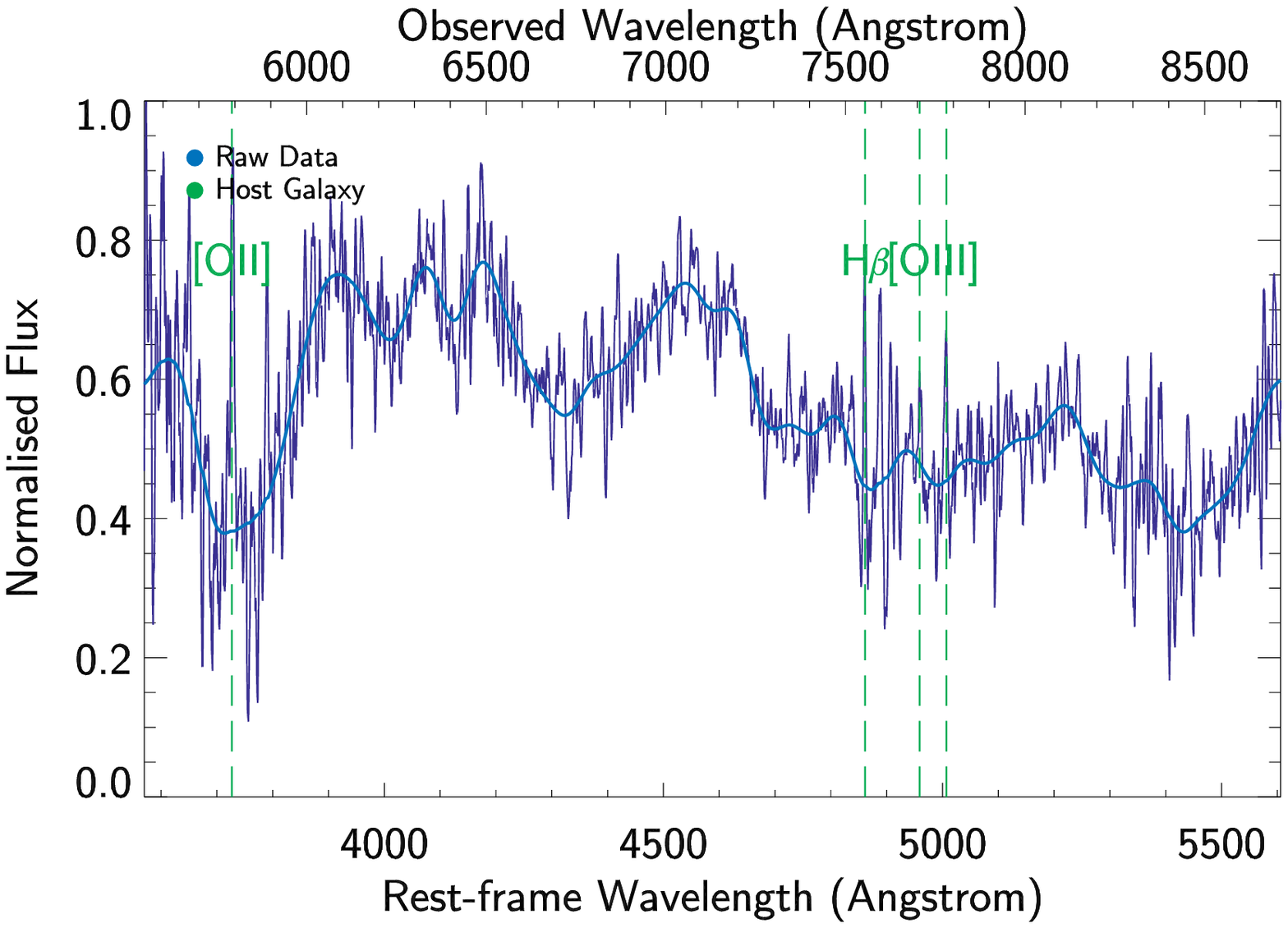}
\label{subfig:08D2ad_data2}
}
\subfigure[08D2bj $z=0.84$]{
\includegraphics[scale=0.40]{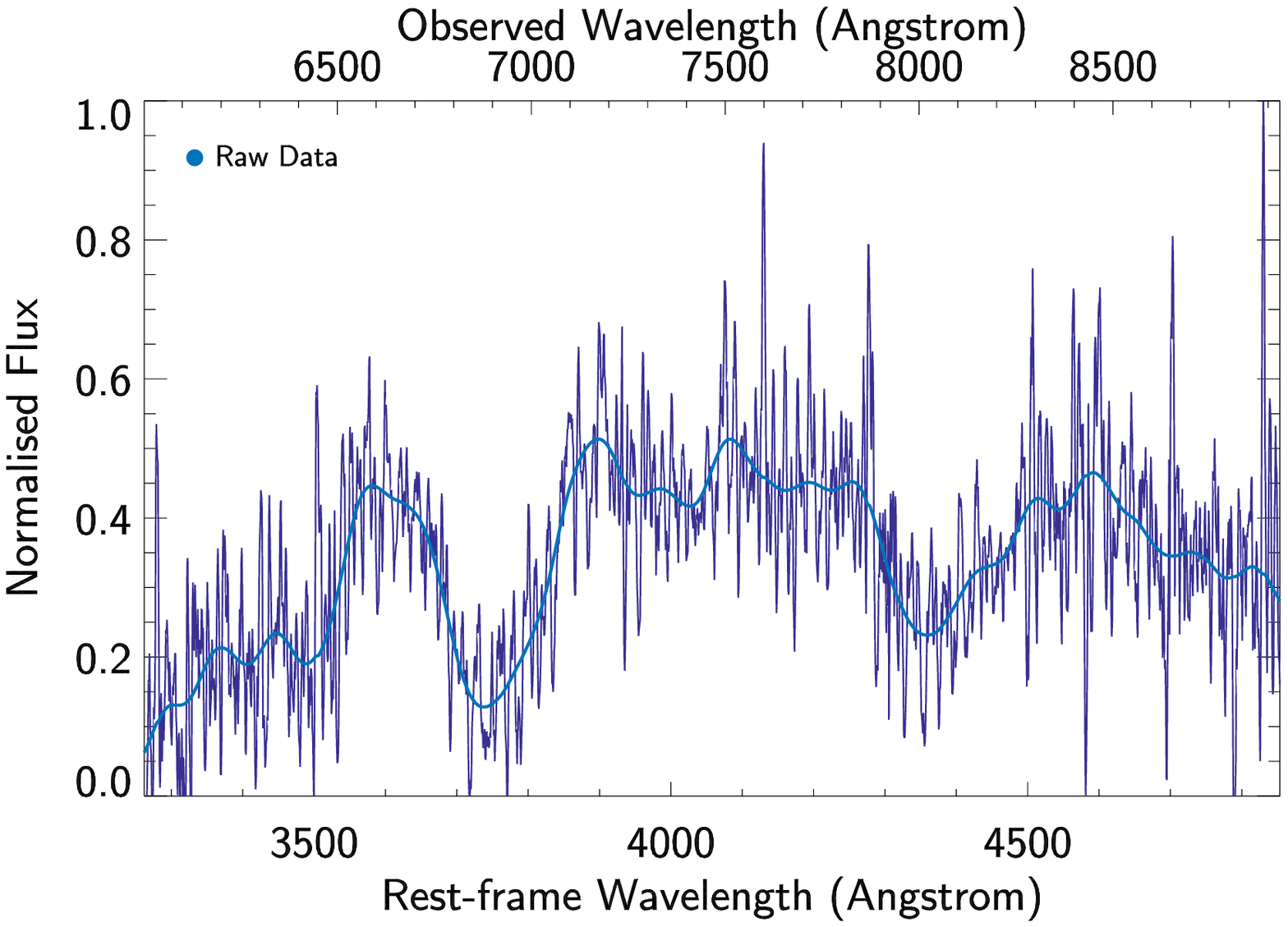}
\label{subfig:08D2bj_data2}
}
\subfigure[08D2dr $z=0.355$]{
\includegraphics[scale=0.40]{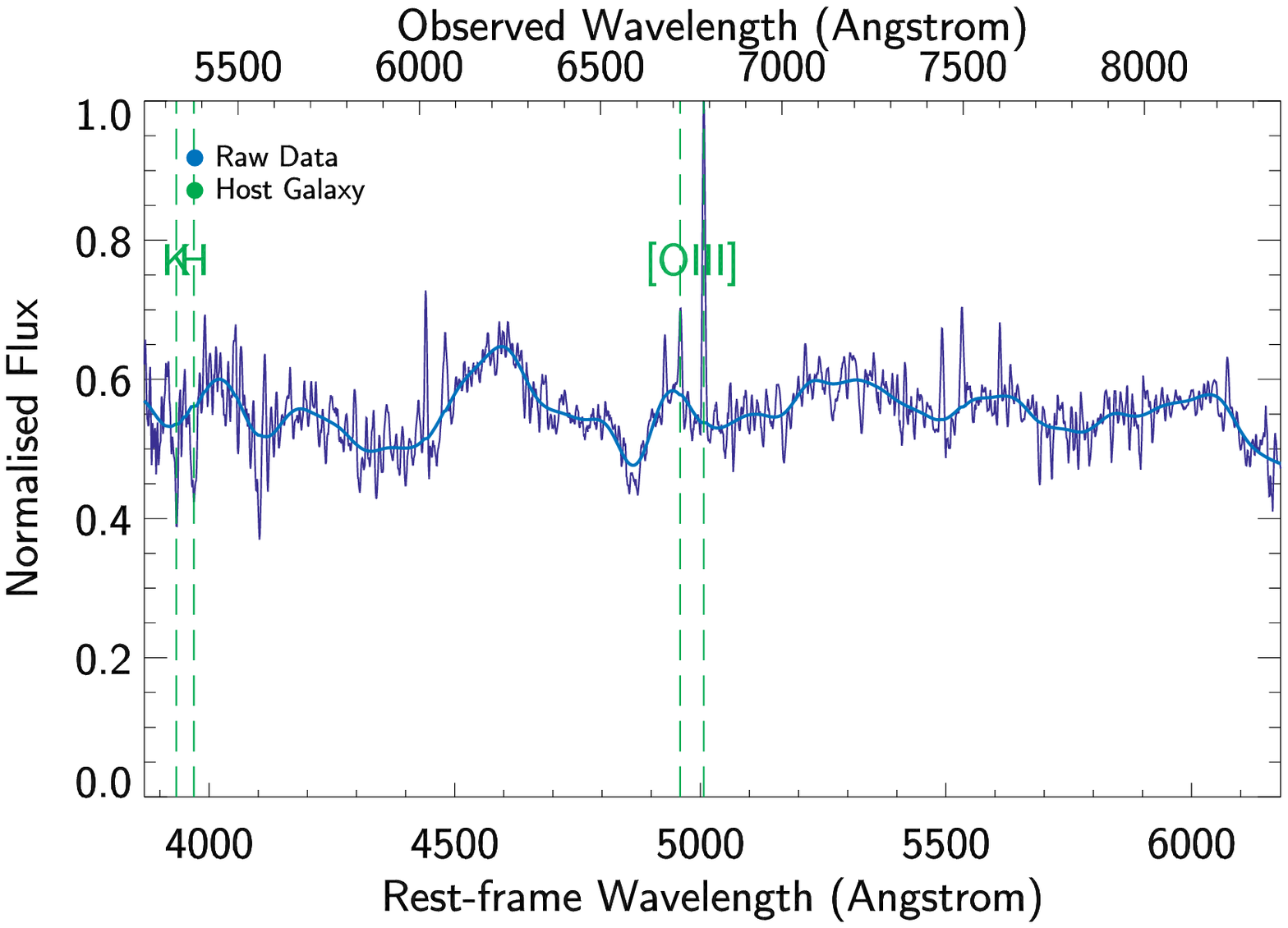}
\label{subfig:08D2dr_data2}
}
\subfigure[08D2dz $z=0.65$]{
\includegraphics[scale=0.40]{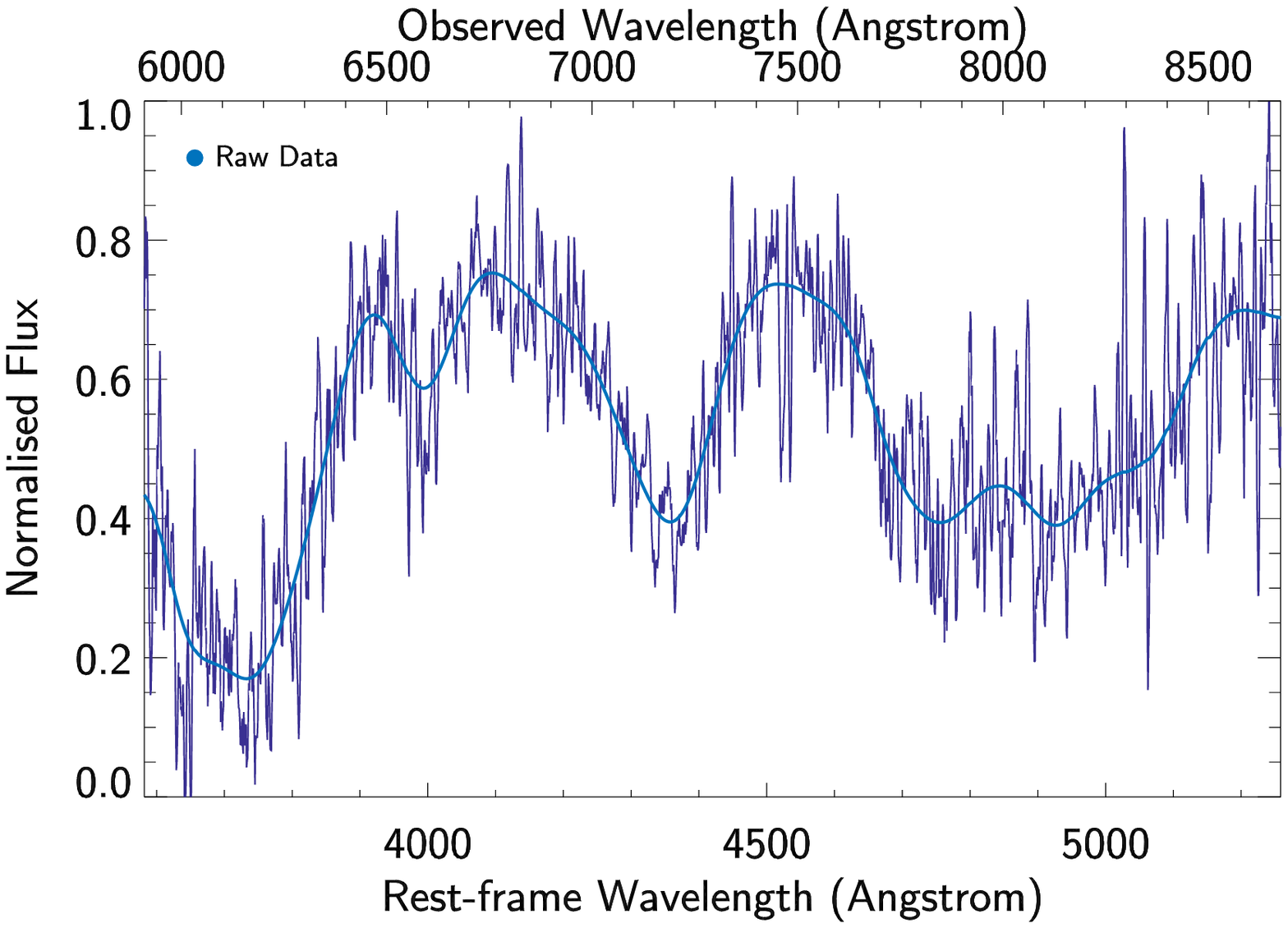}
\label{subfig:08D2dz_data2}
}
\subfigure[08D2gw $z=0.715$]{
\includegraphics[scale=0.40]{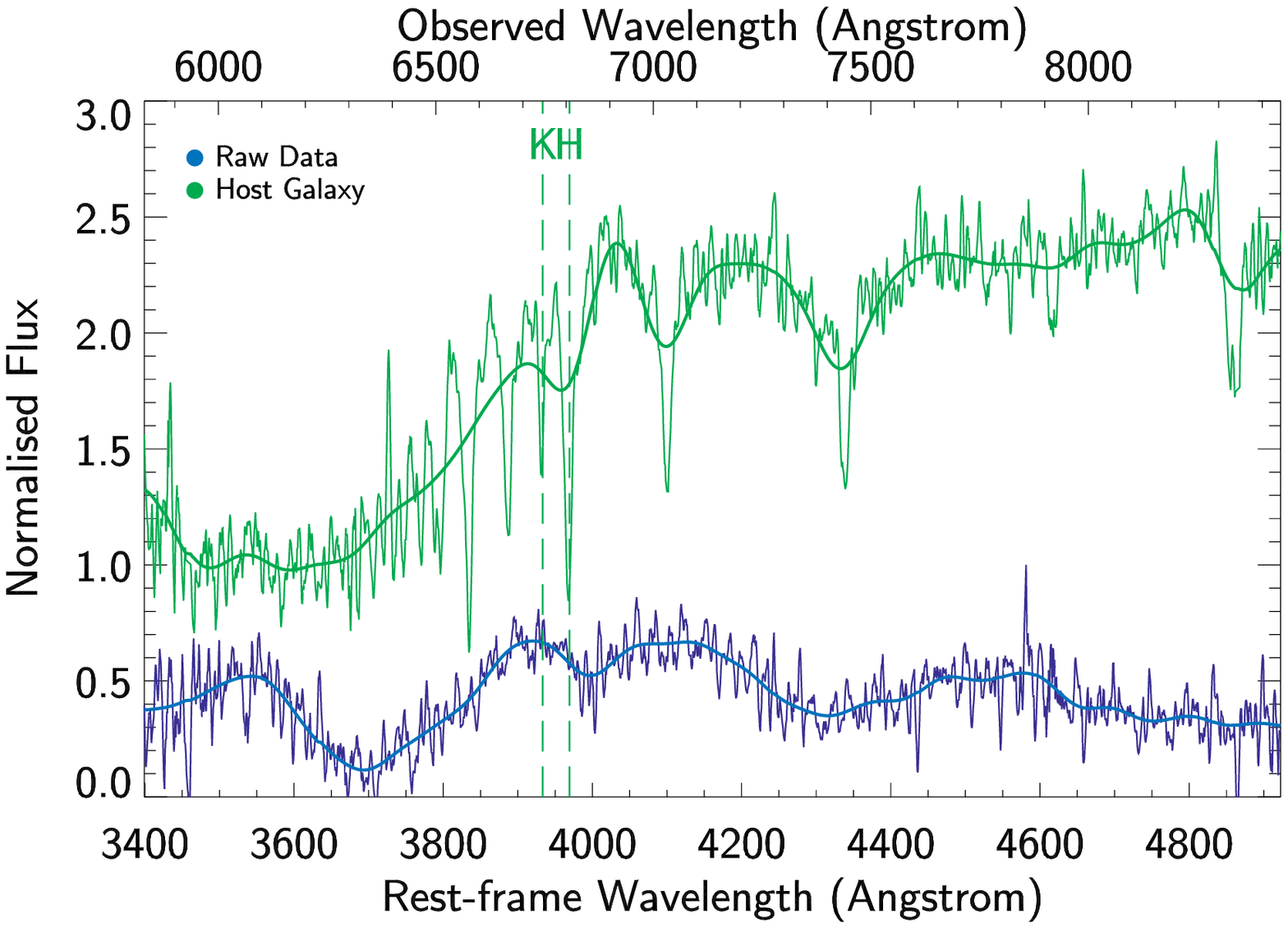}
\label{subfig:08D2gw_data2}
}
\subfigure[08D2hw $z=0.746$]{
\includegraphics[scale=0.40]{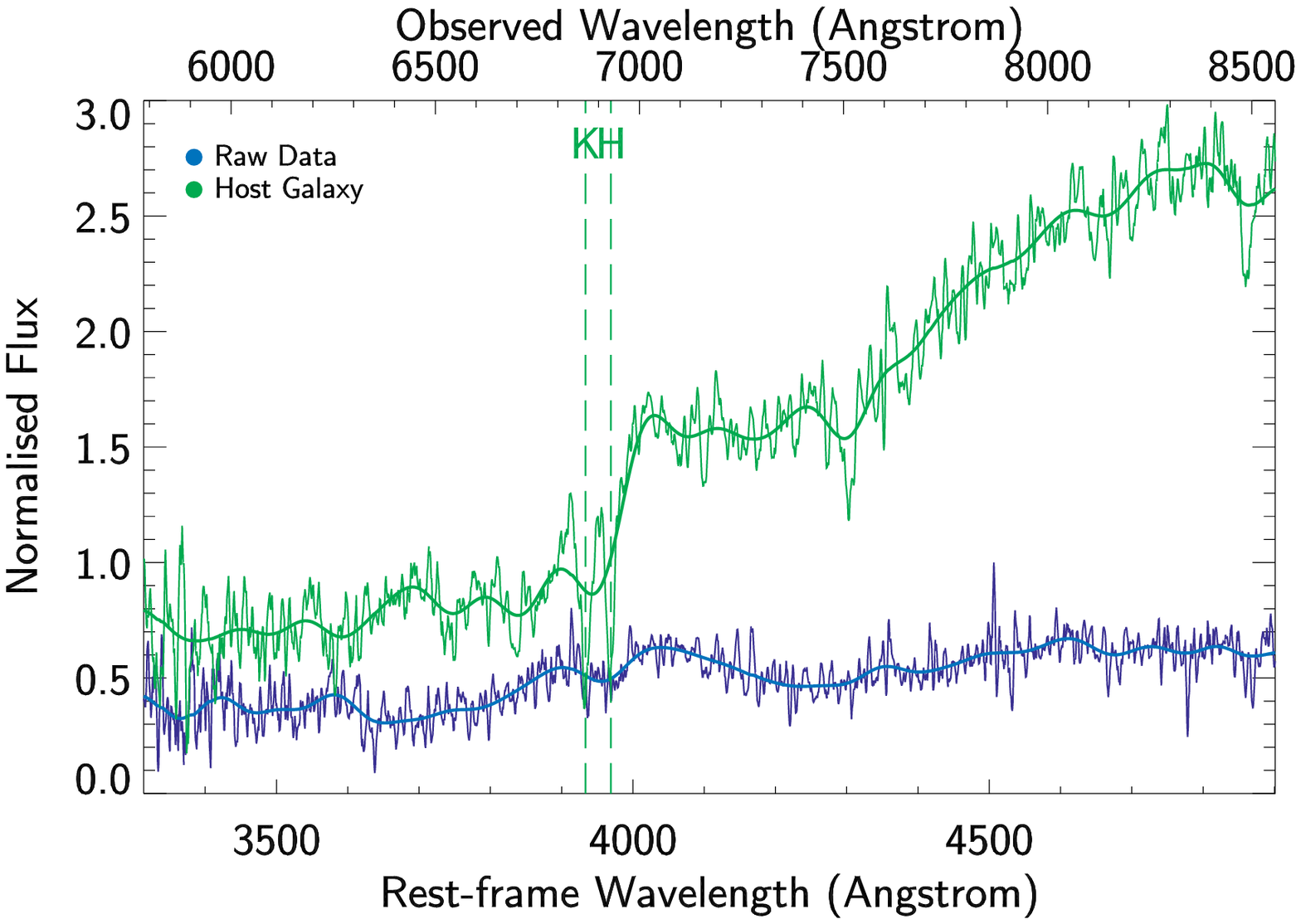}
\label{subfig:08D2hw_data2}
}
\subfigure[08D2id $z=0.833$]{
\includegraphics[scale=0.40]{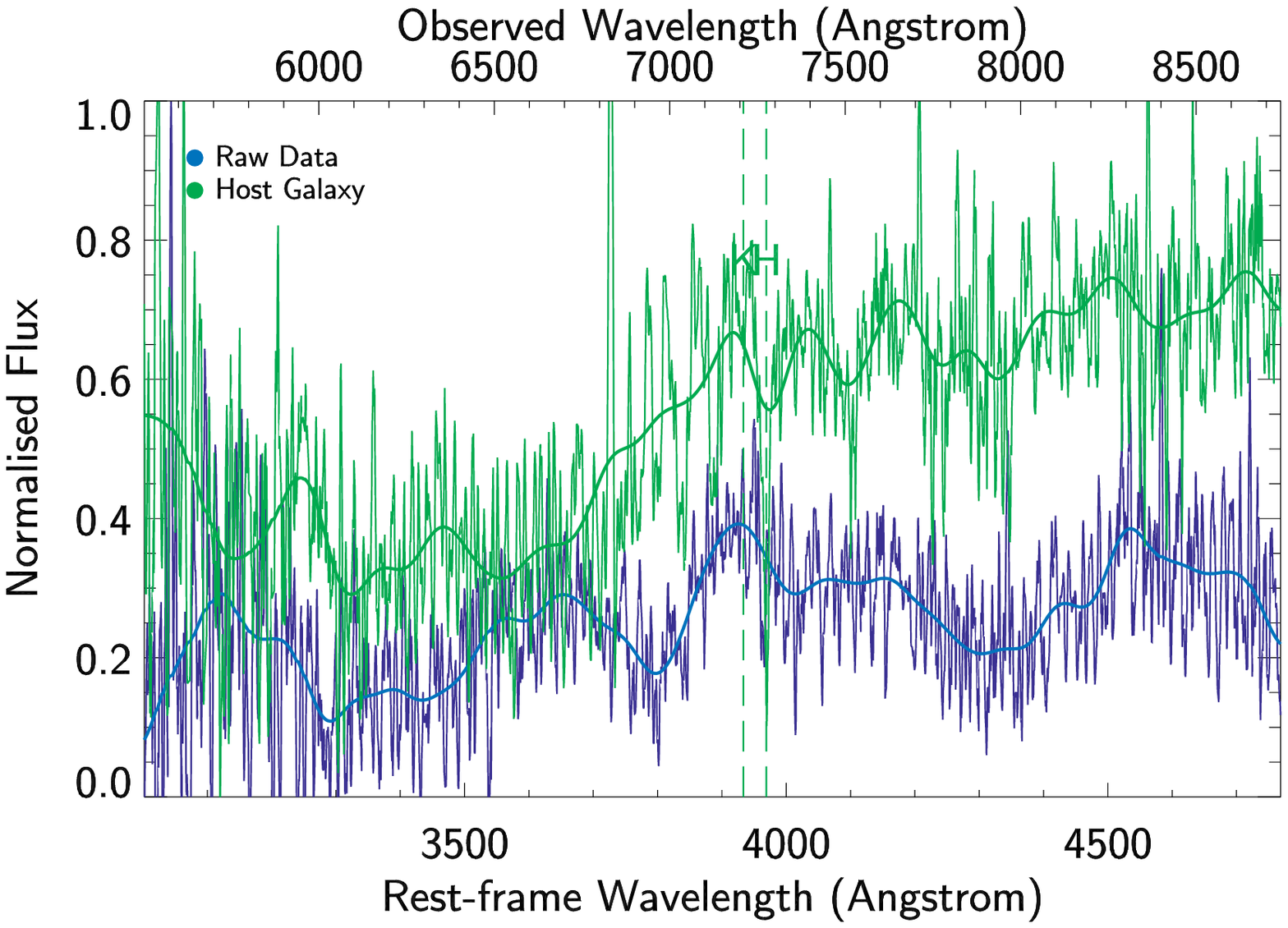}
\label{subfig:08D2id_data2}
}
\end{center}
\caption{Type Ia Supernovae (CI = 4 or 5)}
\end{figure*}

\begin{figure*}
\begin{center}
\subfigure[08D2iq $z=0.709$]{
\includegraphics[scale=0.40]{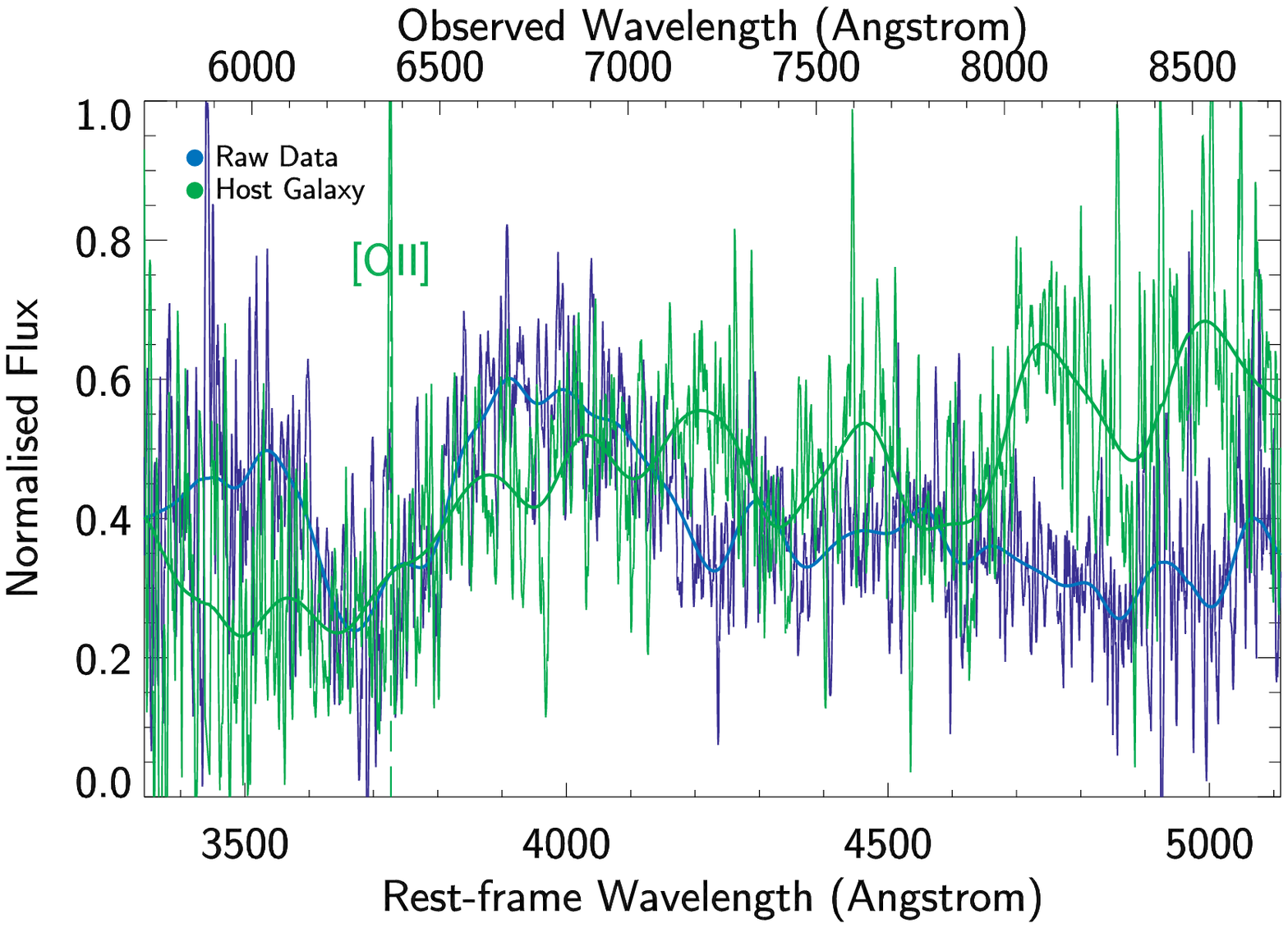}
\label{subfig:08D2iq_data2}
}
\subfigure[08D2kj $z=0.702$]{
\includegraphics[scale=0.40]{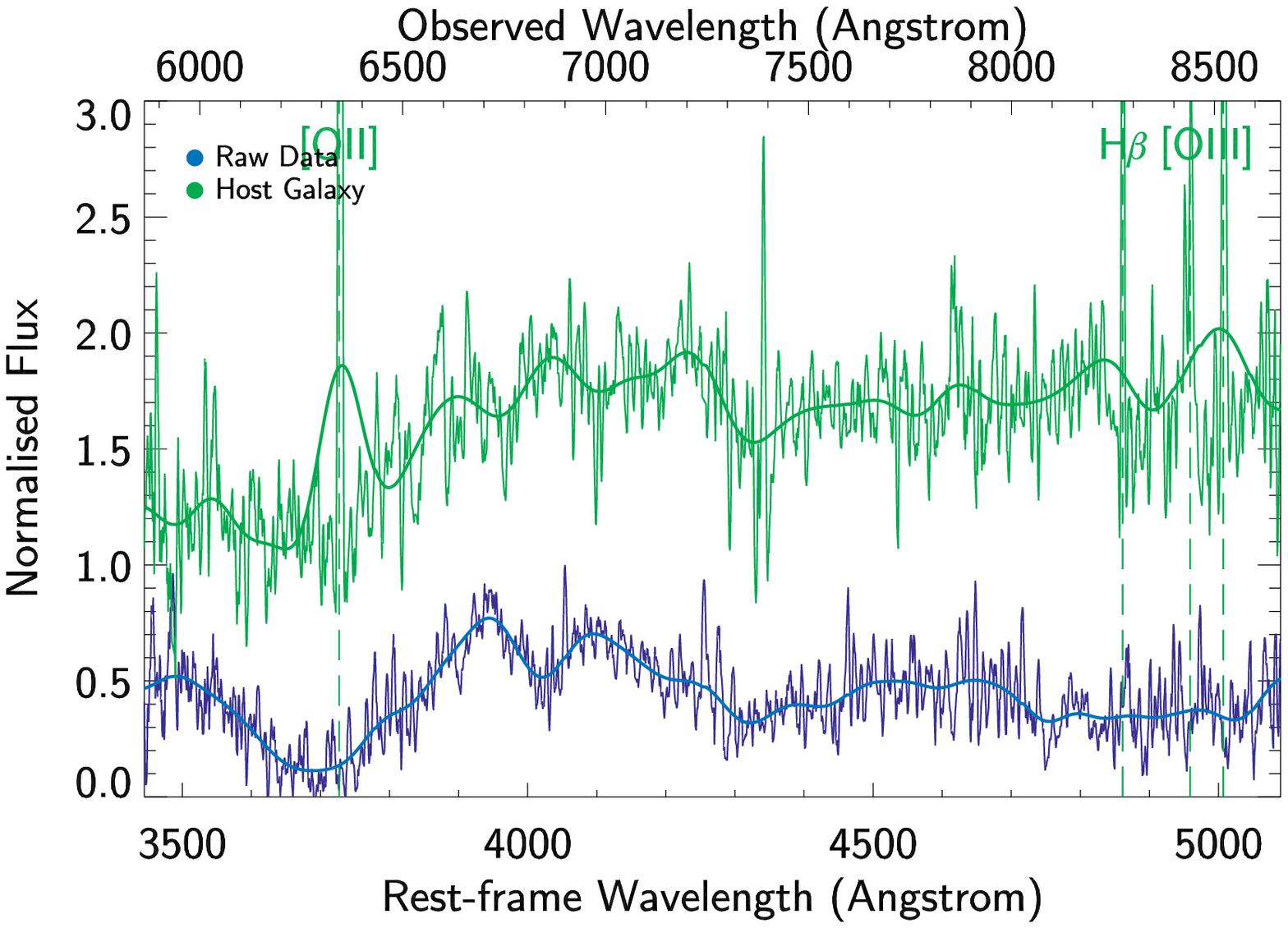}
\label{subfig:08D2kj_data2}
}
\subfigure[08D3bh $z=0.52$]{
\includegraphics[scale=0.40]{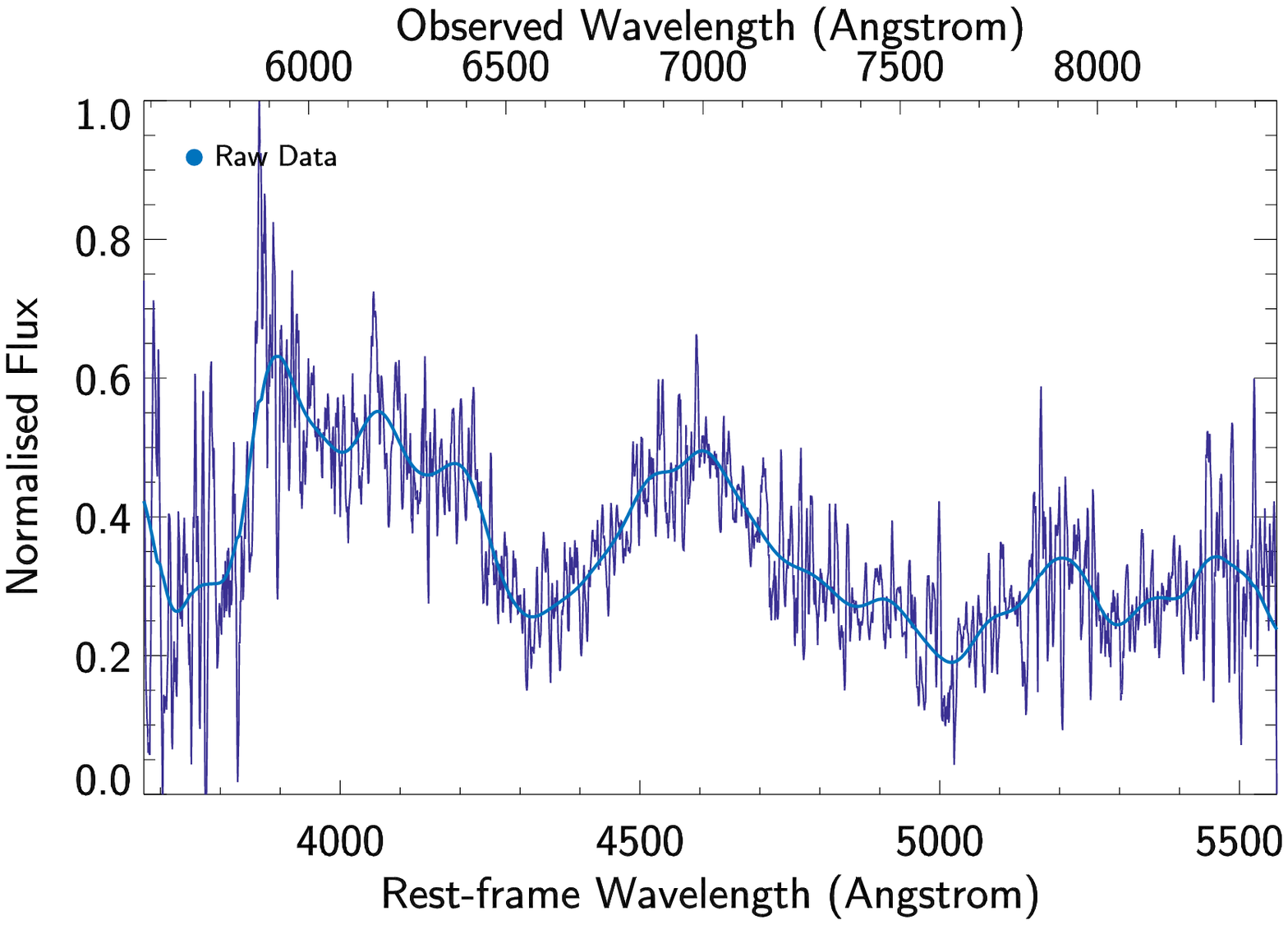}
\label{subfig:08D3bh_data2}
}
\subfigure[08D3dc $z=0.799$]{
\includegraphics[scale=0.40]{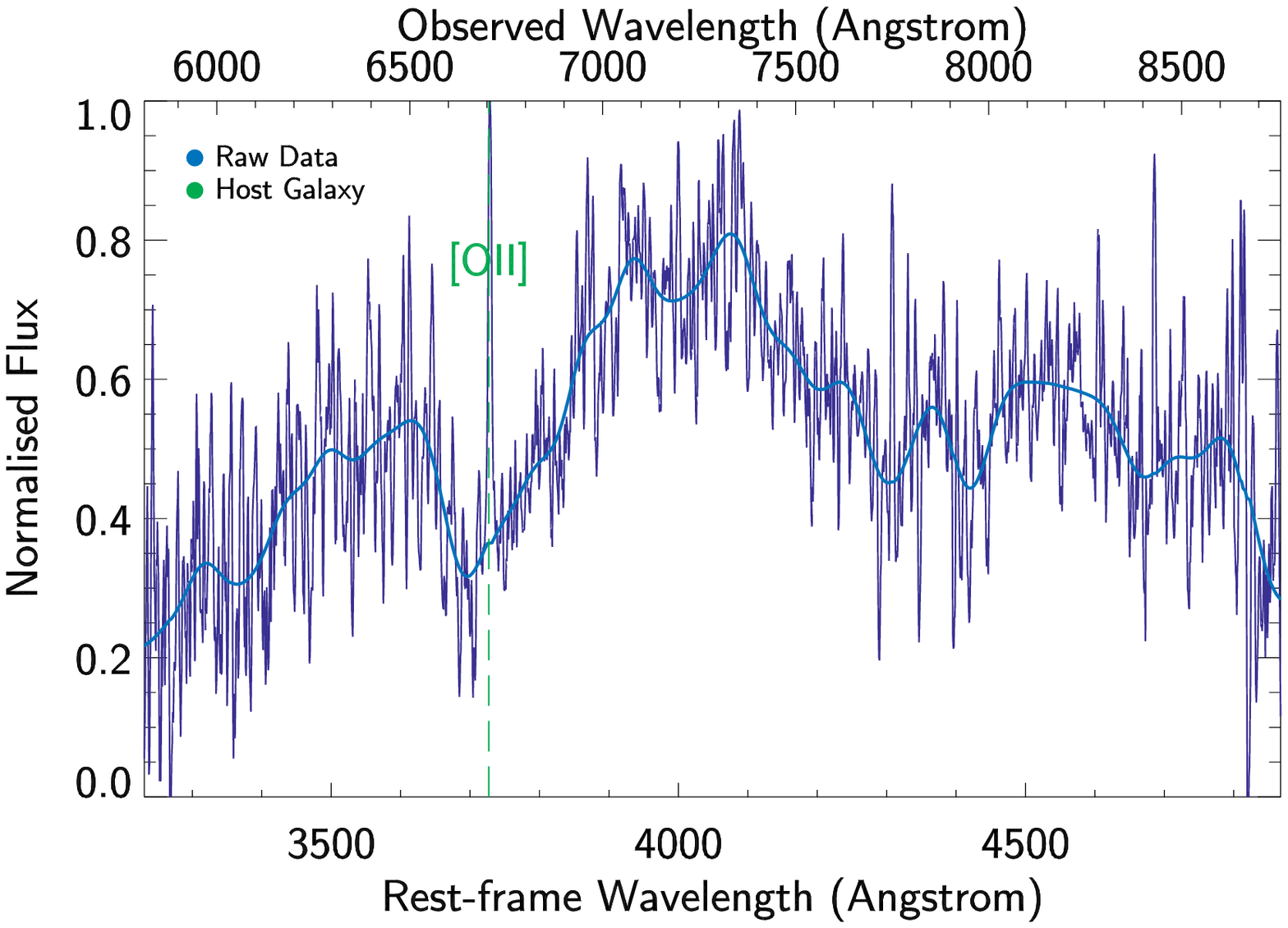}
\label{subfig:08D3dc_data2}
}
\subfigure[08D3gb $z=0.17$]{
\includegraphics[scale=0.40]{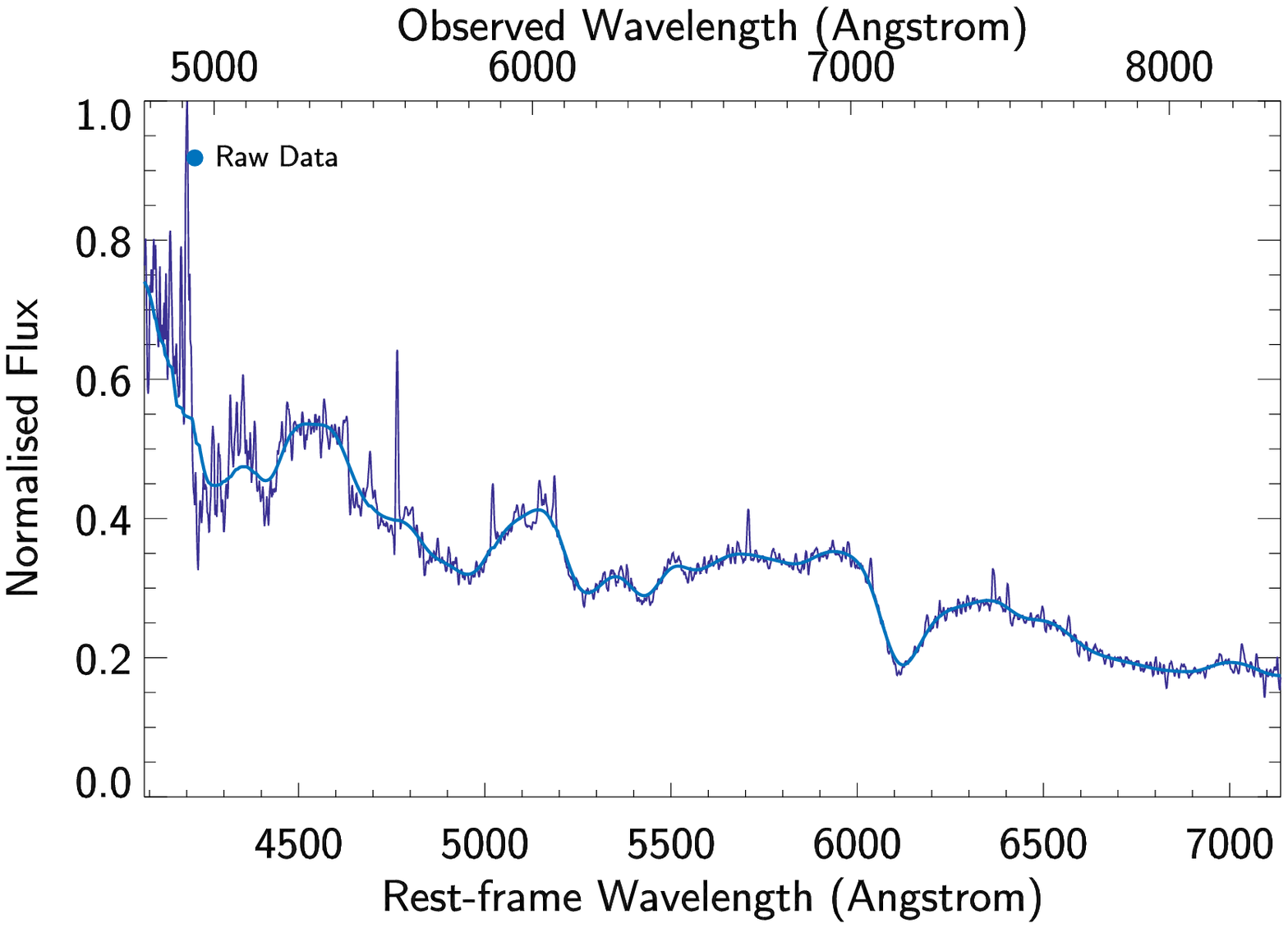}
\label{subfig:08D3gb_data2}
}
\subfigure[08D3gf $z=0.352$]{
\includegraphics[scale=0.40]{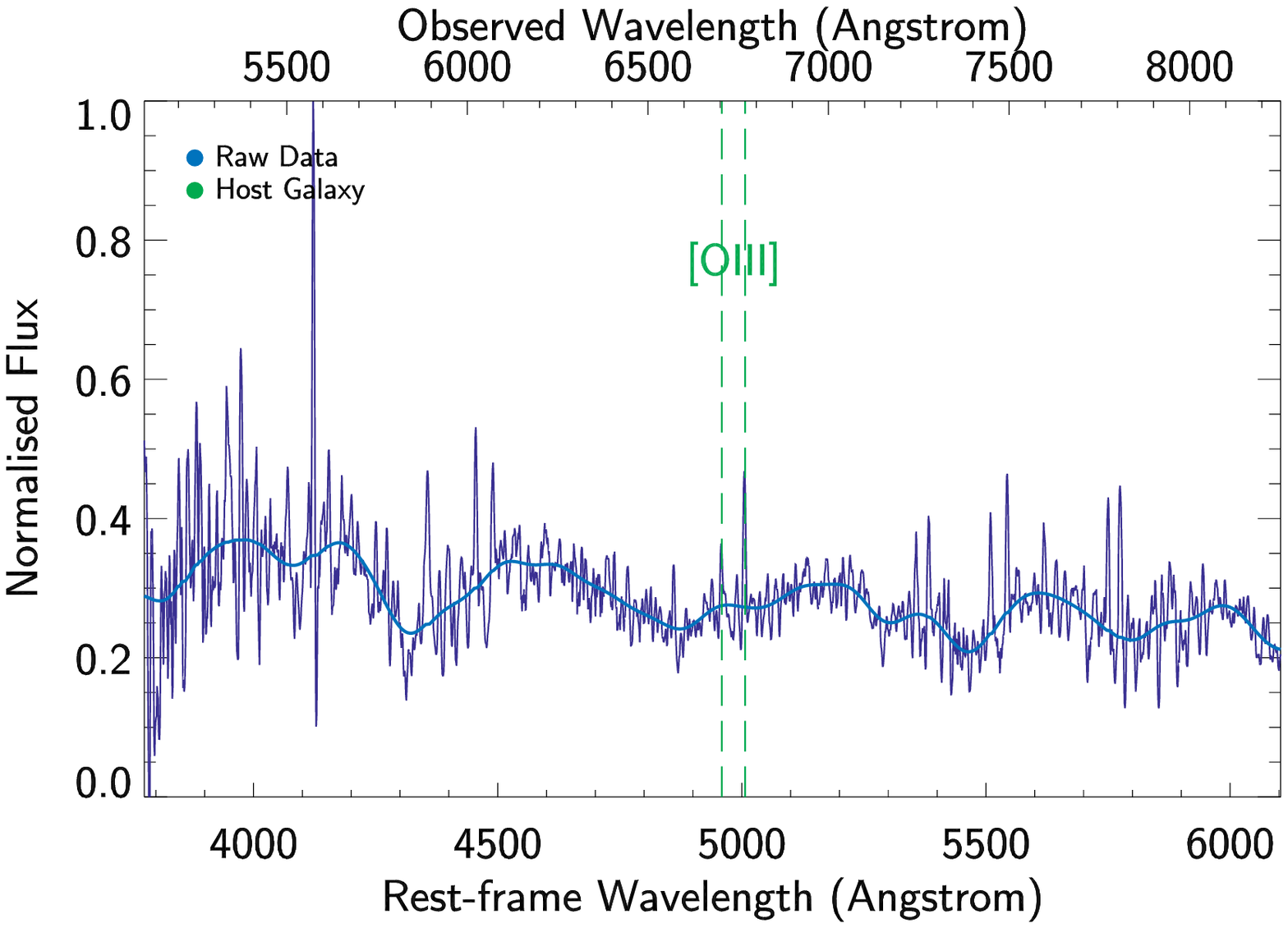}
\label{subfig:08D3gf_data2}
}
\subfigure[08D3gu $z=0.767$]{
\includegraphics[scale=0.40]{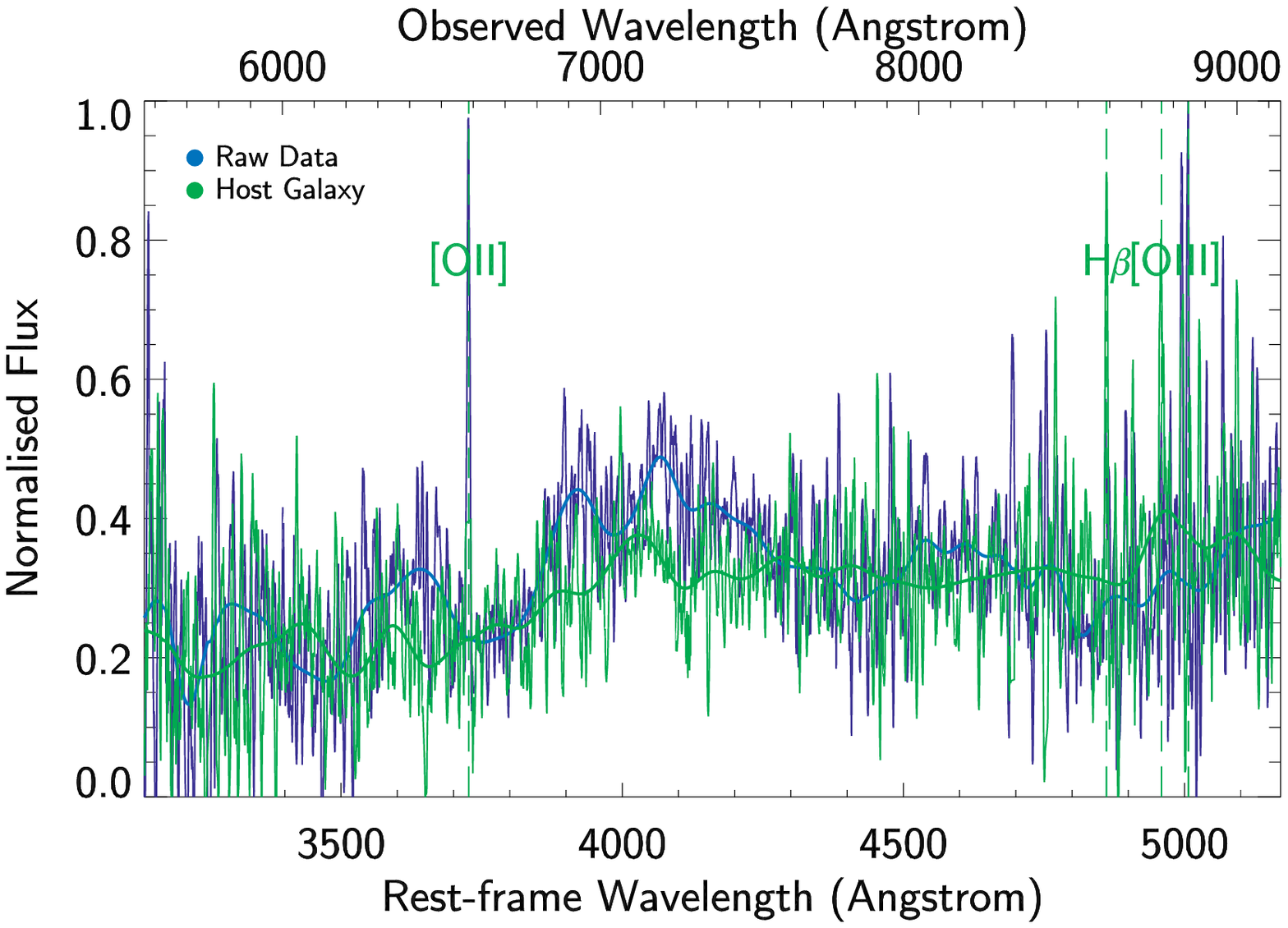}
\label{subfig:08D3gu_data2}
}
\subfigure[08D3hh $z=0.452$]{
\includegraphics[scale=0.40]{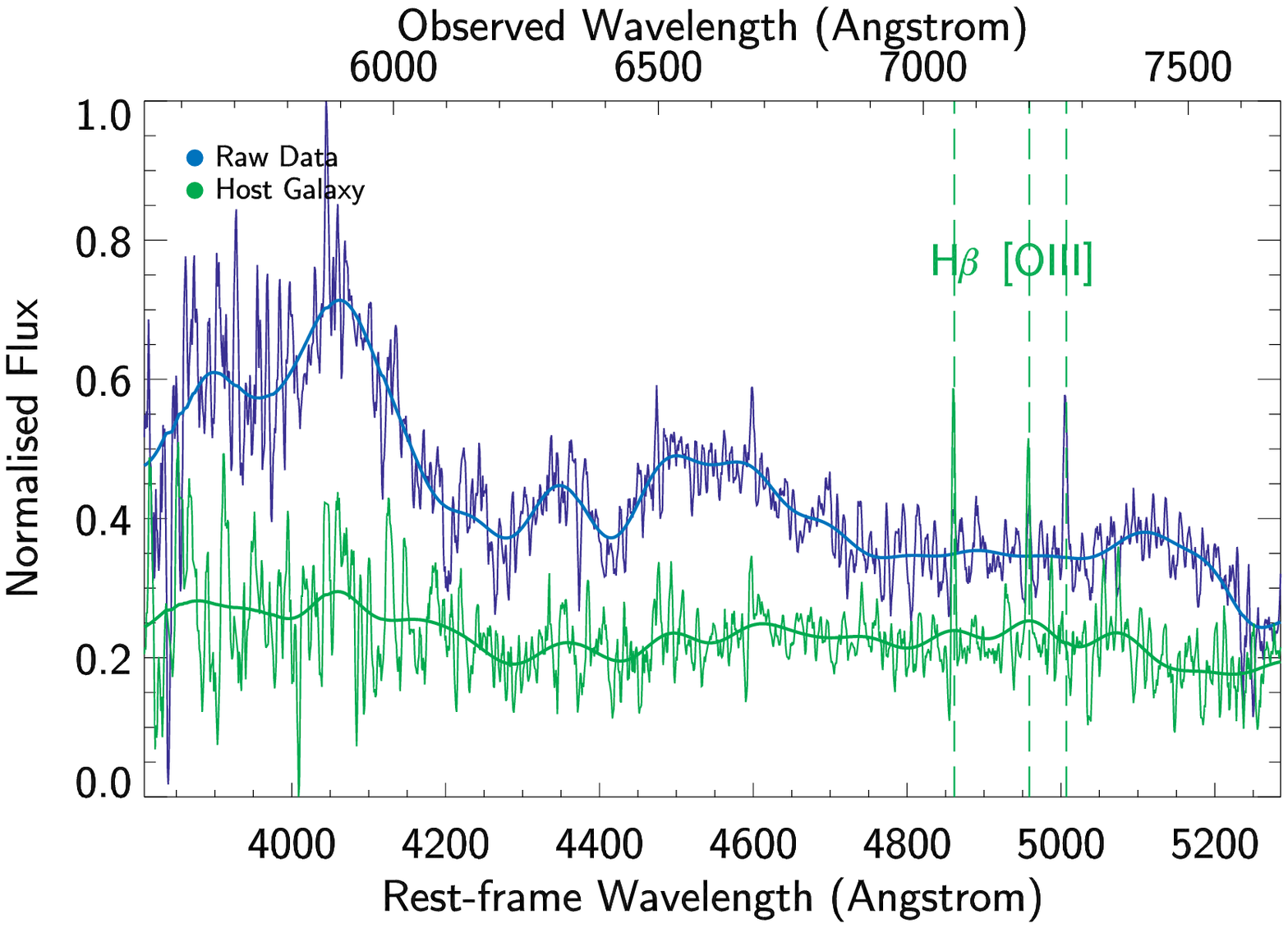}
\label{subfig:08D3hh_data2}
}
\end{center}
\caption{Type Ia Supernovae (CI = 4 or 5)}
\end{figure*}
\newpage

\begin{figure*}
\begin{center}
\subfigure[06D1ck $z=0.90$]{
\includegraphics[scale=0.40]{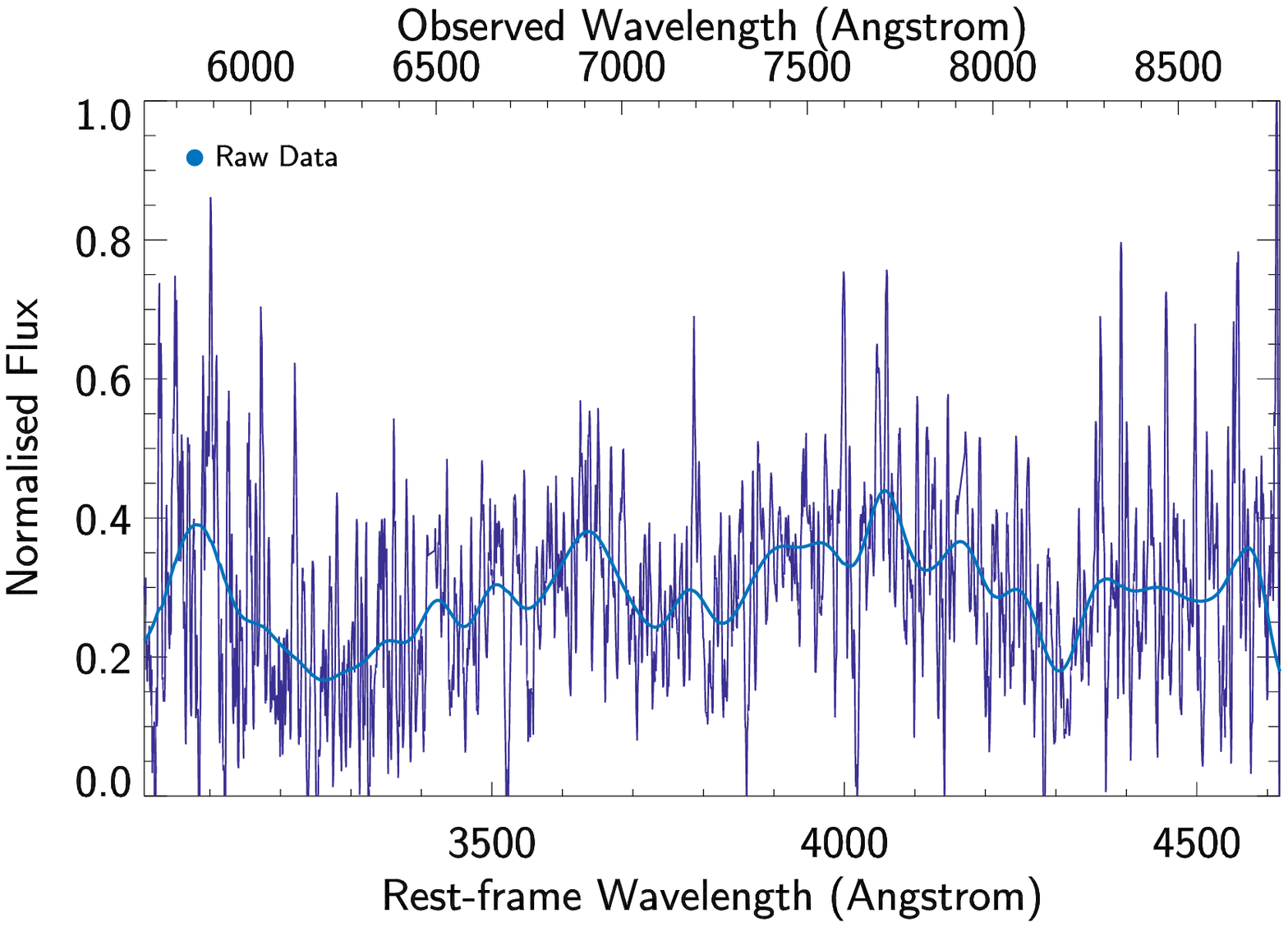}
\label{subfig:06D1ck_data2}
}
\subfigure[06D2iz $z=0.85$]{
\includegraphics[scale=0.40]{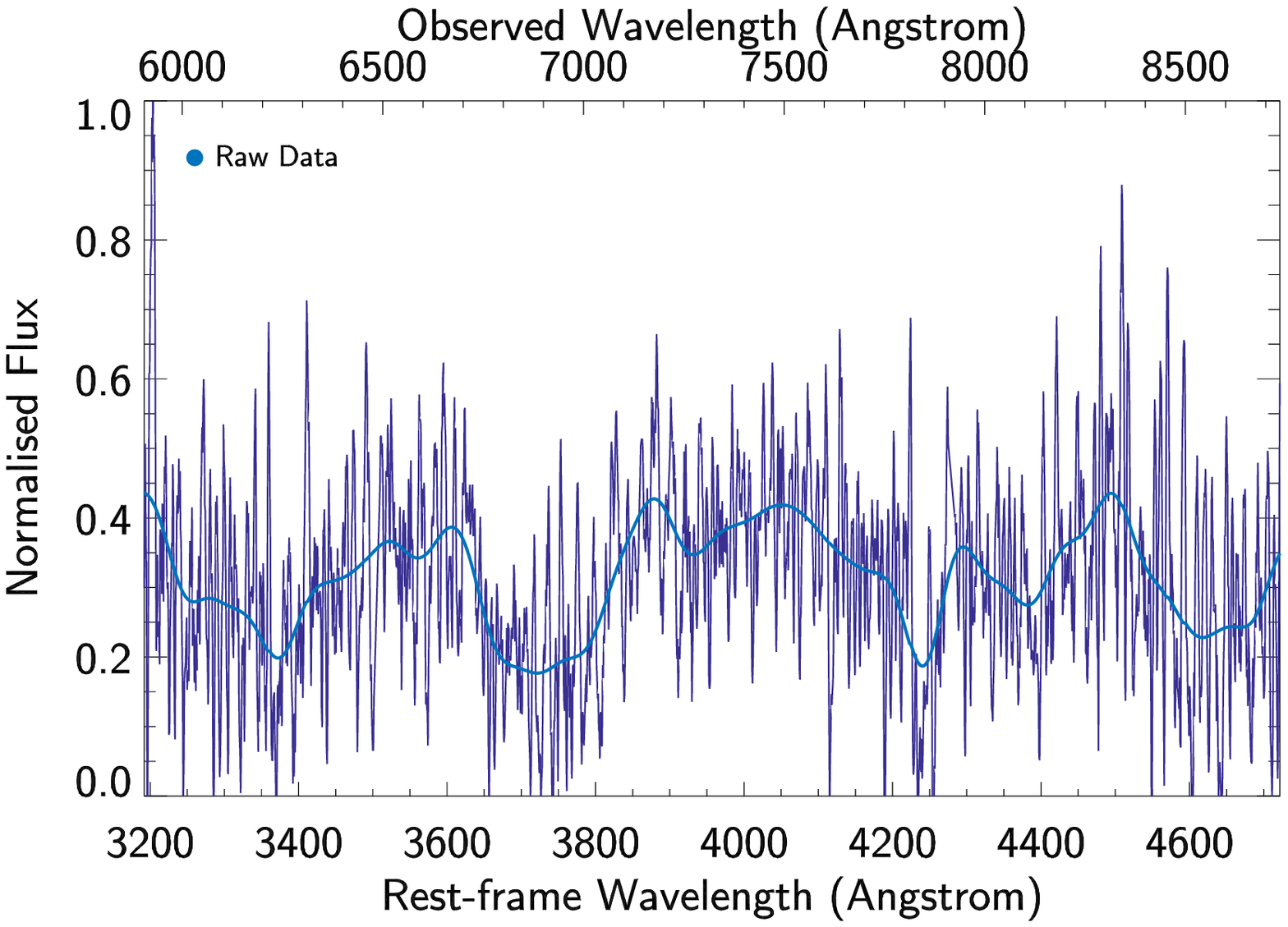}
\label{subfig:06D2iz_data2}
}
\subfigure[06D2ji $z=0.90$]{
\includegraphics[scale=0.40]{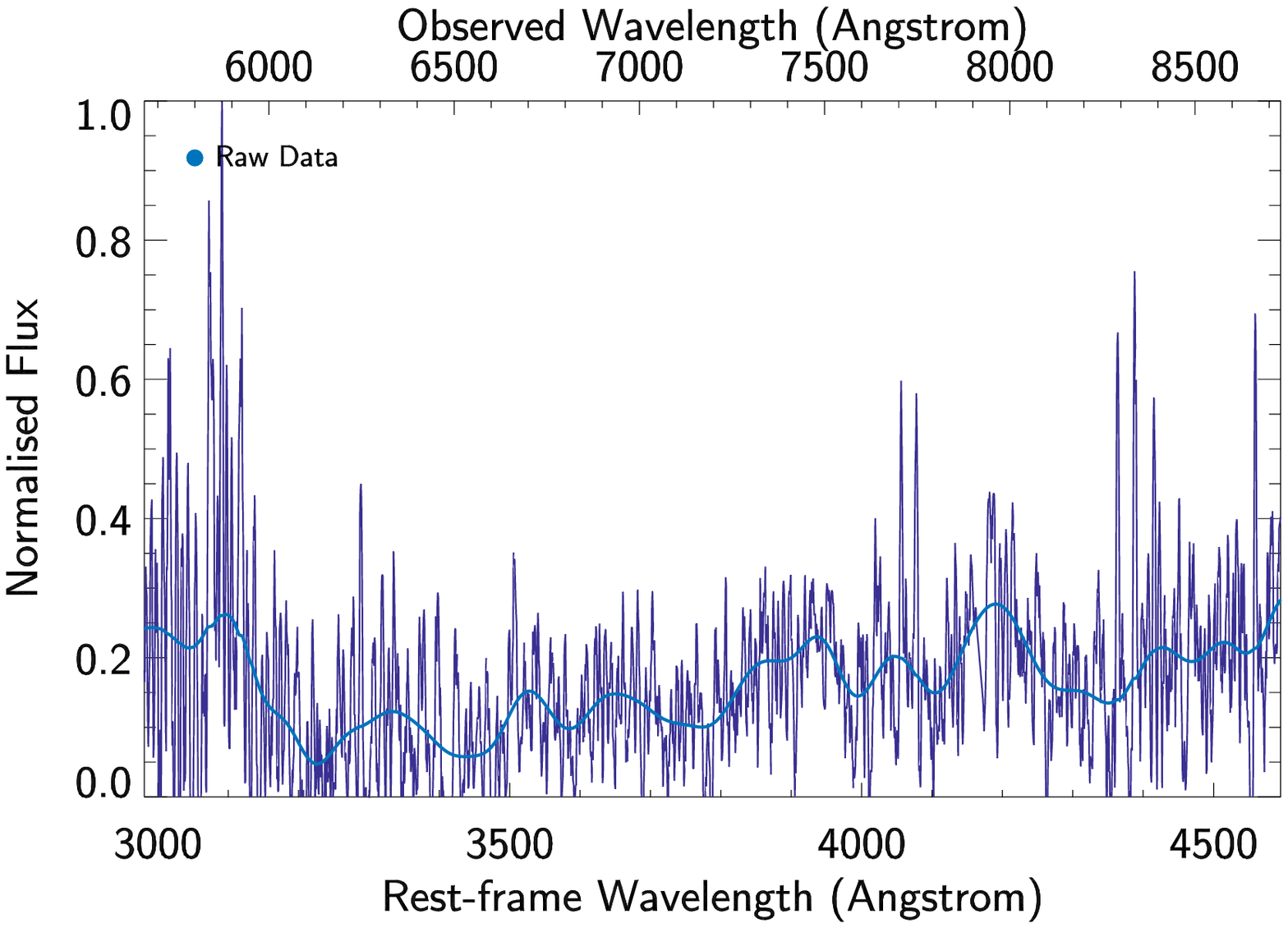}
\label{subfig:06D2ji_data2}
}
\subfigure[06D2ju $z=0.927$]{
\includegraphics[scale=0.40]{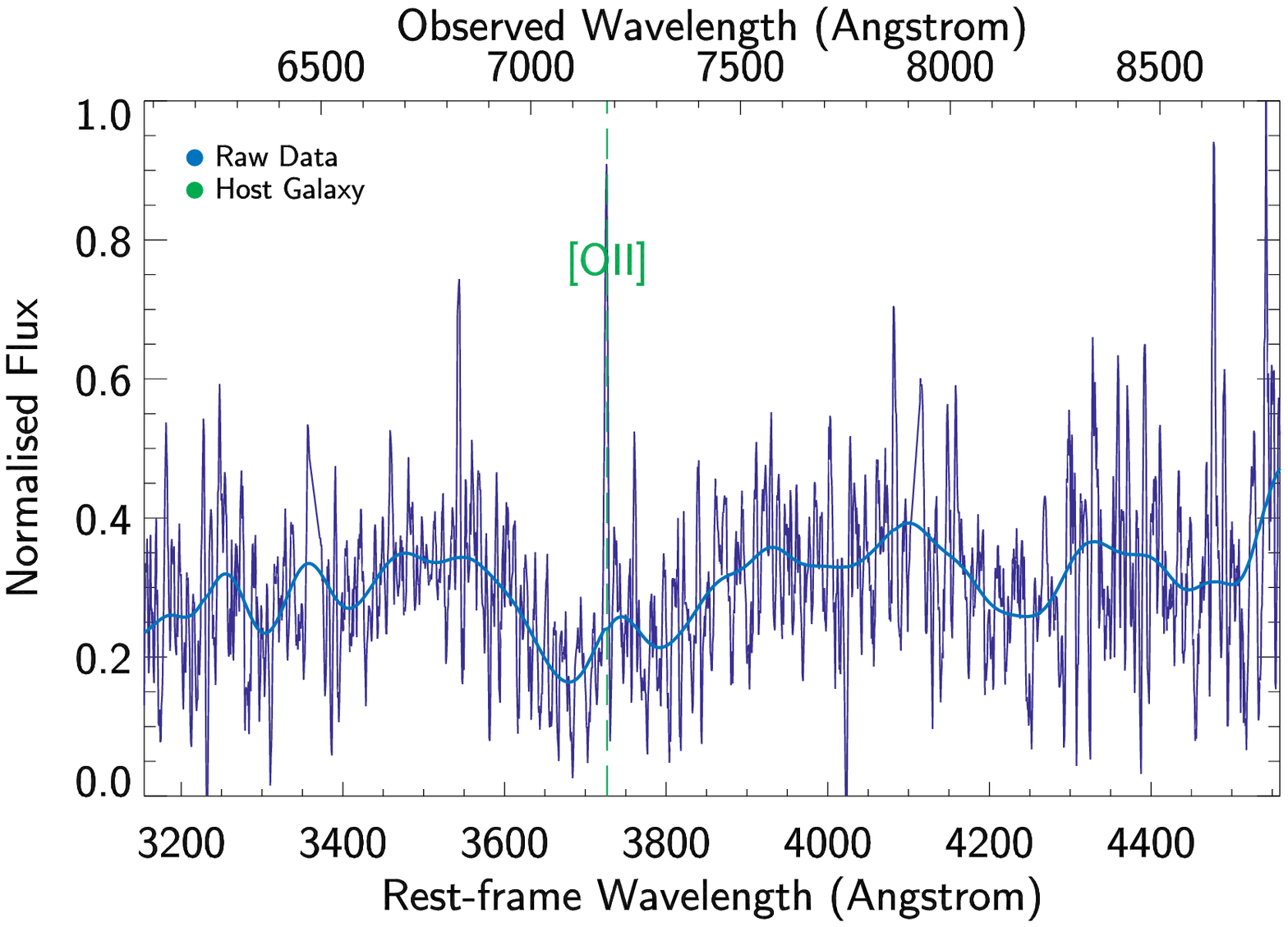}
\label{subfig:06D2ju_data2}
}
\subfigure[06D4fc $z=0.677$]{
\includegraphics[scale=0.40]{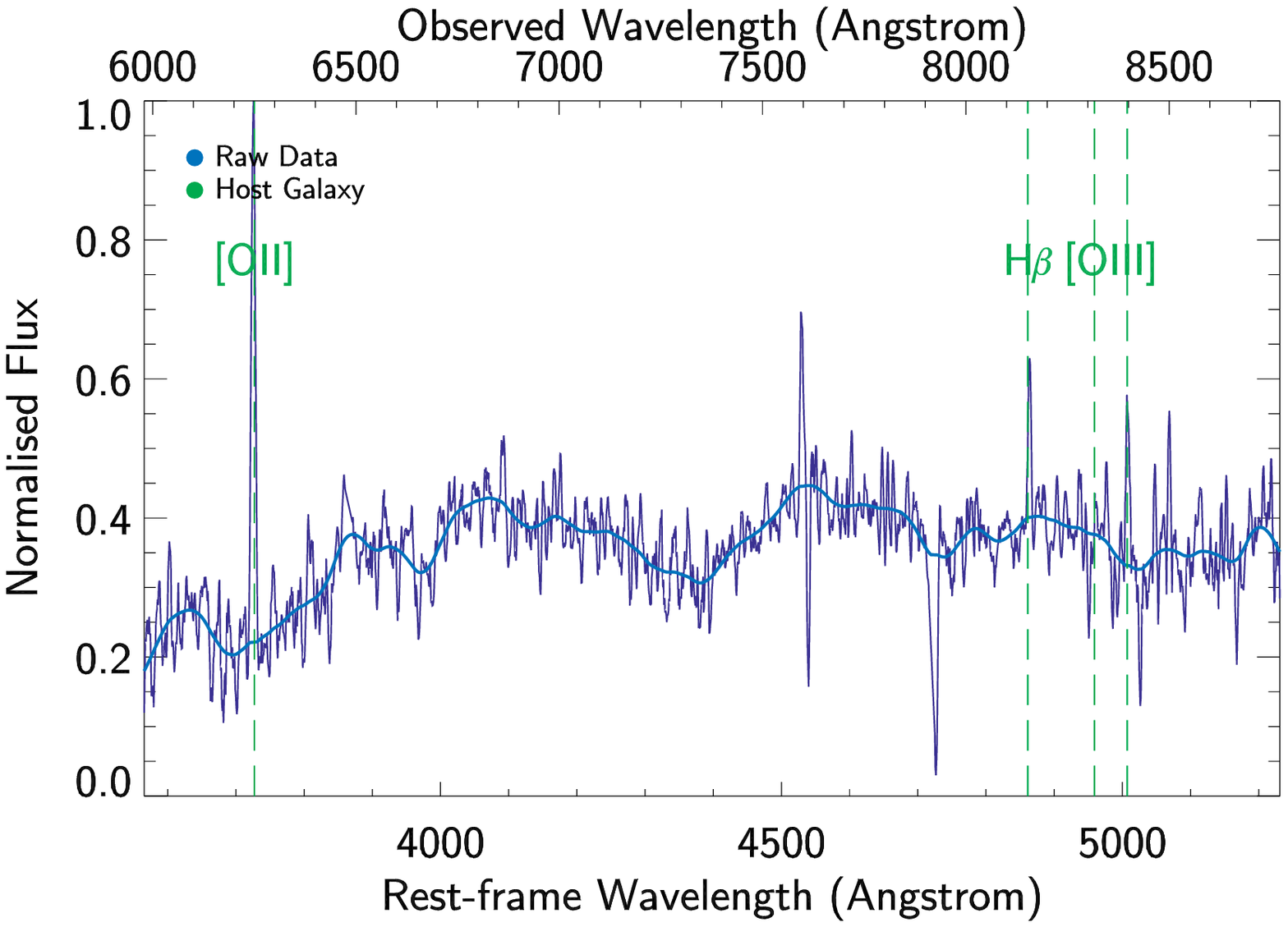}
\label{subfig:06D4fc_data2}
}
\subfigure[07D3bo $z=0.92$]{
\includegraphics[scale=0.40]{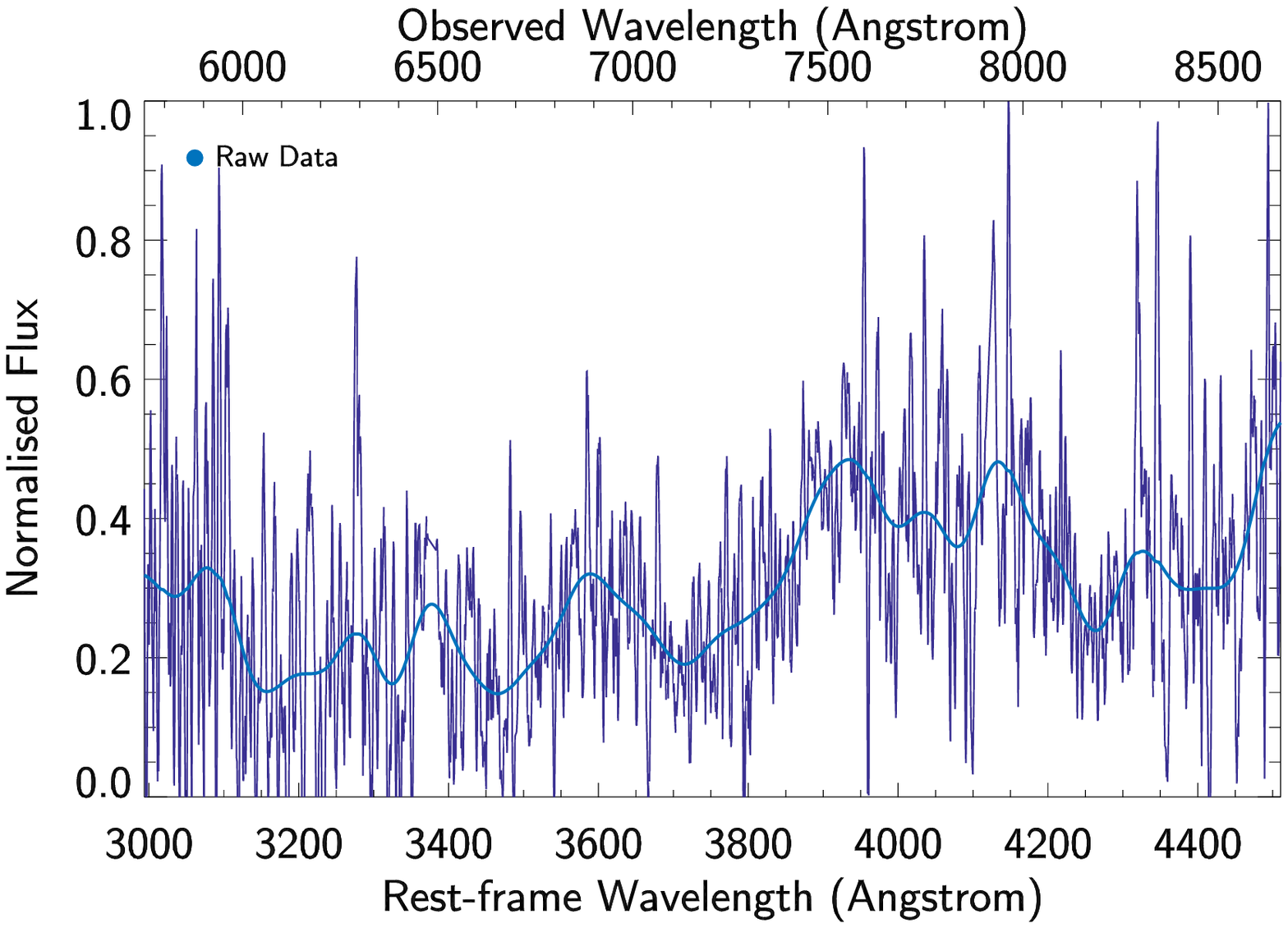}
\label{subfig:07D3bo_data2}
}
\subfigure[07D3bt $z=0.91$]{
\includegraphics[scale=0.40]{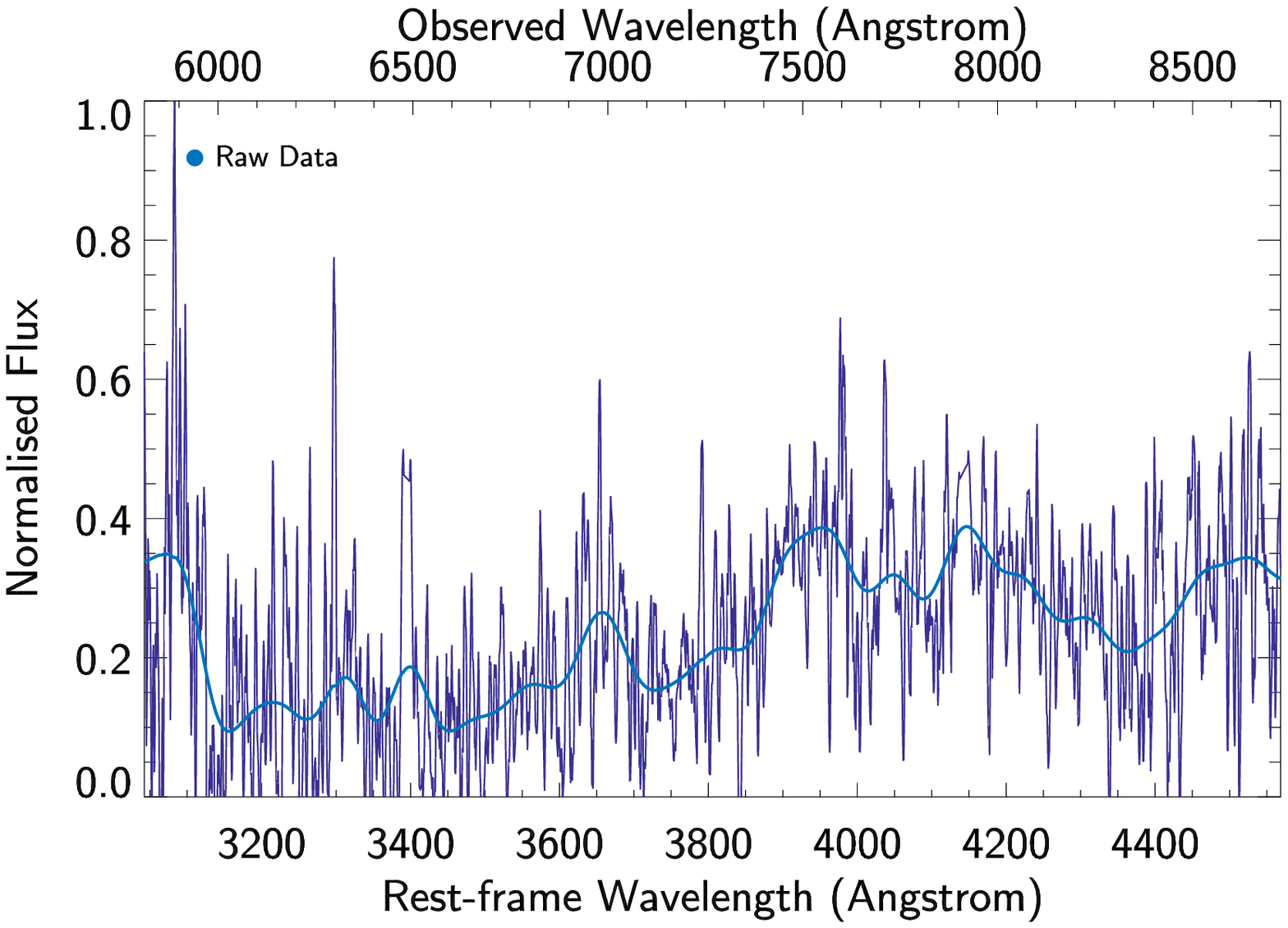}
\label{subfig:07D3bt_data2}
}
\subfigure[07D3do $z=1.02$]{
\includegraphics[scale=0.40]{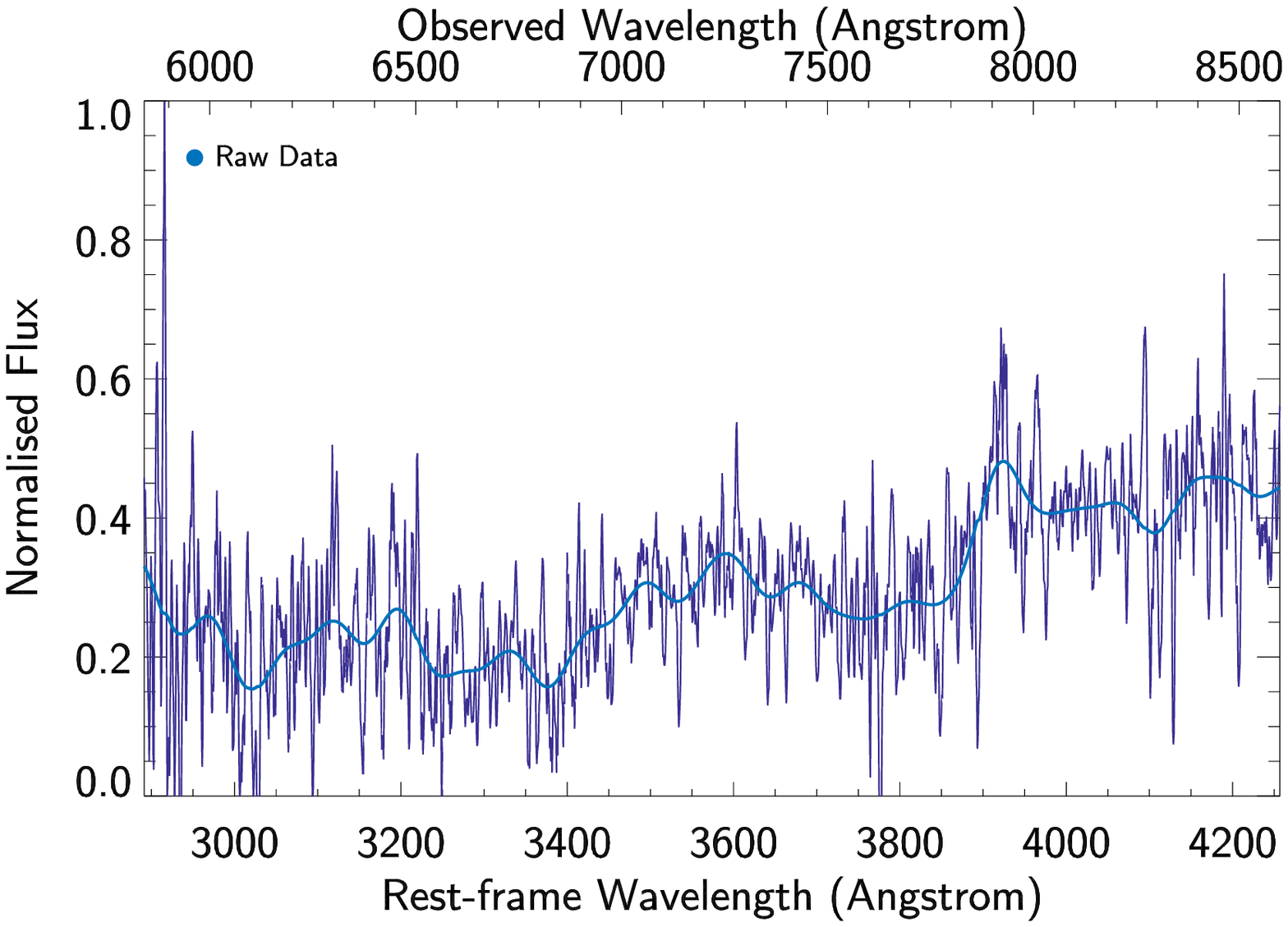}
\label{subfig:07D3do_data2}
}
\end{center}
\caption{Probable Type Ia Supernovae (CI = 3)}
\end{figure*}

\begin{figure*}
\begin{center}
\subfigure[08D1aa $z=0.54$]{
\includegraphics[scale=0.40]{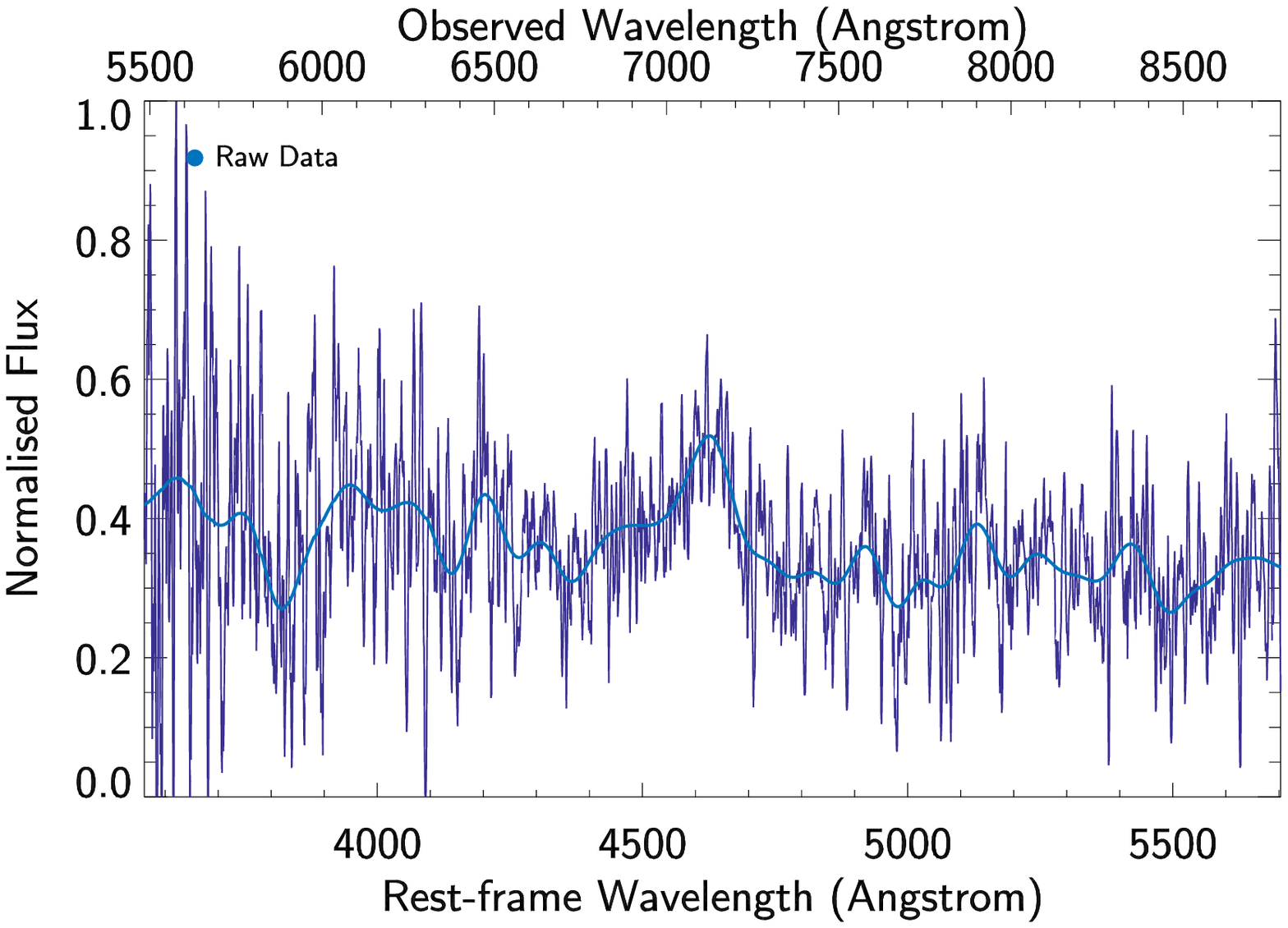}
\label{subfig:08D1aa_data2}
}
\subfigure[08D2ch $z=0.474$]{
\includegraphics[scale=0.40]{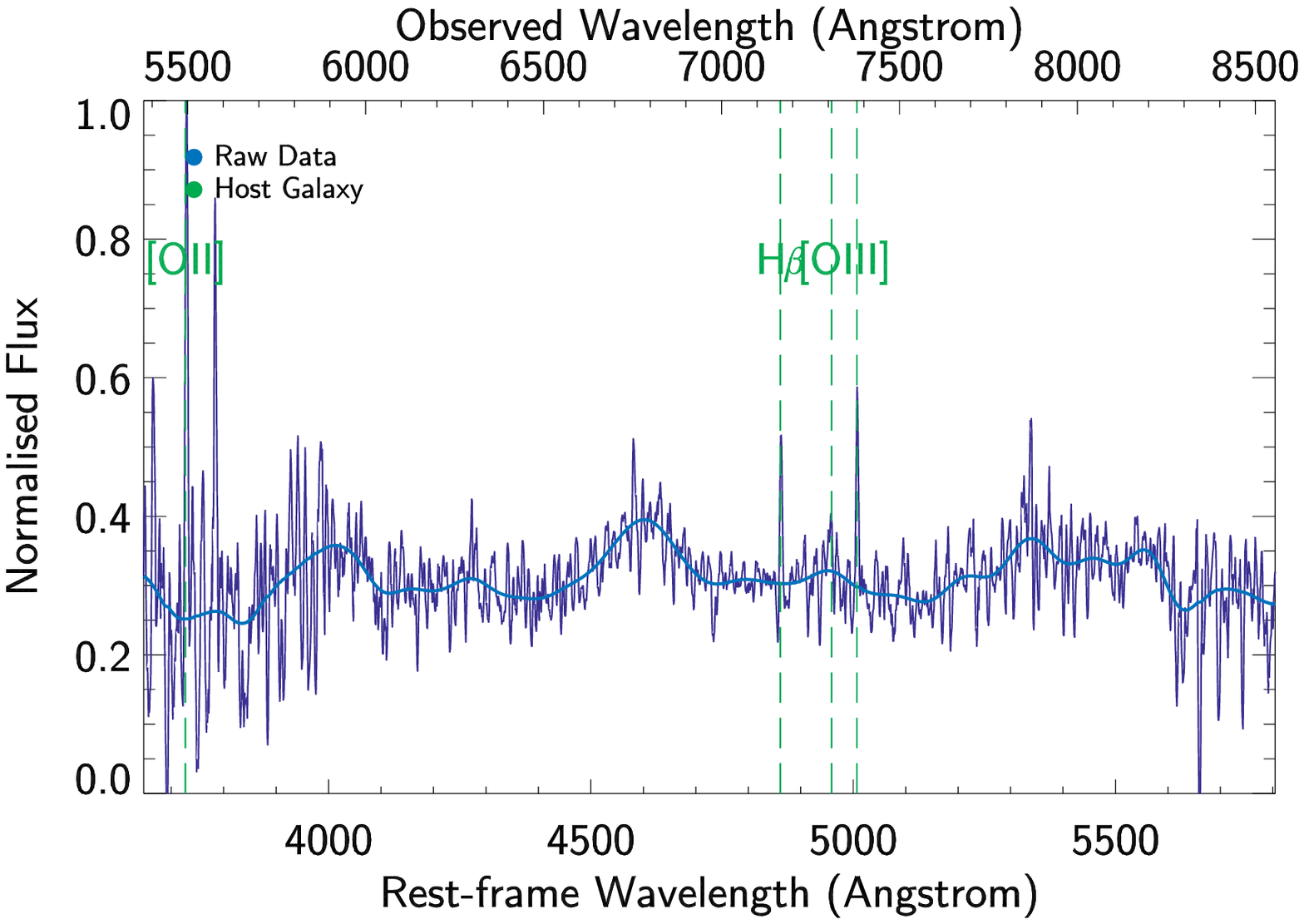}
\label{subfig:08D2ch_data2}
}
\subfigure[08D2cl $z=0.831$]{
\includegraphics[scale=0.40]{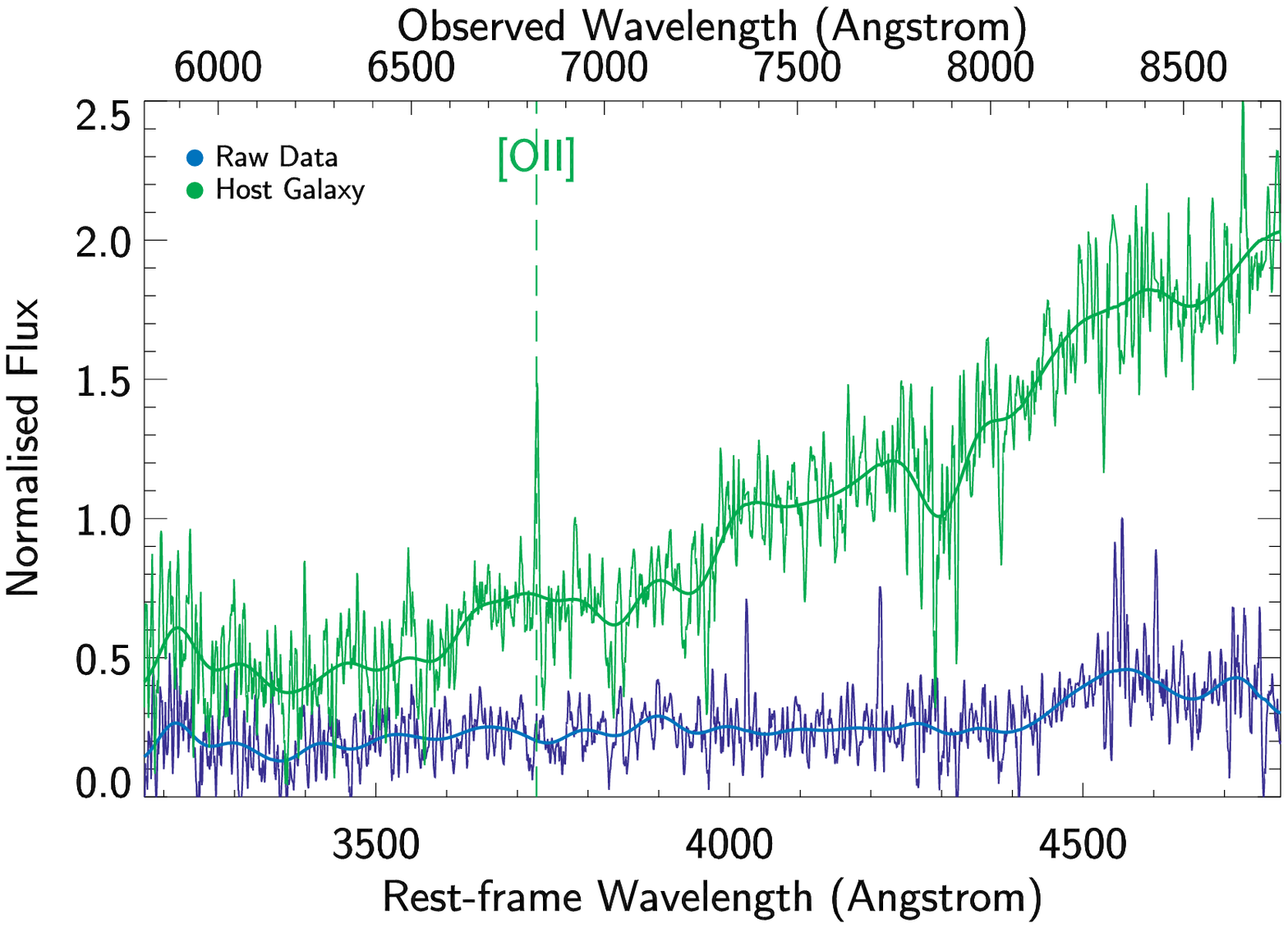}
\label{subfig:08D2cl_data2}
}
\subfigure[08D3dx $z=0.928$]{
\includegraphics[scale=0.40]{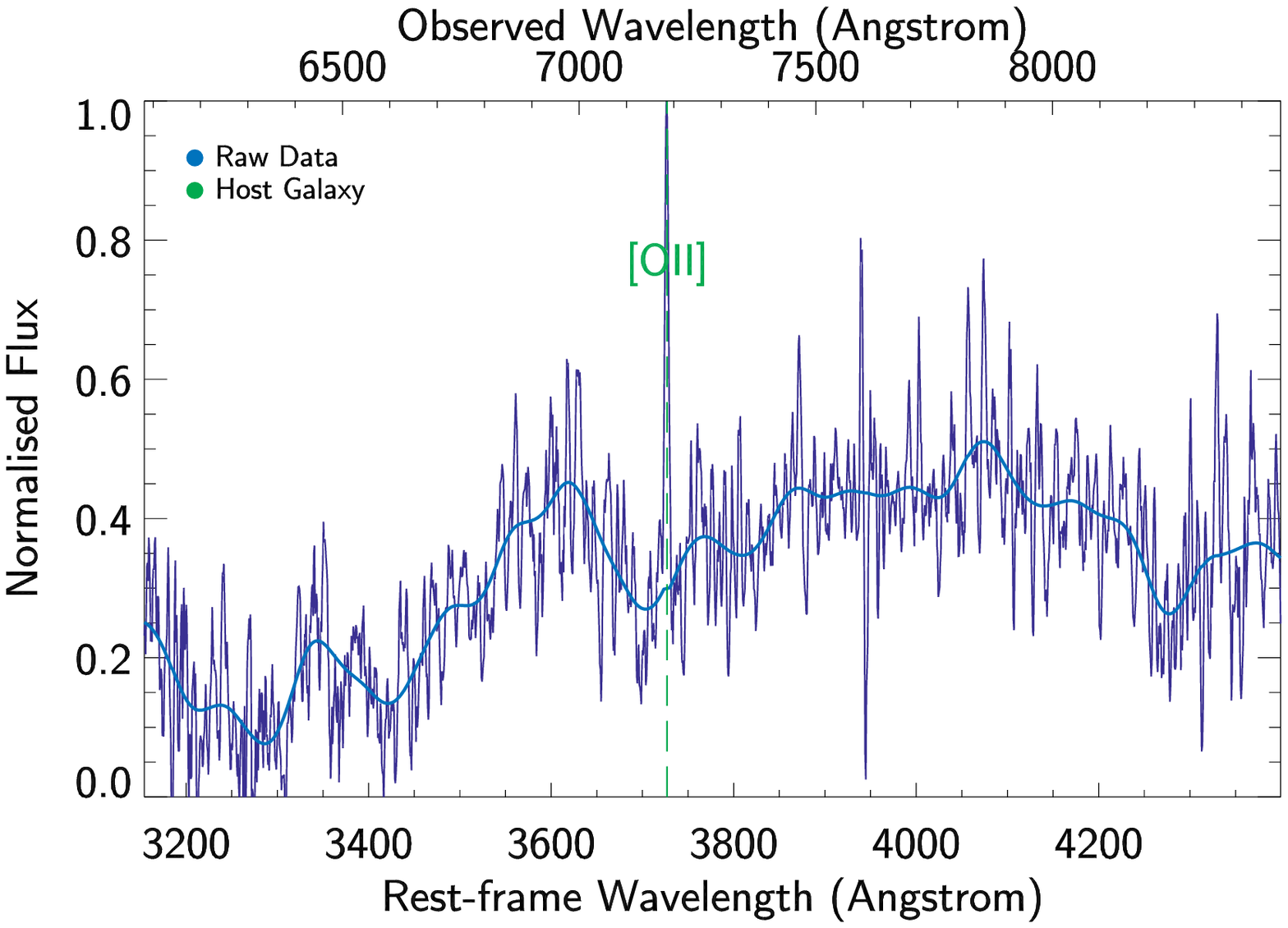}
\label{subfig:08D3dx_data2}
}
\end{center}
\caption{Probable Type Ia Supernovae (CI = 3)}
\end{figure*}

\newpage

\begin{figure*}
\begin{center}
\subfigure[05D4bo $z$ unknown]{
\includegraphics[scale=0.40]{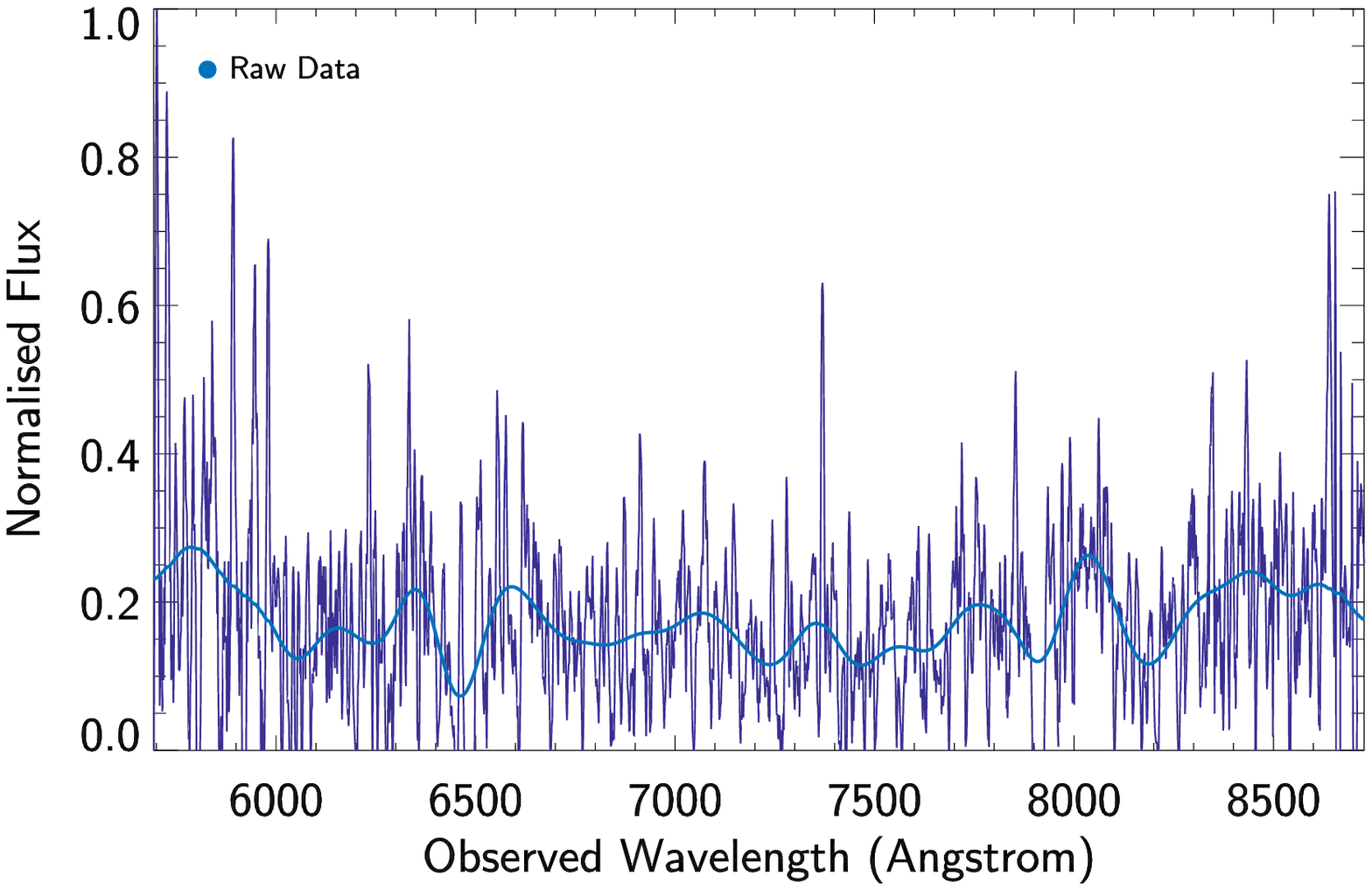}
\label{subfig:05D4bo_data2}
}
\subfigure[06D1gu $z=0.87$]{
\includegraphics[scale=0.40]{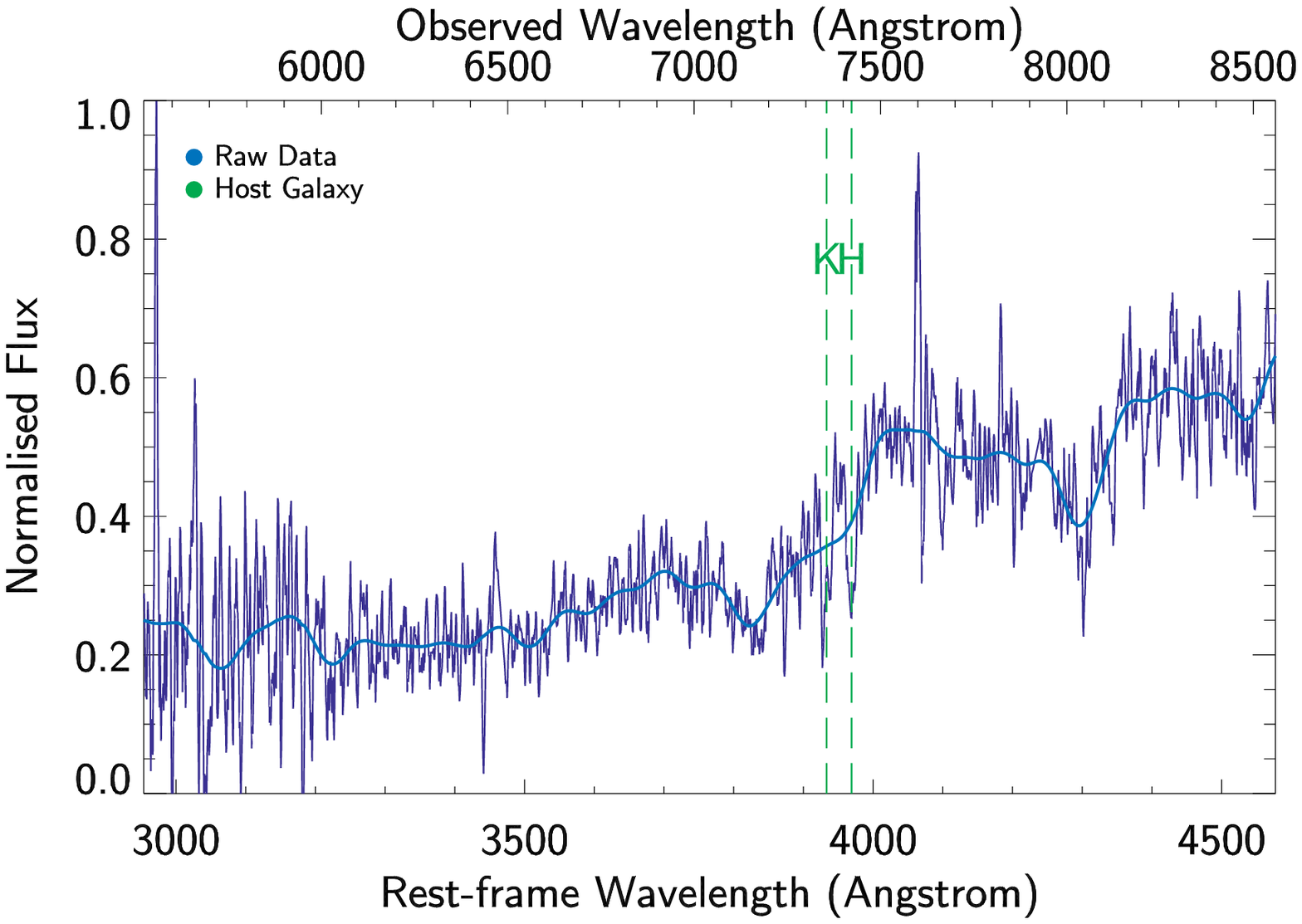}
\label{subfig:06D1gu_data2}
}
\subfigure[06D1hl $z$ unknown]{
\includegraphics[scale=0.40]{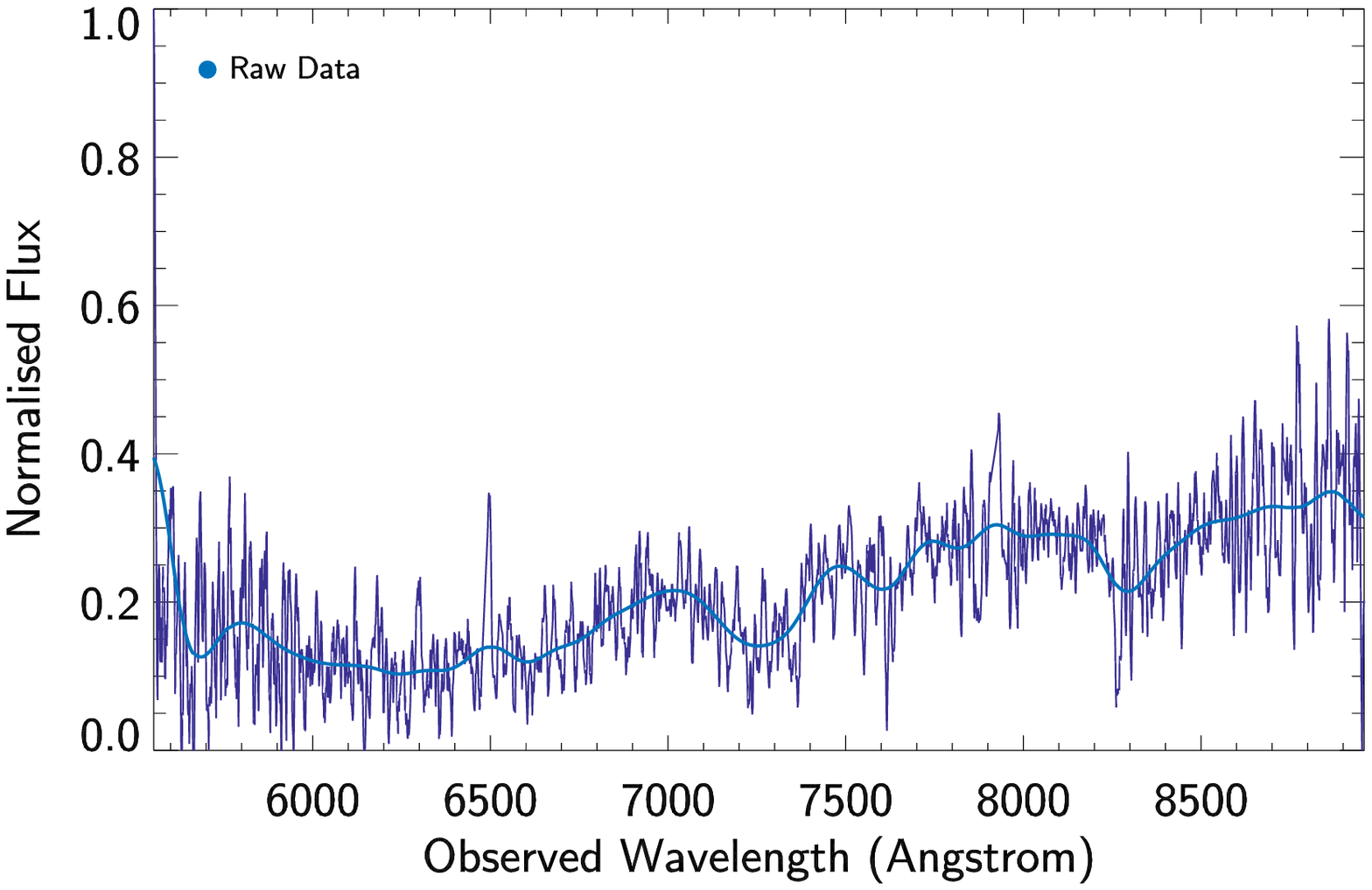}
\label{subfig:06D1hl_data2}
}
\subfigure[06D1jp $z$ unknown]{
\includegraphics[scale=0.40]{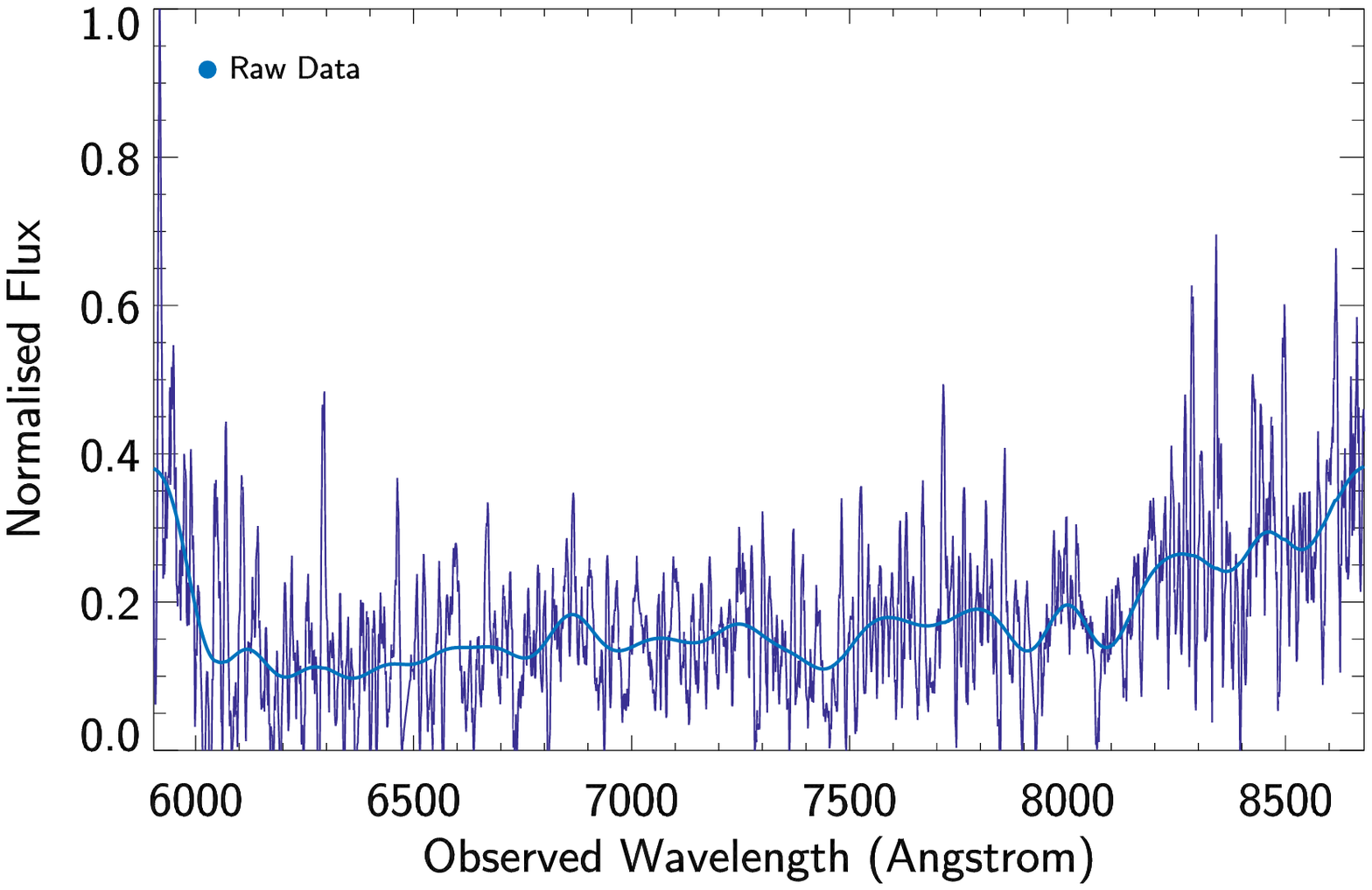}
\label{subfig:06D1jp_data2}
}
\subfigure[06D3fv $z$ unknown]{
\includegraphics[scale=0.40]{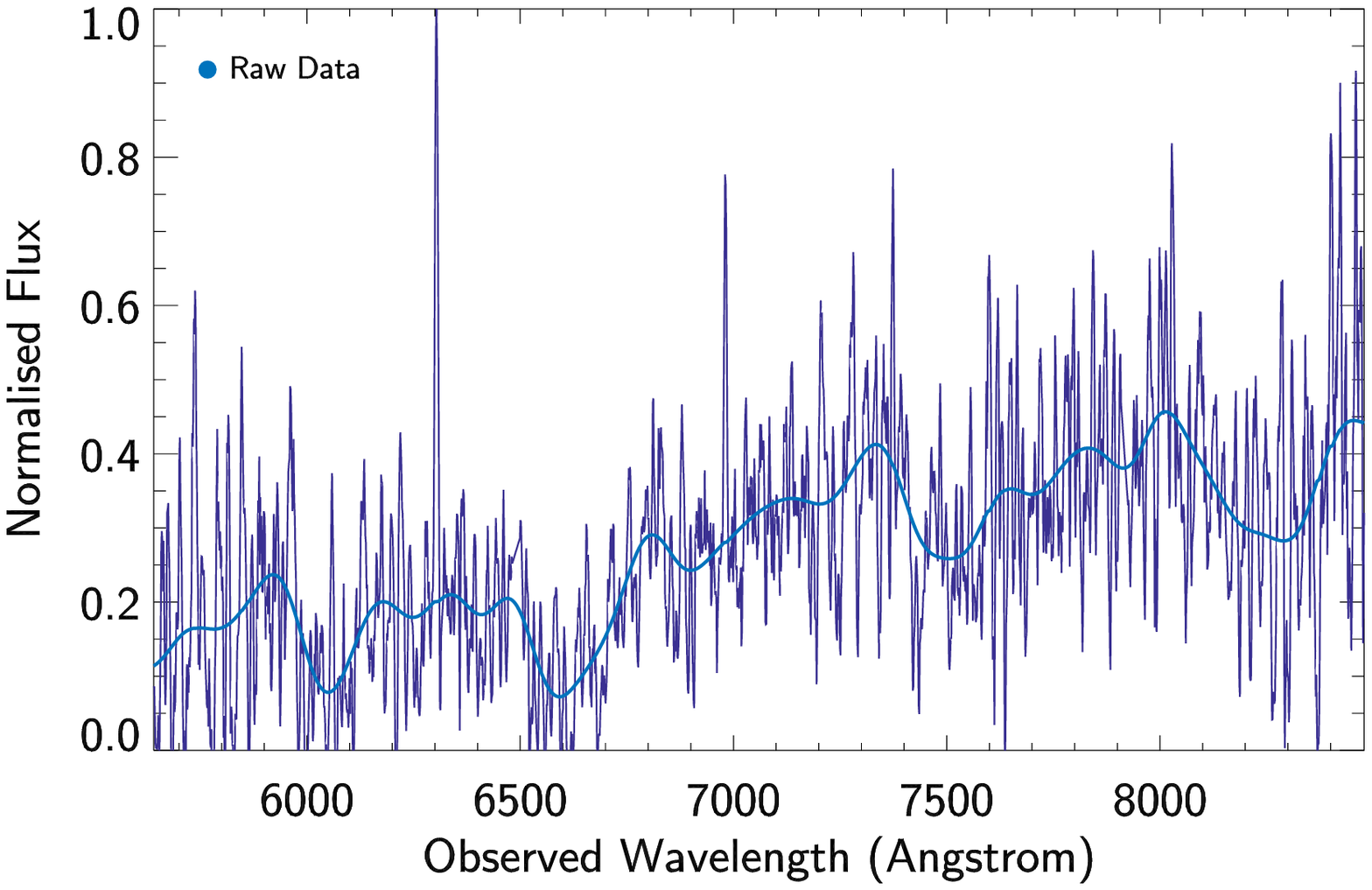}
\label{subfig:06D3fv_data2}
}
\subfigure[06D3hd $z=0.2409$]{
\includegraphics[scale=0.40]{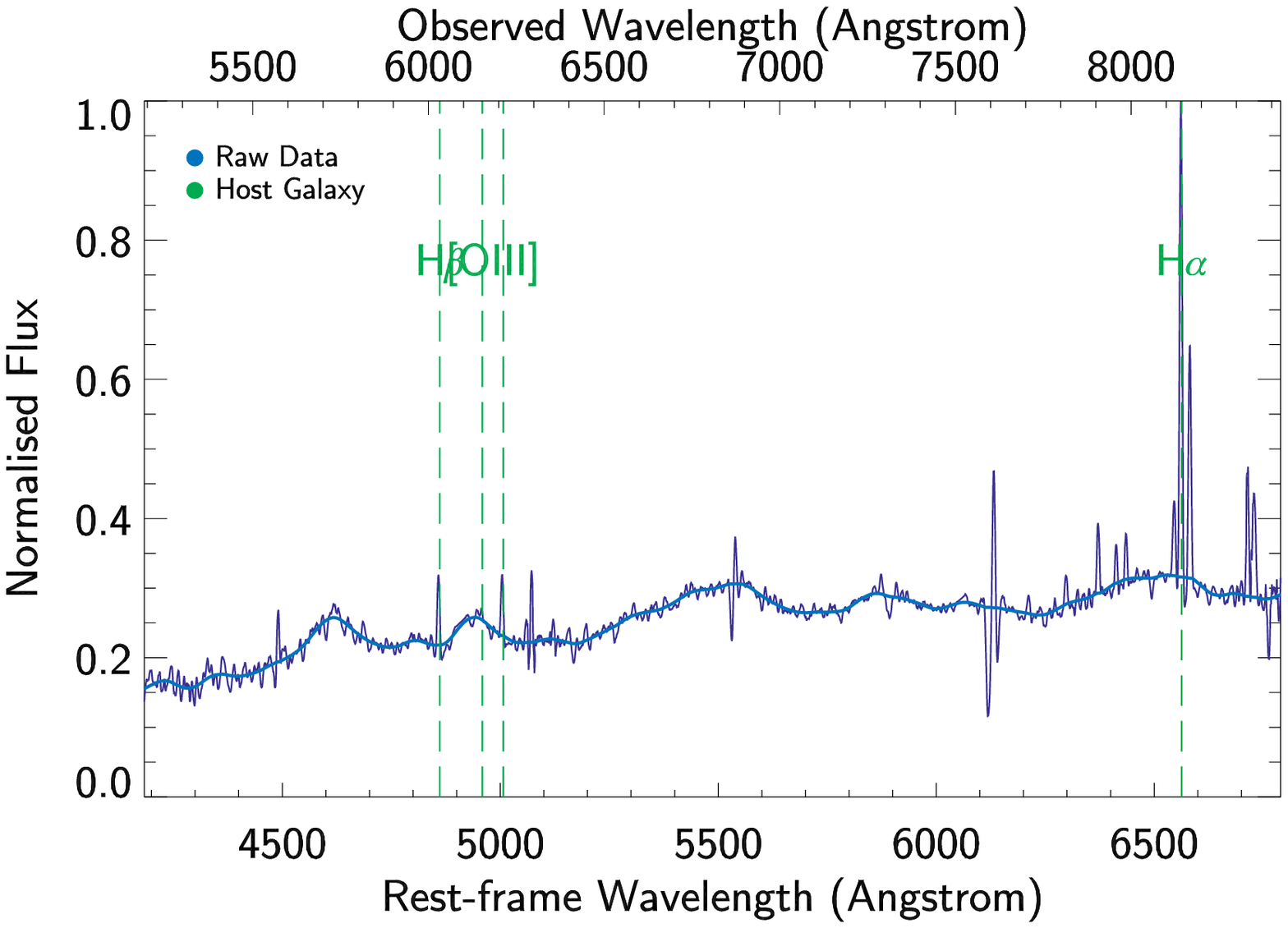}
\label{subfig:06D3hd_data2}
}
\subfigure[06D4gn $z$ unknown]{
\includegraphics[scale=0.40]{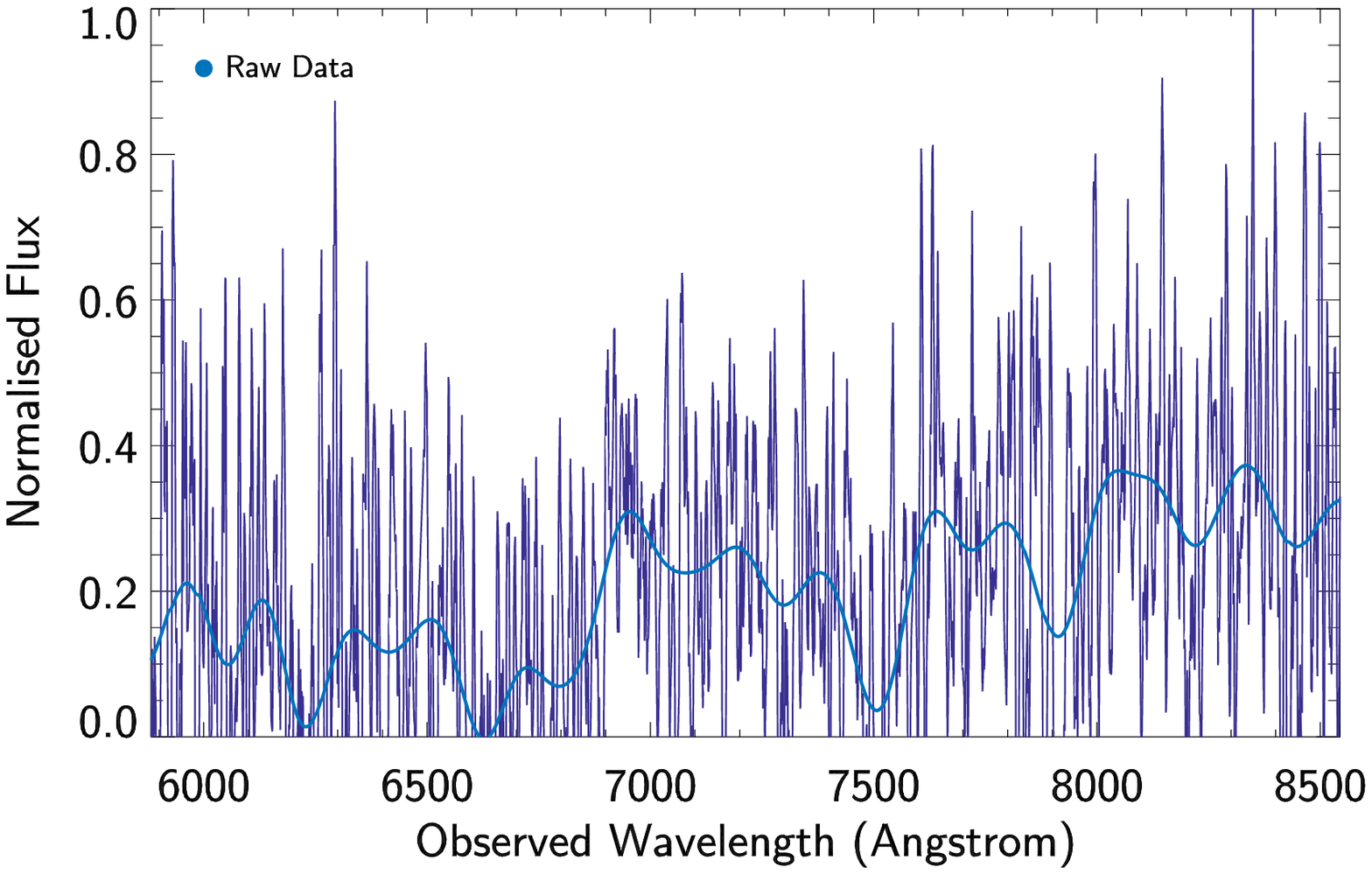}
\label{subfig:06D4gn_data2}
}
\subfigure[07D2bu $z=0.733$]{
\includegraphics[scale=0.40]{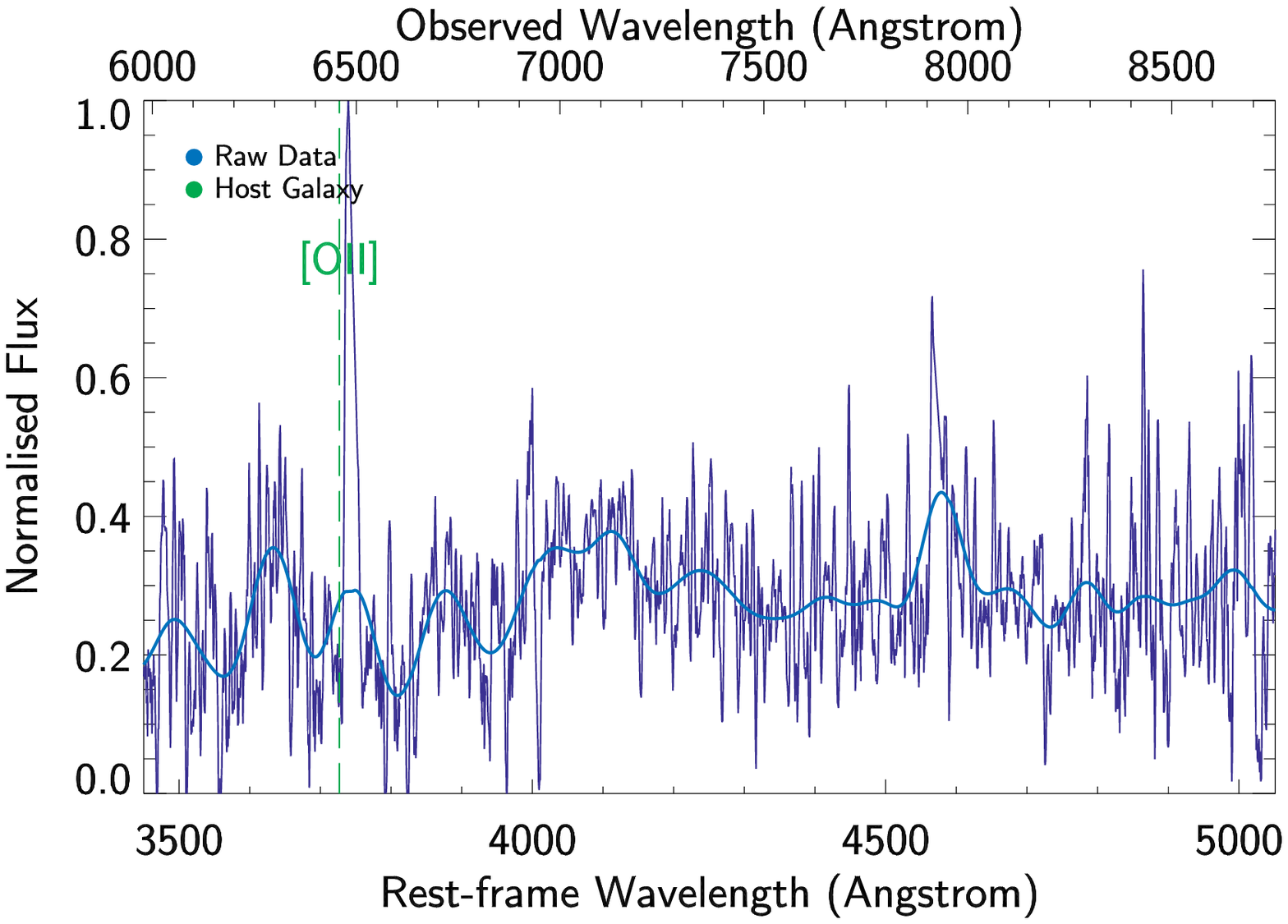}
\label{subfig:07D2bu_data2}
}
\end{center}
\caption{Unknown (CI=2)}
\end{figure*}

\begin{figure*}
\begin{center}
\subfigure[07D2ki $z=0.673$]{
\includegraphics[scale=0.40]{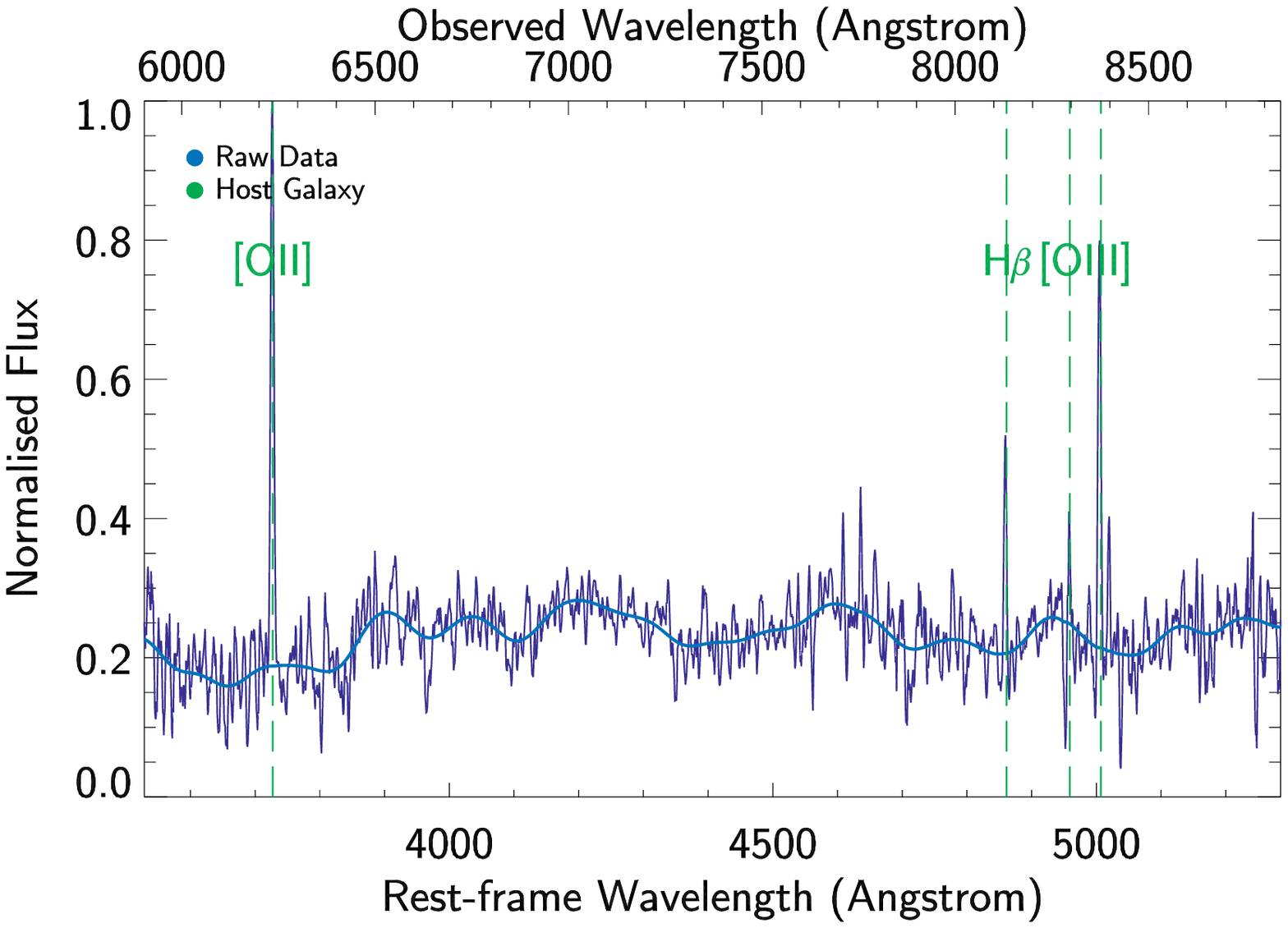}
\label{subfig:07D2ki_data2}
}
\subfigure[07D2kl $z=1.023$]{
\includegraphics[scale=0.40]{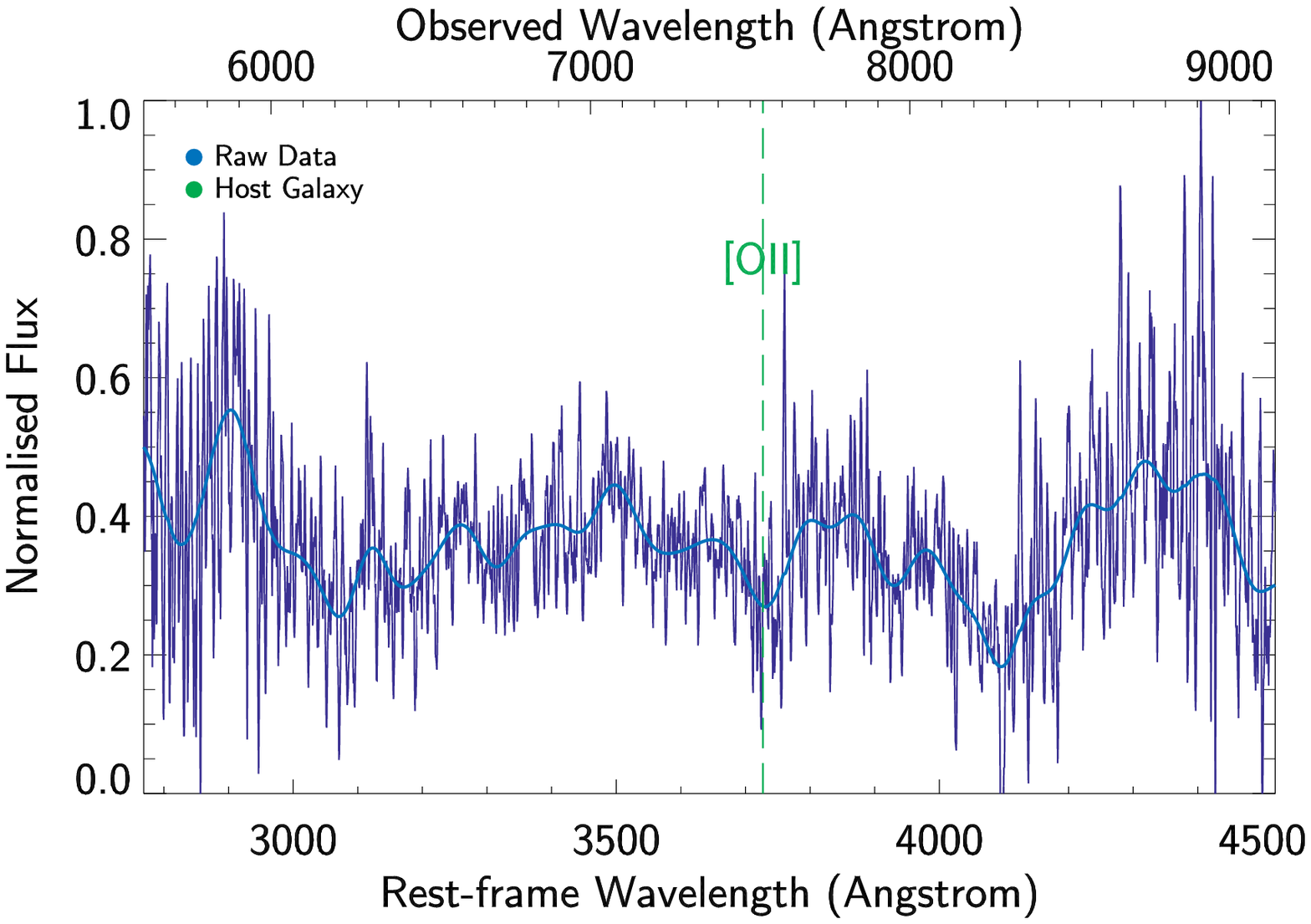}
\label{subfig:07D2kl_data2}
}
\subfigure[07D3bb $z$ unknown]{
\includegraphics[scale=0.40]{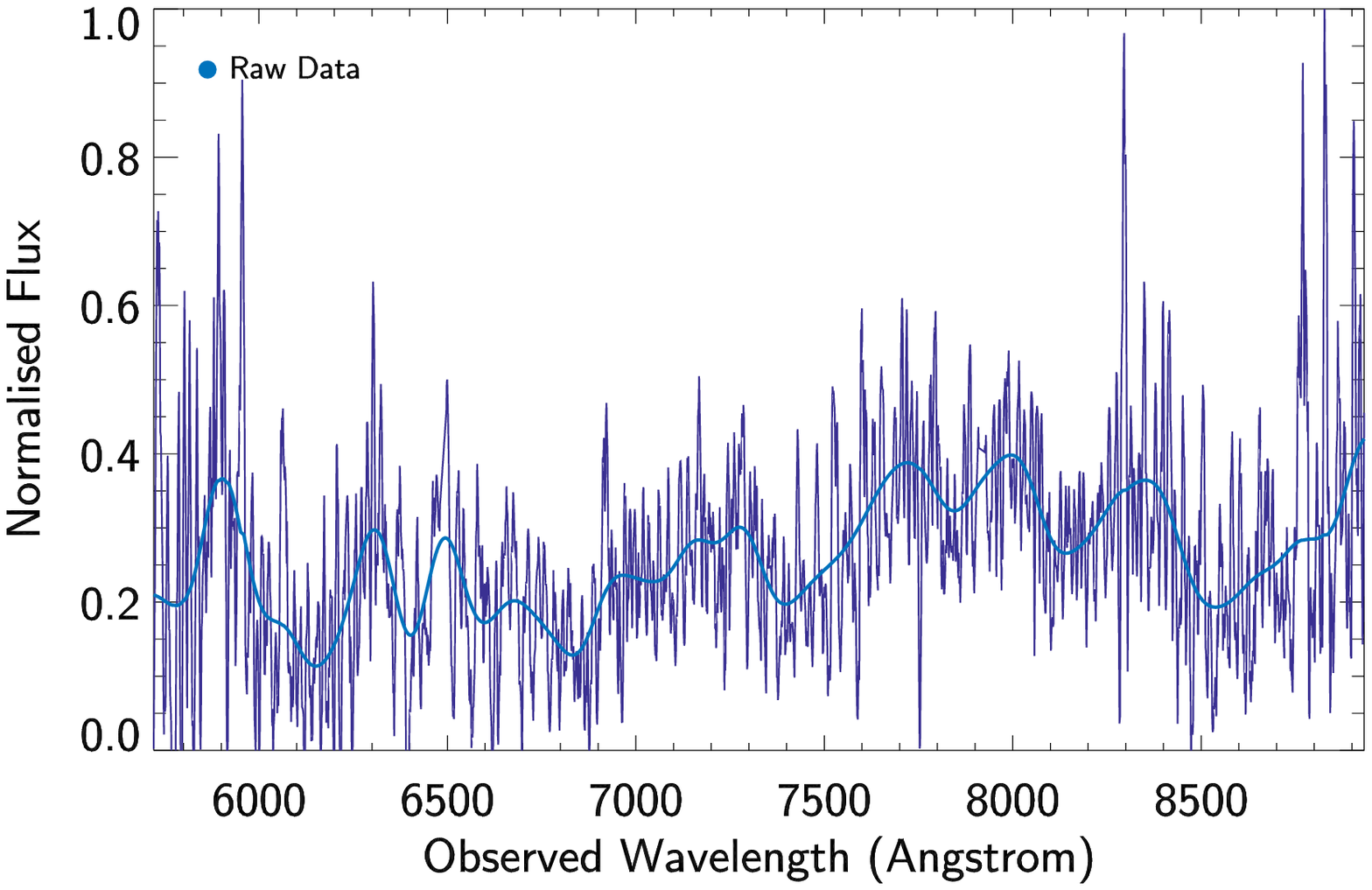}
\label{subfig:07D3bb_data2}
}
\subfigure[07D3bp $z=0.769$]{
\includegraphics[scale=0.40]{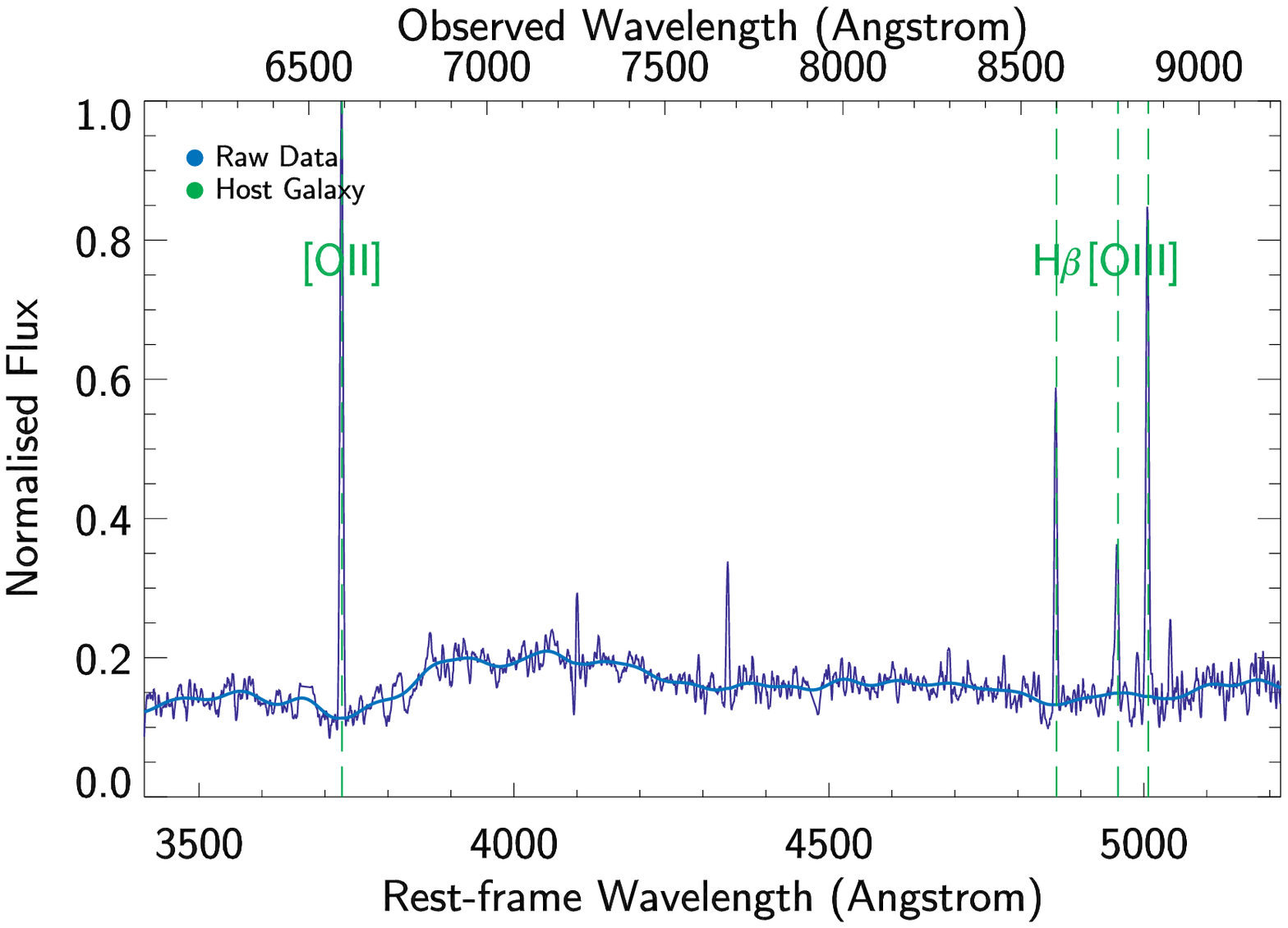}
\label{subfig:07D3bp_data2}
}
\subfigure[07D3fi $z$ unknown]{
\includegraphics[scale=0.40]{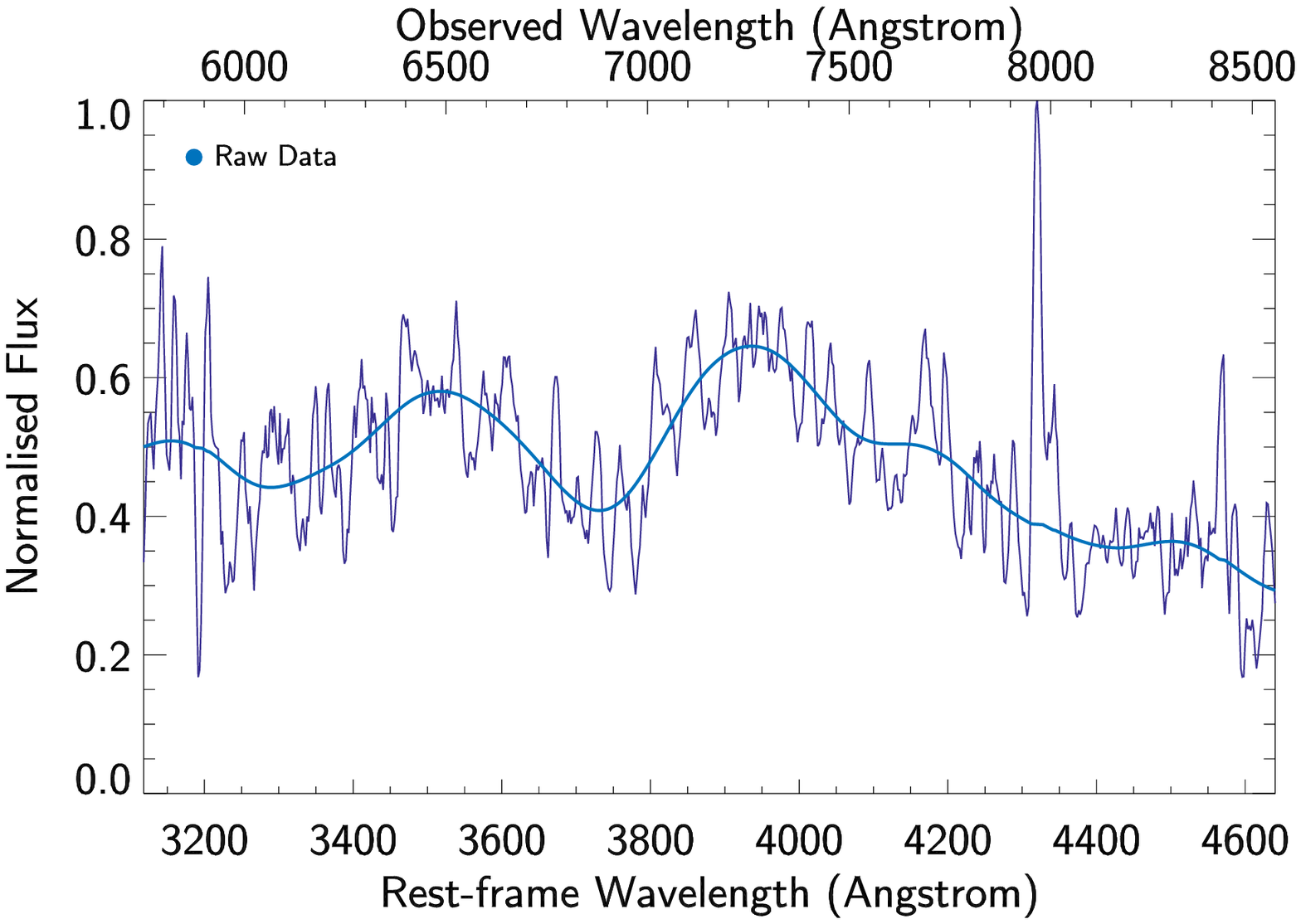}
\label{subfig:07D3fi_data2}
}
\subfigure[08D2au $z=0.563$]{
\includegraphics[scale=0.40]{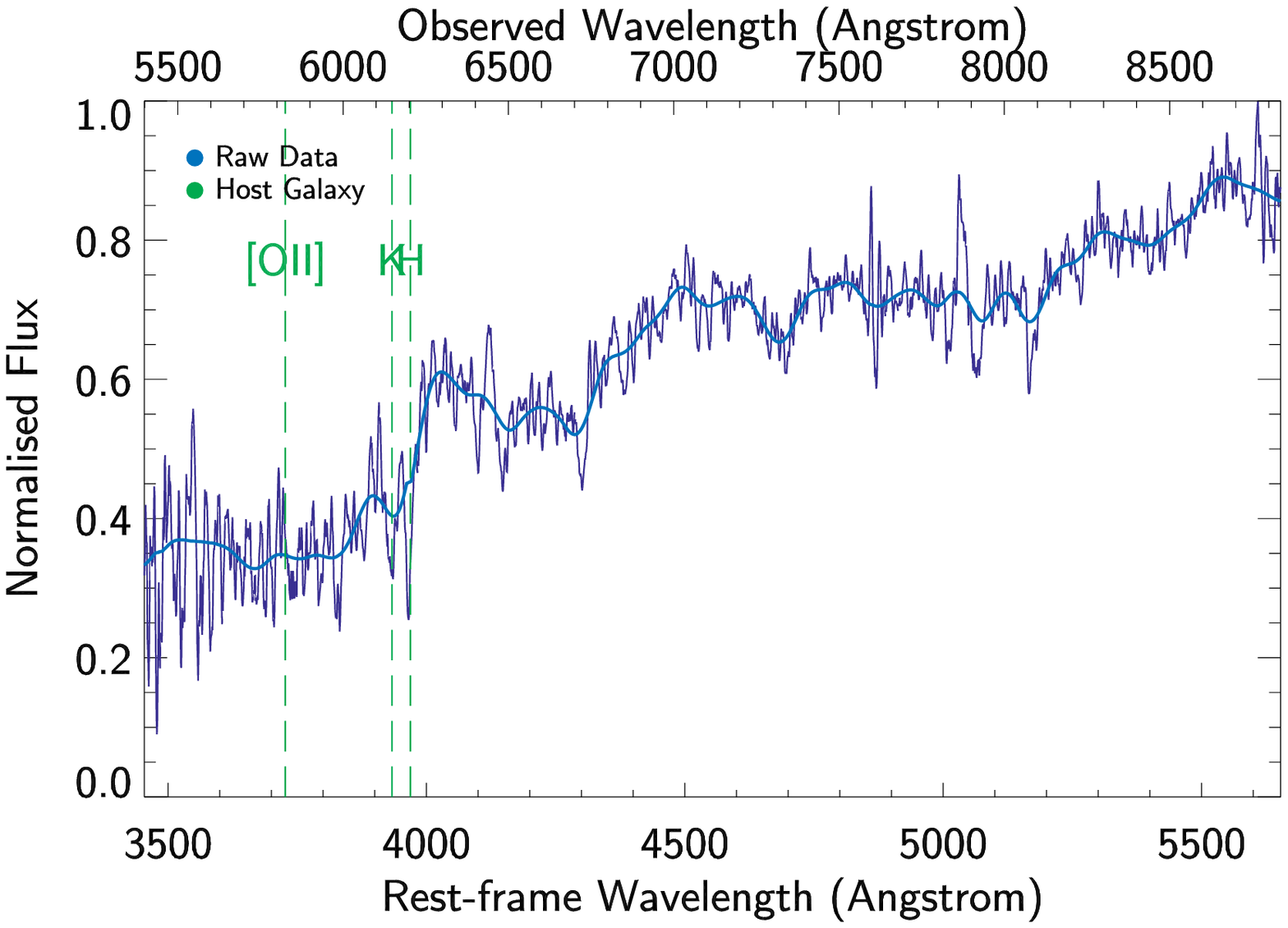}
\label{subfig:08D2au_data2}
}
\subfigure[08D3fp $z$ unknown]{
\includegraphics[scale=0.40]{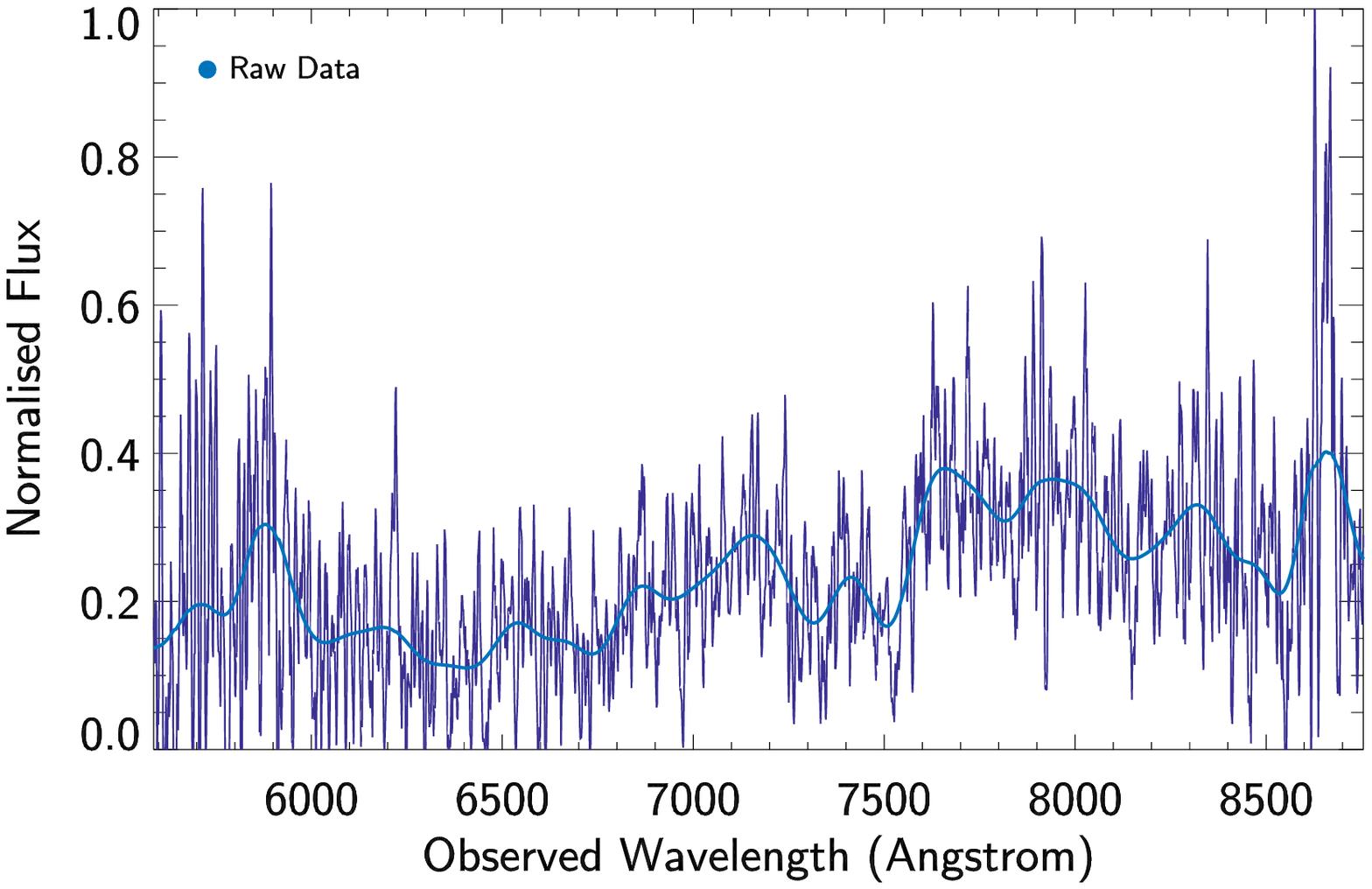}
\label{subfig:08D3fp_data2}
}
\subfigure[08D3fu $z=0.426$]{
\includegraphics[scale=0.40]{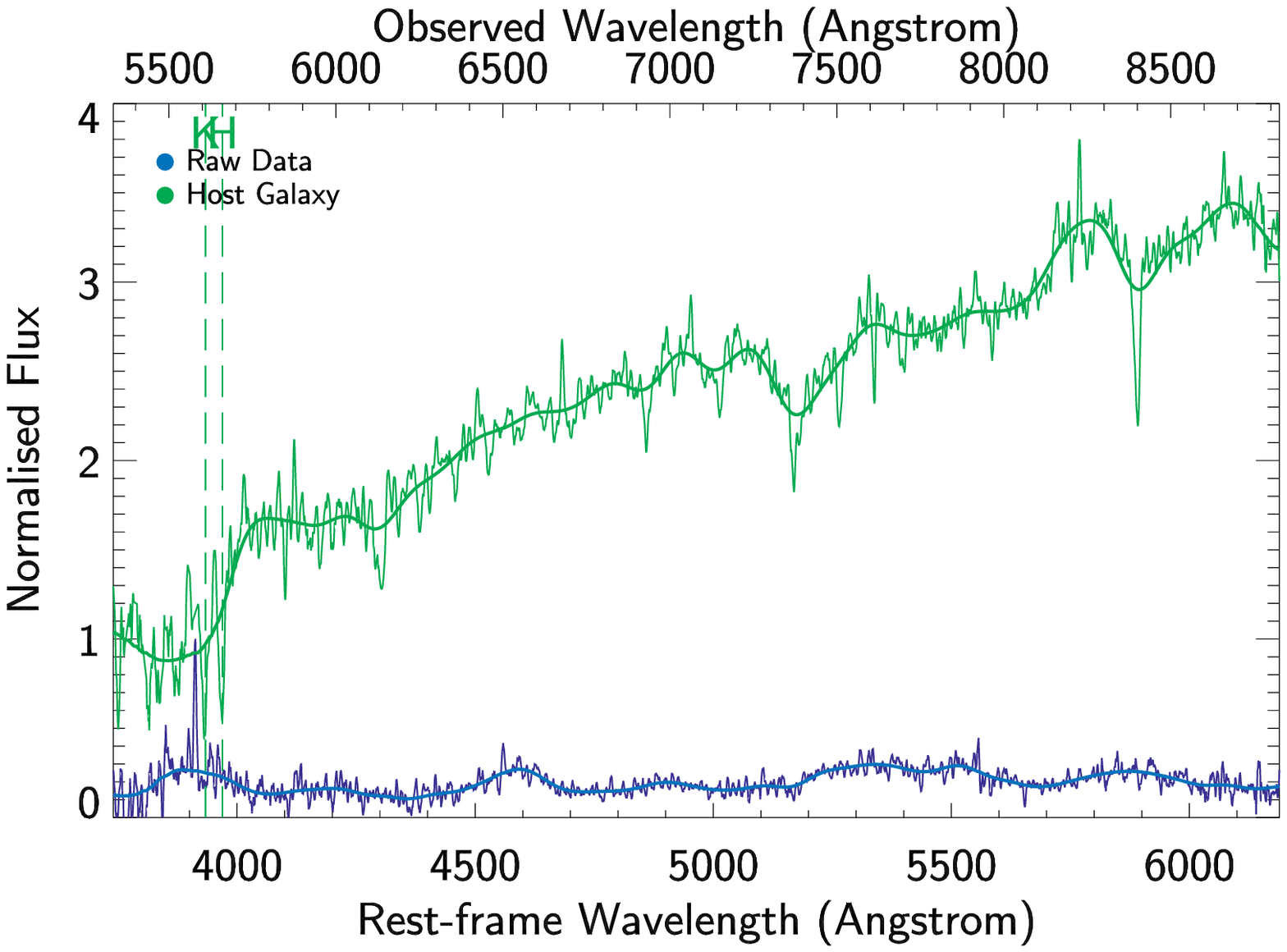}
\label{subfig:08D3fu_data2}
}
\end{center}
\caption{Unknown (CI=2)}
\end{figure*}
\newpage

\begin{figure*}
\begin{center}
\subfigure[06D1dt $z=0.298$]{
\includegraphics[scale=0.40]{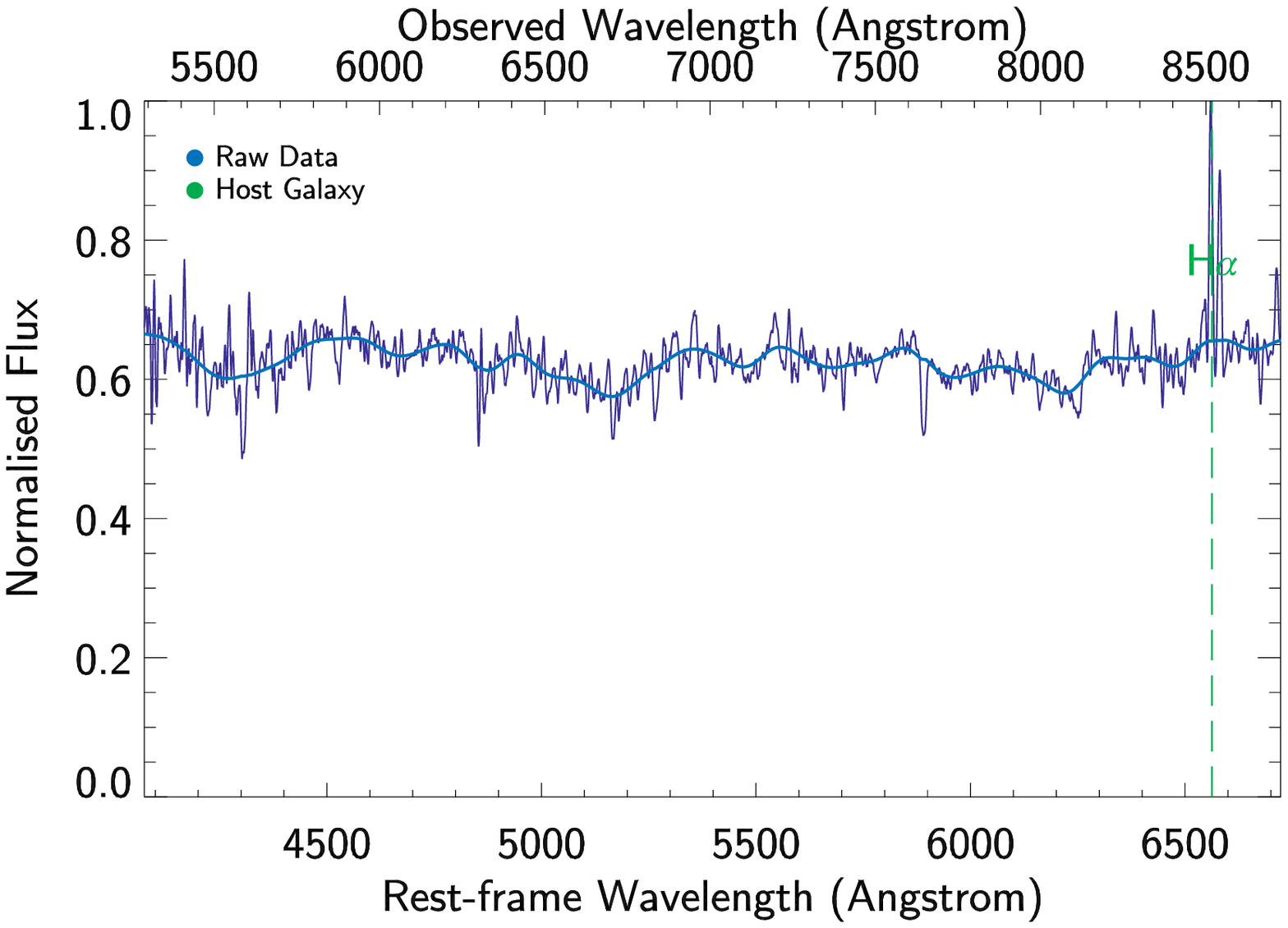}
\label{subfig:06D1dt_data2}
}
\subfigure[06D1gg $z$ unknown]{
\includegraphics[scale=0.40]{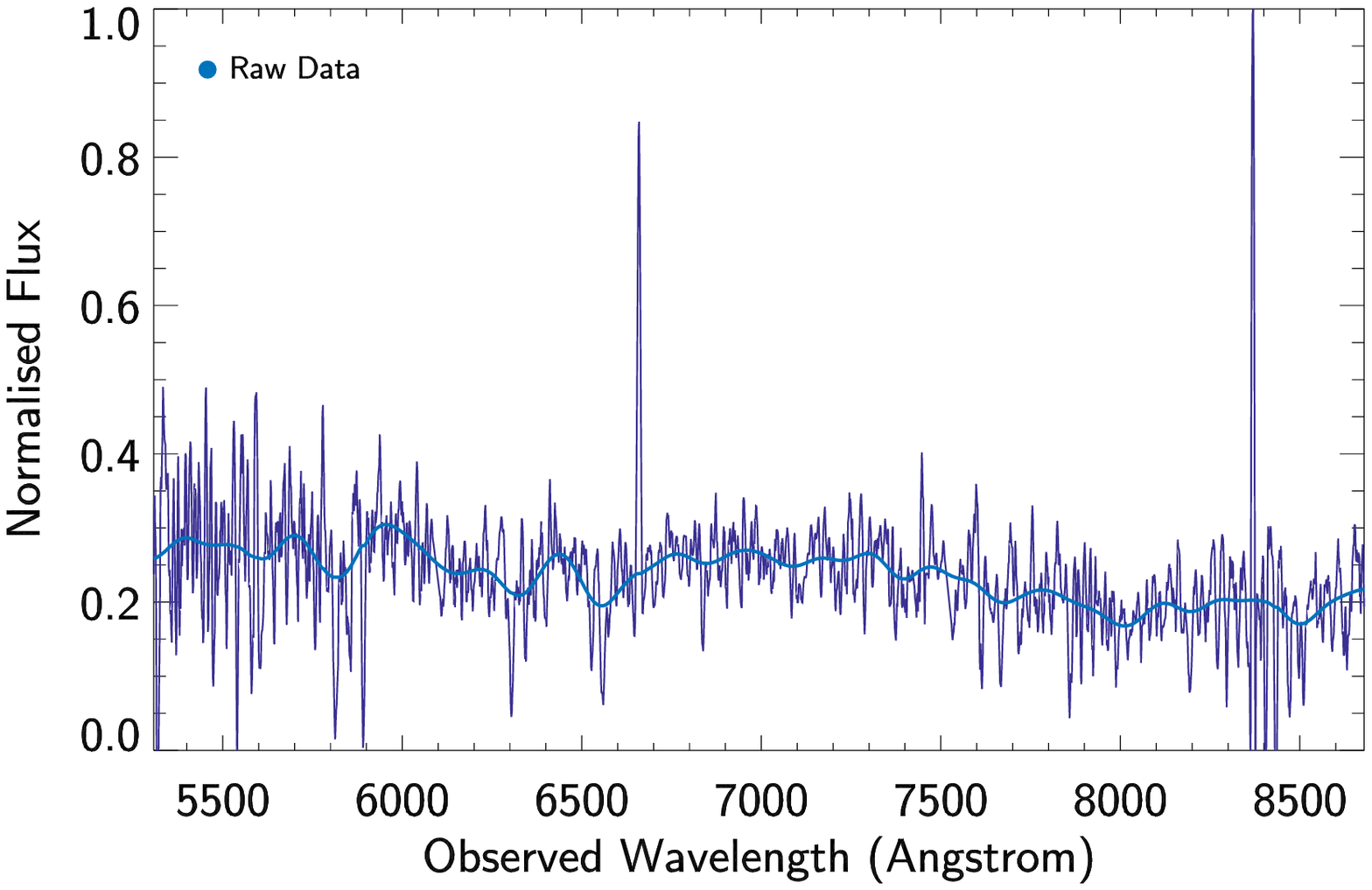}
\label{subfig:06D1gg_data2}
}
\subfigure[06D4hc $z=0.38$]{
\includegraphics[scale=0.40]{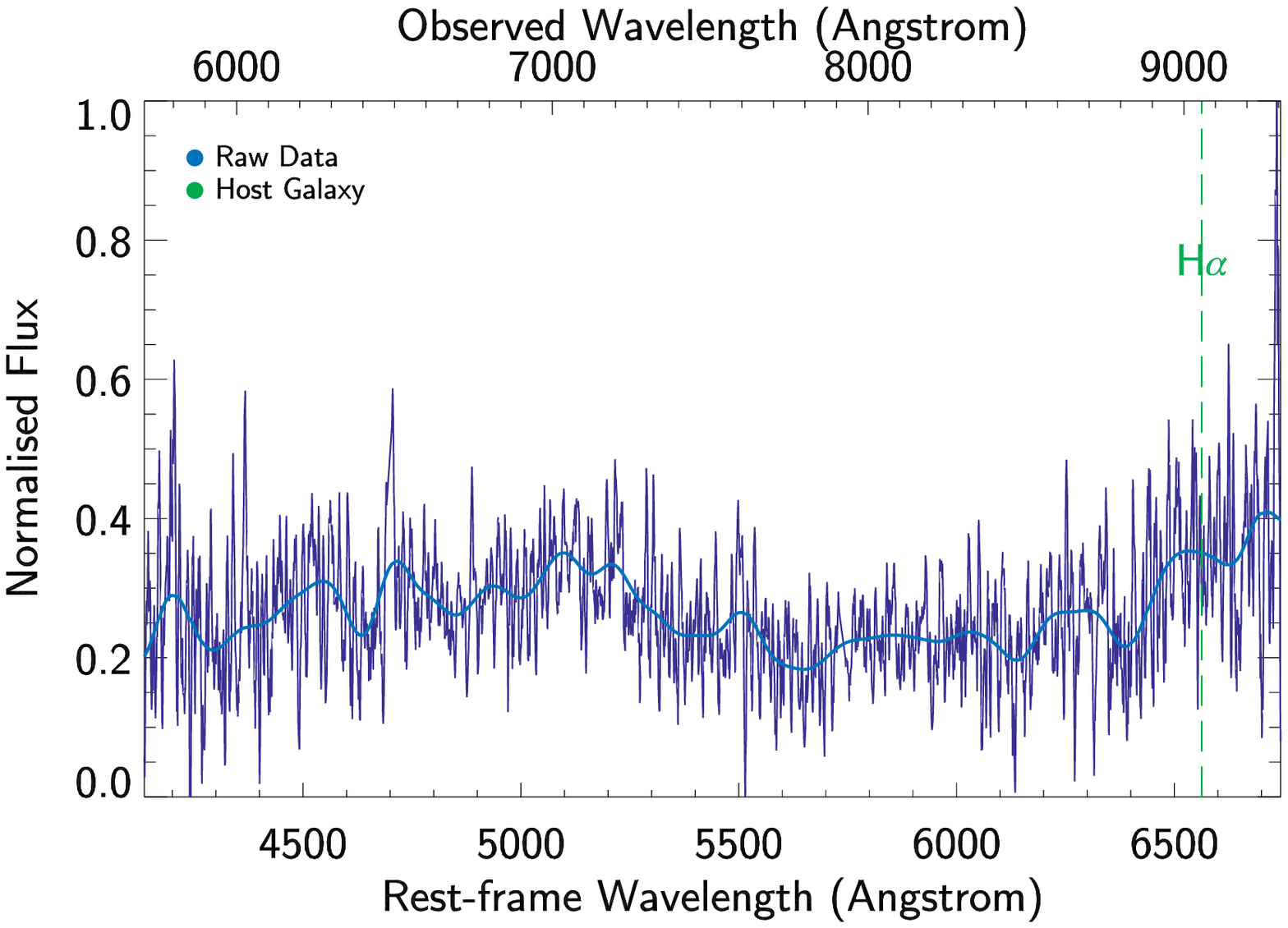}
\label{subfig:06D4hc_data2}
}
\subfigure[07D2ke $z=0.114$]{
\includegraphics[scale=0.40]{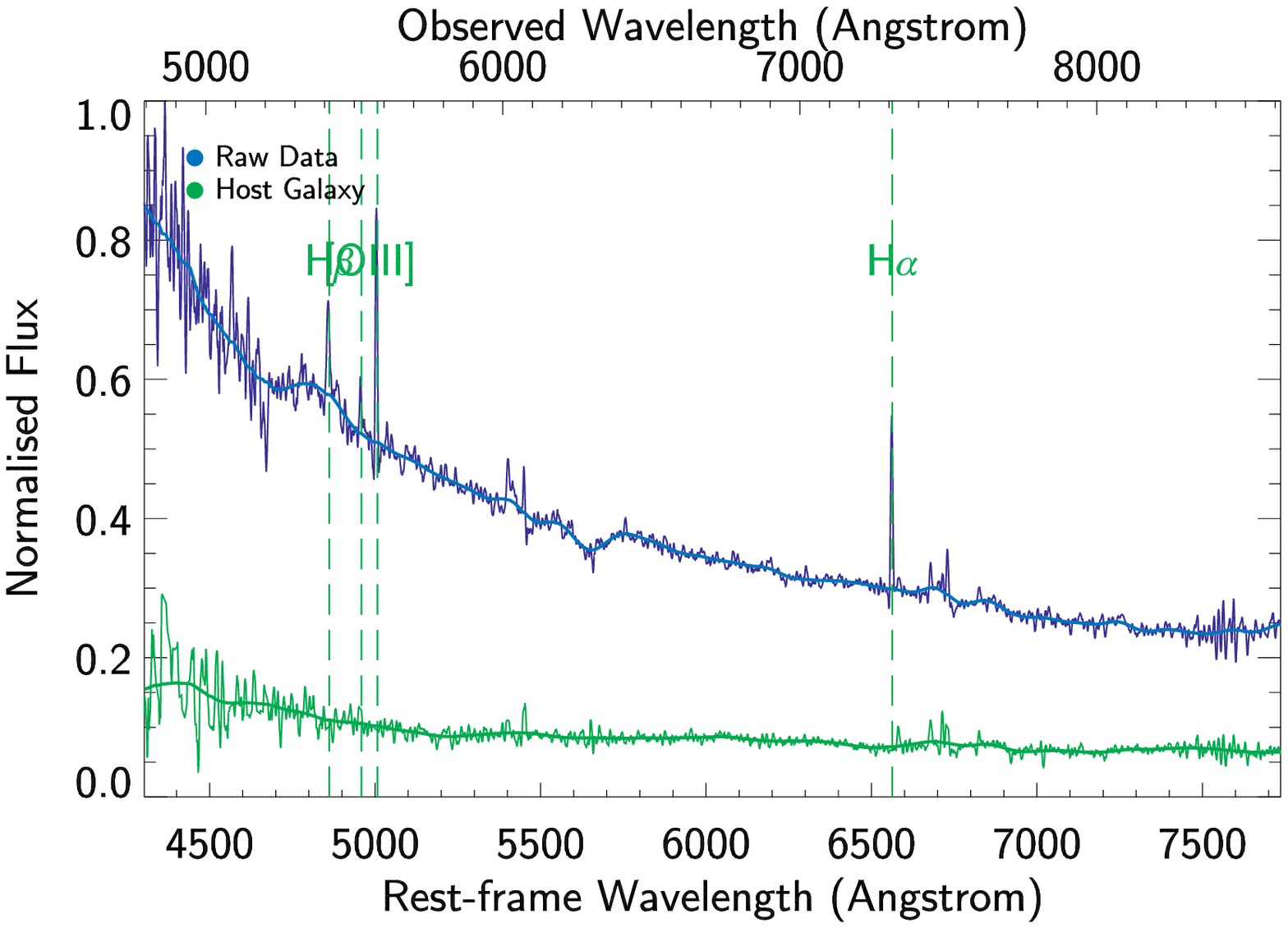}
\label{subfig:07D2ke_data2}
}
\subfigure[07D3ai $z=0.198$]{
\includegraphics[scale=0.40]{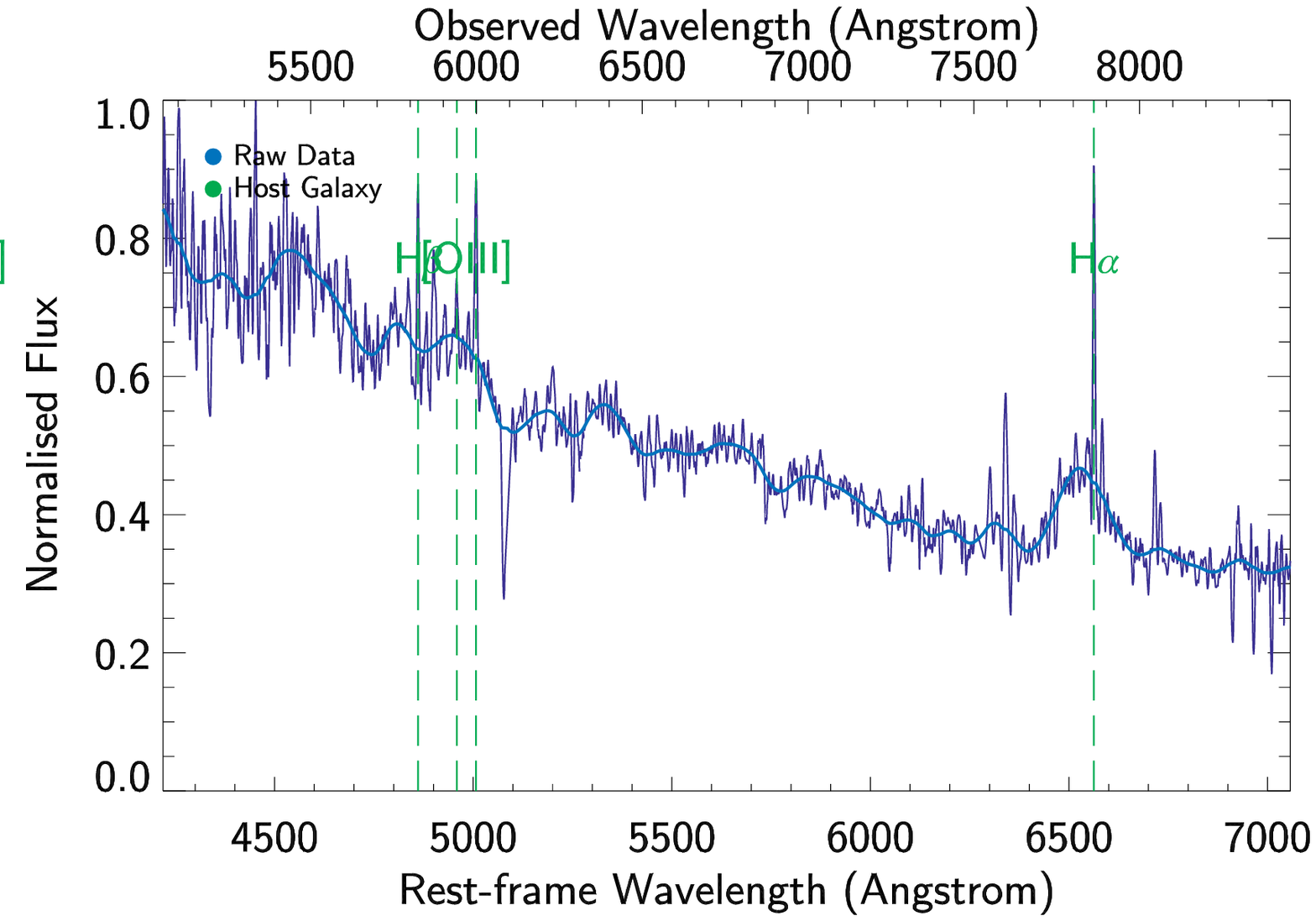}
\label{subfig:07D3ai_data2}
}
\subfigure[07D3am $z=0.214$]{
\includegraphics[scale=0.40]{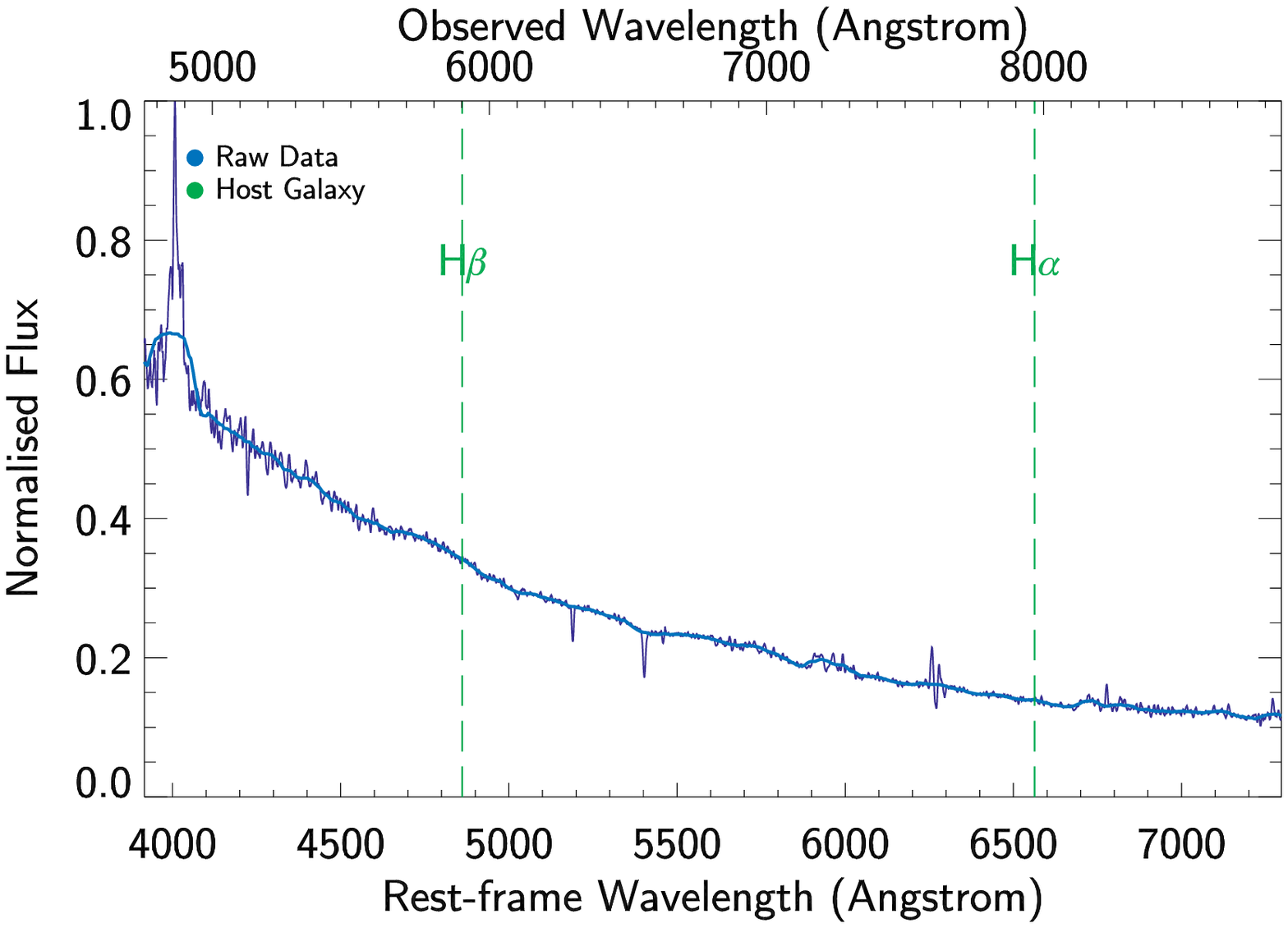}
\label{subfig:07D3am_data2}
}
\subfigure[08D2eo $z=0.125$]{
\includegraphics[scale=0.40]{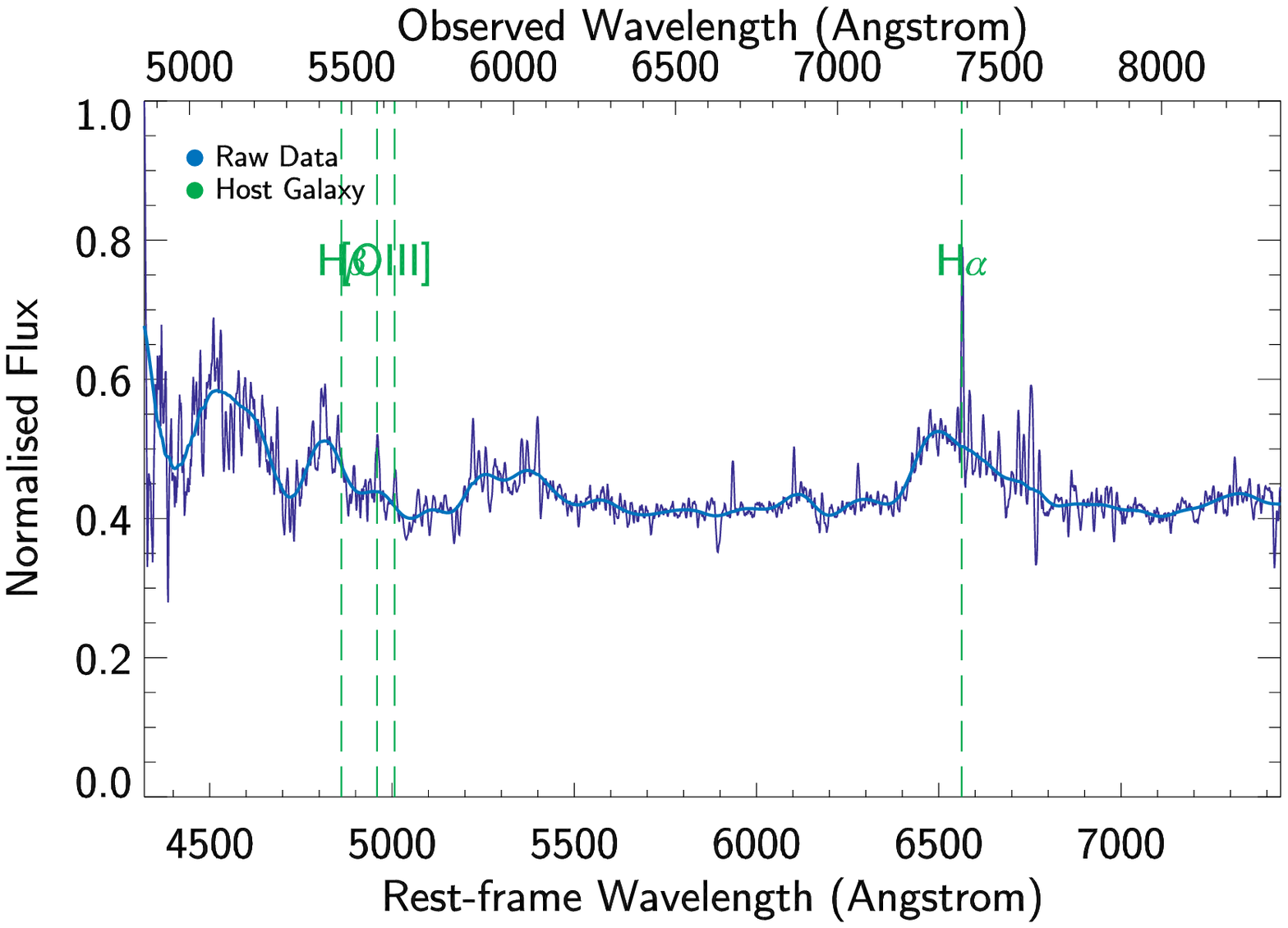}
\label{subfig:08D2eo_data2}
}
\subfigure[08D2fj $z=0.507$]{
\includegraphics[scale=0.40]{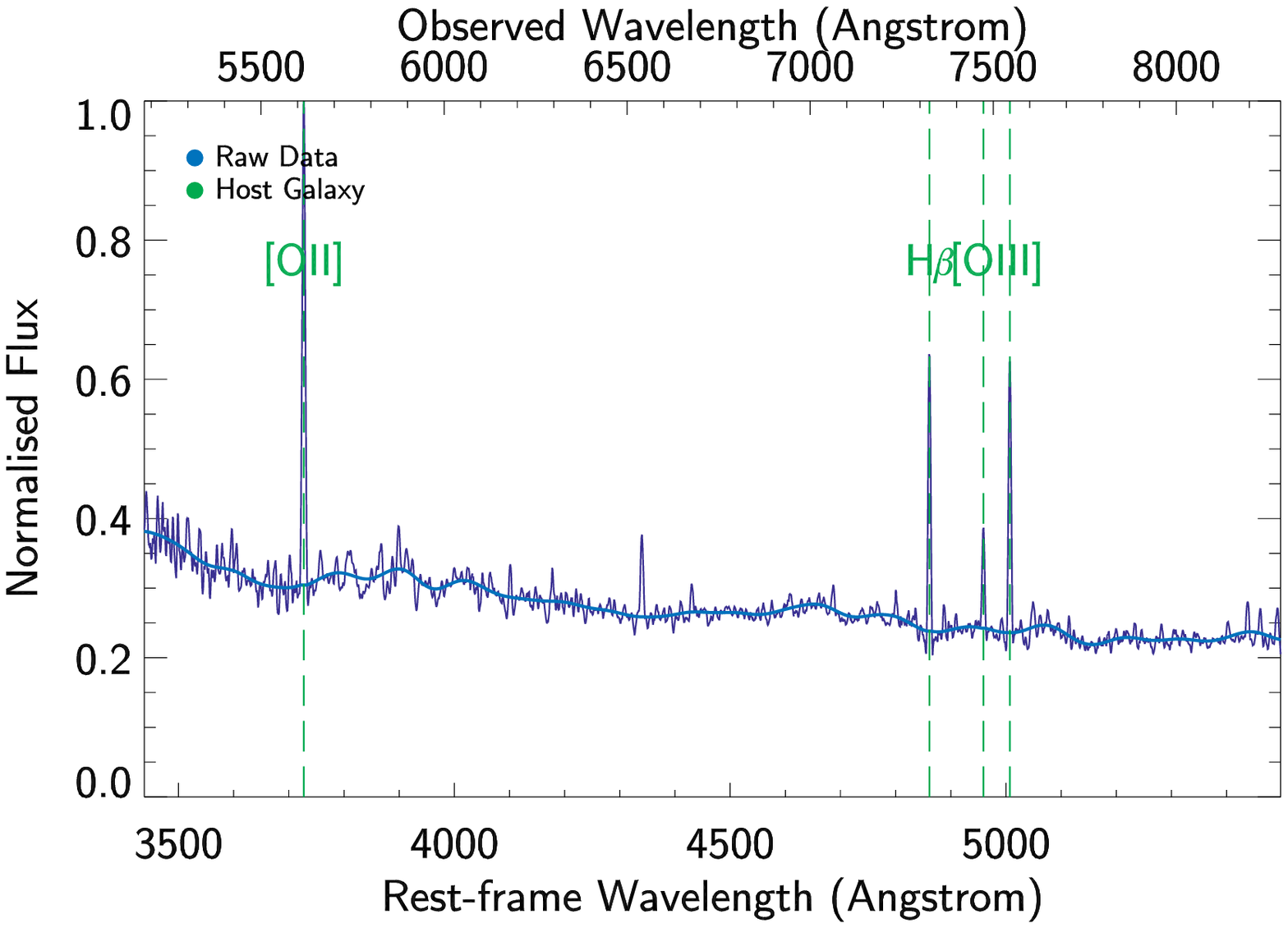}
\label{subfig:08D2fj_data2}
}
\end{center}
\caption{Probably or Definitely Not Type Ia Supernovae (CI=1 or  0)}
\end{figure*}

\begin{figure*}
\begin{center}
\subfigure[08D3au $z$ unknown]{
\includegraphics[scale=0.40]{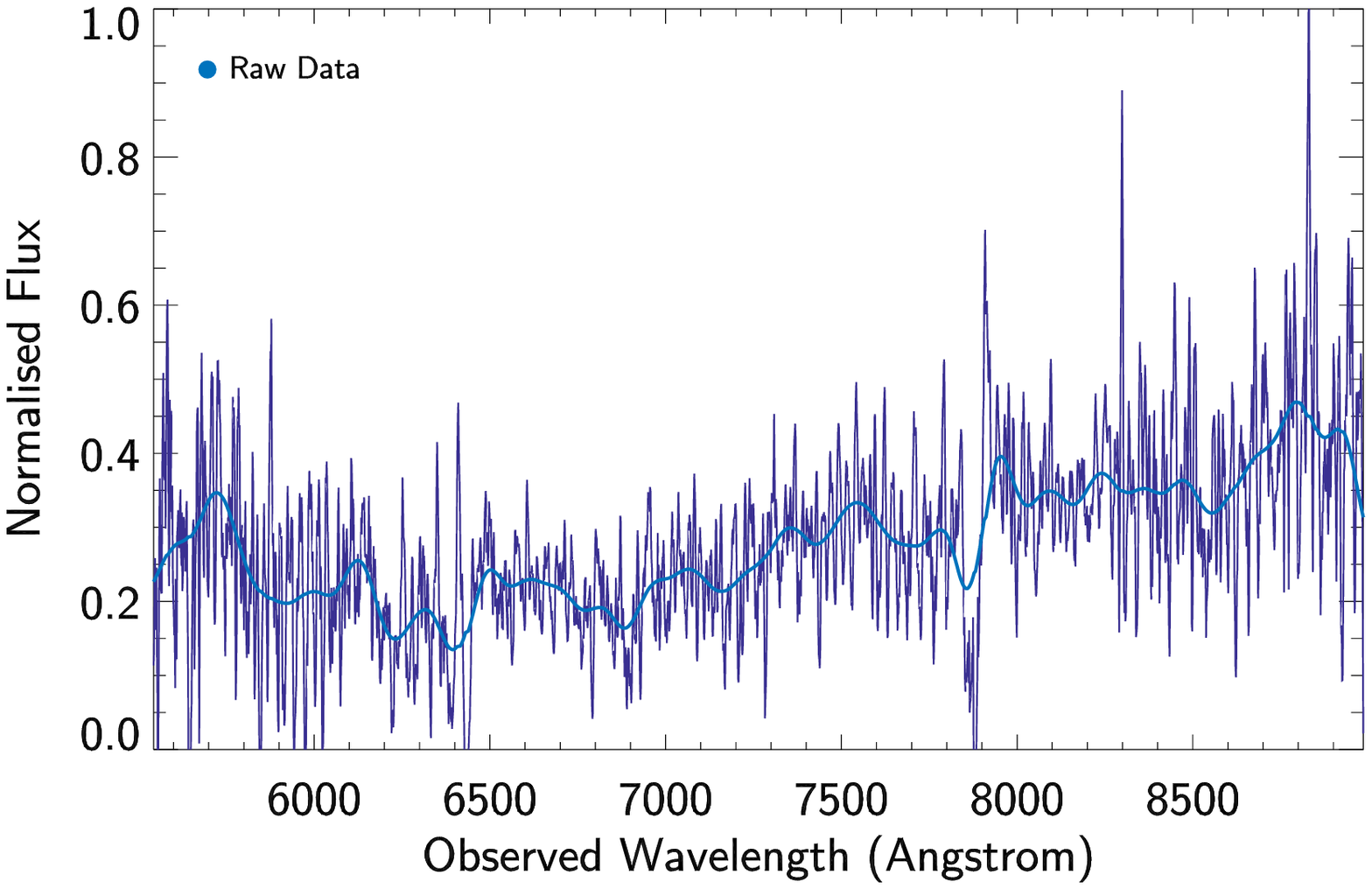}
\label{subfig:08D3au_data2}
}
\end{center}
\caption{Probably or Definitely Not Type Ia Supernovae (CI=1 or  0)}
\end{figure*}

\end{document}